\documentclass[11pt]{article}
\pdfoutput=1
\usepackage{amsmath,amssymb,epsfig,amsfonts}
\usepackage{graphicx}
\usepackage{subfig}
\usepackage[usenames, dvipsnames]{color}
\usepackage{cite}
\usepackage{multirow}
\usepackage{hyperref}
\usepackage{verbatim} 
\usepackage{enumitem}
\usepackage{rotating}
\usepackage{xcolor}
\usepackage{multirow}
\usepackage{bm}
\usepackage[normalem]{ulem}
\usepackage{cleveref}
\usepackage{slashed}

\graphicspath{ {figs/} }

\makeatletter

\newtheorem{definition}{Definition}[section]

\newcommand{\be}{\begin{equation}}
\newcommand{\ee}{\end{equation}}
\newcommand{\ba}{\begin{aligned}}
\newcommand{\ea}{\end{aligned}}

\newcommand{\bea}{\begin{eqnarray}}
\newcommand{\eea}{\end{eqnarray}}

\def\sign{\mathop{\mathrm{sign}}\nolimits}

\def\unit{{1\kern-.65ex {\rm l}}}
\def\1{{1\kern-.65ex {\rm l}}}

\def\bbP{{\mathbb{P}}}

\newcount\hour \newcount\minute
\hour=\time \divide \hour by 60
\minute=\time
\def\now{%
\ifnum \hour<13
  \ifnum \hour=0 \advance \hour by 12 \number\hour:\else \number\hour:\fi%
     \ifnum \minute<10 0\fi%
     \number\minute%
\ A.M.%
\else \advance \hour by -12 \number\hour:%
  \ifnum \minute<10 0\fi%
  \number\minute%
  \ P.M.%
\fi%
}

\makeatother

\def\mb{\mathbb}
\def\mbf{\mathbf}
\def\mc{\mathcal}

\begin{document}

\baselineskip=18pt  
\numberwithin{equation}{section}  
\allowdisplaybreaks  


%
%


\thispagestyle{empty}

\vspace*{0.8cm} 
\begin{center}
{{\Huge {Fibers add Flavor, Part I}:\\ 

\bigskip
\LARGE Classification of 5d SCFTs,
Flavor Symmetries and BPS States}}

 \vspace*{1.5cm}
Fabio Apruzzi$^1$, Craig Lawrie$^2$, Ling Lin$^2$,  Sakura Sch\"afer-Nameki$^1$, Yi-Nan Wang$^1$\\

 \vspace*{1.0cm} 
{\it ${}^1$ Mathematical Institute, University of Oxford, \\
Andrew-Wiles Building,  Woodstock Road, Oxford, OX2 6GG, UK}\\
\smallskip
{\it ${}^2$ Department of Physics and Astronomy, University of Pennsylvania, \\
Philadelphia, PA 19104, USA}\\

\vspace*{0.8cm}
\end{center}
\vspace*{.5cm}

\noindent
We propose a graph-based approach to 5d superconformal field theories (SCFTs)
based on their realization as M-theory compactifications on singular elliptic
Calabi--Yau threefolds.  
Field-theoretically, these 5d SCFTs descend from 6d
$\mathcal{N}=(1,0)$ SCFTs by circle compactification and mass 
deformations. 
We derive a description of these theories in terms of graphs, so-called Combined
Fiber Diagrams, which encode salient features of the partially resolved Calabi--Yau geometry, 
and provides a combinatorial way of characterizing all 
5d SCFTs that descend from a given 6d theory. 
Remarkably, these graphs manifestly capture strongly coupled data of the 5d SCFTs, such as the superconformal flavor symmetry,
BPS states, and mass deformations. The capabilities of this approach are demonstrated by deriving all rank one and rank two 5d SCFTs. 
The full potential, however, becomes apparent when applied to theories with higher rank.
Starting with the higher rank conformal matter theories in 6d, we are led to the discovery of previously unknown flavor symmetry enhancements and new 5d SCFTs.

\newpage

\tableofcontents




\section{Introduction}

Geometry is a well-established tool in the exploration of the landscape of
superconformal field theories (SCFTs). This connection has by now
crystalized into a profound correspondence, where geometric structures have
emerged as central agents in the classification of SCFTs. Hallmarks of this
achievement are the classifications of 6d SCFTs with maximal
\cite{Witten:1995zh} and $\mathcal{N}=(1,0)$ \cite{Heckman:2013pva,
Bhardwaj:2015xxa, Heckman:2015bfa} supersymmetry, as well as rank one and two
5d SCFTs \cite{Intriligator:1997pq, Jefferson:2018irk}, and the Coulomb branch
geometries for rank one 4d $\mathcal{N}=2$ theories \cite{Argyres:2015ffa}.
In these considerations, the geometry not only provides a concrete realization
of the SCFT in an M-/F-theory or string theory background, but, more
importantly, constitutes an organizing principle by which to construct and
enumerate all such theories
systematically. Such classifications may come with caveats in that not all
theories may have purely geometric constructions (see e.g.,
\cite{Bhardwaj:2015xxa}).  However, the geometric classification oftentimes
results in a parallel field theoretic classification, which corroborates the
completeness. 

Ideally, a classification result does not only provide a formal identification
of all theories in a given dimension and amount of supersymmetry, but contains
information about the physics of the strongly coupled SCFTs.  In this paper we
propose a classification approach for 5d SCFTs, with $\mathcal{N}=1$
supersymmetry, which includes not only a systematic way of constructing the
associated geometries in their realization via an M-theory compactification on a Calabi--Yau
threefold, but also allows the reading off of
\begin{enumerate}
	\item the flavor symmetry of the strongly coupled 5d SCFT, and
	\item BPS states of the 5d SCFT.
\end{enumerate}
 The flavor symmetry of the SCFT will be manifestly encoded in the way we
 construct and present the geometries, and literally can be read off. Some BPS states
 can be computed applying a straightforward procedure from the
 geometric information that we provide.  We introduced a graphical
 presentation of this data  in terms of so-called combined fiber diagrams
 (CFDs) in the recent paper \cite{Apruzzi:2019vpe}. The current paper can be
 viewed as providing the geometric derivation of these CFDs.

\begin{figure}
\centering
\subfloat{}{\includegraphics[width=7.5cm]{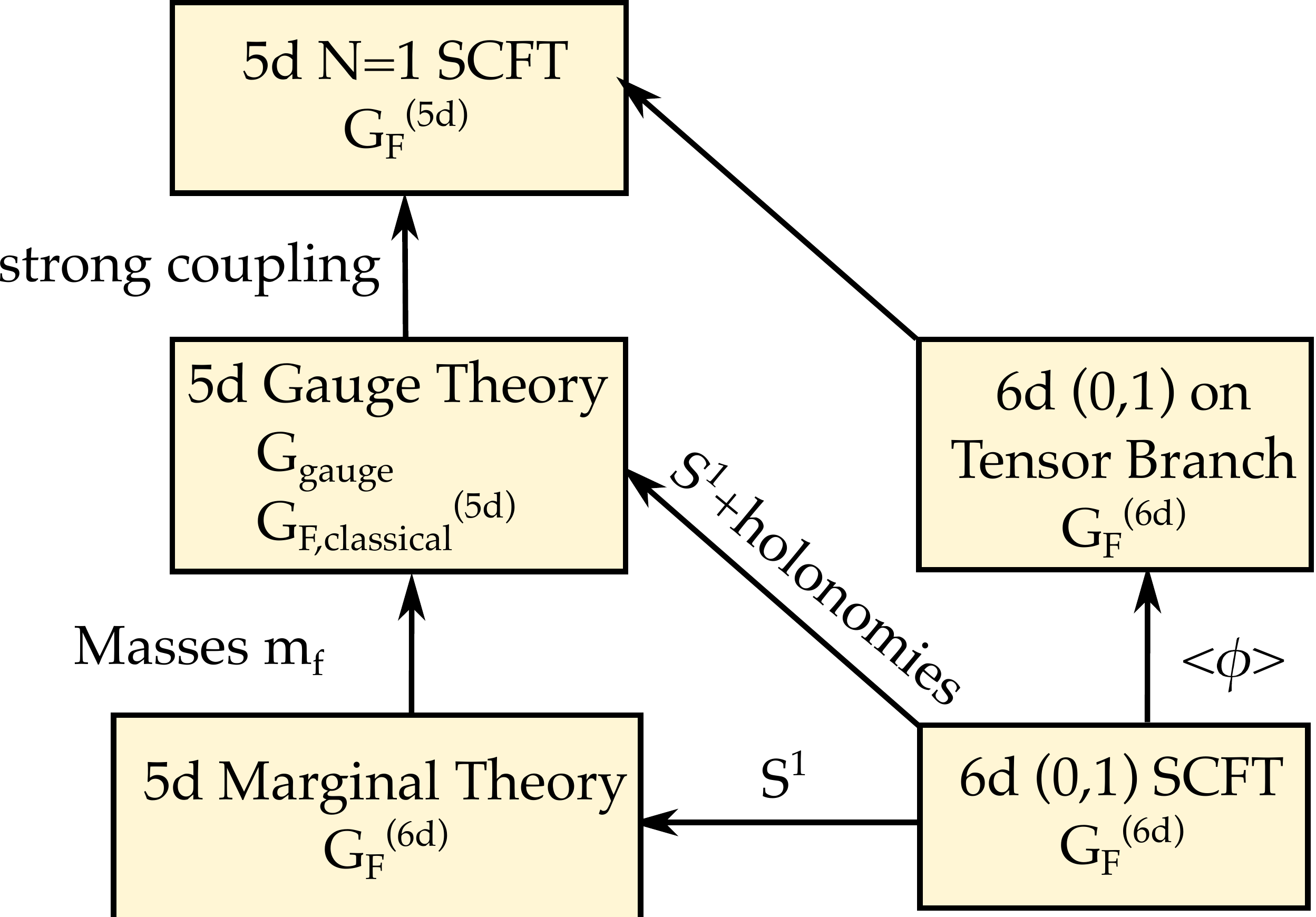}}\qquad 
\subfloat{}{\includegraphics[width=7cm]{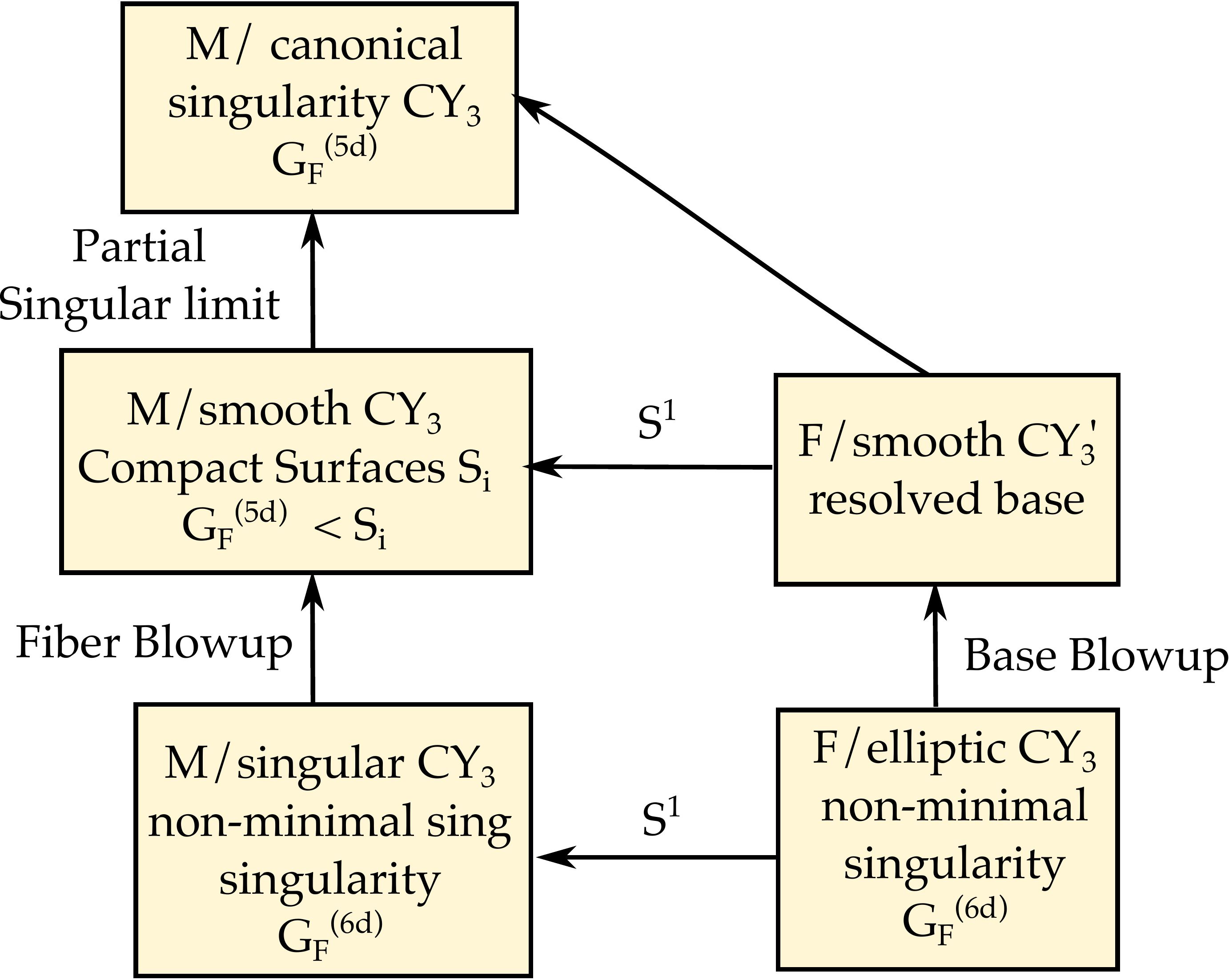}}

 \caption{\, \\
 \underline{(a) Field-Theory overview :}
 Starting from a 6d SCFT, the circle-reduction yields the 5d marginal theory.
 Mass deforming the marginal theory gives rise to 5d gauge theories that flows
 to 5d SCFTs in the UV. Alternatively, one can reduce the 6d theory on a
 circle with holonomies and then flow to said  SCFTs.  A third alternative is
 to take the 6d SCFT onto the tensor branch and reduce to 5d. \\
\underline{(b) Geometry overview:}
The geometric realization in F/M-theory complements the field theoretic
approach. A 6d SCFT is constructed from F-theory on a non-compact
elliptically fibered Calabi--Yau threefold CY$_3$ with non-minimal
singularities.  These have crepant resolutions either by blow-ups in the fiber
(an approach useful e.g. for conformal matter theories) or by blow-ups that
modify the base of the fibration.
Each of these approaches introduces compact surfaces, $S_i$, and the 5d
strongly coupled flavor symmetry is encoded in the geometry of certain curves
associated to the 6d flavor symmetry inside the surfaces $S_i$. We propose a
succinct way of tracking these, so-called flavor curves, in terms of a
graphical tool, making  $G_\text{F}^{(\text{5d})}$ that is encoded in the geometry
manifest --- see figure \ref{fig:OverviewOfTools}.
 \label{fig:Overviews}}
 \end{figure}

The study of 5d SCFTs using M-theory compactifications on singular Calabi--Yau
threefolds goes back to \cite{Morrison:1996xf, Intriligator:1997pq} and was
recently revisited in \cite{Xie:2017pfl,DelZotto:2017pti,  Jefferson:2017ahm,
Jefferson:2018irk, Apruzzi:2018nre, Closset:2018bjz}. This complements the approach using five-brane webs
\cite{Aharony:1997ju,Aharony:1997bh,DeWolfe:1999hj,Bergman:2015dpa,
  Zafrir:2015rga, Zafrir:2015ftn, Ohmori:2015tka,
  Hayashi:2017btw,Hayashi:2018bkd,Hayashi:2018lyv,Seiberg:1996bd,
Brandhuber:1999np, Bergman:2012kr}.  We will follow the M-theory approach by
combining various idea proposed in \cite{Apruzzi:2018nre},
\cite{Hayashi:2014kca}, and \cite{Tian:2018icz}. Conceptually the approach
presented here is
founded in the connection between 6d $\mathcal{N} = (1,0)$ and 5d ${\cal N} = 1$ SCFTs by
  circle-compactification with holonomies in the 6d flavor symmetry.  Such a
  dimensional reduction generically results in 5d theories that flow to
non-trivial UV fixed points. Alternatively, we can think of the resulting
theories as arising from the marginal 5d theory, which is obtained by reducing
the 6d theory on $S^1$ (without holonomies), and subsequently decoupling
hypermultiplets transforming in representations of the (classical) flavor
symmetry, by turning on a mass deformations, and subsequently sending them to
infinity.  The mass deformations can be interpreted as Coulomb branch
parameters for the theory with a weakly gauged classical flavor symmetry.
These form part of an extended Coulomb branch $\mathcal{K}(\phi^i, m_f) $ of
the 5d gauge theory, which in addition to the vacuum expectation values,
$\phi^i$, of the scalars in the vector multiplet also encodes these masses for
hypermultiplets $m_f$.  This connection is depicted on the left-hand side of
figure \ref{fig:Overviews}.

\begin{figure}
\centering
\includegraphics[width=8cm]{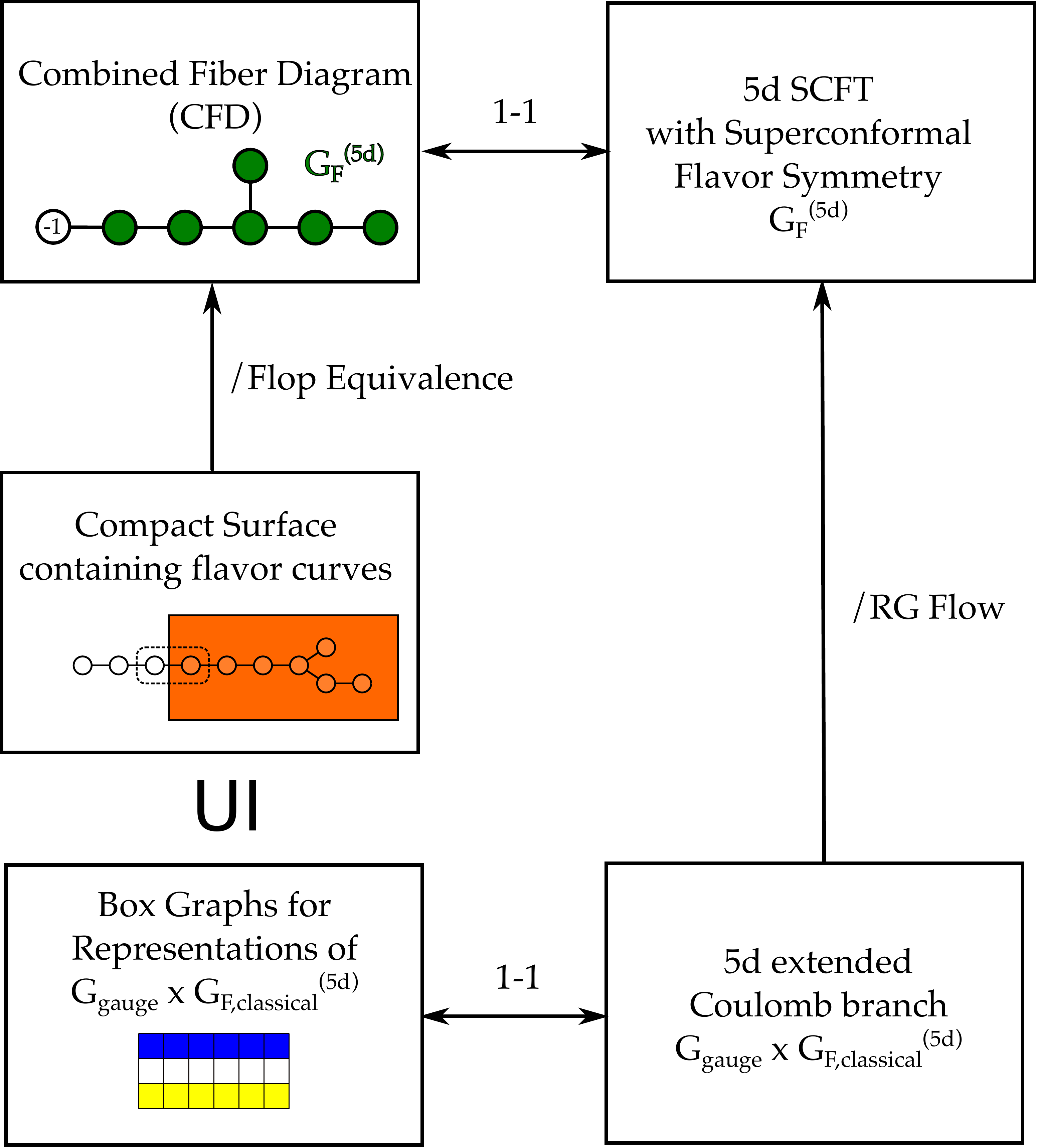}
 \caption{
This diagram shows schematically the approaches taken in the present paper and
in the companion paper \cite{Apruzzi:2019enx}, where box graphs and Coulomb branch
phases will be discussed. The CFDs were initially introduced in
\cite{Apruzzi:2019vpe} and encode (among other things) the strongly-coupled
flavor symmetry $G_\text{F}^{(5d)}$ of the 5d SCFT. The present paper provides
the geometric foundation for and a derivation of CFDs using the structure of
``flavor curves'' within compact surfaces in the resolution of Calabi--Yau
threefold singularities. In \cite{Apruzzi:2019enx} the focus will be on a
derivation using weakly-coupled gauge theory descriptions, whenever these
exist.  
 \label{fig:OverviewOfTools}}
 \end{figure}

It is this extended Coulomb branch, which we would like to map out
systematically, and thus determine the distinct strongly coupled UV fixed points.
This can be either achieved by analysing the Coulomb branch structure, or
using the M-theory realization in terms of singular Calabi--Yau threefolds. We
will consider both approaches --- see figure \ref{fig:OverviewOfTools}: 
\begin{enumerate}
\item[] \underline{Part I: Geometry}\\
  In M-theory the extended Coulomb branch is parametrized in terms of
  the relative extended K\"ahler cone of the singular Calabi--Yau
  threefold which underlies the 5d marginal theory (and thereby the parent 6d
  $\mathcal{N}=(1,0)$ theory that the marginal theory flows to in the UV).
   This K\"ahler cone parametrizes the distinct
  resolutions of singularities of the Calabi--Yau threefold. The
  main result in this paper is to find a succinct parametrization of these in
  terms of what we call {\it combined fiber diagrams (CFDs)}
  \cite{Apruzzi:2019vpe}, which allow us to determine the distinct 5d SCFTs
  descending from a given 6d theory.  We illustrate the power of this approach
  by determining the rank one and rank two classification\footnote{The rank
  refers to the dimension of the Coulomb branch, or equivalently, to the rank
of the weakly-coupled gauge group, if such a description exists.}, and further
we study a large class of examples of 5d SCFTs that are descendants of
conformal matter theories. A systematic analysis of the higher rank theories
will appear in subsequent work \cite{Apruzzi:2019kgb}.

\item[] \underline{Part II: Box Graphs and Coulomb branch phases}\\
In the companion paper \cite{Apruzzi:2019enx} we systematically explore the
Coulomb branches using the combinatorial device that was called \textit{box
graphs} in \cite{Hayashi:2014kca}.  This analysis emphasizes the weakly
coupled 5d gauge theory phases that flow to the SCFT, and complements the
classification obtained from the geometric approach. 
\end{enumerate}

At this point we should clarify the distinctly new aspects of the current
proposal towards the classification of 5d SCFTs.  We develop a new approach to the
geometric classification, which provides a purely combinatorial derivation of
all relevant geometries. This approach not only easily extends to higher rank
SCFTs, it, at
the same time, encodes strongly coupled data such as the (generically enhanced)
flavor symmetry for the SCFT, as well as mass deformations, which trigger
flows between 5d SCFTs. It is due to these characteristics that we pursue the
current approach.  Related work, e.g. \cite{Jefferson:2018irk}, determines the
geometries relevant for the construction of the rank two 5d SCFTs, however the
characterization of the geometries does not keep track of the strongly-coupled
data in the way that our approach manifestly achieves.  When applied to rank
two theories, the geometries we find are birationally equivalent to the ones
in \cite{Jefferson:2018irk}.  However, we choose a description that tracks the
flavor symmetries and systematically encodes the relevant deformations at
every step of the classification. This seems a natural way to organize
the classification of 5d SCFTs, and as we will show, it comes with the added
benefit of providing a very efficient algorithm for the classification.

That 5d gauge theories often experience a non-trivial flavor symmetry enhancement
at their superconformal fixed point was first noted in
\cite{Seiberg:1996bd,Intriligator:1997pq}.  To detect these enhanced
symmetries usually requires an analysis of the spectrum of operators charged
under the instantonic $U(1)_T$ symmetry, or the computations of protected
quantities such as the superconformal index \cite{Kim:2012gu,Bergman:2013aca,
Zafrir:2014ywa, Mitev:2014jza, Hwang:2014uwa, Gaiotto:2015una, Tachikawa:2015mha,
Yonekura:2015ksa, Zafrir:2015uaa, Bergman:2016avc}. In some cases, the flavor
symmetry at strong coupling can be also understood by analyzing string
junctions between 7-branes in a $(p,q)$-fivebrane web, see
\cite{Hayashi:2015fsa} for rank two examples. An alternative approach consists
of the study of the Higgs branch at infinite coupling by compactifying the 5d
theory on a $T^2$, which is a 3d $\mathcal{N}=4$ theory. 
The Coulomb branch of the mirror describes the Higgs branch at infinite coupling of the 5d SCFTs \cite{Ferlito:2017xdq, Cabrera:2018jxt}.

Our proposal in turn encodes the flavor enhancement from the get-go: the
strongly coupled flavor symmetry is manifest in the geometries
and in the graphical description in terms of CFDs, that we propose. This
approach is particularly easy to implement for -- though not limited to --
theories that descend from 6d theories, where the geometric realization
manifestly encodes the 6d superconformal flavor symmetry.  Examples are the
conformal matter theories \cite{DelZotto:2014hpa}, which have a particularly
nice characterization in terms of non-flat resolutions of the (non-compact)
elliptically fibered Calabi--Yau threefold.  In the context of F-theory, such
fibrations have been systematically studied in \cite{Lawrie:2012gg} for
Kodaira fibers, with examples in codimension two and three appearing in
\cite{Candelas:2000nc,Braun:2011ux,Braun:2013nqa,Cvetic:2013uta,Baume:2015wia,Anderson:2016ler,Buchmuller:2017wpe,Anderson:2017aux,Apruzzi:2018nre,Tian:2018icz,Huang:2018gpl,Dierigl:2018nlv,Achmed-Zade:2018idx}.
Unlike the more commonly studied resolutions of minimal collisions of elliptic singularities \cite{Miranda3, Esole:2017hlw, Esole:2019asj, Esole:2018csl, Esole:2018mqb},
which result in complex one-dimensional fibers, non-minimal singularities require insertions of complex
surfaces, $S_i$, into the fiber in order to resolve the singularity.  Such
fibrations, where the fibers are not equidimensional and thus have
higher dimensional fiber components, are called non-flat as the projection
defining the fibration is not a flat morphism. Similarly we shall refer to a
resolution of singularities whose resulting smooth geometry is a non-flat
fibration as a non-flat resolution.

We should stress, however, that our approach using CFDs to characterize 5d
SCFTs is not limited to geometries that admit a non-flat resolution, but can
be used to characterize any crepant resolution of the elliptic Calabi--Yau
threefold that underlies a given 6d SCFT. We will encounter examples that are
not based on non-flat resolutions in the context of rank two theories. 

Throughout, the geometries that we will consider have non-minimal singularities
in codimension two, i.e., over points, in the base, $B$, and the singular
elliptic Calabi--Yau threefolds $Y$ have a Weierstrass model,
\begin{equation}
  y^2 = x^3 + f (u,v) \, x + g(u,v) \,,
\end{equation}
where $(u,v)$ are local coordinates on the base $B$.  The non-minimal
singularity at $u=0=v$ will correspond to collisions of two non-compact curves
in the base, $u=0$ and $v=0$, above which the fiber has standard minimal
Kodaira singularities, associated to some Lie algebras $\mathfrak{g}_\nu$,
$\nu=u, v$.  F-theory compactified on $Y$ gives a 6d SCFT, with flavor
symmetry $\mathfrak{g}^\text{(6d)}_\text{F}$
\cite{DelZotto:2014hpa,Bertolini:2015bwa,Merkx:2017jey},  and it is this
flavor symmetry, and the remnants of this symmetry that percolate down to 5d,
which we will encode in our characterization of the resolution geometries, and
in their graphical presentation in terms of CFDs.

There are two approaches to resolve an elliptically fibered Calabi--Yau threefold with non-minimal singularities: 
One approach, most commonly used in F-theory, is to blow up the non-minimal
locus $u=v=0$ in the base successively, until the resulting fibration only has
minimal Kodaira singularities.  We refer to this, in reference to its
interpretation in the field theory, as the \textit{tensor branch} geometry.
This introduces compact surfaces, $S_i$, which are ruled over the blow-up
curves in the base.

Alternatively, it is useful in certain cases to blow up the fiber of the
elliptic threefold $Y$ (without changing $B$).  This approach is particularly
useful for 5d theories that descend from 6d conformal matter theories, where
the 6d flavor symmetry can be realized such that $\mathfrak{g}^\text{(6d)}_\text{F} = \mathfrak{g}_u \oplus \mathfrak{g}_v$. 
The resolution of the codimension one Kodaira singular fibers gives rise to
non-compact ruled surfaces, the so-called Cartan divisors,  
\be\label{CartDiv}
	\mathbb{P}^1_{l} \hookrightarrow D_{l}^{\mathfrak{g}_\nu} \rightarrow \{\nu =0 \} \,,
\ee
which are fibered over the codimension one loci $\nu = 0$ (associated to the simple roots of $\mathfrak{g}_\nu$). 
Over codimension two, these rational curves can become reducible, in addition however, due to the   
non-minimal singularity, the fiber resolution also introduces
\textit{compact} surfaces $S_i$ over the codimension two locus $u=0=v$. The
resulting smooth model is therefore not fibered by complex one dimensional
curves, but includes higher dimensional surface components -- the hallmark of
a non-flat fibration. 

The two approaches are, of course, birationally equivalent.  We argue that
whichever way one chooses to resolve the singularity, it will be key to retain
the information about the intersection between the compact surfaces $S_j$
(either from base blow-ups or non-flat resolutions) and the non-compact Cartan
divisors (\ref{CartDiv}),  in order to manifestly encode the flavor symmetries of
the 5d SCFTs.
 
In gauge theoretic terms, a resolved geometry corresponds to a point in the
extended Coulomb branch of the 5d marginal theory.  The compact surface
components, $S_i$, that resolve the non-minimal singularity realize the gauge
group of the effective field theory, while the non-compact divisors
$D_{l}^{\mathfrak{g}_\nu}$ furnish the flavor symmetry.  The flavor symmetry
is determined by fibral $\mathbb{P}^1$s, that are contained within the surface
components $S_i$ -- we will refer to these curves as {\it flavor curves}.  In
the singular limit, which corresponds to taking the volume of all $S_i$ to
zero, and thus the gauge coupling to infinite, the flavor curves also collapse
to zero volume, and contribute to the flavor symmetry at the strongly coupled
point.  The flavor symmetries we find in this way are in agreement with all
known flavor enhancements in 5d SCFTs at rank one and two and predict new
flavor symmetries at higher rank.

The CFDs encode essentially the generators of the Mori cone of the compact
(reducible) surfaces. From these we can furthermore determine a set of  BPS
states, which arise from wrapped M2-branes over complex curves. These become
massless particles in the SCFT limit.  For genus $0$ curves, their spectrum can
be straightforwardly computed in our constructions, as we will demonstrate for
spin $0$ and spin $1$ states in all rank one and rank two theories we construct.

In the present paper, we will have less of a focus on the weakly coupled gauge
theory interpretations of the geometries.  It should be stressed, however,
that the geometries equally contain the gauge theory data
\cite{Intriligator:1997pq}.  Using the  concept
of box graphs introduced in \cite{Hayashi:2014kca}, we systematically map out the Coulomb branch phases 
in the companion paper \cite{Apruzzi:2019enx}. This agrees with the present geometric analysis, whenever a weakly coupled gauge theory description is available.

\subsection{Strategy}
\label{sec:strategy}

We now summarize our strategy. The starting point is a 6d SCFT from which we
determine upon circle reduction the 5d marginal theory, to which we associate
three quantities: a resolved Calabi--Yau threefold, an effective gauge theory
description and a graph, called a {\it CFD}.  From this data, we determine all
5d descendant SCFTs, which can be obtained from the marginal theory by
relevant deformation and RG-flow, as well as key physical features, by applying a
straightforward algorithm. 

We support our proposal using a geometric and a gauge theoretic approach, respectively:
\noindent
\textbf{Part I: Geometry}\\
A given marginal theory is associated to a crepant resolution of an elliptic
Calabi--Yau threefold with a non-minimal singularity.  From such a smooth
geometry, one can extract an object we refer to as a  {\it marginal combined
fiber diagram}, which allows us to track all descendant SCFTs in terms
of so-called descendant CFDs. Each descendant CFD similarly is associated to a
different crepant resolution of a Calabi--Yau threefold. Each CFD corresponds to a 5d SCFT,  and
encodes the superconformal flavor symmetry as well as the relevant
deformations that trigger flows to another SCFTs.

The CFD is a graph with vertices that are curves inside the reducible surface
$\mathcal{S} = \bigcup_k S_k$ with self-intersection (inside ${\cal S}$) of
$-2$ or higher.  The $(-2)$-curves correspond to the $\bbP^1$-fibers of
$D^{\mathfrak{g}_\nu}_l$ and give rise to the non-abelian part of the flavor
symmetry of the SCFT -- the above-mentioned flavor curves.  Vertices that
correspond to $(-1)$-curves can be removed, which corresponds to flop
transitions that map the curve out of ${\cal S}$; they correspond physically to the
hypermultiplets that can decouple via mass deformations. Such CFD-transitions
allow charting the entire tree of descendant SCFTs systematically.  Finally,
the CFDs  contains higher self-intersection curves which are related to
abelian flavor factors and play a key role in determining the BPS states.

To obtain a comprehensive description it is thus key to determine the CFD
associated to the marginal theory, from which all descendants are obtained by
CFD-transitions.  For example, the unique rank one marginal theory originating
from an $S^1$-compactification of the 6d rank one E-string has the CFD
\begin{equation}
\includegraphics*[width=6cm]{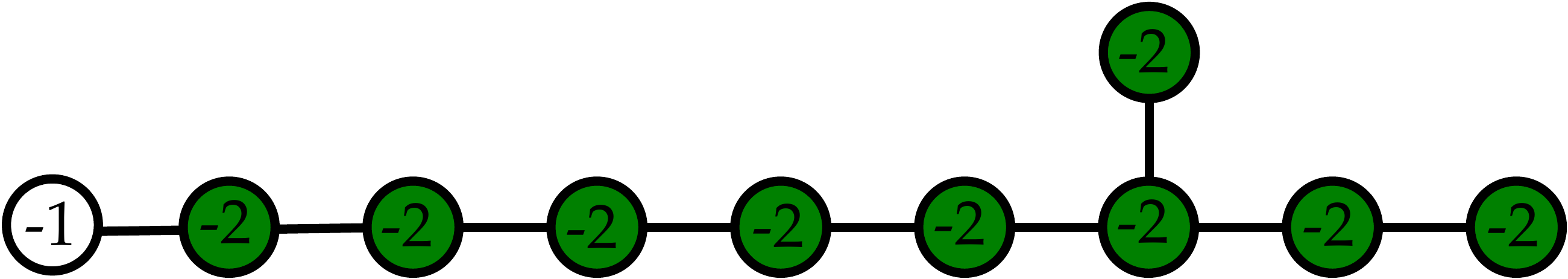} \,,
\end{equation}
where the $(-2)$-curves are the colored vertices, and correspond precisely to the flavor curves. 
This geometry is obtained by studying the curves in the compact surface, which is a generalized del Pezzo surface, gdP$_9$. This is a rational elliptic surface and 
the $(-1)$-curve is the zero-section, i.e., the copy of the
base $\mathbb{P}^1$ in the surface; 
the $(-2)$-curves intersecting in the affine $E_8$ Dynkin diagram correspond to the geometric realization of the flavor symmetry of the 5d marginal theory inherited from the E-string. 

For higher rank 5d theories, the corresponding CFDs should be thought of as an
equivalence class of geometries, related to each other by flops which do not
correspond to mass deformations, but simply to different gauge theory
descriptions that yield the same UV fixed point. In this way the CFDs are more
effective in their characterization of the SCFTs than any given resolution of
a singular geometry.

\noindent
\textbf{Part II: Coulomb Branch and Box Graphs}\\
Another path to support out CFD-approach is to consider the effective gauge
theory description.  To a marginal theory one can also associate in general
multiple gauge theoretic or quiver descriptions\footnote{Geometrically, these
correspond to rulings of the surface components $S_k$.} To each gauge or
quiver description, we can again encode the extended Coulomb branch
diagrammatically, in terms of so-called \textit{box graphs}
\cite{Hayashi:2014kca, Hayashi:2013lra, Braun:2014kla, Braun:2015hkv, Lawrie:2015hia}, which
were used to characterize crepant resolutions of minimal singularities in
elliptic fibrations. These are representation-theoretic ways of characterizing
the different Coulomb branch phases, and in the present application, encode in
particular the mass deformations that trigger flows to other SCFTs. 
 
To exemplify this, consider again the rank one marginal theory in 5d, i.e. the
circle-reduction of the rank one E-string theory. This admits an
$SU(2)_\text{G}$ gauge theory description with eight fundamental flavor
hypermultiplets, transforming under a classical flavor  symmetry,
$SO(16)_\text{F}$.  The matter fields in the $({\bf 2}, {\bf 16})$
representation of $SU(2)\times SO(16)$ can be represented in terms of a
representation graph, that encodes whether the corresponding weight or its
negative is in the Coulomb branch (indicated by blue/yellow coloring); this
comprises the box graph for this gauge theory description
\begin{equation}
({\bf 2}, {\bf 16}):\qquad \ba  \includegraphics*[width=4cm]{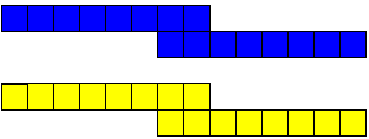} \ea \ .
\end{equation}
Here, each diagram is a ${\bf 16}$ of $SO(16)$, and the simple root of $SU(2)$
maps between the two diagrams.  In the companion paper \cite{Apruzzi:2019enx} we
show how  this characterization of the gauge theory phases encodes the mass
deformations and analogous box graph descriptions for all descendant 5d theories.
We also determine how the classical flavor symmetry is enhanced to the
superconformal one.  In summary, we will show in \cite{Apruzzi:2019enx} that for
5d SCFTs, which admit a weakly-coupled gauge theory description, the box
graphs encode equivalent information to the geometric approach of the present
paper. Furthermore, as in the geometric case,  equivalence classes of box
graphs, which are inequivalent gauge theory phases that yield the same UV
fixed point, can be packaged together into a CFD. In this way we also able to
provide an independent derivation of the tree of SCFTs/CFDs, using box graphs
and transitions between box graphs. 


The present paper is organized as follows: section \ref{sec:Setup} contains
background on 5d gauge theories, SCFTs and their realization in M-theory, as
well as resolutions of singular elliptic fibrations -- all central to our
subsequent endeavors.  These ideas are combined in section
\ref{sec:Resolutions}: we begin in section \ref{sec:Dictionary} by detailing
the dictionary between M-theory on resolutions of elliptic threefolds and 5d
SCFTs.  The proposal is exemplified by determining all rank one geometries
by considering non-flat resolutions of the elliptic fibration in section
\ref{sec:EstringGeo}. In sections \ref{sec:NonFlatRes} and
\ref{sec:HigherRankBU}, we determine the resolutions for the marginal theories
(which are the essential starting points for our subsequent analysis of the
descendant theories) and some examples of explicit
resolutions for rank two and higher. In section \ref{sec:CFD} we propose a
succinct way of encoding salient properties of the crepant resolutions
into graphs, the combined fiber diagrams (CFDs). We define CFDs associated to
crepant resolutions, and then demonstrate the result by determining all
rank one theories and rank two theories, using this proposal.
Furthermore, we determine the marginal theories for the $(E_n, E_n)$  and
$(E_8, SU(n))$ conformal matter theories and determine their descendants. In
section \ref{sec:BPS} we discuss the BPS states that are encoded in the CFDs
and compute them for rank one and rank two theories.  We conclude in section
\ref{sec:Conclusions}. There are various appendices containing summaries and
details of our analysis.  Most noteworthy are the summary tables for the rank
two theories in appendix \ref{sec:origin}, specifically appendix
\ref{app:SUMMARY}, which summarizes for each rank one and two theory the
corresponding CFDs,  the strongly-coupled flavor symmetry, the weakly-coupled
gauge theory description (if it exists), and the spin $0$ and spin $1$ BPS states.
The RG-flow trees, which are determined by considering the CFDs and their
transitions, starting with the marginal CFDs are shown in figure
\ref{fig:Rank1CFDTree} for rank one, and in figures \ref{fig:D10FibsAll},
\ref{fig:E8FibsAll}, and \ref{fig:Model3FibsAll} for rank two.

\section{M-theory, 5d SCFTs, and Resolutions of Elliptic Fibrations}
\label{sec:Setup}

Due to their intrinsically strongly-coupled nature, 5d SCFTs have been traditionally studied via their low-energy effective gauge theory phase.
As we will review now, the Coulomb branch structure of this infrared (IR) description motivates the approach of geometric engineering in M-theory.
Moreover, we discuss -- in the framework of M-/F-theory duality -- how the geometry can also capture essential physical features of theories that descend from circle reductions of 6d ${\cal N}= (1,0)$ SCFTs.

\subsection{5d SCFTs, Gauge Theories, and Coulomb Branches}
\label{sec:Gauge-Coulomb}

Gauge theories in 5d can be viewed as effective low-energy descriptions of an SCFT in the UV, which is deformed by the following relevant operator
\begin{equation} \label{eq:gkterm}
\int d^5x \, g_{\mathcal O}\, \mathcal O(x) \sim\frac{1}{g^2_{\rm YM}} \int d^5x\, {\rm Tr}\left(F^{\mu \nu} F_{\mu \nu}\right) \,.
\end{equation}
The explicit form of this operator is itself part of the low-energy effective
description, as $g_{\rm YM}$ has negative scaling dimension, which means that
the theory is non-renormalizable. In other words the gauge theory description
is not consistent at all scales. From the gauge kinetic term we infer the potential existence of an RG-fixed point at  $g_{\rm
YM} \rightarrow \infty$, which is a  strongly coupled UV SCFT. 
Moreover, there can be several operators like
\eqref{eq:gkterm},  leading to different IR gauge theories, which are dual in
the UV.

A 5d $\mathcal{N}=1$ gauge theory with gauge group $G_\text{gauge}$ and associated gauge Lie algebra $\mathfrak{g}$ has the following multiplets:
\begin{align}
\begin{split}
  \hbox{Vector multiplet in } {\bf Ad} (\mathfrak{g}): & \quad\mathcal{A} = (A_{\mu} , \phi^i, \lambda)\,, \cr 
  \hbox{Hypermultiplets in } {\bf R}_{\mathfrak{g}}: & \quad \mathbf{h} = ( h \oplus h^c , \psi) \,.
\end{split}
\end{align}
The hypermultiplets transform in representations of the classical flavor symmetry $G_\text{F,cl} $ of the effective gauge theory.
Moreover, any 5d gauge theory has instanton operators which are charged under an abelian global symmetry $U(1)_T$ associated to the current 
\begin{align}
	J_T= \frac{1}{8\pi^2}  \star \text{Tr} (F \wedge F) \,.
\end{align}
If $G_\text{gauge}$ is semi-simple, each simple gauge factor has a topological $U(1)_T^{(n)}$.
Non-perturbative effects can lead to an enhanced superconformal flavor symmetry, which we denote by 
\begin{align}\label{eq:flavor_enhancement_general}
	G_\text{F} \supset G_\text{F,cl} \times \prod_{n=1}^N U(1)^{(n)}_T \, ,
\end{align}
with $\text{rank}(G_\text{F}) = \text{rank}(G_\text{F,cl}) + N$.

The Coulomb branch of a 5d gauge theory is parametrized by the vacuum
expectation values (vevs), $\langle \phi^i \rangle$, of the real scalars in
the vector multiplets. These generically break the gauge group
$G_\text{gauge}$ to its Cartan $U(1)^r$.  The effective Lagrangian that
governs the dynamics on the Coulomb branch is 
\be
\mathcal L_\text{eff}= G_{ij} \, d\phi^i \wedge \star d \phi^j + G_{ij} \, F^i \wedge \star F^j + \frac{c_{ij \ell}}{24 \pi^2} \, A^i \wedge F^j \wedge F^\ell + \ldots \,,
\ee
where the couplings are determined by derivatives of the prepotential
$\mathcal{F}$, which is a real cubic function of the vevs of $\phi^i$
\begin{align}
  \begin{split}
    & G_{ij} = \frac{\partial^2\mathcal F}{\partial \phi^i \partial \phi^j} \, ,\\
    & c_{ij \ell} = \frac{\partial^3\mathcal F}{\partial \phi^i \partial \phi^j\partial \phi^\ell} \, .
  \end{split}
\end{align}
Since the Chern--Simons term is not gauge invariant, its coefficient must be integer, $ c_{ij\ell} \in \mathbb Z$ to avoid axial anomalies \cite{Ganor:1996pc,Intriligator:1997pq}.
The prepotential has a classical and one-loop contribution
\begin{align} \label{eq:prep}
\ba
\mathcal{F}=& \mathcal{F}_\text{classical} + \mathcal{F}_\text{1-loop} \cr 
=& \left( \frac{1}{2g_{YM}^2}\, C_{ij} \phi^i \phi^j + \frac{ k}{6} \, d_{ij \ell} \phi^i \phi^j \phi^\ell \right)
 +\frac{1}{12} \left( \sum_{\alpha \in \Phi_{\mathfrak{g}}} |\alpha_i \, \phi^i|^3 - \sum_{{\bf R}_f} \sum_{ \lambda \in \mathbf W_{{\bf R}_f}} |\lambda_i \, \phi^i + m_f|^3\right) \, ,
\ea
\end{align} 
where $d_{ij\ell}= \frac{1}{2} {\rm tr}_\text{fund}\left( T_i (T_j T_\ell +
T_\ell T_j)\right)$, $C_{ij} = \text{tr} (T_i T_j)$, and $k$
the half-integer quantized classical Chern--Simons level.  The roots of
$\mathfrak{g}$ are $\Phi_{\mathfrak{g}}$, and weights of the representation
${\bf R}_f$ by $\mathbf W_{{\bf R}_f}$.  Finally, $m_f$ are the masses of the
hypermultiplets. 

The prepotential determines different phases of the gauge theories as well as
the existence of a UV fixed point.  At $g_{YM}\rightarrow \infty$ the
effective Lagrangian description breaks down due to the appearance of
infinitely many light states. The phases of the gauge theory form the
extended Coulomb branch
\begin{equation} \label{eq:extKC}
\mathcal{K}(\phi^i,m_f) \,,
\end{equation} 
which includes the vevs of scalars $\phi^i$ but also the masses of
hypermultiplets. This can be thought of as the Coulomb branch of the theory,
where we in addition weakly gauge the flavor symmetry $G_{\text{F, cl}}$.
These mass parameters can be interpreted as Coulomb branch parameters, where
we have weakly gauged the classical flavor symmetry $G_\text{F,cl}$.

Our approach in this paper is based on the observation that these gauge theory
structures are equivalently realized in terms of M-theory compactifications on
Calabi--Yau threefolds
\cite{Cadavid:1995bk,Ferrara:1996hh,Witten:1996qb,Ferrara:1996wv,Morrison:1996xf,Intriligator:1997pq}.
In particular, the non-abelian gauge degrees of freedom arise from M2-branes
wrapping collapsed holomorphic curves at singularities.
In this correspondence, the Coulomb branch \eqref{eq:extKC} is identified with
the so-called extended K\"ahler cone associated to a singular  Calabi--Yau
threefold.  This cone is defined by the union of all the K\"ahler cones
associated to the possible crepant resolutions of the singularity.  Each of
these resolutions is geometrically given by a collection ${\cal S} = \bigcup_j
S_j$ of compact surfaces $S_j$, where the surfaces are in one-to-one
correspondence with the Cartans of the vector multiplets and with the Coulomb
branch scalars.  On the other hand, each mass parameter $m_f$ is associated
with a non-compact surface $D_f$, and for fixed choices of $m_f$ there are
subcones 
\begin{equation} \label{eq:kinK}
K(\phi^i)|_{{\rm fixed}\, m_f}\subset \mathcal K(\phi^i,m_f) \,.
\end{equation}
The origins of these subcones define different singular limits in which the
compact surfaces $S_j$ collapse to zero volume, which in the gauge theory
correspond to distinct 5d UV fixed points. 

Before we discuss in detail the dictionary between Calabi--Yau geometry and 5d
physics, we point out that the extended Coulomb branch $K(\phi^i,m_f)$ has an
elegant and systematic description in terms of so-called box graphs
\cite{Hayashi:2014kca}, which encode the extended Coulomb branch in terms of
representation-theoretic data of 
\begin{equation}
G_\text{gauge} \times G_\text{F,cl} \, .
\end{equation}
We will explore this gauge-theoretic approach in the companion paper
\cite{Apruzzi:2019enx}.


\subsection{Geometric Engineering of 5d Gauge Theories and SCFTs}
\label{sec:geometry_to_gauge-theory}

The gauge theoretic content of the last section has a counterpart in the
realization of 5d SCFTs in M-theory on singular Calabi--Yau threefolds
\cite{Intriligator:1997pq}. Associated to a singular Calabi--Yau $Y$ is a
resolution that retains the Calabi--Yau condition, a so-called crepant
resolution. The process of resolving the singularity introduces compact
divisors, i.e., complex surfaces, $S_i$, into the geometry. The space of
crepant resolutions of the singularity should be thought of as playing the
role of the extended Coulomb branch of the gauge theory. The precise
dictionary is as follows:
\begin{itemize}
\item \underline{Cartan subgroup of the gauge group:} \\
A compact divisor $S$ is dual to a $(1,1)$-form $\omega^{(1,1)}$, which in turn can be used to dimensionally reduce the M-theory three-form $C_3 = \omega^{(1,1)}\wedge A$, where $A$ is a $U(1)$-gauge field. The gauge coupling is set by the volume of $S$.
The number of compact surface components $S_i$, $i= 1, \cdots, r$, sets the rank of the weakly-coupled gauge theory.  
\item \underline{Non-abelian gauge symmetry:} \\
The enhancement to a non-abelian gauge group results from rulings of the
compact surface components.
A surface $S$ is ruled, if it is fibered by rational curves $f$ over a curve $\Sigma$, 
\be
f \hookrightarrow S \rightarrow \Sigma \,,
\ee
with the intersection numbers
\begin{equation}
  S \cdot f = -2 \,,\qquad  f\cdot_S f = 0 \, .
\end{equation}
This allows collapsing the surface along the fibers $f$ to curve $\Sigma$
worth of singularities, inside of the Calabi--Yau threefold. M2-branes wrapped on $f$ become massless in this limit
and furnish the W-bosons for the enhancement to a  non-abelian gauge symmetry.
It is possible for $r$ surfaces $S_i$ to intersect pairwise along curves
$\Sigma_{ij} = S_i \cap S_j$ that are (multi-)sections of the rulings $f_i
\hookrightarrow S_i \rightarrow \Sigma_i$ on both surfaces.  In this case,
collapsing all $f_i$ leads to a simple rank $r$ non-abelian gauge group
$G_\text{gauge}$ determined by the Cartan matrix
\be
C^{G_{\rm gauge}}_{ij} = - S_i \cdot f_j  \, .
\ee
The genus $g(\Sigma_{ij})$ of the intersection curve is computed by
\begin{equation} \label{eq:genusform}
S_j S_{j+1}^2+S_{j+1} S_{j}^2=2g(\Sigma_{ij} )-2 \,.
\end{equation}
The fibers that rule the configuration of surfaces $\mathcal S = \bigcup_j S_j$ will be denoted by
\begin{equation}\label{eq:ruling_reducible_surface_notation}
	\{f\} \equiv  \bigcup_j f_j   \hookrightarrow \mathcal S \,.
	\end{equation}
\item \underline{Matter:} \\
If a ruling $f_i \hookrightarrow S_i$ has reducible fibers, such that the
rational curves splits (i.e., $f_i = \sum_l C_i^{(l)}$ in homology), M2-branes
wrapping $C_i^{(l)}$ give rise to charged matter states which become massless
in the singular limit when $S_i$ collapses to $\Sigma_i$. 
\item \underline{Quiver Theories:} \\
A weakly-coupled description of an SCFT can also be in terms of a quiver gauge
theory. For instance a quiver with gauge groups $G_1 \times G_2$ is realized,
if there are surfaces  $S_{G_{i}}$, $i=1,2$ (contributing to the gauge groups
$G_i$ in the above fashion), intersecting each other along $\Sigma$, which is
a special fiber of the ruling on both surfaces $S_{G_i}$.  In this case,
collapsing all fibers of these rulings leads to massless states from M2-branes
on $\Sigma$, which are charged as bifundamentals of $G_1\times G_2$.

\item \underline{Prepotential:}\\
From the M-theory compactification is a polynomial in the K\"ahler parameters
$\phi^i$ dual to the divisors $S_i$ we can define the geometric prepotential
\begin{equation} \label{eq:geomprep}
\mathcal F_{\rm geo} = \frac{1}{6}c_{ij \ell}\phi^i \phi^j \phi^\ell \qquad \text{with} \qquad c_{ij\ell}=S_i \cdot S_j \cdot S_\ell \, .
\end{equation} 
This is to be identified with the cubic part of the field theoretic
prepotential \eqref{eq:prep}, which receives classical and one-loop
contributions
\begin{align}\label{eq:matching_prep}
	c_{ij\ell} = S_i \cdot S_j \cdot S_\ell \stackrel{!}{=} k \, d_{ij\ell} + c_{\text{one-loop}, \, ij\ell} \, .
\end{align}
In practice, the prepotential matching allows us to compute the classical
Chern--Simons level $k$ from geometry, see appendix
\ref{subsubsec:rank_2_prep} for the rank two theories.

\item \underline{Dualities:}\\ 
A surface can have multiple rulings, which give rise to dual gauge theory descriptions of the same SCFT. 
For example, in the rank two theories, two surfaces $S_1$ and $S_2$ intersecting along $\Sigma$ can have  three distinct types of rulings $f_i^{(n)} \hookrightarrow S_i$, $n=1,2,3$, corresponding to the following weakly-coupled gauge theory descriptions:
\begin{enumerate}
	\item $\Sigma$ is a section for both surfaces, i.e., $\Sigma \cdot_{S_1} f_1^{(1)} = \Sigma \cdot_{S_2} f_2^{(1)} = 1$.
	This gives rise to an $SU(3)$ theory.
	\item $\Sigma$ is a section for $S_1$ and a bi-section for $S_2$, i.e., $2 \, \Sigma \cdot_{S_1} f_1^{(2)} = \Sigma \cdot_{S_2} f_2^{(2)} = 2$. This realizes an $Sp(2)$ theory.
	\item $\Sigma$ is a special fiber in both surfaces, i.e., $\Sigma \cdot_{S_i} f_i^{(3)} = 0$. This correspond to an $SU(2) \times SU(2)$ quiver description.
\end{enumerate}
\item \underline{Coulomb branch phases:}\\
In general, there can be different configurations of surfaces $S_k$ that give rise to the same gauge theory upon flop transitions along suitable rulings within the reducible surface $\cal S$. 
This is the geometric incarnation of different chambers within the Coulomb branch of a 5d ${\cal N}=1$ gauge theory.
As one crosses the interfaces between two such chambers, the masses of certain states undergoes a sign change.
This is distinct from flopping curves in or out of $\mathcal{S}$, which in contrast gives rise to different SCFTs.

\item \underline{Relevant Deformations and Flows to new SCFTs:}\\
While flop transitions that map curves out of $\cal S$ correspond to mass
deformations in the gauge theory, physically, it involves sending one of the flavor masses, $m_f$, to $\pm \infty$ and 
effectively decouples the corresponding hypermultiplet. Geometrically, this is
due to the curve being flopped out of $\mathcal{S}$, and thereby not getting
collapsed in the singular limit (i.e., the associated state remains massive).
This decoupling process yields an effective theory with a different UV fixed
point.  In case the effective description has an appropriate gauge theory
interpretation, the classical Chern--Simons level shifts as
\begin{equation}\label{CSShift}
k\rightarrow k +\sign(m_f) \frac{1}{2}\,.
\end{equation}

\item \underline{Absence of weakly-coupled gauge theory description:}\\
The geometric description is slightly more general as it captures SCFTs that
do not allow a weakly-coupled description. This occurs e.g. in rank one for
the ``$E_0$ theory''  \cite{Morrison:1996xf}, with the number of such theories
increasing in higher rank. Geometrically, this occurs, when two surfaces
$S_{1,2}$ only have rulings such that $\Sigma =S_1 \cap S_2$ is a fiber of one
ruling, but a (multi-)section of the other. There is no ``classical''
non-abelian gauge theory phase, because collapsing either surface along its
rulings results in light non-perturbative states from the forced collapse of
the other surface.  Such geometries are nevertheless consistent M-theory
compactifications and give rise to non-trivial strongly interacting limits,
corresponding to 5d SCFTs.

Importantly, the geometric description of mass deformations puts theories
without effective gauge descriptions on the same footing as theories with
gauge theory descriptions.  While certain states that can be decoupled may be
non-perturbative states from a gauge theoretic perspective, the geometry
uniformly characterizes these as curves that can be flopped out of the
reducible surface $\cal S$.  Such curves play a vital role in determining all
descendant SCFTs obtainable from a given SCFT via mass deformations.

\end{itemize}

\subsection{SCFTs and Flavor from Geometry}\label{sec:flavor_symmetry_from_geo_general}

The SCFT limit of a gauge theory, where $\phi^i \rightarrow 0$ with fixed
$m_f$, corresponds in geometry to collapsing the reducible compact surface
$\mathcal{S}= \bigcup_j S_j$ to a point.  In this limit, non-perturbative
states from M2-branes wrapping (multi-)sections of the rulings on $S_j$ become
part of the spectrum, signaling a breakdown of the effective description.

The existence of such a strongly coupled UV fixed point poses certain convexity constraints on the prepotential \eqref{eq:prep} as a function on the extended Coulomb branch.
Translated into geometry, this implies conditions on the singular limits of surface configuration $\cal S$ \cite{Intriligator:1997pq}. Essentially, the condition is that the singularity one obtains from the collapse is not too severe, and allows for a crepant resolution.
Such a singularity is called a canonical singularity. 
Importantly, these conditions are purely geometric and thus also apply to cases without an effective gauge theory description.

Moreover, the geometric perspective easily accommodates the feature of UV dualities, as a canonical singularity can have different resolutions corresponding to different effective gauge descriptions of the same SCFT.
This is reflected by certain details, e.g., the intersection curves between different surfaces, being vital to the gauge theory description, but not to the SCFT.
In contrast, the contractible curves inducing mass deformations have to be kept track across different, dual phases.

A key feature of the UV fixed points is that the flavor symmetry of the theory
can enhance compared to the effective description as a gauge theory. While the
classical flavor symmetry $G_\text{F,cl}$ of an effective gauge theory is
easily inferred from the spectrum of hypermultiplets, the superconformal
flavor group $G_\text{F}$ is oftentimes difficult to determine, as it is an
intrinsically strongly-coupled datum. 
In cases when the SCFT has a gauge theory phase in the IR, the intricate
enhancement \eqref{eq:flavor_enhancement_general} can be computed via the
superconformal index \cite{Kim:2012gu}.
However, such methods fail when no effective description is available.

A central point that we will make in this paper is that the geometry does in
fact track the flavor symmetries via the collapse of non-compact divisors
intersecting the compact surfaces $\cal S$ \cite{Xie:2017pfl,
Apruzzi:2018nre}.
Concretely, if a non-compact divisor $D \subset Y$ is ruled, $F \hookrightarrow D \rightarrow W$, then collapsing $F$ generates orbifold singularities fibered along a non-compact curve isomorphic to $W$.
More generally, if the generic fibers of multiple non-compact divisors intersect in a Dynkin diagram, then the above collapse produces a curve of the corresponding ADE singularities.

In particular, when $\cal S$ collapses along the ruling $\{f\}$
\eqref{eq:ruling_reducible_surface_notation}, a subset of generic fibers $F$
contained in $\cal S$ may be forced to collapse to a curve, leading to
singularities of ADE-type -- this determines the classical flavor symmetry
$G_\text{F,cl}$. When furthermore collapsing $\mathcal{S}$ to a point, this
flavor symmetry can enhance to $G_{\text{F}}$, which generically is larger
than $G_\text{F,cl}$. 

The classical flavor symmetry $G_\text{F,cl}$ depends on the choice of ruling $\{f\} \hookrightarrow {\cal S}$, and may only be manifest in certain geometric phases.
However, the superconformal flavor symmetry $G_\text{F}$ only depends on the curves in $\cal S$ that collapse in the singular limit. We will refer to these as {\it flavor curves}. 
A particularly nice way of extracting these flavor curves arises for 5d theories that descend from 6d conformal matter theories, where the 6d flavor symmetry is manifest in the elliptic fibration. By considering non-flat resolutions of such singularities, the flavor curves have a very simple presentation and are manifest in the resolved geometries.


\subsection{5d Marginal Theories}
\label{sec:marginal_theories_general}

A special class of 5d gauge theories that arise by circle-reduction of 6d
SCFTs are so-called marginal theories.  These are effective theories whose
UV-completion is not an honest 5d SCFT, but rather the 6d SCFT itself.  This
limiting theory, dubbed 5d Kaluza--Klein (KK) theory in
\cite{Jefferson:2018irk}, is the M-theory compactification on the fully
singular elliptic fibration, i.e., the geometric limit where all compact
($S_j$) and non-compact ($D_i^{(\nu)}$) exceptional divisors are collapsed to
points and curves, respectively.

The marginal theories play in important role in our classification program as they contain the relevant information about the 5d SCFTs that descend from a given 6d SCFT.
Geometrically, the marginal theory can be characterized most straightforwardly by blowing up the base to remove the non-minimal point $u=v=0$, possibly multiple times, until all fiber singularities are of minimal Kodaira type.
In 6d, these blow-ups correspond to giving the scalars of ${\cal N}=(1,0)$ tensor multiplets a vev, thus moving the 6d SCFT onto its  tensor branch.
A subsequent fiber resolution yields a smooth Calabi--Yau threefold with a flat fibration.
Note that this resolution process also introduces compact surfaces $S_j$, which are now ruled (or elliptically fibered) over the compact rational curves that were introduced into the base to blow up $u=v=0$.

In this smooth space, one can in principle read off all necessary physical data such as the effective gauge descriptions and their mass deformations.
There are ongoing efforts, for example in
\cite{Bhardwaj:2018yhy,Bhardwaj:2018vuu}, towards classifying the smooth geometries associated with marginal theories arising from 6d SCFTs classified by F-theory.

A key feature of the marginal geometry is that the compact surfaces ${\cal S} = \bigcup_j S_j$ contain \emph{all} codimension one fibers $F_i^{(\nu)}$, i.e., the maximal set of flavor curves. 
This means that when ${\cal S}$ is collapsed to a point, the flavor symmetry from codimension one singularities, as discussed in section \ref{sec:flavor_symmetry_from_geo_general}, would formally be of \emph{affine} type.
The ``affinization'' is due to the presence of the Kaluza--Klein $U(1)$ arising in the circle reduction, and is the indicator that the UV physics corresponding to this singular geometry is only appropriately described as a 6d theory.

Once one knows the 5d marginal theory and its associated ``marginal'' geometry, mass deformations corresponding to flops generate all descendant theories that have a genuine 5d SCFT limit.
Geometrically, these flops move flavor curves $F_i$ out of $\cal S$, such that when we collapse the latter, not all codimension one divisors $D_i$ are forced to collapse to curves. 

The geometries associated to the 5d marginal theories that come from 6d $(G_1,
G_2)$ conformal matter theories can be particularly elegantly characterized in
terms of non-flat resolutions of elliptic fibrations. In this case, the flavor
curves that are the fibers of the Cartan divisors associated to the two flavor
groups $G_1$ and $G_2$ in 6d, can be straightforwardly identified inside the
non-flat fibers.  This  is based on the idea of \cite{Apruzzi:2018nre}, where
non-flat fibrations were used in this context.  Here, we will construct the
marginal geometries for conformal matter theories from non-flat resolutions,
where the non-flat fiber is a reducible surface $\mathcal{S}$.  An advantage
of this approach is that  $\cal S$ is \emph{by construction} contractible, as
it arises from crepant resolutions of a singularity in a Calabi--Yau
threefold.  Thus, the singular limit of blowing down $\cal S$ gives
rise to a well-defined 5d SCFT in M-theory.

\subsection{Singular Elliptic Calabi--Yau Threefolds}
\label{sec:elliptic_threefolds_general}

Given our motivations, we focus on singular Calabi--Yau threefolds that are elliptic fibrations over a non-compact base two-fold $B$.
In the context of F-theory \cite{Vafa:1996xn}, these manifolds define 6d theories. We assume first of all that the base is smooth\footnote{We will discuss models where the base has codimension two singularities momentarily.}.
In complex codimension one in the base, singularities in the fiber are of
so-called minimal type, and are classified by the classical results of Kodaira
and N\'{e}ron, and give rise to gauge fields in 6d.  In codimension two,
minimal singularities give rise to 6d bifundamental matter in hypermultiplets
\cite{Morrison:1996na,Morrison:1996pp,Katz:1996xe}.
The case of interest in this paper is, however, when the fibration has a codimension two  non-minimal singularity.
These are indicators of strongly coupled sectors in F-theory, and form the
geometric foundation of the recent classification of 6d ${\cal N}=(1,0)$ SCFTs
\cite{Heckman:2013pva, Heckman:2015bfa ,Bhardwaj:2015xxa}. In 6d the canonical
way to study these singularities is by blowing up the base, i.e., removing the
codimension two non-minimal locus by inserting rational curves, until the
singularities all become minimal. 

In M-theory on such Calabi--Yau threefolds, there is an alternative approach, which allows resolutions of the singular fiber, keeping the base unchanged. Fiber-resolutions of non-minimal singularities have the key feature that they insert higher-dimensional fiber components. For codimension two non-minimal singularities in Calabi--Yau threefolds these are complex surface components. In this section we summarize the background of how to study such non-flat resolutions. 

Consider an elliptically fibered Calabi--Yau threefold  $Y$, defined over a complex base surface $B$, 
\be
\mathbb{E} \hookrightarrow Y \stackrel{\pi}{\longrightarrow} B\,,
\ee
with projection $\pi$ and generic fiber given by an elliptic curve $\mathbb{E}$. We will assume that this has a section, i.e., a map from the base to the fiber, and thus a Weierstrass model
\be
y^2 = x^3 + fx + g \,.
\ee 
We will generally work with the Tate model of the elliptic fibration, 
\be\label{TateMod}
y^2 + b_1 x y + b_3 y = x^3 + b_2 x^2 + b_4 x + b_6 \, ,
\ee
which is (largely) equivalent to the Weierstrass model, but comes with some added computational benefits  \cite{Bershadsky:1996nh, Katz:2011qp}.

The $b_i$ are sections of line bundles over the base $B$, i.e., they depend on
the base coordinates. 
The generic fiber above a point in the base will be smooth, and singularities in codimension one occur whenever the discriminant $\Delta = 4 f^3 + 27 g^2$ vanishes.
Let $(u,v)$ denote local coordinates in the base. 
In general, the discriminant locus $\{\Delta =0\}$ can have multiple components.
For our purposes, it will suffice to consider two codimension one irreducible components of $\{\Delta =0\}$, which locally are described by $u=0$ and $v=0$. 
The singularity type in the fiber is determined by the vanishing orders of $b_i$ as functions of $u$ and $v$.
We will use the notation for resolutions of singularities as introduced in \cite{Lawrie:2012gg, Hayashi:2014kca}.

\subsubsection{Codimension One: Kodaira Fibers}
\label{sec:cod-1fiber}

Denote by $W_u = \{u=0\}$ an irreducible component of the discriminant locus.
To each set of vanishing orders in codimension one $\hbox{ord}_{u}(b_i)$ there
is an associated Kodaira singular fiber. These are determined by resolving the
singularity, which amounts to introducing new (exceptional) rational curves
(i.e., $\mathbb{P}^1$s) in the fiber. Except for sporadic outliers, a Kodaira
fiber is characterized in terms of an affine simple Lie algebra
$\widehat{\mathfrak{g}}$, with a rational curve $F_i$ associated to each
simple root $\alpha_i$ 
\be
\alpha_i \quad \leftrightarrow \quad F_i \cong \mathbb{P}^1\,,\qquad i=0, \ldots, \, \hbox{rank} (\mathfrak{g}) \,.
\ee
Fibering $F_i$ over $W_u$ results in divisors, the so-called Cartan divisors,
\be
F_i \hookrightarrow D_{i}   \rightarrow W_u \,,
\ee
which are in one-to-one correspondence with the Cartans of $\widehat{\mathfrak{g}}$. 
For simply-laced Lie algebras $\mathfrak{g}$, the fibral curves $F_i$ intersect with the Cartans in the negative Cartan matrix 
\be \label{eq:cod1fib}
D_{i} \cdot F_j = - {\mathcal C}^{\widehat{\mathfrak{g}}}_{ij} \,.
\ee
An example is shown in figure \ref{fig:NonFlatFiberPic1}, where in codimension
one above $u=0$ the fiber is of type $II^*$ or $\widehat{\mathfrak{g}} =
\widehat{\mathfrak{e}_8}$.

For a non-simply-laced Lie algebra $\mathfrak{g}$, the $F_i$ can be multiple copies of $\mb{P}^1$, and the associated $D_i$ is reducible as well. In this case, the intersection matrix is given by~\cite{Park:2011ji}
\be \label{eq:cod1fib-ns}
D_{i} \cdot F_j = -\frac{2\langle\alpha,\alpha\rangle_{\rm max}\langle \alpha_i,\alpha_j\rangle}{\langle\alpha_i,\alpha_i\rangle\langle\alpha_j,\alpha_j\rangle}\, ,
\ee
where $\alpha_i$ is the root that corresponds to the curve $F_i$ and $\langle\alpha,\alpha\rangle_{\rm max}$ is the maximal length of the roots in the Lie algebra $\mathfrak{g}$. We will show an example of $\mathfrak{g}_2$ in section~\ref{sec:HigherRankBU}.


\subsubsection{Codimension Two: Flat and Non-flat Resolutions}
\label{sec:cod-2fiber}

Consider next the situation when two such codimension one singular fibers
collide. Let $W_u$ and $W_v$ be components of the discriminant along $u=0$ and
$v=0$, respectively, with associated Kodaira fibers of type
$\widehat{\mathfrak{g}}_{u/v}$.  There are essentially two cases to
distinguish: 
\be
\ba
\hbox{Minimal:} & \quad \hbox{ord}_{u=v=0} (f)<4\quad \hbox{or}\quad \hbox{ord}_{u=v=0} (g)<6\cr 
\hbox{Non-Minimal:} & \quad \hbox{ord}_{u=v=0} (f, g, \Delta ) \geq (4,6,12)  \,.
\ea
\ee
In the Tate form, non-minimality is characterized by $\hbox{ord}_{u=v=0} (b_i;
\Delta) \geq (1,2,3,4,6; 12)$.  In the minimal case the fiber corresponds to
what normally is interpreted as bifundamental matter, and above the
codimension two locus $u=v=0$ there is again a collection of rational curves,
intersecting in a (possibly reduced) Kodaira singular fiber.  A fiber
resolution (and the associated smooth fibration) giving such codimension two
fibers is called flat (the fiber dimension stays at complex one dimension).
An in-depth analysis of the possible fiber types can be found in
\cite{Esole:2011sm, Marsano:2011hv, Krause:2011xj, Lawrie:2012gg,
Hayashi:2014kca, Braun:2014kla, Braun:2015hkv, Esole:2017kyr}.  Here we will consider a simple example. 

The simplest case arises for $u=0$ with an $I_n$ fiber, i.e., $\mathfrak{g} = \mathfrak{su}(n)$, and $v=0$ with an $I_1$ (effectively a $\mathfrak{u}(1)$).
The singular fiber in this case above $u=0$ is given in terms of a ring of $n$ rational curves $F_i$, $i=0, \ldots, n-1$. 
At $u=v=0$  one of these curves will become reducible and splits into two rational curves
\be\label{FCC}
F_i \rightarrow C_i^{+} + C_{i+1}^{-} \,.
\ee
The fiber in codimension two corresponds in this situation to the Kodaira fiber $I_{n+1}$, and the $n+1$ rational curves intersect in the $\widehat{\mathfrak{su}(n+1)}$ affine Dynkin diagram.
Moving away from the locus $u=v=0$ along the divisor $u=0$ can be thought of as a Higgsing $\mathfrak{su}(n+1) \rightarrow \mathfrak{su}(n) \oplus \mathfrak{u}(1)$. The curves $C_i^{\pm}$ carry charges under the Cartan divisors $D_{j}$ which in this case correspond to the fundamental ${\bf n}$ representation, consistent with this Higgsing. The sign $\pm$ indicates whether the curve carries $\pm \lambda$ as charge vector, where $\lambda$ is a weight of the fundamental representation. 

\begin{figure}
\centering
\includegraphics[width=10cm]{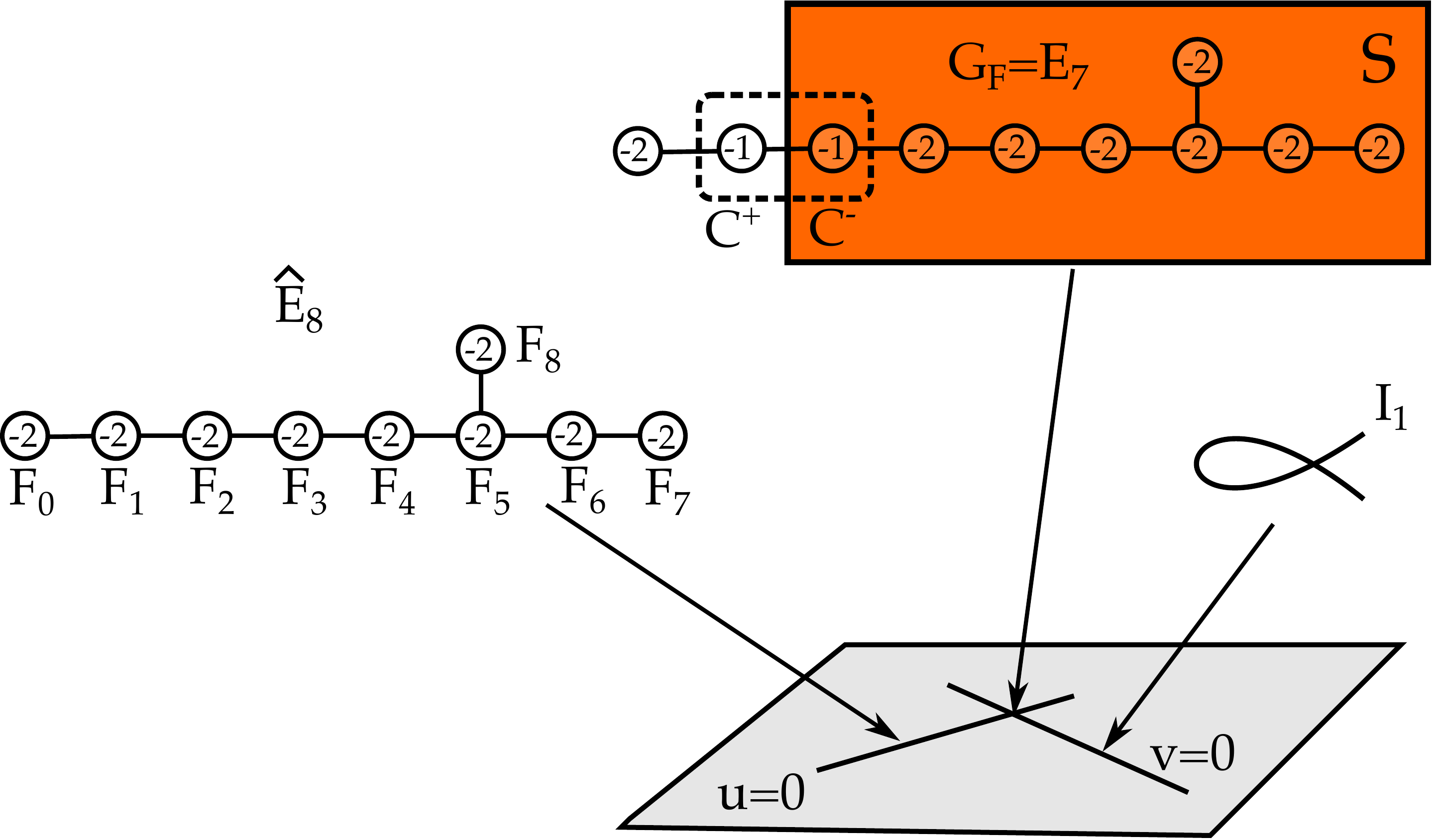}
\caption{Rank one E-string: Depicted is the collision of two codimension one singularities of type $E_8$ and $I_1$, respectively, as well as the codimension two fiber including the non-flat component $S$. 
In codimension one, the fiber above $u=0$ is a collection of $(-2)$-curves
$F_i \cong \mathbb{P}^1$ that are in one-to-one correspondence with the affine
roots of $E_8$. At the collision point $u=v=0$, one of these rational curves
splits (here $F_1 \rightarrow C^+ + C^-$). The non-flat fiber component $S$ is
a complex surface above the codimension two locus, which contains some of the
$(-2)$-curves of the codimension one fiber -- the so-called flavor curves. In
the situation depicted, $S$ contains $F_2, \cdots, F_8$, which intersect in
the $E_7$ Dynkin diagram -- we will refer to these curves in the following as
flavor curves.
Indeed, we will argue that they encode the flavor symmetry of the associated strongly coupled SCFT to be $E_7$.   \label{fig:NonFlatFiberPic1}}
\end{figure}

The situation of interest in this paper occurs when  the collision is {\it non-minimal}. 
In this case the codimension two fiber has a resolution in terms of a non-flat fiber, i.e., the fiber above $u=v=0$ is not only a collection of curves, but also complex surfaces and 
so overall is not of complex dimension one. Alternatively we can resolve the base by successively blowing up the locus $u=v=0$. As this is more commonly done, we will focus here on describing the non-flat fiber resolution.
An example is shown in figure \ref{fig:NonFlatFiberPic1}, where $S$ is the
non-flat surface component above $u=v=0$, arising from colliding an $\widehat{E_8}$ and an $I_1$ fiber.\footnote{For $I_1$ the codimension one fibration is generically smooth, and shown as a node.}
The surface $S$ contains a collection of curves that are part of the codimension one singular fiber (in the example shown in figure \ref{fig:NonFlatFiberPic1}, this is a subset of curves, which form an $E_7$ sub-Dynkin diagram of the codimension one affine $E_8$ singular fiber).
Non-flat fiber resolutions in codimension two have been studied systematically in \cite{Lawrie:2012gg}, and more recently in \cite{Tian:2018icz, Apruzzi:2018nre}.

The codimension one fibral curves $F_i$ can split in the non-flat case as
well, i.e., the rational curves in codimension two will be a collection of
curves associated to the roots $F_j$ and weights $C_{i}^\pm$ -- in figure
\ref{fig:NonFlatFiberPic1}, $F_1$ splits into the two $(-1)$-curves $C^{\pm}$,
of which one is contained in the non-flat surface.  In general there can be
multiple surface components $S_k$ (the number of distinct values of $k$
depends on the precise rank enhancement in codimension two), which can in part
contain the fibral curves.

As we have argued in section \ref{sec:geometry_to_gauge-theory}, the compact surface components $S_j$ obtained in this way will in general give rise to the 5d gauge group, whereas the $(-2)$-curves $F_i$  of the codimension two fiber contained in $\mathcal{S}= \bigcup_j S_j$ determine the flavor symmetry of the SCFT that is obtained in the singular limit. We will refer these curves as {\it flavor curves}. 
While we come back to the associated 5d physics in more detail in the next section, we point out that the resolution process very easily produces many physically inequivalent (in 5d) geometries related via flop transitions.
Crucially, these transitions also involve flops between compact divisors $S_j$ \emph{and} non-compact divisors $D_i$, which from the perspective of $\cal S$ is a contraction, i.e., a mass deformation in 5d.
In that way, fiber resolutions of elliptic threefolds with non-minimal singularities naturally produce \emph{different} 5d SCFTs associated to the same 6d theory.

While most of our geometric examples are obtained from a non-flat resolutions, we stress that this is merely a convenient method to find smooth phases of a non-minimal elliptic singularity.
For higher rank examples discussed in sections \ref{sec:HigherRankBU}, we employ a combination of base and fiber blow-ups, resulting in compact surfaces $S_j$ which are both flat and non-flat.
In fact, flatness of the $S_j$ with respect to the elliptic fibration is irrelevant for the geometric realization of 5d SCFTs.
The only crucial part of our proposal is that we keep track of the flavor curves inside the compact surface ${\cal S} = \bigcup_j S_j$.

For 5d SCFTs descending from 6d conformal matter theories, this is particularly simple, as the flavor curves are geometrically manifest.
For other types of 6d SCFTs, such as models on fibrations with a singular base, making the flavor symmetry geometrically manifest require in general additional, non-trivial tunings of the generic Weierstrass model for these SCFTs \cite{Bertolini:2015bwa,Merkx:2017jey}.
Moreover, certain 5d marginal theories may arise from outer-automorphism twists of 6d SCFTs, which geometrically corresponds to taking discrete quotients of the tensor branch geometries.
Making the flavor curves explicitly manifest in such examples will require a detailed analysis beyond the scope of this paper.
However, we will propose in section \ref{sec:rank_two_CFDs} a way to encode the SCFT-relevant data of rank two marginal theories that are quotient geometries.
The results which agree perfectly with previous works \cite{Jefferson:2017ahm,Jefferson:2018irk,Hayashi:2018lyv} will serve as additional evidence for the efficacy of our proposal.

Importantly, these results are based on intuitions drawn from non-flat resolutions.
Because of their importance, we will now briefly explain technical details of the fiber resolution procedure, following \cite{Lawrie:2012gg}.


\subsubsection{Resolution of Singular Fibers}
\label{sec:ResSing}

Consider an elliptic fibration with singular locus given by $x_1=x_2=x_3=0$.
We will denote the blow-up by the shorthand notation similar to 
\cite{Lawrie:2012gg} 
\be\label{eq:one_single_blowup}
\{x_1, x_2, x_3, u_i\}:\qquad   x_j= u_i x'_j\,,\quad  j=1,2,3 \,,
\ee
where $x_i$ are placeholders for coordinates $x,y,u, u_k$, $k<i$, appearing in a (partially resolved) Tate model. The coordinate $u_i$ corresponds to the exceptional divisor of the resolution. 
After the blow-up, the new coordinates $[x'_1, x'_2, x'_3]$ are treated as projective coordinates on a $\mathbb{P}^2$ and thus cannot vanish simultaneously. The  singular locus $x_1=x_2=x_3=0$ is hence replaced by the exceptional divisor $u_i=0$, which corresponds to the aforementioned $\mathbb{P}^2$.
The blow-up requires also performing a proper transform to keep the canonical class invariant, which amounts to dividing the Weierstrass equation by $u_i^2$. 

Likewise, small resolutions will be denoted by $\{x_1, x_2, u_i\}$, which maps $x_1 = x'_1 u_i$ and $x_2 = x'_2 u_i$, where again the last entry $u_i$ corresponds to the exceptional section of the blow-up, where now $[x'_1, x'_2]$ are coordinates on an exceptional $\mathbb{P}^1$. The proper transform of the Weierstrass equation in this case is division by $u_i$.
More generally, we also allow weighted blow-ups where the coordinates are
transformed as $x_j\rightarrow u_i^{a_j} x_j$, $j=1,2,3$. They will play a
role in the following in certain resolution sequences, for example the one in appendix \ref{app:D5d5}. When describing blow-ups, we will in the following always simply write $x_i$ instead of $x_i'$, following a common abuse of notation. 

For the situation at hand where we collide two codimension one singularities, the sequence of blow-ups contains the blow-ups of the codimension one fibers along $u=0$ and $v=0$, respectively. In the case of a non-minimal codimension two collision, we also need blow-ups of the type 
\be
\{u_i, v_j, \delta_k\} :\qquad   u_i \rightarrow \delta_k u_i \,,\ v_j \rightarrow \delta_k v_j \,,
\ee
where $u_i$ and $v_j$ are exceptional sections of the respective codimension one resolutions. 
Here, $\delta_k =0$ corresponds to the non-flat fiber component $S_k$. 

We will denote a complete resolution sequence, which smooths the Calabi--Yau
threefolds, generally by $BU$. After such a resolution the proper transform of $u$ and $v$ are 
\be\label{uvtotaltrans}
\ba
u&= U\left(\prod_i u_i^{m^{(u)}_i}\right)\cdot\left(\prod_{j=1}^r\delta_j^{\xi^{(u)}_i}\right)\cr
v&= V\left(\prod_i v_i^{m^{(v)}_i}\right)\cdot\left(\prod_{j=1}^r\delta_j^{\xi^{(v)}_i}\right),
\ea
\ee
where $U=0$ and $V=0$ correspond to the Cartan divisor  $D_{0}$ associated to the affine node of $\widehat{\mathfrak{g}}_{u}$ and $\widehat{\mathfrak{g}}_{v}$, respectively. 
While $m_i$ correspond to the well-known affine Dynkin labels of $\widehat{\mathfrak{g}}$, the $\xi_i$ coefficients are uniquely determined by the tensor branch information of the conformal matter theory. Suppose that the base coordinates $u$ and $v$ are described by 2d toric rays $(1,0)$ and $(0,1)$ on a local $\mb{C}^2$ coordinate patch, then the surface component $\delta_i$ will correspond to exactly the ray $(\xi^{(u)}_i,\xi^{(v)}_i)$ on the 2d base. 

Different blow-up sequences correspond to distinct resolutions of the Calabi--Yau threefold, and have different 5d physical characteristics as we will explain in the next section. 
One particularly central quantity to characterize the physics, that is encoded in the the geometry of the surfaces $S_k$, is the intersection matrix $S_k \cdot D_{i}^{(\nu)}\cdot D_{j}^{(\mu)}$. 
However, to determine the flavor curves, i.e., the fibral curves are contained in the surfaces $S_k$, it is enough to quote the reduced intersection matrix 
\be\label{nkinu}
n_{k, i, \nu}= S_k \cdot D_{i}^{(\nu)}\cdot D_{i}^{(\nu)}\,.
\ee
We can rewrite the total transform (\ref{uvtotaltrans}) in terms of the linear relation in integer homology involving the resolution divisors
\be
\label{eq:homology_relation_vert_div}
  \pi^{-1}(W_\nu) = \sum_i m_i^{(\nu)}\,D_{i}^{(\nu)} + \sum_i \xi_i^{(\nu)} \, S_i \,,
\ee
where $\pi$ is the projection to the base. This linear relation is independent of the specific resolution. 

Note that if we intersect (\ref{eq:homology_relation_vert_div}) for $\nu=u$ with $D_j^{(v)}\cdot D_j^{(v)}$, we obtain the relation
\be
\ba
   \sum_i {m_i^{(u)}}\,D_{i}^{(u)}\cdot D_j^{(v)}\cdot D_j^{(v)}+ \sum_i \xi_i^{(u)} \, S_i\cdot D_j^{(v)}\cdot D_j^{(v)}&=\pi^{-1}(W_{u})\cdot  D_j^{(v)}\cdot D_j^{(v)} \, .
\ea
\ee
Because $W_u$ intersects $W_v$ transversely in the base, we have $D_j^{(v)} \cdot \pi^{-1}(W_u) \cong F_j^{(v)}$, and the right-hand side can be simplified with \eqref{eq:cod1fib}.
If $F_j^{(v)}$ is a flavor curve, i.e., fully wrapped inside the surfaces $S_j$, then $D_j^{(v)} \cdot D_i^{(u)} = 0$ for any $i$, as otherwise a component of $F_j^{(v)}$ would sit inside $D_i^{(u)}$.
Hence, the criterion for $F_j^{(v)}$ to be a flavor curve is
\be\label{eq:wrap-condition}
\sum_i\xi_i^{(u)}n_{i,j,v} = D_j^{(v)} \cdot F_j^{(v)} = -\frac{2\langle\alpha,\alpha\rangle_{\rm max}}{\langle\alpha_j,\alpha_j\rangle}\quad \text{(no summation)}\, ,
\ee
which includes the case of non-simply laced algebras, see (\ref{eq:cod1fib-ns}).



\section{5d SCFTs from Conformal Matter Theories}
\label{sec:Resolutions}

We begin our analysis by studying a particular class of 5d SCFTs, which descend from 6d conformal matter theories. This will motivate the graphical approach using CFDs in the next section, which goes beyond the class of conformal matter descndants.
Our starting point is a 6d ${\cal N} = (1,0)$ SCFT given in terms of a singular elliptic fibration with a non-minimal singularity in the fiber over $u=v=0$, which is given in terms of the collision of two non-compact codimension-one loci $W_\nu \subset B$, $\nu = u,v$ with Kodaira fibers of affine type $\widehat{\mathfrak{g}}_\nu$.\footnote{The base in these cases is always smooth. We discuss 6d SCFTs realized with singular bases in section \ref{sec:rank_two_CFDs}.
Moreover, we will not consider elliptic fibrations with non-trivial Mordell--Weil groups. }
These 6d SCFTs, so-called conformal matter theories, have flavor symmetry $G_\text{F}^\text{(6d)} = G_u \times G_v$, which is manifest in the elliptic fibration.

As we explained in section \ref{sec:elliptic_threefolds_general}, these singularities admit a  crepant resolution in terms of a non-flat fibration. 
The non-flat fiber components are compact surfaces that are intersected by non-compact divisors $F^{(\nu)}_i \hookrightarrow D_i^{(\nu)}$ that resolve the codimension one singularities over $W_\nu$.
Any such geometry defines via M-theory an effective 5d theory on its Coulomb branch, which comes from a circle reduction with appropriate holonomies of the 6d SCFT.
By determining all inequivalent non-flat resolutions, we can thus map out the entire network of descendant 5d SCFTs.

Since the 5d theory arises from an $S^1$-reduction of the 6d theory, we expect for the classical and the superconformal flavor symmetries, $G_\text{F,cl}$ and $G_\text{F}$, respectively, the relation
\begin{equation}\label{eq:flavor_symmetry_hierarchy}
	G^\text{(6d)}_\text{F} \supset G_\text{F} \supset G_\text{F,cl}\,.
\end{equation}
Following the discussion in section \ref{sec:flavor_symmetry_from_geo_general}, the 5d flavor symmetries are then determined by the collapse of the flavor curves $F_i^{(\nu)}$.
This, and other physical properties largely follow from the discussion in the previous sections.
As certain aspects are more manifest in the setup we discuss, we give a short summary adapted to this description.

\subsection{M-theory on Non-flat Fibrations: A Dictionary}
\label{sec:Dictionary}

Denote by $S_j$, $j=1, \ldots, r$ the non-flat fibers above $u=v=0$ and let again
\be
\mathcal{S} = \bigcup_{j=1}^r S_j
\ee
be the reducible surface component.

\subsubsection*{Weakly Coupled Gauge Description}

The rank of the weakly coupled gauge group $G_\text{gauge}$ is given by the number of independent surface components in the non-flat fiber, i.e., rank$(G_\text{gauge})=r$. 
The precise gauge group is given by determining a ruling of the surfaces $f_j \hookrightarrow S_j \rightarrow \Sigma_j$, which allows a partial collapse $S_j \rightarrow \Sigma_j$, and the pairwise intersection pattern amongst the $S_j$, see section \ref{sec:geometry_to_gauge-theory}.

By matching the prepotential \eqref{eq:matching_prep} with the triple intersection numbers $S_i \cdot S_j \cdot S_m$, one can determine discrete data such as the Chern--Simons level $k$ for {$SU(K) \subset G_\text{gauge}$} or the number $M$ of mass deformations.
If $G_\text{gauge} = \prod_{n=1}^N G_{\text{G};n}$, where $G_{\text{G};n}$ are simple gauge factors, then
\begin{align}\label{eq:mass_deformation_definition}
	M = \text{rank}(G_\text{F,cl}) + N = \text{rank}(G_\text{F}) \, .
\end{align}

Field theoretically, the classical flavor symmetry $G_\text{F,cl}$ is entirely determined by the hypermultiplet spectrum.
It can be verified geometrically, following section \ref{sec:flavor_symmetry_from_geo_general}, from the collapsed codimension one fibers when the compact surfaces are blow-down to curves, $S_j \rightarrow \Sigma_j$ \cite{Apruzzi:2018nre}.
Since the weakly coupled phase will play an underpart in this work, we will refer to the companion paper \cite{Apruzzi:2019enx}, where the classical flavor symmetry $G_\text{F,cl}$ becomes part of the main cast.

\subsubsection*{Enhanced Flavor Symmetry from Flavor Curves}

The central new aspect of our approach is that the superconformal flavor symmetry $G_\text{F}$ is manifest from the geometry.
This information can be determined, without explicitly knowing the effective theory, just from the numbers $n_{j, i, \nu}$, $\nu=u, v$, defined in (\ref{nkinu}).
These numbers are an intersection-theoretic description of whether a codimension one fiber $F^{(\nu)}_i$ is contained inside ${\cal S}$, i.e., is a flavor curve or not.
The numbers $n_{j,i,\nu}$ compute the degree of the normal bundle of the curve $S_j \cap D_i^{(\nu)}$ inside $S_j$. 

For a Cartan node $F_i^{(\nu)}$ of $\mathfrak{g}_\nu$ inside $\mathcal{S}$ define (recall that $F_i$ is the fibral curve in the Cartan divisor $D_i$)
\be\label{nFDef}
\ba
n(F_i^{(u)}) & =\sum_{j=1} \xi^{(v)}_j \, n_{j,i, u} = \sum_{j =1 } \xi^{(v)}_j \, S_j \cdot (D_i^{(u)})^2 \, ,\cr 
n(F_i^{(v)}) & =\sum_{j=1} \xi^{(u)}_j \, n_{j,i,v} = \sum_{j=1} \xi^{(u)}_j \, S_j \cdot (D_i^{(v)})^2 \, .
\ea
\ee
where $\xi_j^{(\nu)}$ are defined in (\ref{uvtotaltrans}). 

For a simply-laced Lie algebra $\mathfrak{g}_\nu$, a curve $F_i^{(\nu)}$ by \eqref{eq:wrap-condition} is a flavor curve if $n(F_i^{(\nu)})=-2$ --- and the associated root contributes to the flavor symmetry. 
 There are two instances to consider:
\begin{itemize}
	\item If $n_{j, i, \nu}=-2$, then the fibral curve $F^{(\nu)}_i$ is irreducible in codimension two and is contained in $S_j$. 

	\item If for $j \not= \ell$, $n_{j, i, \nu}=n_{\ell, i, \nu}= -1$, then the fibral curve $F_i$ is reducible, but is fully contained in $\mathcal{S}$. 
	Its irreducible components of self-intersection $-1$ are contained in $S_j$ or $S_{\ell}$, respectively, and the root associated to $F^{(\nu)}_i$ is part of $G_\text{F}$. 

\end{itemize}

If on the other hand there is only one $j$ with $n_{j, i, \nu}=-1$, but $n_{j, k, \nu}= 0$ for all $k \not= i$, then the curve $F^{(\nu)}_i$ is reducible, and only one of the split components (denoted by $C^\pm$ in (\ref{FCC})) is contained in $\mathcal{S}$ with self-intersection $(-1)$.
In this case the curve $F_i^{(\nu)}$ does not contribute to the non-abelian part of the strongly coupled flavor symmetry.

For the non-simply-laced case, the $F_i^{(\nu)}$ is a flavor curve if and only if (see (\ref{eq:wrap-condition}))
\be
n(F_i^{(\nu)})=-\frac{2\langle\alpha,\alpha\rangle_{\rm max}}{\langle\alpha_i,\alpha_i\rangle} \,.
\ee

The Dynkin diagram of the non-abelian flavor symmetry $G_\text{F,na}\subset G_\text{F}$ is read off from the graph that is formed by the flavor curves and is a subgraph of the affine Dynkin diagrams of $\hat{\mathfrak{g}}_u$ and $\hat{\mathfrak{g}}_v$.
The abelian part $U(1)^s \subset G_\text{F}$ can be in most cases computed from knowing the mass deformations $M = \text{rank}(G_\text{F})$:
\begin{align}
\label{U(1)counting}
	s = \mathrm{rank}(G_\text{F})-\mathrm{rank}({G}_\text{F,na}) = M - \mathrm{rank}({G}_\text{F,na}).
\end{align}
Geometrically, these $U(1)$s correspond to $s$ independent linear combinations $D_a$ involving non-compact divisors, that intersect $\cal S$ non-trivially and are orthogonal to $G_\text{F,na}$.
In practice, we can compute $M$ as the rank of the intersection matrix between all independent divisors and curves of the form $D_i^{(\nu)}\cdot S_j$, after subtracting the rank of 5d gauge group $r=\mathrm{rank}(G_\text{gauge})$; see appendix \ref{app:BlowupsRank2E} for explicit examples.
Note that since the intersections are all occurring locally inside $\cal S$, we can also identify a $U(1)$ generator $D_a$ with the curves $D_a \cap {\cal S}$, and determine the $U(1)$ charges as intersection numbers between curves in the surface $\cal S$.
Such a representation of a flavor $U(1)$ in terms of curves is in general not unique, due to linear relations amongst non-compact divisors.

Note that while we have assumed that the geometry realizes the \emph{full} 6d flavor symmetry as $G^\text{(6d)}_\text{F} = G_u \times G_v$, it can be $G_u \times G_v$ is only a subgroup of $G^\text{(6d)}_\text{F}$.
In the latter case, the geometrically determined putative superconformal flavor group $G_\text{F}$ can also be a subgroup of the actual, larger 5d flavor group.
In this case, the full flavor symmetry is obtained by computing the BPS states of a certain spin (see section~\ref{sec:BPS}), which will recombine into representations of a large flavor symmetry group $G_\text{F}$. If this combination is indeed possible, the 5d superconformal flavor group should enhance to $\widetilde{G}_\text{F}$. 
Such examples are discussed in appendix~\ref{app:Rank1E6SU3} and ~\ref{app:D5d5}. 
E.g., the rank one 5d SCFTs are described either by $(E_8,I_1)$ or $(E_6,SU(3))$ collisions. 
In the latter case, we only see the full flavor symmetry by considering the BPS states, whereas in the former it is manifest in geometry.
This also reflects the situation in 6d where the flavor symmetry is $G_\text{F}^\text{(6d)} = E_8 \supset E_6 \times SU(3)$.
For the remaining part of this paper, we will assume that the singular elliptic fibration we start with already has the 6d flavor symmetry manifest.

\subsubsection*{Relevant Deformations and SCFT-Trees}

By shrinking a $(-1)$-curve that is contained in only one surface $S_j \subset {\cal S}$, the triple intersection number $S_j^3$ increases by one.
Field theoretically, this corresponds to a relevant deformation of the SCFT (or mass deformation of the gauge theory) that triggers an RG-flow.
Each contraction (or flop) of this type lowers $M = \text{rank}(G_\text{F})$ by one, thus the resulting SCFT is a different one. What appears to be a curve contraction on $\cal S$ corresponds in the full \emph{smooth} Calabi--Yau threefold to a birational map, where the collapsed $(-1)$-curve is flopped into a \emph{non-compact} divisor.
In this way we will determine all descendant SCFTs from a given marginal theory by flop transitions.

 \subsubsection*{UV-Dualities}

There are also transitions in which a $(-1)$-curve $C$ inside $S_j$ is flopped into an adjacent $S_k$ as $C'$.
In this case the limiting SCFT does not change, as this does not change the
overall structure of flavor curves (whether a flavor curve is irreducible or
reducible in codimension two is immaterial as long as it is fully contained in
the reducible surface $\mathcal{S}$).
Such transformations neither change $G_{\text{F}}$ nor $M$.  In a field theory context, these flops between surface components inside $\mathcal{S}$ correspond to different weakly coupled gauge theory phases. 
For example, it can in rank two examples happen that a geometry with both a weakly coupled $SU(3)_\text{G}$ and $SU(2)_\text{G} \times SU(2)_\text{G}$ description only allows for an $SU(3)_\text{G}$ interpretation after such a flop (see~\cite{Apruzzi:2019enx}  for more details).
Nevertheless, these gauge theories are dual to each other in the sense of flowing to the same UV theory.


\subsection{Rank one Classification from Non-Flat Resolutions}
\label{sec:EstringGeo}

To see our proposal at work we start with the rank one 5d SCFTs, which are known to arise from circle reductions with flavor holonomies of the 6d rank one E-string theory. The weakly-coupled description is given by an $SU(2)$ gauge theory with $N_F$ fundamental hypermultiplets, which has classical flavor symmetry $SO(2N_F)$ that at the UV fixed point enhances to $E_{N_F+1}$. 
Geometrically, these SCFTs correspond to different M-theory compactifications on non-compact Calabi--Yau threefolds with canonical singularities, stemming from the collapse of a del Pezzo surface dP$_n$ or the Hirzebruch surface $\mathbb{F}_0 \cong \bbP^1 \times \bbP^1$ \cite{Seiberg:1996bd, Morrison:1996xf}.

It has been known that there are many other equivalent ways to engineer the same SCFTs, e.g., M-theory on a Calabi--Yau threefold with an $\mathbb{F}_0$ is considered to be equivalent to one with $\mathbb{F}_2$. The key difference in our description is that we will use such equivalences to make the superconformal flavor symmetry of the UV fixed points manifest within the geometry.

The 6d E-string theory is obtained from an $(E_8 ,I_1)$ (i.e., $(II^*,I_1)$) non-minimal collision of Kodaira singularities, where the $E_8$ associated with the $II^*$ fibers encodes the 6d flavor symmetry group. 
The 5d rank one theories descend via flops from the so-called marginal geometry, from which all other theories are obtained by flop transitions. This corresponds to a particular resolution of this non-minimal singularity in terms of a non-flat fibration, where the surface component $S$ contains all fibral curves of the affine $E_8$ of the $II^*$ fiber.

\begin{figure}
\centering
\includegraphics[width=6cm]{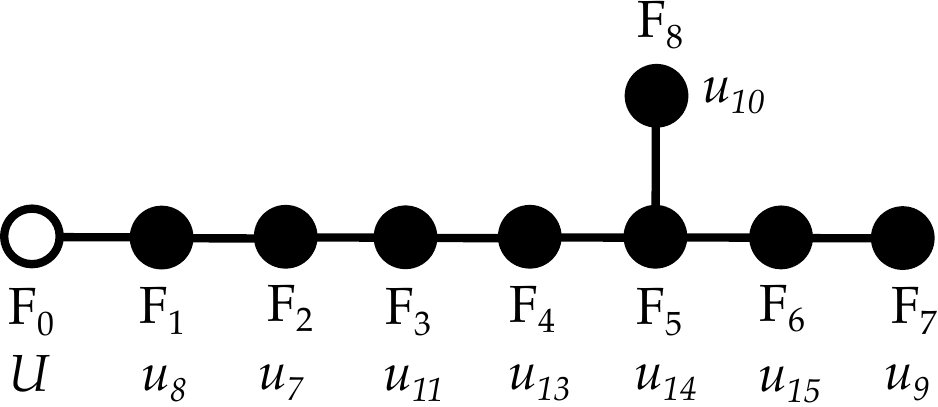}
\caption{Affine $E_8$ Dynkin diagram with labels corresponding to the exceptional sections $u_i$ and associated to the Cartan divisors $D_{k}$ that correspond to the simple roots $\alpha_k$. The labels $u_i$ refer to the exceptional sections of the blow-up. \label{fig:E8Dynkin}}
\end{figure}

The marginal theory has a surface omponent that is  a generalized del Pezzo surface gdP$_9$. 
While a dP$_n$ contains a number of $(-1)$-curves as Mori cone generators, a gdP$_n$ has rational  generators with self-intersection $(-2)$ as well.
Contracting the $(-1)$-curve in a gdP$_n$ maps it to a gdP$_{n-1}$ surface. A discussion of the geometry of these surfaces can be found in appendix \ref{app:gdP}.
It is from the $(-2)$-curves that are contained within the surfaces, that we read off the superconformal flavor symmetry:
in the case of gdP$_n$ surfaces, an $E_n$ Dynkin diagram worth of rational $(-2)$-curves shrink in the UV limit, and furnish the flavor symmetry.

\subsubsection*{Non-flat resolutions of \boldmath{$(E_8,I_1)$}}
\label{sec:E8-I1}

The collision of a $II^*$ Kodaira fiber transversally with an $I_1$ fiber has a simple description in terms of a Tate model for the elliptic fibration (\ref{TateMod}) with vanishing orders
\be
\hbox{ord}_u(b_i) = (1,2,3,4,5)\,,\qquad  \hbox{ord}_v(b_i)= (0,0,0,0,1) \,,
\ee
which takes the form 
\be\label{rank-1Tate}
y^2 + b_1 u x y + b_3 u^3 y = x^3 + b_2 u^2 x^2 + b_4 u^4 x  + b_6 u^5 v\,.
\ee
At the codimension two locus in the base $u=v=0$ the vanishing orders of the $b_i$ are $(1,2,3,4,6)$, which is equivalent to the non-minimality condition in the Weierstrass model 
\be\label{WeierNonMin}
	\hbox{ord}_{u=v=0}(f,g,\Delta)=(4,6,12) \,.
\ee

We now turn to deriving the resolutions of the singular model (\ref{rank-1Tate}). In any resolution, the Cartan divisors $D_{i}$ associated to the $E_8$ affine roots are given in terms of the sections 
\be\label{CartansRank1}
(U, u_{8}, u_7, u_{11}, u_{13}, u_{14},u_{15}, u_{9}, u_{10}) \equiv (D_{0}, \cdots,  D_{8}) \,,
\ee
with the ordering shown in figure \ref{fig:E8Dynkin}. These intersect with the fibral curves $F_i$ in  the negative affine $E_8$ Cartan matrix,
\be
F_i \cdot D_j =  - \mathcal{C}^{\widehat{E_8}}=\left(
\begin{array}{ccccccccc}
 -2 & 1 & 0 & 0 & 0 & 0 & 0 & 0 & 0 \\
 1 & -2 & 1 & 0 & 0 & 0 & 0 & 0 & 0 \\
 0 & 1 & -2 & 1 & 0 & 0 & 0 & 0 & 0 \\
 0 & 0 & 1 & -2 & 1 & 0 & 0 & 0 & 0 \\
 0 & 0 & 0 & 1 & -2 & 1 & 0 & 0 & 0 \\
 0 & 0 & 0 & 0 & 1 & -2 & 1 & 0 & 1 \\
 0 & 0 & 0 & 0 & 0 & 1 & -2 & 1 & 0 \\
 0 & 0 & 0 & 0 & 0 & 0 & 1 & -2 & 0 \\
 0 & 0 & 0 & 0 & 0 & 1 & 0 & 0 & -2 \\
\end{array}
\right)\,.
\ee
This is depicted in figure \ref{fig:E8Dynkin}.

\paragraph{gdP\boldmath{$_9$}:}  
In the first model, the entire affine $E_8$ worth of rational curves is contained within the compact surface $S$. 
This requires blowing up first the non-minimal locus $u=v=0$ in the base 
\be\label{BaseBU}
u\rightarrow U\delta\ ,\quad  v\rightarrow V\delta \,.
\ee
We will argue in section \ref{sec:CFD} that after this blow-up, the compact surfaces contain all fibral curves of the codimension one fibers. 

After this blow-up, the locus $u=v=0$ is removed and replace by the curve $\{\delta=0\} \subset B_2$. The Tate model still has the same form
\be
y^2+b_1 Uxy+b_3 U^3  y=x^3+b_2 U^2 x^2+b_4 U^4  x+b_6 U^5 V\, ,
\ee
but no longer has any non-minimal locus.

\begin{figure}
\centering
\includegraphics[height=3.5cm]{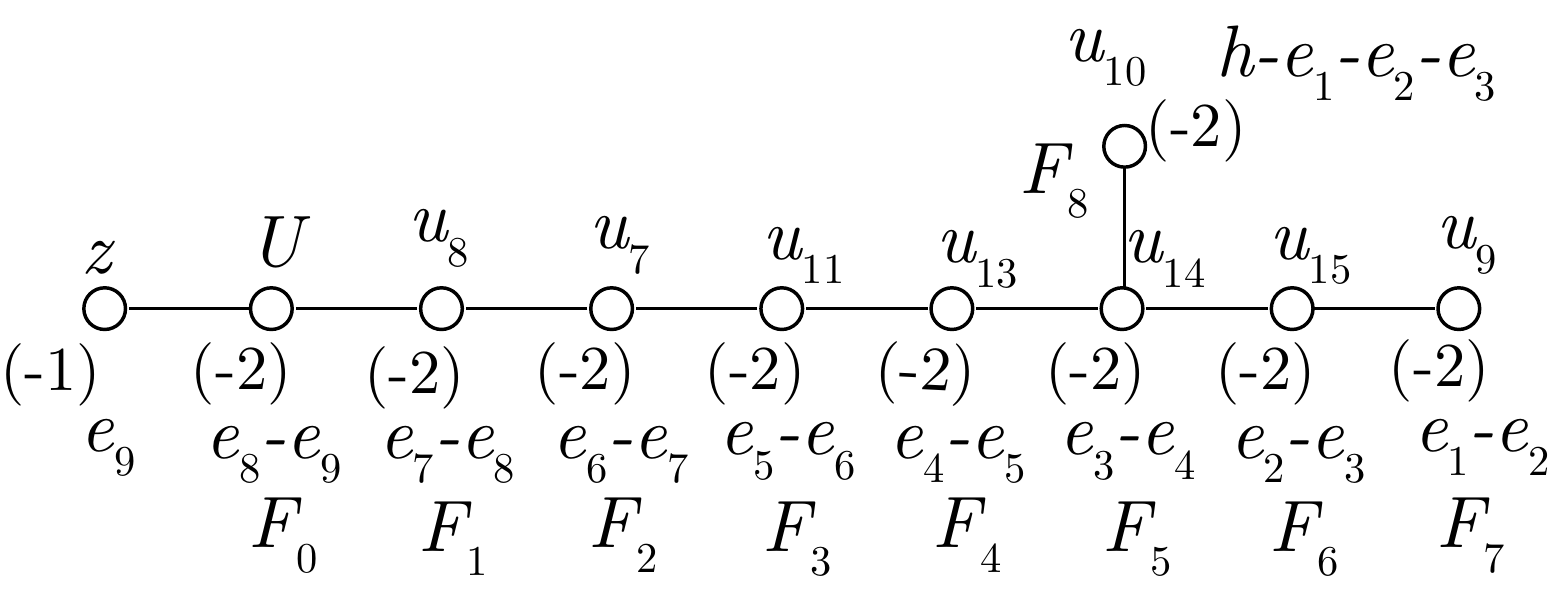}
\caption{
The negative curves on the vertical divisor gdP$_9$, where the numbers in the brackets  denote the self-intersection numbers 
and the letter $u_i$ or $z$ indicates the intersection of that divisor in the resolved Calabi--Yau with $S=\hbox{gdP}_9$. 
We also denote the expression of each curve in terms of the standard basis of the Picard group $h$ and $e_i$ of a rational surface.}\label{fig:gdP9}
\end{figure}

The model is still singular. The following blow-up sequence generates all the $E_8$ Cartan divisors in (\ref{CartansRank1})
 \be\label{gdP9BU}
\text{gdP}_9:\quad \ba
&\{\left\{x,y,U,u_1\right\},\left\{x,y,u_1,u_2\right\},\left\{y,u_2,u_3\right\},\left\{y,u_1,u_3,u_4\right\} , \cr 
&   \left\{y,u_1,u_5\right\}, \left\{u_1,u_3,u_6\right\},
   \left\{u_1,u_4,u_7\right\},\left\{u_1,u_5,u_8\right\},\left\{u_2,u_3,u_9\right\},  \left\{u_3,u_4,u_{10}\right\},
   \cr 
& \left\{u_4,u_6,u_{11}
   \right\},
   \left\{u_3,u_6,u_{12}\right\},\left\{u_6,u_{10},u_{13}\right\},\left\{u_{10},u_{12},u_{14}\right\},\left\{u_3,u_{12},u_{15}\right\} \} \,.
\ea
\ee

The configuration of curves on the gdP$_9$ is shown in figure~\ref{fig:gdP9}, and we can read off the following intersection numbers:
\be
S \cdot D_i \cdot D_i = (-2,-2,-2,-2,-2,-2,-2,-2,-2)\ ,\ (i=0,\dots,8).
\ee
Furthermore we can compute
\be
S^3=0\quad \Rightarrow \quad 
h^{1,1}(S)=10-S^3=10\,,
\ee
consistent with the fact that $S$ is a nine-fold blow-up of a $\bbP^2$. 

We can now either apply different blow-up sequences and obtain other surface components $S$, which contain a different subset of the $E_8$ fibral curves --- this is detailed in appendix \ref{app:Rank1Res} --- or we apply consecutively flop transitions to the $(-1)$-curves. 
The resulting tree of geometries connected by blow-ups or flops is summarized in table \ref{tbl:rank1Res}. 
This lists both the geometry of the surface $S$, which are generalized del Pezzo surfaces gdP$_n$ or Hirzebruch surfaces $\mathbb{F}_n$.
Furthermore we determine the intersection numbers $n_i\equiv S\cdot D_{i} \cdot D_{i}$ i.e., (\ref{nkinu}), which determine the flavor symmetry. 
In the present context, whenever a fibral curve $F_i$ with $F_i^2= -2$ is contained in the surface $S$, it contributes to the strongly coupled flavor symmetry. 
The curves for which $n_i=-1$ are reducible in codimension two and split. The associated root is not part of the flavor symmetry $G_\text{F}$.

The key here is that the geometry manifestly encodes the strongly coupled flavor symmetry, as well as the complete flop chain descending from the marginal theory, which in the SCFT language corresponds to mass deformations and subsequent RG-flows to another UV fixed point. Needless to say, this is in complete agreement with the known rank one theories and their strongly coupled flavor symmetries \cite{Seiberg:1996bd, Morrison:1996xf}. 

\begin{table}\centering
\begin{tabular}{|c|c|c|c|}\hline
Geometry of $S$ & Intersections $S\cdot D_{i}\cdot D_{i}$& Codim 2 Fiber  & SCFT Flavor $G_\text{F}$\cr \hline 
gdP$_9$ (Marginal)&  (-2,-2,-2,-2,-2,-2,-2,-2,-2)&\includegraphics[width=4cm]{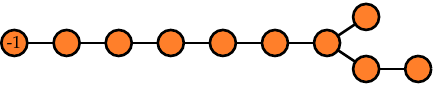} & \cr \hline 
gdP$_8$&  (-1,-2,-2,-2,-2,-2,-2,-2,-2)&\includegraphics[width=4cm]{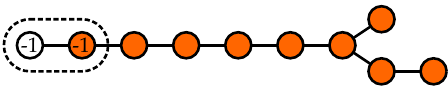} & $E_8$ \cr \hline 
gdP$_7$&  (0,-1,-2,-2,-2,-2,-2,-2,-2)&\includegraphics[width=4cm]{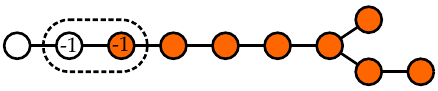} & $E_7$ \cr \hline 
gdP$_6$&  (0,0,-1,-2,-2,-2,-2,-2,-2)&\includegraphics[width=4cm]{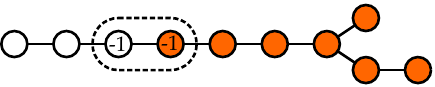} & $E_6$ \cr \hline 
gdP$_5$&  (0,0,0,-1,-2,-2,-2,-2,-2)&\includegraphics[width=4cm]{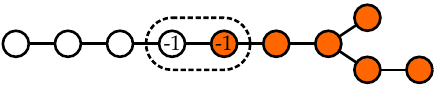} & $SO(10)$ \cr \hline 
gdP$_4$&  (0,0,0,0,-1,-2,-2,-2,-2)&\includegraphics[width=4cm]{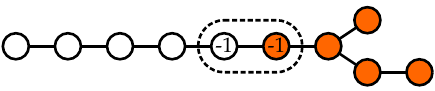} & $ SU(5)$ \cr \hline 
gdP$_3$&  (0,0,0,0,0,-1,-2,-2,-2)&\includegraphics[width=4cm]{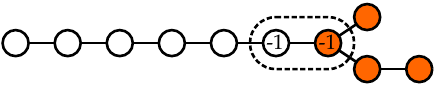} & $SU(2)\times SU(3)$ \cr \hline 
gdP$_2$&  (0,0,0,0,0,0,-1,-2,-1)&\includegraphics[width=4cm]{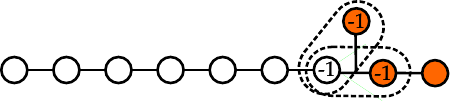} & $SU(2)\times U(1)$ \cr \hline 
$\mathbb{F}_1$&  (0,0,0,0,0,0,0,-1,0)&\includegraphics[width=4cm]{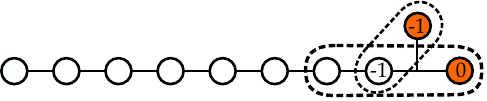} & $ U(1)$ \cr \hline 
gdP$_1 \cong \mathbb{F}_2$&  (0,0,0,0,0,0,0,-2,0)&\includegraphics[width=4cm]{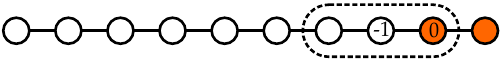} & $ SU(2)$ \cr \hline 
$\mathbb{P}^2$ & (0,0,0,0,0,0,0,0,1) & \includegraphics[width=4cm]{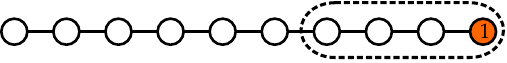} & -  \cr\hline
\end{tabular}
\caption{Geometries for rank one 5d SCFT:  the table contains the geometry of the non-flat surface component $S$ in the elliptic fibration, the intersection of $S$ with the Cartans $D_i$, $i=0, \cdots, 8$ of the $E_8$ fiber; the codimension two fiber, where the dotted lines show the splitting of the codimension one rational curves of the $II^*$ fiber that become reducible. Futhermore, the marked nodes are contained in the non-flat fiber $S$. Full colored nodes that remain irreducible in codimension two correspond to $(-2)$ curves in the $II^*$ fiber and contribute to the flavor symmetry of the SCFT, $G_\text{F}$. The last column gives the flavor symmetry, which can be read off from the fiber. }\label{tbl:rank1Res}
\end{table}

We conclude the discussion of rank one geometries by noting that there are alternative starting points, or marginal theories. 
E.g., the rank one collision $(E_6, SU(3))$ also gives rise to the 6d E-string.
In appendix \ref{app:Rank1E6SU3} we consider resolutions of this collision.
The maximal flavor symmetry present at the superconformal point is also correctly obtained from these models. 
However, it may not be manifest in these cases as the full enhancement, and only becomes apparent by combining the BPS states into representations of a higher rank group.
This is in particular the case for the models with $E_8$ and $E_7$ flavor symmetry, which in this alternative description would not be manifest, and only is seen by computing the BPS states and their representations, which combine into multiplets of the higher rank flavor symmetries. In this sense, the marginal theory we consider in this section, i.e., $(E_8,I_1)$, captures all of these symmetries complete manifestly within the geometric resolution, and is thus preferred.


\subsection{Rank two:  Non-Flat Resolutions}
\label{sec:NonFlatRes}

\begin{figure}
\centering
\includegraphics*[width= 7.5cm]{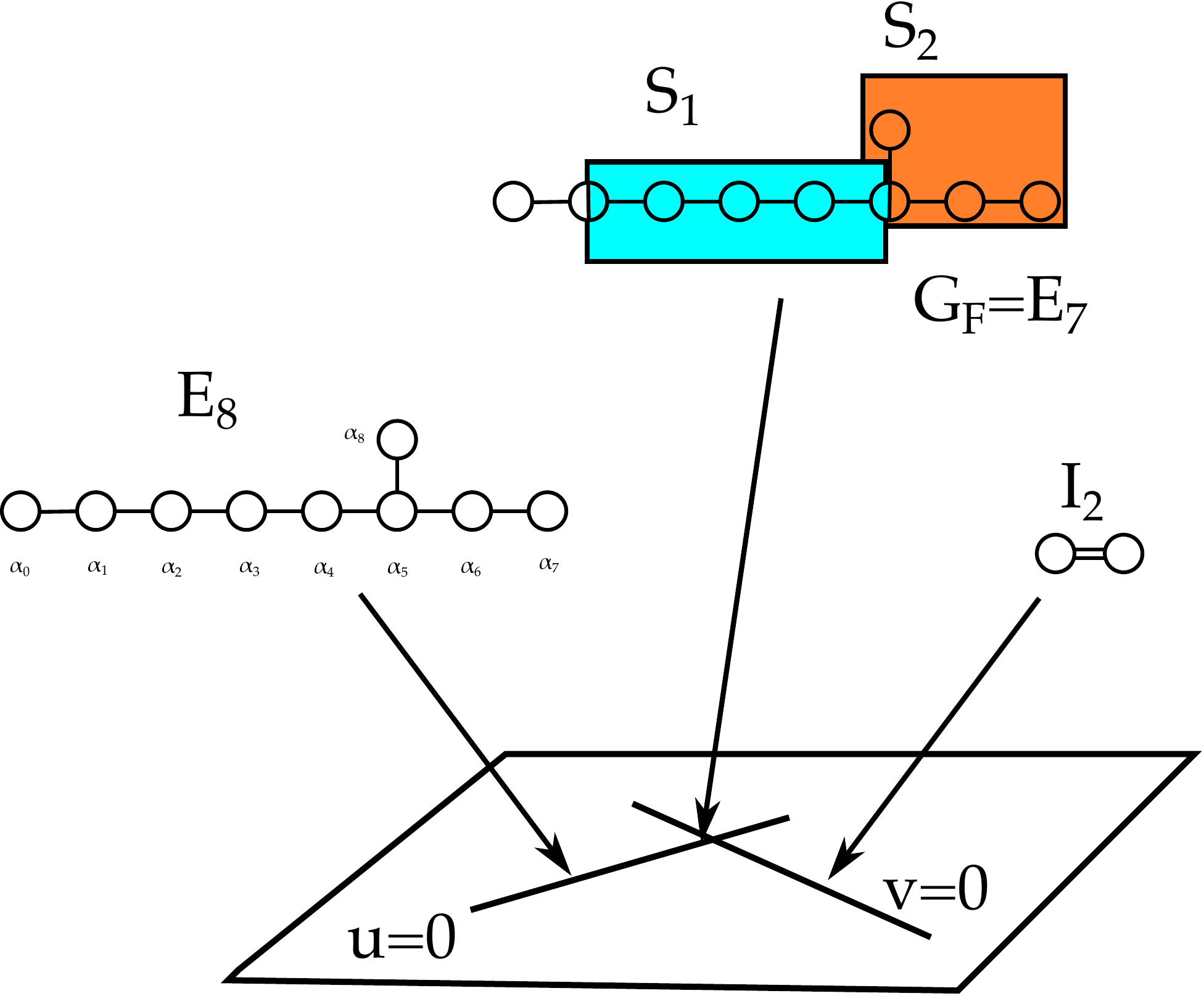}
\caption{Cartoon of a codimension two fiber for the rank two  E-string. $S_i$ are the two surface components of the non-flat fiber in codimension two. The rational curves (circles) that are fully contained in the surfaces contribute to the non-abelian flavor symmetry, which in the shown example is $G_\text{F}= E_7$. 
\label{fig:ExampleSketchRankTwo}}
\end{figure}

The next application of our method is to the geometries that result in rank two theories. The non-flat fiber resolutions have two surface components, which in M-theory give rise to the Cartans of the gauge group (if there is weakly coupled 
gauge theory description). An example is shown in figure \ref{fig:ExampleSketchRankTwo}. 
As for rank one, we extract the flavor symmetries of the strongly coupled SCFT from the rational curves that are contained in the surface components of the fiber. 

We will provide a systematic exploration of all rank two theories in section \ref{sec:CFD}. Here, we will focus on two aspects: determine the geometries associated to the marginal theories and to point out some new features that occur in higher rank, e.g., the existence of resolutions that result in the same superconformal theories, i.e., the geometric avatar of different gauge theory phases for the same SCFT.

This will be exemplified in concrete resolution geometries, but we will pass to a more streamlined description in 
 section~\ref{sec:CFD}, where equivalence classes of resolutions will be characterized  in terms of CFDs (combined fiber diagrams). 
The equivalence classes will in particular contain resolution geometries, which correspond to the same SCFTs and are simply different resolutions associated to gauge theory phases for the same UV fixed point.  
The CFDs, as we emphasized before, encode the essential information of the non-flat resolutions, such as the rational curves associated to the flavor symmetries hat are contained in the non-flat fiber components, and allow for a systematic determination of all 5d descendants of a 6d SCFT.

The two main marginal theories (whose descendants map out most of the rank two SCFTs) are the rank two E-string and the $(D_5,D_5)$ conformal matter theory. The latter is equivalently represented in terms of a collision between an $D_{10}$ with a transverse $I_1$, which is what we will consider in the following. They are given in terms of 
the Tate models (\ref{TateMod}) with vanishing orders 
\be\label{Rank2TateVan}
\ba
(D_{10} , I_1):\qquad & \hbox{ord}_{U=0} (b_i)= (1, 1, 5, 5, 10)\,, \qquad 
\hbox{ord}_{V=0} (b_i)=  (0, 1, 0, 0, 0) \cr 
(E_8 , I_2):\qquad & \hbox{ord}_{U=0} (b_i)= (1, 2, 3, 4, 5)\,, \ \,  \qquad 
\hbox{ord}_{V=0} (b_i)=  (0, 0, 1, 1, 2) \,. 
\ea
\ee


\subsubsection[Non-flat Resolutions of \texorpdfstring{$(E_8,SU(2))$}{(E8,SU(2))}]{Non-flat Resolutions of \boldmath{$(E_8,SU(2))$}}\label{subsec:non-flat_res_E8SU2}

Consider the rank two  E-string geometry that is the codimension two non-minimal collision of $E_8$ along $u=0$, with an $I_2$ ($SU(2)$) singular fiber  along $v=0$. Our focus will be on determining the blow-up for the marginal theory, as well as providing some examples. 
The explicit resolutions can be found in the appendix \ref{app:BlowupsRank2E}.

The geometry for the marginal theory is obtained as a non-flat resolution in appendix \ref{app:TopCFDsRank2E}, where the full derivation is explained. Here we only summarize it in terms of the reduced triple intersection matrix
\be\label{E8SU2MarginalRes}
\ba
\begin{array}{c|ccccccccc|cc|cc}
 S_i\cdot D_j^2  & D_0 & D_1 & D_2& D_3& D_4& D_5& D_6& D_7& D_8& D_0^{SU(2)} & D_1^{SU(2)} & S_1 & S_2 \\ \hline
 S_1 & -2 & -1 & 0 & 0 & 0 & 0 & 0 & 0 & 0 & -2 & 0 & 6 & 0 \\
 S_2 & 0 & -1 & -2 & -2 & -2 & -2 & -2 & -2 & -2 & 0 & -1 & -2 & 1 \\
\hline
n(F_j)&  -2 & -2 & -2 & -2 & -2 & -2 & -2 & -2 & -2 & -2 & -2 & - & - \\
\end{array}
\ea
\ee

Here $n(F_j)$ is defined in (\ref{nFDef}).
The Cartan divisors of the affined $E_8$ and $SU(2)$ are intersected with the two surface components $S_1$ and $S_2$, as $n(F_j)$. Note that in the case of $(E_8,SU(2))$, there are non-trivial multiplicities (see (\ref{uvtotaltrans}))
\be
\xi^{E_8}_1=1\,,\quad  \xi^{E_8}_2=2\,,\quad  \xi^{SU(2)}_1=\xi^{SU(2)}_2=1\,,
\ee
hence the entries $S_2\cdot (D_i^{SU(2)})^2$ need to be multiplied by $\xi_2=2$ in the bottom line of (\ref{E8SU2MarginalRes}), in order to read off the correct flavor curves.

 We see that the curves at the intersection of the non-compact Cartan divisors and surface components are all $(-2)$ curves, so that all fibral curves associated to the affine roots of $E_8\times SU(2)$ are contained within the reducible surface $\mathcal{S}$ --- the hallmark of the marginal geometry.

We can now determine all descendant geometries/SCFTs by flops, which will be the subject of  section \ref{sec:CFD}. 
Here we should consider a few more examples of non-flat resolutions to point out some new effects that occur in higher rank. 
Again the details of the resolutions are explained in appendix \ref{app:ExamplesE8SU2}. The key characteristic is the reduced triple intersection number. We consider three example blow-ups, for which this is given by
\be
\label{Triples}
\ba
&BU_1^{(E_8, SU(2))} : \cr 
&
\begin{array}{r|ccccccccc|cc|cc}
S_i \cdot D_j^2&D_0 & D_1 & D_2& D_3& D_4& D_5& D_6& D_7 &D_8& D^{SU(2)}_0& D_1^{SU(2)} & S_1 & S_2\cr \hline
S_1  & 0 & -1 & -2 & -2 & -2 & -1 & 0 & 0 & 0 & 0 & 0 & 4 & -4 \cr 
S_2  &0 & 0 & 0 & 0 & 0 & -1 & -1 & 0 & -2 & 0 & 1 & 2 & 7 \cr \hline
n(F_j)  & 0 & -1 & -2 & -2 & -2 & -2 & -1 & 0 & -2 & 0 & 2 &  -& -\cr
\end{array}
\cr 
&BU^{(E_8, SU(2))}_2: \cr 
& 
\begin{array}{c|ccccccccc|cc|cc}
S_i\cdot D_j^2&D_0 & D_1 & D_2& D_3& D_4& D_5& D_6& D_7 &D_8& D^{SU(2)}_0& D_1^{SU(2)} & S_1 & S_2\cr \hline
S_1 &  0 & 0 & 0 & 0 & 0 & 0 & 0 & 0 & 0 & 0 & 0 & 8 & 0\cr 
S_2   & 0 & -1 & -2 & -2 & -2 & -2 & -1 & 0 & -2 & 0 & 1 & -2 & 3  \cr \hline
n(F_j)  & 0 & -1 & -2 & -2 & -2 & -2 & -1 & 0 & -2 & 0 & 2 &  -& -\cr
\end{array}
\,.
\cr 
&BU^{(E_8, SU(2))}_3: \cr 
&
\begin{array}{c|ccccccccc|cc|cc}
S_i \cdot D_j^2 &D_0 & D_1 & D_2& D_3& D_4& D_5& D_6& D_7&D_8& D^{SU(2)}_0& D_1^{SU(2)} & S_1 & S_2\cr \hline
 S_1 &0 & -1 & -2 & -2 & -2 & -1 & 0 & 0 & 0 & 0 & 0 & 4 & -4 \cr 
S_2 & 0 & 0 & 0 & 0 & 0 & -1 & -2 & -1 & -2 & 0 & 0 & 2 & 6 \cr \hline
n(F_j) &0 & -1 & -2 & -2 & -2 & -2 & -2 & -1 & -2 & 0 & 0 & - & -
\end{array}
\ea
\ee

\begin{figure}
\centering
  \subfloat[BU$_1^{(E_8, SU(2))}$: Wrapping of codimension one Fiber by surfaces $S_i$.]{\includegraphics*[width= 6cm]{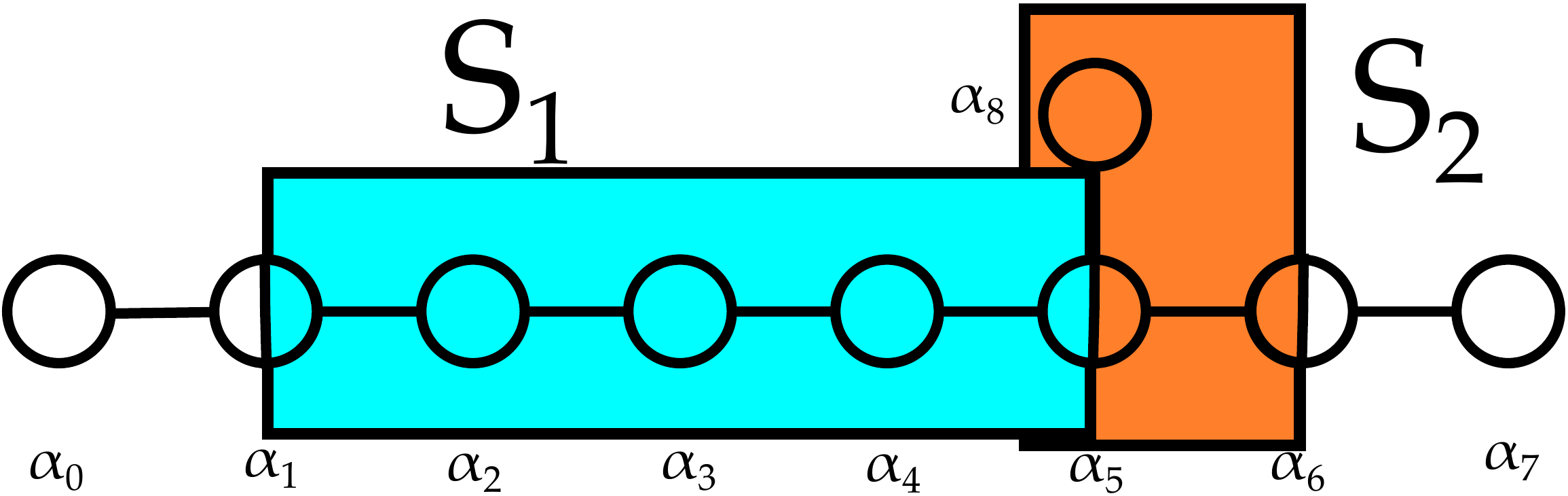}}
  \qquad  \qquad
  \subfloat[BU$_1^{(E_8, SU(2))}$: Codimension two Fiber.]{{\includegraphics[width=7cm]{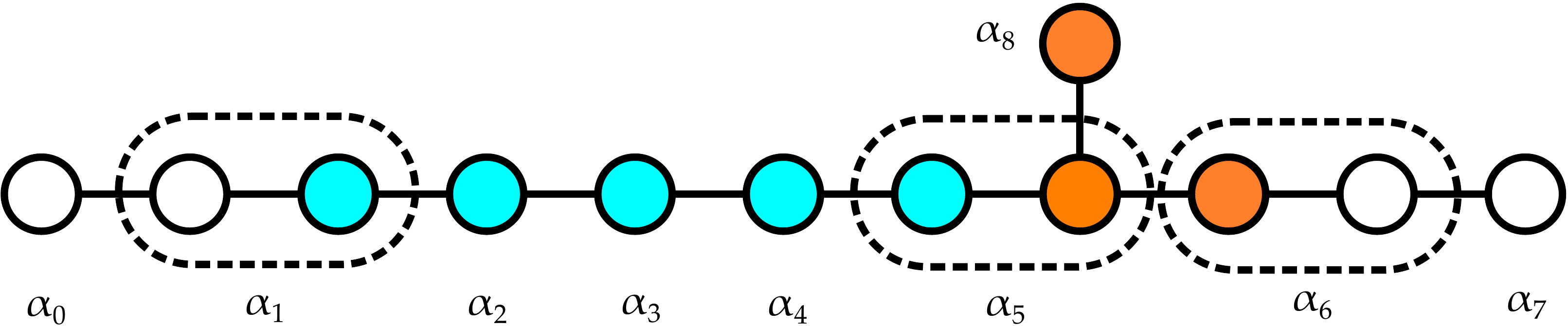}}}\\

 \subfloat[BU$_2^{(E_8, SU(2))}$: Wrapping of codimension one Fiber by surfaces $S_i$.]{\includegraphics*[width= 6cm]{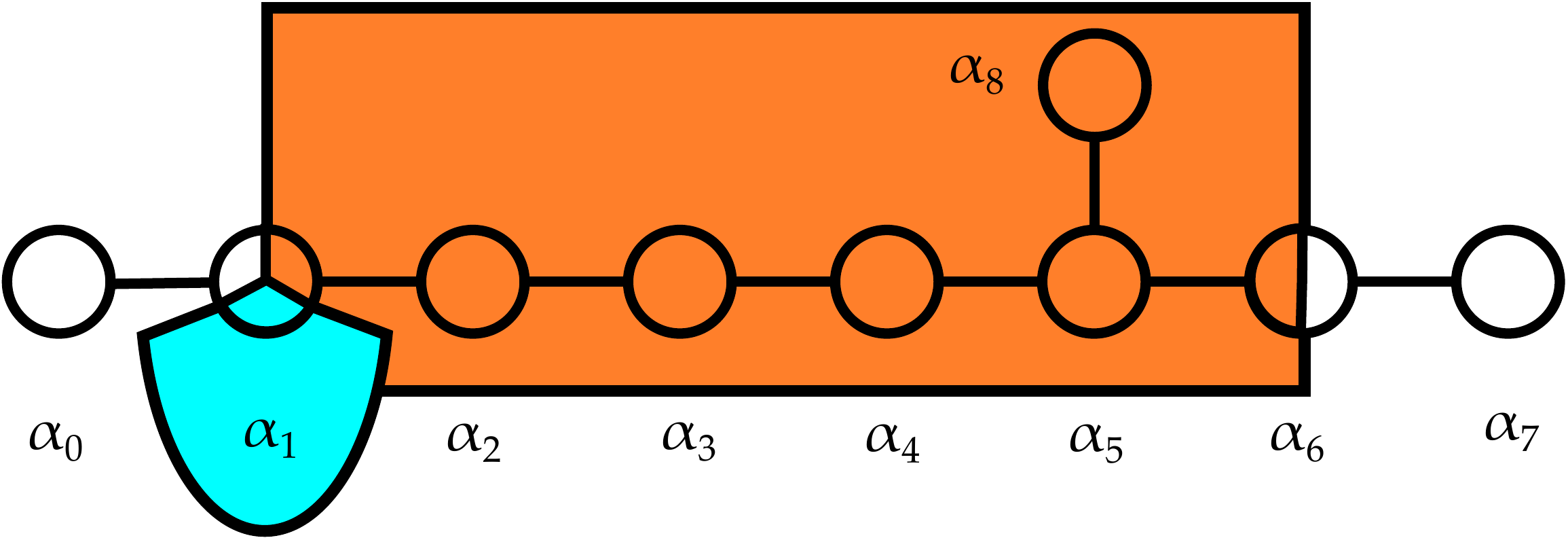}}
  \qquad  \qquad
  \subfloat[BU$_2^{(E_8, SU(2))}$: Codimension two Fiber.]{{\includegraphics[width=7cm]{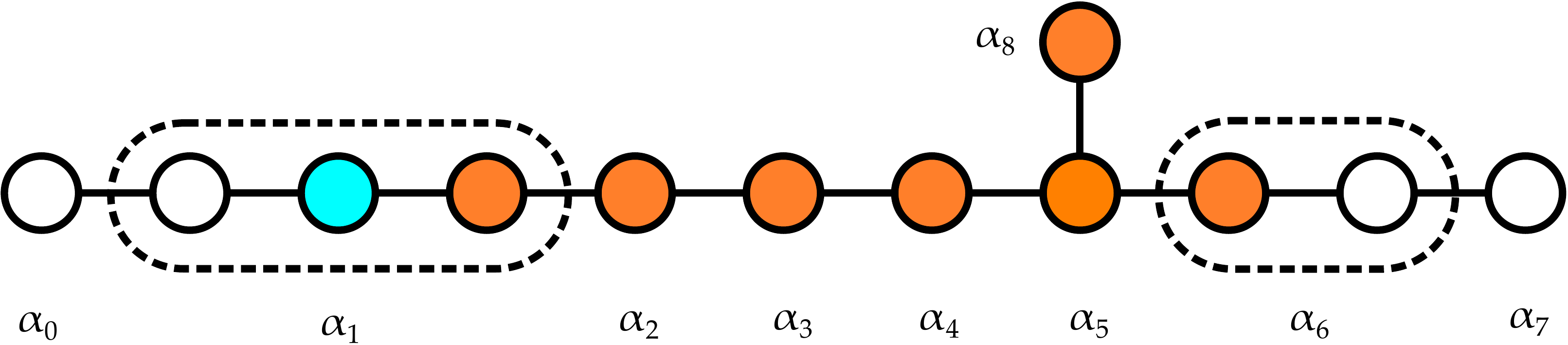}}}\\

 \subfloat[BU$_3^{(E_8, SU(2))}$: Wrapping of codimension one Fiber by surfaces $S_i$.]{\includegraphics*[width= 6cm]{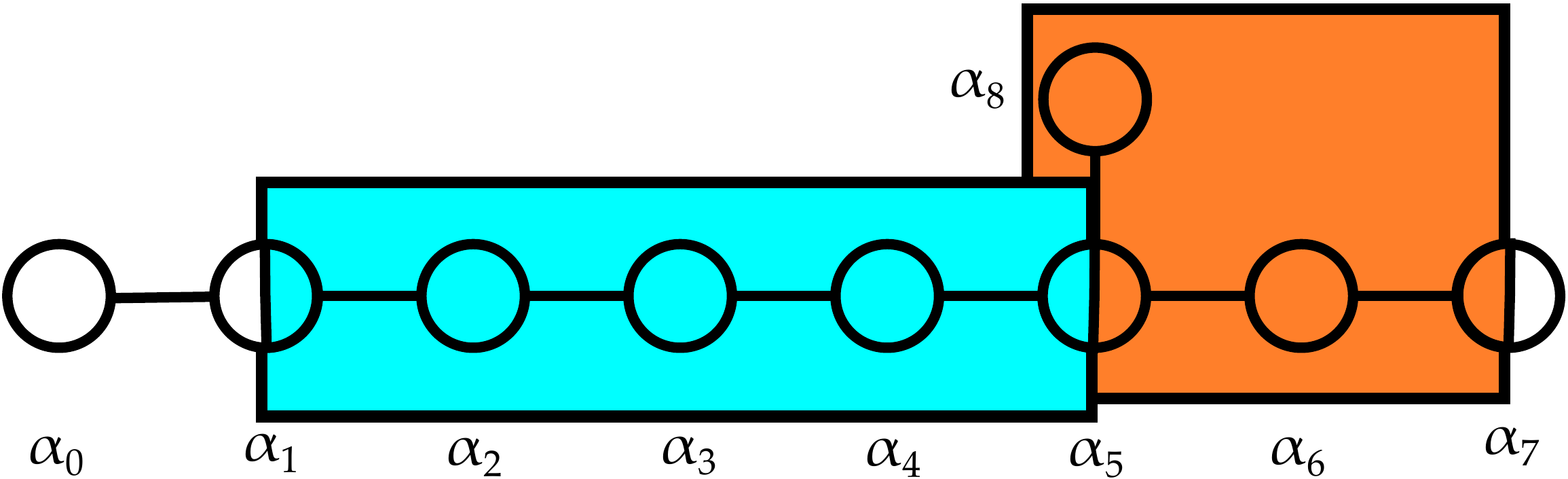}}
  \qquad  \qquad
  \subfloat[BU$_3^{(E_8, SU(2))}$: Codimension two Fiber.]{{\includegraphics[width=7cm]{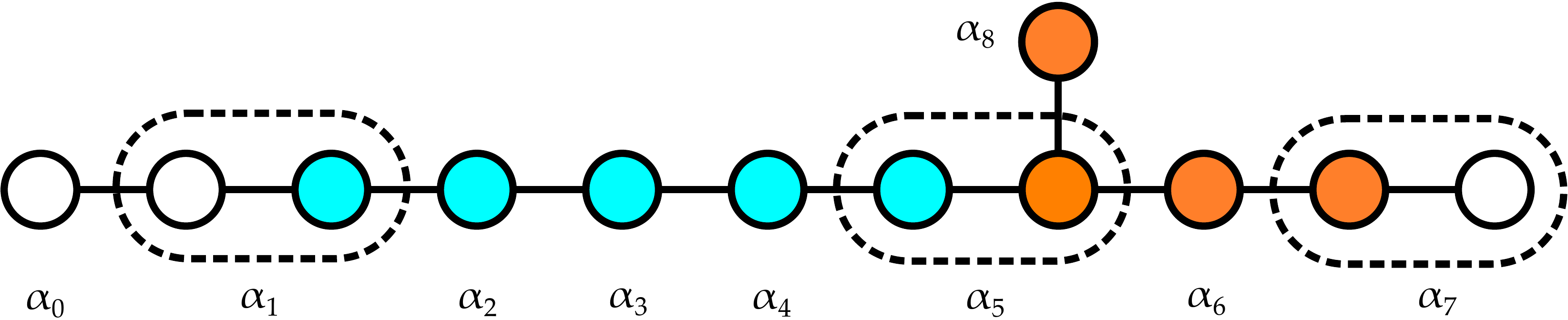}}}\\

\caption{Example Blowups for the Rank two E-string $BU_{i}^{(E_8, SU(2))}$. We show both the fiber of the $E_8$ and the curves that are contained in the surfaces components $S_1$ (turquois) and $S_2$ (orange) of the non-flat fiber, as well as, in the second row of each model, the actual codimension two fiber with all the irreducible fiber components and their intersections. 
\label{fig:ExampleBUE8Rank2}}
\end{figure}

The codimension two fibers for these blow-ups are shown in figure \ref{fig:ExampleBUE8Rank2}.
By considering the triple intersection number with the reducible surface $\mathcal{S}$, we can read off the strongly-coupled flavor symmetries as follows
\be
\ba
G_\text{F}^{BU_1}= G_\text{F}^{BU_2}= & \, SU(6)\times U(1) \cr 
G_\text{F}^{BU_3}= & \, SO(12)\times U(1) 
\ea\,.
\ee
Note that the resolutions $BU_1$ and $BU_2$ result in the same UV fixed point, although the geometric resolutions are distinct. The resolution geometries describe two distinct weakly-coupled gauge theory descriptions, of the same SCFT. It is this equivalence between resolutions that we will modd out by in the subsequent discussion of CFDs  in section \ref{sec:CFD}, and condense the geometric description to one that does not have such redundancies in the characterization of 5d SCFTs.


\subsubsection[Non-flat Resolutions of \texorpdfstring{$(D_{10},I_1)$}{(D10,I1)}]{Non-Flat Resolutions of \boldmath{$(D_{10},I_1)$}}\label{subsec:non_flat_res_D5d5}

The second class of non-flat resolutions in rank two  that we will consider is  the collision $D_{10}$ with a non-generic $I_1$ Kodaira fiber with the following vanishing orders in the Tate model 
\be
\hbox{ord}_{U=0} (b_i)= (1, 1, 5, 5, 10)\,, \qquad 
\hbox{ord}_{V=0} (b_i)=  (0, 1, 0, 0, 0) \,. 
\ee
Here we tuned the vanishing order of $b_6$ to trivially satisfy the split condition of the Kodaira fiber for $SO(20)$. 

First we summarize what we find in appendix \ref{app:TopCFDsD10} for the marginal geometry. The reduced triple intersection matrix is\footnote{For $(D_{2k},I_1)$ (or equivalently, $(D_k,D_k)$) theories, the multiplicities $\xi_i=1$ for all $i$.}
\be\label{D10MarginalBU}
\begin{array}{c|ccccccccccc|c|cc}
S_i   D_j^2 & D_0 & D_1 & D_2 & D_3 & D_4 & D_5 & D_6 & D_7 & D_8 & D_9 & D_{10} & D_0^{I_1} & S_1 & S_2 \\ \hline
S_1 & 0 & 0 & 0 & 0 & 0 & 0 & 0 & 0 & 0 & 0 & 0 & 0 & 8 & 4 \\
 S_2 & -2 & -2 & -2 & -2 & -2 & -2 & -2 & -2 & -2 & -2 & -2 & 0 & -6 & -2\\ \hline
n(F_i) & -2 & -2 & -2 & -2 & -2 & -2 & -2 & -2 & -2 & -2 & -2 & 0 & - & -\\

\end{array}\,.
\ee
In this resolution, the rational curves of the $D_{10}$ affine fiber are all contained within one surface component already. This geometry will define the marginal CFD in section \ref{sec:CFD}. 

\begin{figure}
\centering
\subfloat[BU$_1^{(D_{10}, I_1)}$: Wrapping of codimension one Fiber by surfaces $S_i$.]{\includegraphics*[width= 8cm]{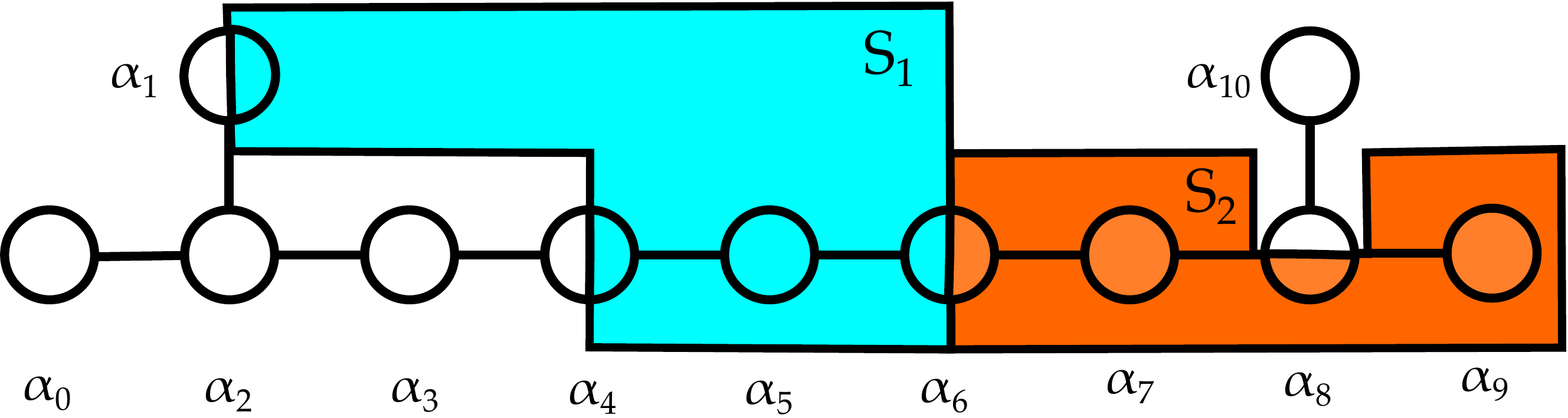}}\\
  \subfloat[BU$_1^{(D_{10}, I_1)}$: Codimension two Fiber where dotted lines indicate the splitting of the codimension one fiber components.]{{\includegraphics[width=8cm]{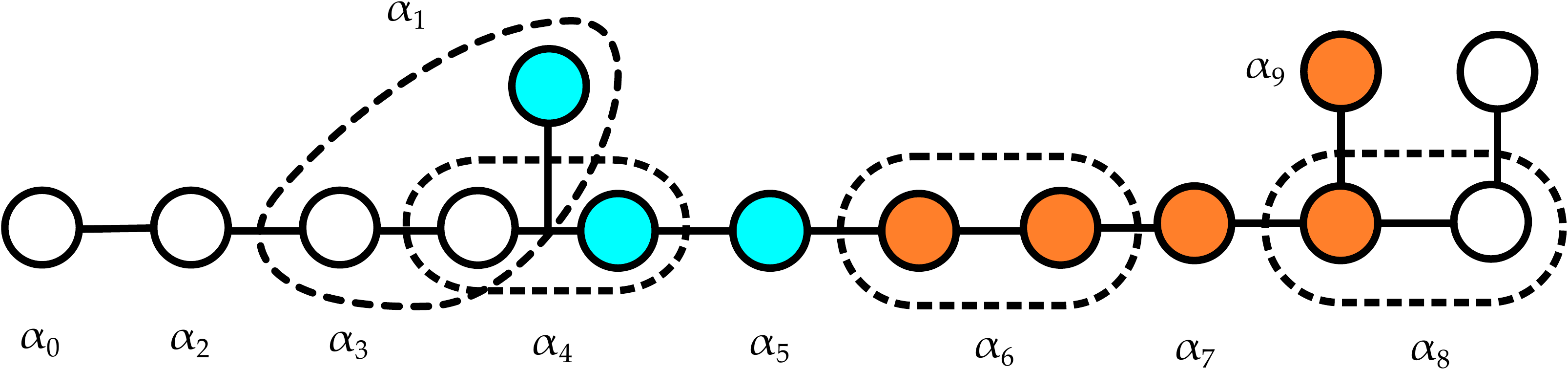}}}\\
\caption{Example blow-up $BU_1^{(D_{10}, I_1)}$. The first row shows the wrapping of the codimension one fiber. The second row gives the full codimension two fiber in terms of irreducible curves. }\label{fig:D10Example}
\end{figure}

Let us consider also an example resolution of a non-marginal theory --- the details are given in appendix \ref{app:BUD10Ex}. 
The reduced triple intersection numbers are
\be
\begin{array}{r|ccccccccccc|c|cc}
 S_i \cdot D_j^2  &
  D_0 & D_1 & D_2& D_3 & D_4 & D_5 & D_6& D_7 & D_8& D_9 &D_{10} & D_{0}^{I_1}&S _1 & S_2\\ \hline
S_1  & 0 & 0 & 0 & 0 & -1 & -2 & -1 & 0 & 0 & 0 & 0 & 4 & 6 & -2 \\
S_2  & 0 & 0 & 0 & 0 & 0 & 0 & -1 & -2 & -1 & -2 & 0 & 2 & 0 & 6 \\\hline
n(F_j)& 0 & 0 & 0 & 0 & -1 & -2 & -2 & -2 & -1 & -2 & 0 & 6 & - & -\\
\end{array}
\ee
from which we can read off the strongly-coupled flavor symmetry 
\be
G= SU(4) \times SU(2) \times U(1) \,.
\ee
The corresponding codimension two fiber is shown in figure \ref{fig:D10Example}.
Note that the abelian part of the flavor symmetry is obtained by considering the full triple intersection matrix, as discussed in appendix \ref{app:BUD10Ex}. 

An alternative starting point is the non-minimal $(D_5,D_5)$ collision, which will be discussed in appendix \ref{app:D5d5}. The  models obtained from the resolution of this collision will give rise to equivalent models, however to see the full flavor symmetry, one needs to compute the BPS states and repackage them in terms of higher rank groups. Our $(D_{10},I_1)$ starting point does not require this, and manifestly encodes the flavor symmetries geometrically.


\subsection{Higher Rank Conformal Matter}
\label{sec:HigherRankBU}

Starting with any non-minimal codimension two collision of two codimension one singular fibers of type $\mathfrak{g}_u$ and $\mathfrak{g}_v$ can be analyzed in the fasion described in this section. 

An infinite class of such 5d theories at arbitrary rank descending from 6d 
\be
(D_k, D_k) \,,\qquad \hbox{or} \qquad (D_{2k}, I_1)
\ee 
minimal conformal matter were discussed in \cite{Apruzzi:2019vpe}, as well as $(E_6, E_6)$ conformal matter for which we determined all descendant SCFTs. 

A systematic exploration of higher rank will appear in \cite{Apruzzi:2019kgb}. Here we will give a higher rank conformal matter example, $(E_8, G_2)$ for which we also determine the complete set of daughter 5d SCFTs in section \ref{sec:CFD}. A more systematic analysis of the higher rank cases will follow in that section as well, where we determine a more combinatorial way of generating all the geometries. 

Consider the collision of $II^*$ and $I_3$ Kodaira fibers associated to the  $(E_8, SU(3))$ conformal matter, which has  vanishing orders
\be
\hbox{ord}_{u=0} (b_i)= (1, 2, 3, 4, 5)\,, \qquad 
\hbox{ord}_{v=0} (b_i)=  (0, 1, 1, 2, 3) \,. 
\ee
The 5d SCFTs obtained from this have rank 4 and the 
resolution is given in  appendix \ref{app:E8SU3}. 
The reduced triple intersection matrix of the Cartan divisors $D_i^{\mathfrak{g}}$ of $E_8$ and $SU(3)$, respectively, with the four non-flat fiber components $S_i$ are 

{\small
\be
\ba
&BU^{(E_8, SU(3))}: \cr 
& 
\begin{array}{c|ccccccccc|ccc|cccc}
S \cdot D_j^2 &D_0 & D_1 & D_2& D_3& D_4& D_5& D_6& D_7 &D_8& D^{I_3}_0& D_1^{I_3} & D_{2}^{I_3}  & S_1 & S_2 & S_3 & S_4\cr \hline
S_1 & 0 & 0 & 0 & 0 & 0 & 0 & 0 & 0 & 0 & 0 & -1 & -1 & 6 & -2 & -2 & 0 \cr 
S_2 & 0 & 0 & 0 & 0 & 0 & 0 & 0 & -1 & 0 & 0 & -1 & -1 & -2 & 4 & -4 & -2 \cr 
S_3 & 0 & -1 & -2 & -1 & 0 & 0 & 0 & 0 & 0 & 0 & 0 & 0 & 0 & 0 & 6 & -2 \cr 
S_4 & 0 & 0 & 0 & -1 & -2 & -2 & -2 & -1 & -2 & 0 & 0 & 0 & 0 & 0 & 0 & 4 \cr\hline
n(F_j)& 0 & -1 & -2 & -2 & -2 & -2 & -2 & -2 & -2 & 0 & -2 & -2 & - & - & - & -
\end{array}
\ea
\ee
}

Note that although the $(E_8, SU(3))$ conformal matter has the following $\xi_i$ coefficients
\be\label{E8SU3-xi}
\xi^{E_8}_1=\xi^{E_8}_2=1\,,\quad  \xi^{E_8}_3=2\,,\quad  \xi^{E_8}_4=3\,,\quad  \xi^{SU(3)}_i=1\,,\quad i=1,\cdots,4\,,
\ee
they only affect the triple intersection numbers $S_3\cdot (D_i^{I_3})^2$ and $S_4\cdot (D_i^{I_3})^2$, which are zero in this case. 

\begin{figure}
\centering
\subfloat[BU$^{(E_8, SU(3))}$: Wrapping of codimension one Fiber by surfaces $S_i$.]{\includegraphics*[width= 5cm]{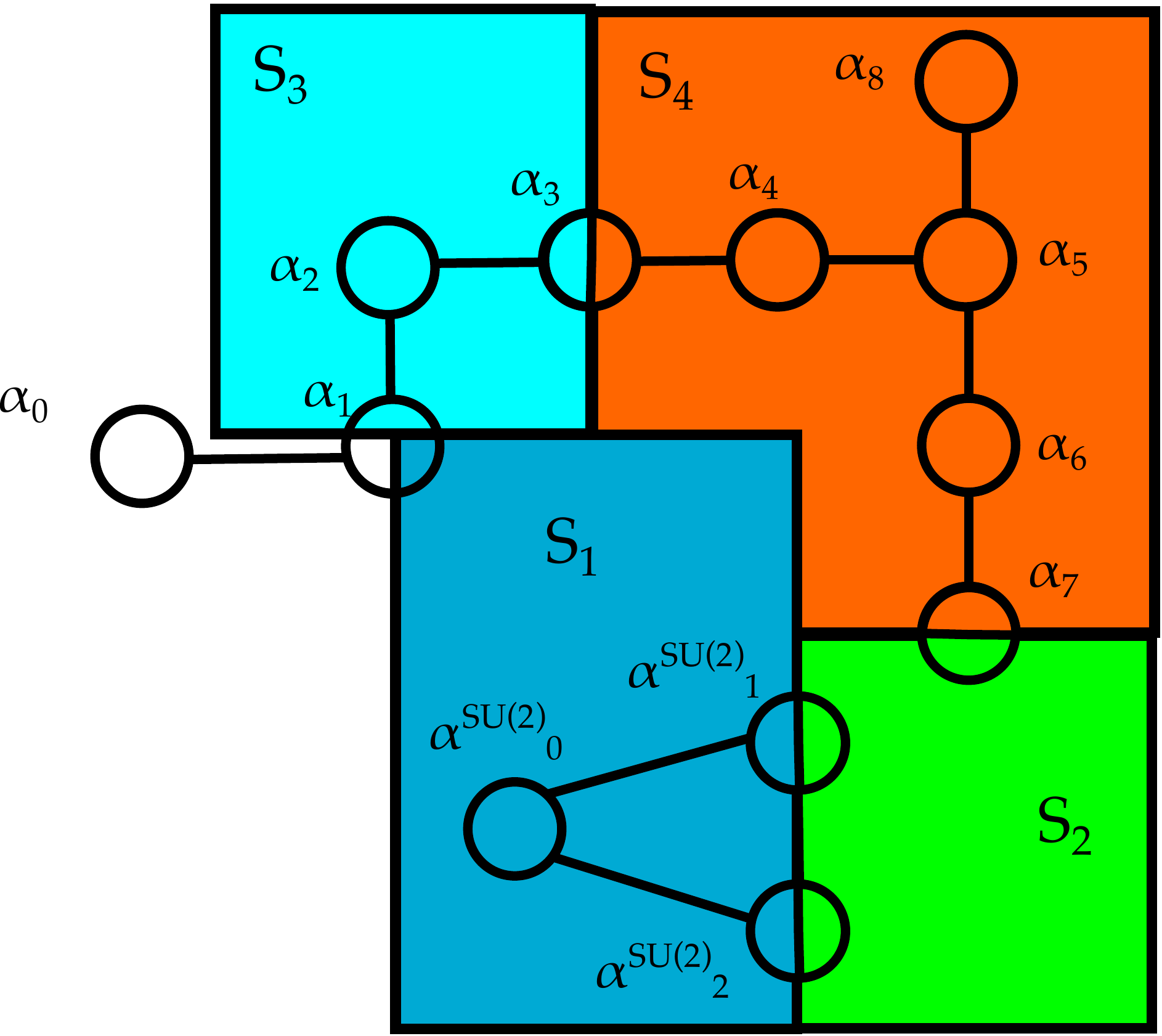}}\qquad 
 \subfloat[BU$^{(E_8, G_2)}$: Wrapping of codimension one Fiber by surfaces $S_i$.]{\includegraphics*[width= 5cm]{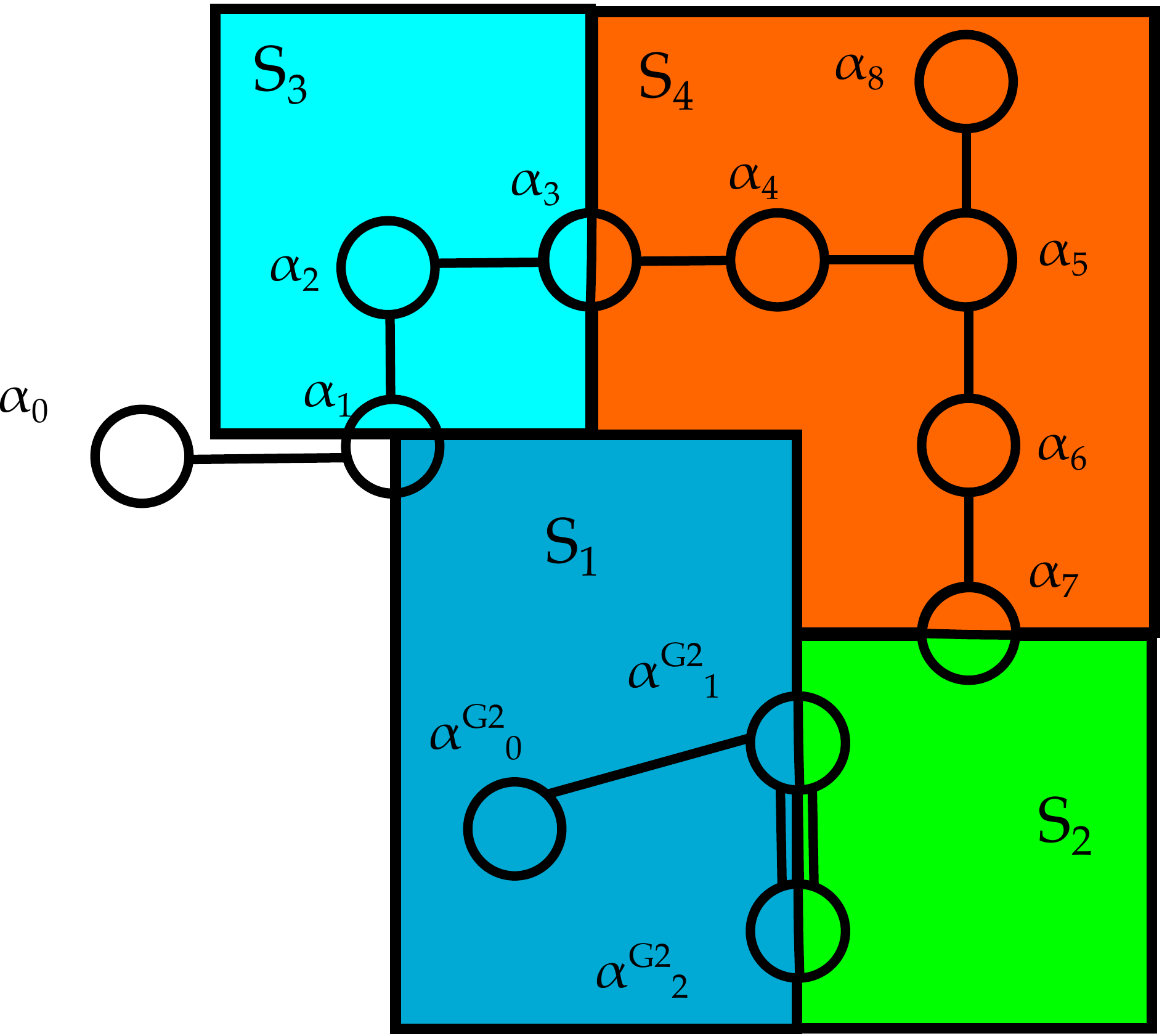}}\\
  \subfloat[BU$^{(E_8, SU(3))}$: Codimension two Fiber where dotted lines indicate the splitting of the codimension one fiber components.]{{\includegraphics[width=10cm]{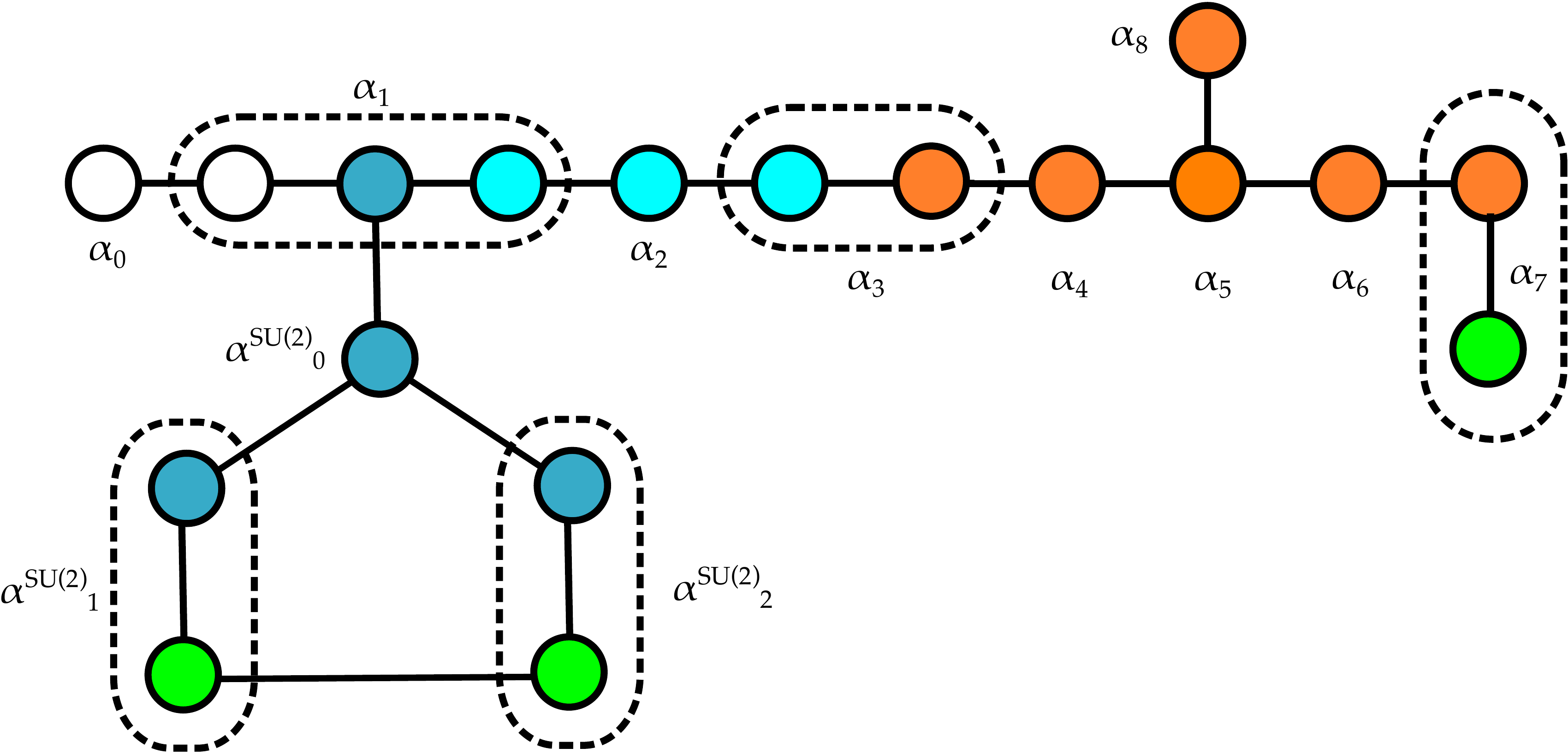}}}\\
  \subfloat[BU$^{(E_8, G_2)}$: Codimension two Fiber where dotted lines indicate the splitting of the codimension one fiber components.]{{\includegraphics[width=10cm]{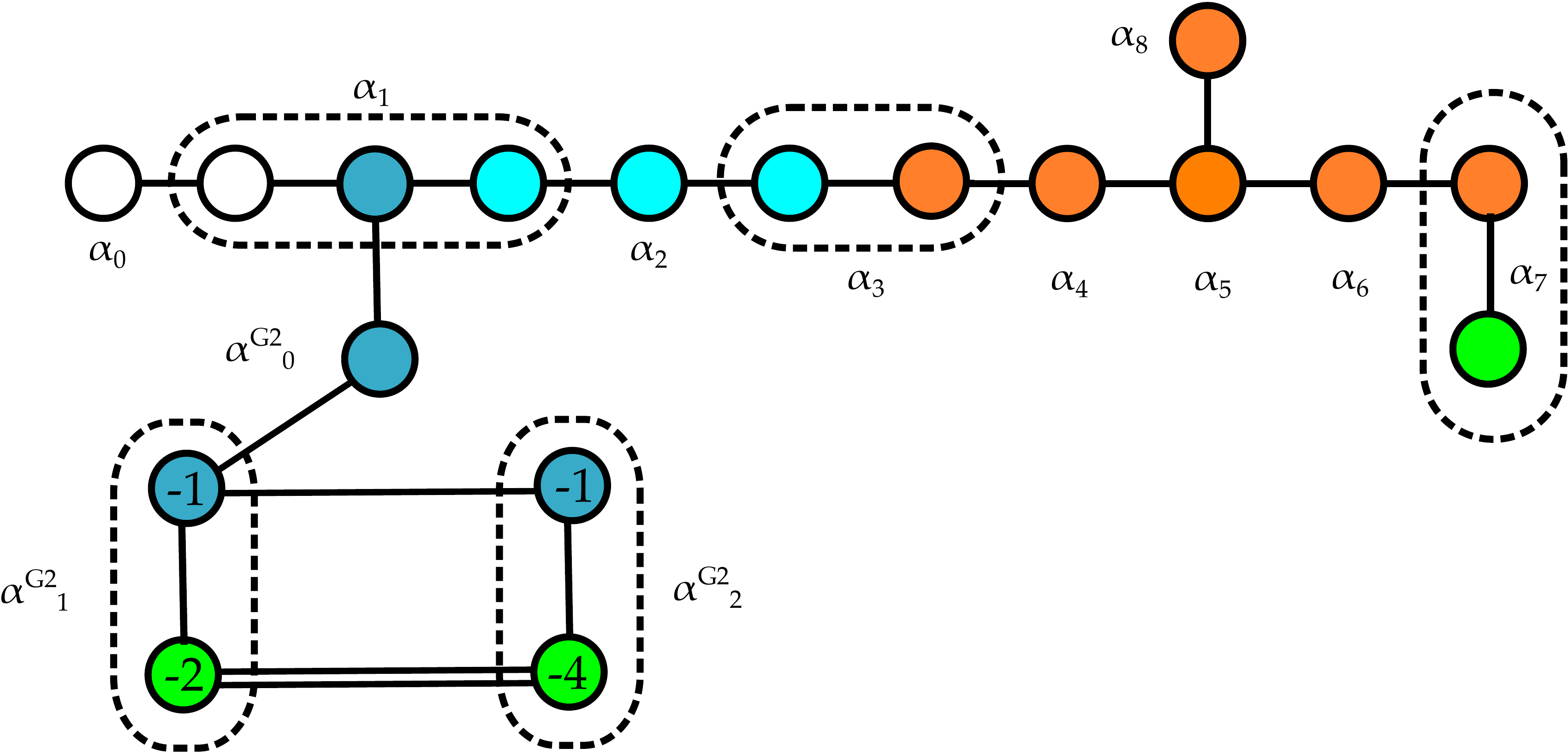}}}\\

\caption{Example blow-up for $(E_8, SU(3))$ and $(E_8, G_2)$. The gauge theory description has rank 4 gauge group, corresponding to the four surface components $S_i$. The first image shows the affine $E_8$ and $SU(3)$ and how the $S_i$ wrap these (half-wrapping corresponding to self-intersection $(-1)$, full wrapping to $(-2)$), see (\ref{BUE8SU3}).  }\label{fig:E8SU3Example}
\end{figure}

The wrapped components of the fiber and codimension two fiber is shown in figure \ref{fig:E8SU3Example}.
The strongly coupled flavor symmetry for the SCFT from this point of view is 
\be
G_\text{F} \supset E_7 \times SU(3) \,.
\ee
In fact there is an alternative starting point, where the full 6d flavor symmetry $(E_8, G_2)$ is manifest \cite{Mekareeya:2017jgc}.

With some minor changes we can generalize this to $(E_8, G_2)$, i.e., the collision of $II^*$ with $I_1^{ns}$ (non-split $I_1^*$). 
The vanishing orders  change to 
\be
\hbox{ord}_{v=0} (b_i)=  (0, 1, 1, 2, 3) \,. 
\ee
The reduced triple intersection matrix for this case is, with the coefficients $\xi_i$ as in (\ref{E8SU3-xi}) and (\ref{nFDef})

{\small
\be
\ba
&BU^{(E_8, G_2)}: \cr 
&
\begin{array}{c|ccccccccc|ccc|cccc}
S \cdot D_j^2 &D_0 & D_1 & D_2 & D_3 & D_4 & D_5 & D_6 & D_7  &D_8 & D^{G_2}_0& D_1^{G_2} & D_{2}^{G_2}  & S_1 & S_2 & S_3 & S_4\cr \hline
S_1 & 0 & 0 & 0 & 0 & 0 & 0 & 0 & 0 & 0 & 0 & -1 & -2 & 6 & -2 & -2 & 0 \\
S_2 & 0 & 0 & 0 & 0 & 0 & 0 & 0 & -1 & 0 & 0 & -1 & -4 & -2 & 4 & -4 & -2 \\
S_3 & 0 & -1 & -2 & -1 & 0 & 0 & 0 & 0 & 0 & 0 & 0 & 0 & 0 & 0 & 6 & -2 \\
S_4 & 0 & 0 & 0 & -1 & -2 & -2 & -2 & -1 & -2 & 0 & 0 & 0 & 0 & 0 & 0 & 4 \\\hline
n(F_j) & 0 & -1 & -2 & -2 & -2 & -2 & -2 & -2 & -2 & 0 & -2 & -6& - & - & - & -
\end{array}
\ea
\ee
}
The fiber (and containments within the non-flat fibers) is shown in figure \ref{fig:E8SU3Example}. 
So here the flavor symmetry at the strongly coupled point is 
\be
G_\text{F} = E_7 \times G_2 \,.
\ee
Clearly the realization in terms of the collision with $G_2$ captures the full flavor symmetry in 5d. 
In the next section we will discuss using our proposed graphical presentation the higher rank generalizations to $(E_8, SU(m))$, $m>3$,  as well as the  $(E_n, E_n)$ conformal matter theories systematically.
In these cases the maximal flavor symmetry is manifest already in the 6d realizations.


\section{5d SCFTs from Graphs}
 \label{sec:CFD}

We now turn to reformulating the geometric description of 5d SCFTs that descend from 6d SCFTs in a succinct way, which we already introduced in our recent paper \cite{Apruzzi:2019vpe}, in terms of a  graph-theoretic tool, the {\it combined fiber diagram (CFD)}. A CFD characterizes a 5d SCFT, its superconformal flavor symmetry and mass deformations.
Furthermore, it enables a systematic and comprehensive derivation of all descendant SCFTs from a given marginal theory. 
 

\subsection{Combined Fiber Diagrams (CFDs)}
\label{sec:CFD-def}

A CFD was defined in \cite{Apruzzi:2019vpe} as a graph, whose vertices $C_i$ are curves and whose edges are given by intersection numbers  $m_{i,j}= C_i\cdot C_j$ between the curves. Furthermore, vertices carry labels $(n_i, g_i)$, which are the self-intersection number and genus of the curve associated to the vertex. 

We will now explain how a CFD can be associated to any crepant resolution of an elliptically fibered Calabi--Yau threefold $Y$. 

Denote by $B$, the non-compact base of $Y$, where there is a local coordinate patch $(u, v)$. Furthermore, consider the elliptic fibration with Kodaira fibers of type $\mathfrak{g}_u$ above $u=0$ and $\mathfrak{g}_v$ above $v=0$ ($\mathfrak{g}_\nu$ can either refer to the algebra associated to the fiber or the fiber type). 
At $u=v=0$ let there be a non-minimal singularity, i.e., in terms of the Weierstrass model ord$(f, g, \Delta)_{u=v=0} \geq (4,6,12)$. 
Consider a crepant blow-up, $BU: \tilde{Y} \rightarrow Y$ of the threefold that introduces $r$ compact surface components $S_i$  with 
\be
\mathcal{S}= \bigcup_{i=1}^r S_i \,.
\ee
This corresponds to a rank $r$ 5d theory. There are a number of non-compact divisors that intersect $\mathcal{S}$, including the Cartan divisors $D_i^{(\nu)}$, $\nu=u,v$, for the codimension one singularities. 

Denote by $F_{k}$ all the fibral curves that are complete intersections between $\mathcal{S}$ and non-compact divisors, and are entirely contained within $\mathcal{S}$, we denote these by 
\be\label{MathCalFDef}
\mathcal{F}= \left\{C =  D_i^{(\nu)}\cdot\mathcal{S}; \ C \subset \mathcal{S} \right\} \,.
\ee 
These are the flavor curves. 
They are (a subset of) the rational curves associated to simple roots of the Lie algebras $\mathfrak{g}_u$ and $\mathfrak{g}_v$. Note that a rational curve that is reducible in codimension two is contained in $\mathcal{F}$, if all its irreducible components are all contained in $\mathcal{S}$ (i.e., they may not be contained in one single surface component, but in the reducible surface $\mathcal{S}$).
For the precise conditions of a fully wrapped Cartan divisor, see the ``Enhanced Flavor Symmetry'' section in section~\ref{sec:Dictionary}. The CFD is effectively the set of Mori cone generators of the reducible surface $\mathcal{S}$. More precisely:

{\definition {\bf CFD associated to a Crepant Resolution}\\

Given a resolution $BU$ of a non-minimal singular Weierstrass model, with the compact (reducible) surface $\mathcal{S} = \cup_k S_k$, the associated $\hbox{CFD}_{BU}$ is a graph whose vertices are the curves $C\subset \mathcal{S}$, including:

\begin{enumerate}

\item{The flavor curves $F_i\subset\mathcal{F}$ generating the non-abelian flavor symmetry, which are marked (usually colored green). }

\item{The rational $(-1)$-curves with normal bundle $\mc{O}(-1)\oplus \mc{O}(-1)$ that can be flopped outside of $\mathcal{S}$.}

\item{The other curves generating the Mori cone of  $\mathcal{S}$, which are not intersection curves between $S_i$ and $S_j$.}

\end{enumerate}

We label each vertex by a pair of integers $(n,g)$, encoding its self-intersection number $n$ 
and the genus $g$. 
For the  $\mc{O}(-1)\oplus\mc{O}(-1)$ curve in case 2., they are labelled by $(n,g)=(-1,0)$. For the flavor curves $F_i$ in case 1., they are usually labeled as $(n,g)=(-2,0)$ unless there is a surface component $S_j$ with a non-trivial multiplicity $\xi_j>1$, which contains both the flavor curve $F_i$ and a $(-1)$-curve connected to $F_i$.

Two vertices $C_i$ and $C_j$ are connected by
\be
m_{i,j} = C_i \cdot C_j
\ee
edges if they intersect at $m_{i,j}$ points in $\mathcal{S}$.
}

%

\noindent
\underline{\it Conventions:} The  label $g_i$ is generically omitted when $g_i=0$. Furthermore, marked vertices have, unless noted otherwise,  $(n_i, g_i)= (-2, 0)$, which again is often omitted. 

The marked vertices  form a sub-graph, which we identify with the Dynkin diagram of the non-abelian part of the superconformal flavor symmetry. 

As examples consider the rank two  E-string resolutions $BU^{(E_8, SU(2))}_1$ and $BU^{(E_8, SU(2))}_2$, see (\ref{BU3}) and (\ref{BU34}), respectively. The reduced intersection matrices (\ref{Triples}) show that these models differ only 
 in terms of how the $(-2)$ and $(-1)$ curves are distributed among the two surface components $S_i$, but the overall structure of $(-2)$ and $(-1)$ curves in the fiber that are contained in some non-flat surface are the same. Equivalently, the intersections with $\mathcal{S}$ and the Cartan divisors is the same in both cases. 
Hence, the associated CFDs (and SCFTs) are identical:
\be\label{CFDBU3}
\includegraphics[height=2.7cm]{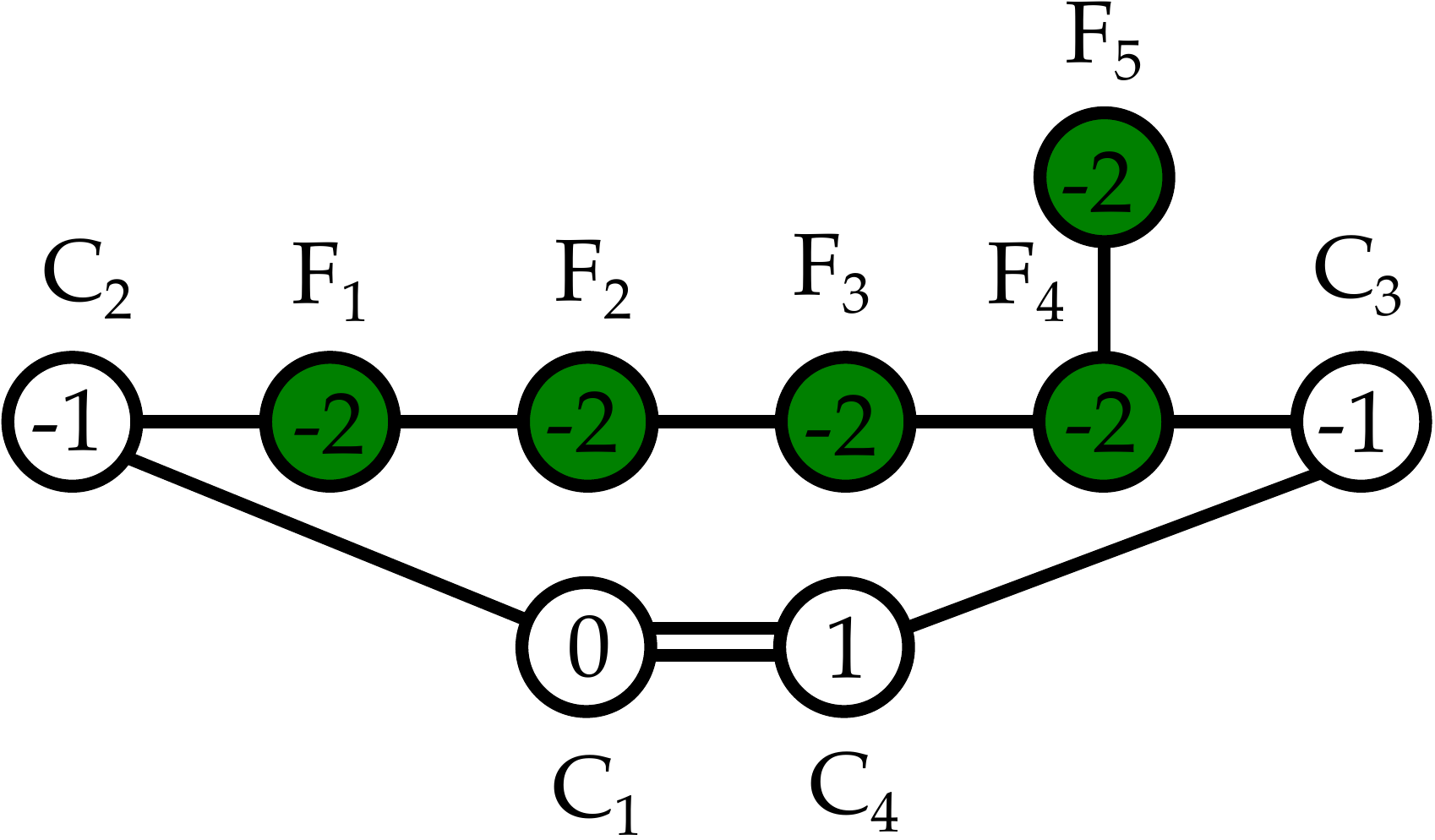} \,.
\ee
The marked, green vertices are the $(-2)$-curves that are contained in $S_1 \cup S_2$, and determine the non-abelian part of the SCFT flavor symmetry to be $SU(6)$.
The white curves have self-intersections as indicated by the numbers in the nodes. 
In triple intersection matrices for $S_i$ (\ref{BU3Triple1}), the corresponding $(-2)$- and $(-1)$-curves are boxed in. The curves $C_1$ and $C_4$ with self-intersection $0$ and $1$ inside the $\mc{S}$ are double-boxed, which are originally components of the fiber that come from the $I_2$ singular fiber above $v=0$. 
Note that the curve $C_4$ can be expressed as a linear combination of other curves:
\be
C_4=2C_2+2F_1+2F_2+2F_3+2F_4+F_5+C_3\,,\label{C4linear}
\ee
and is not an indepedent Mori cone generator.

The advantage of the description in terms of CFDs compared to the fibers is two-fold: 
a CFD encodes the non-abelian part of the flavor symmetry manifestly in terms of the marked vertices, which form a sub-graph, which is the  Dynkin diagram of the non-abelian part of the flavor symmetry of the SCFT, $G_{F,na}$. For the abelian part $U(1)^s$, the number of $U(1)$ factors should still be computed by (\ref{U(1)counting}), with the number of mass deformations $M$. For the descendants of a given marginal CFD, this effectively equals to the number of white nodes minus a fixed number, see the details in section \ref{sec:rank_two_CFDs}.

Furthermore, we can transition between CFDs, which correspond in the geometry to 
contractions of rational curve with self-intersection $C_i\cdot C_i = -1$ (which do not correspond flavor curves). 
 In (\ref{CFDBU3}) there are two $(-1)$-curves that can be contracted. The two descendant CFDs are obtained by the standard rules of blowing down $(-1)$-curves on a complex surface. In terms of the CFDs, which `flops' of curves from within $\mathcal{S}$ to outside of $\mathcal{S}$ can be described in terms of simple graph-theoretic rules.

{\definition\label{def:Flops}
 {\bf CFD Transitions}\\
 Let $C_i$ be an unmarked vertex with label $(n_i, g_i)= (-1, 0)$ of a CFD. We can define a new CFD, CFD$'$ that is obtained by removing the vertex $C_i$ and updating the graph as follows:
 \begin{enumerate}
 \item Let $C_j$ be a vertex in the original CFD with label $(n_j, g_j)$ with $C_i\cdot C_j= m_{i,j}$, then in CFD$'$, the vertex $C_j$ is labeled by 
\be
 {n}'_j = n_j + m_{i,j}^2  \,,\qquad {g}'_j = g_j+ \frac{m_{i,j}^2 - m_{i,j}}{2} \,.
 \ee
\item If $C_i\cdot C_j = m_{i,j}$ and $C_i\cdot C_k = m_{i,k}$, and $C_j \cdot C_k = m_{j,k}$ then in CFD$'$ 
\be
m_{j, k}' =m_{j,k}+m_{i,j} m_{i,k} \,.
\ee
\item  If $C_i$ intersects multiple curves $C_i\cdot C_j = m_{i,j}$, then rule 2. applies pairwise. 
\end{enumerate}
}

\begin{figure}
\centering
\includegraphics*[width= 10cm]{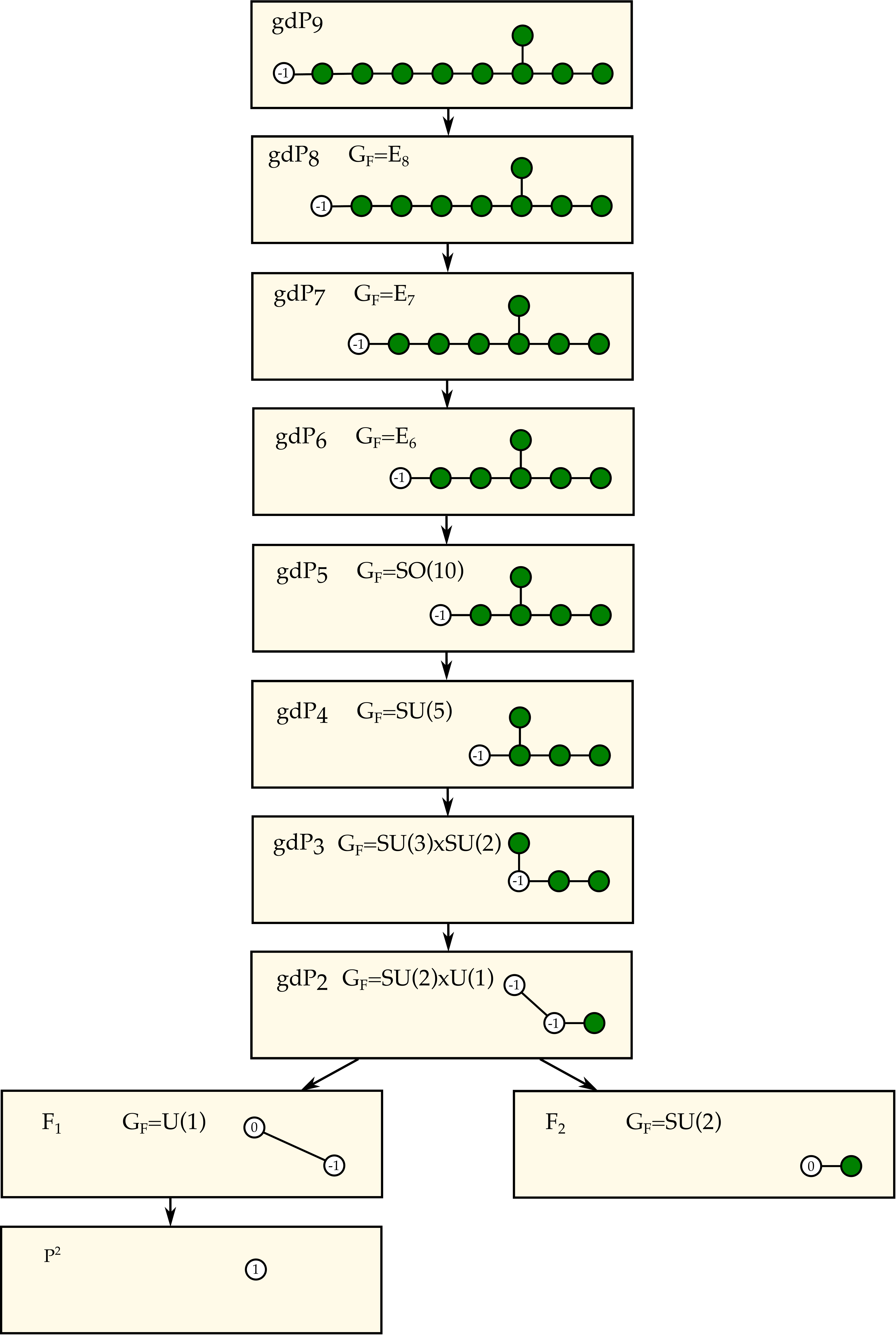}
\caption{Rank one CFD tree starting with the marginal CFD, given by the gdP$_9$, including all $(-2)$-curves (shown in green) as well as one $(-1)$-curve. The transitions are dictated by the rules in Definition \ref{def:Flops}.  The strongly coupled flavor symmetry is denoted by $G_\text{F}$ and the non-abelian part of it is read off from the $(-2)$-curves in the CFDs. \label{fig:Rank1CFDTree}}
\end{figure}

In particular if $n_j=-2$ with $m_{i,j}= 1$, then $n_{j}'= -1$.
These rules determine the complete chain of descendant CFDs (and associated SCFTs). For instance for 
(\ref{CFDBU3}) the two descendants are 
\be
\includegraphics[height=1.7cm]{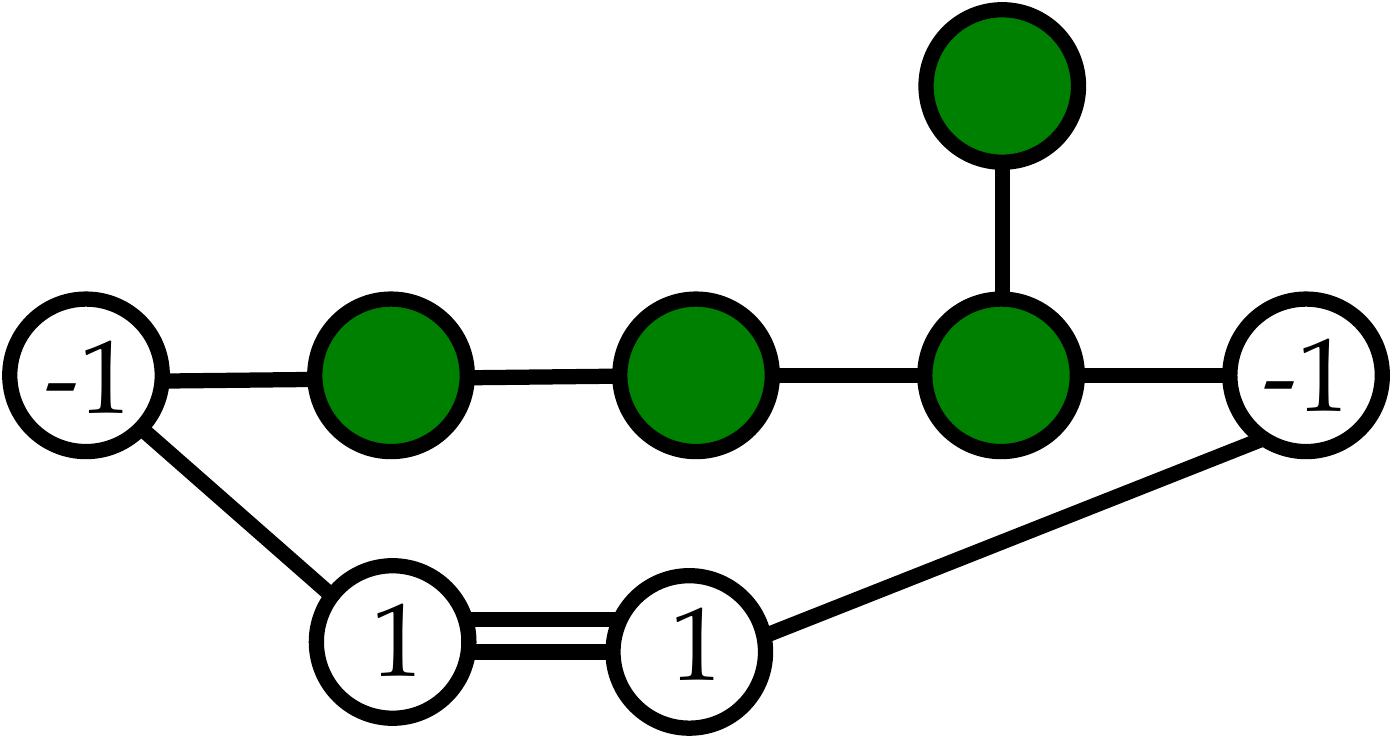} \qquad \hbox{and }\qquad 
\includegraphics[height=1.7cm]{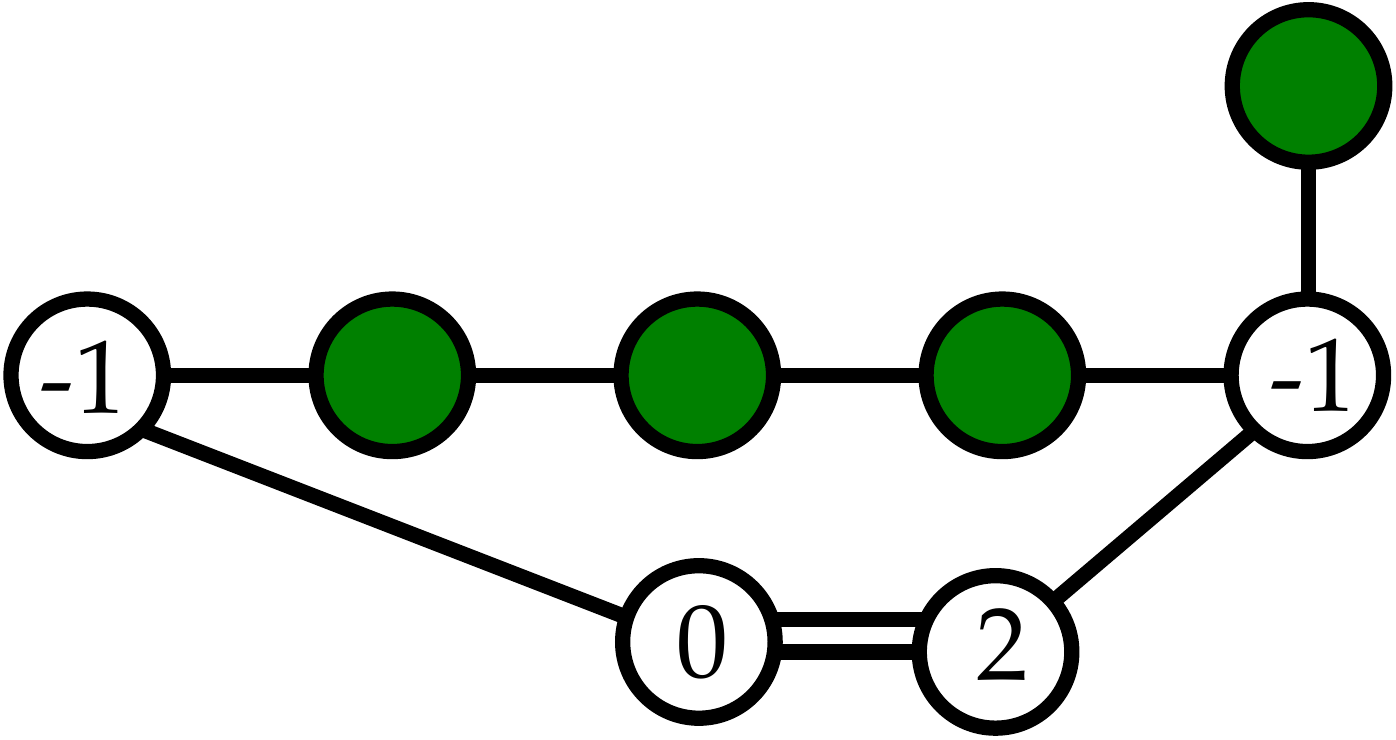}\,.
\ee
These have $SU(5)\times U(1)$ and $SU(4)\times SU(2) \times U(1)$ strongly coupled flavor symmetry, respectively. We continue the CFD-transitions until a descendant CFD is reached, which does not have any vertices with $(n_i, g_i)=(-1, 0)$. These are the endpoints of the CFD tree (and the RG-flows).

The descendant CFDs that we obtain in this way are irreducible in the sense, that the corresponding SCFTs do not factorize into SCFTs of lower ranks. Geometrically, this means that the CFD-transitions only correspond to flops of  $(-1)$-curves, which are not at the intersection of two surface components. By construction, we do not include these curves into the CFDs.

\subsection{Starting Points: Marginal CFDs}
\label{sec:marginal-CFD}

In light of these extremely simple rules governing CFDs and their transitions, it is important to determine 
the CFD for the marginal theories, from which all other descendant CFDs/SCFTs can be obtained by this simple, graph-theoretic operation. 
In this section, we explain how  for a given codimension two non-minimal collision, one can compute the CFD for the marginal theory. 
The main characteristic of the marginal (or top) CFD is that its marked vertices form affine Dynkin diagrams for the codimension one singularities \cite{Apruzzi:2019vpe}. 

Consider the codimension two collision, as before, with $(\mathfrak{g}_u, \mathfrak{g}_v)$. To compute the marginal geometry and associated CFD, we follow the same strategy as in section \ref{sec:Resolutions}. Before resolving the singularities in codimension one, we blow up the non-minimal locus on the base
\be
\{u,v,\delta_1\}:\qquad u\rightarrow U \delta_1 \,,\qquad v\rightarrow V \delta_1 \,.
\ee 
From a 6d point of view this corresponds to a partial tensor branch. After this single blow-up in the base, we perform the resolution of the codimension one singular fibers as usual \cite{Lawrie:2012gg}.
After the blow-ups, we have the following set of divisors: the vertical divisor  
\be
S_1: \qquad \delta_1=0 
\ee
as well as the Cartan divisors for the codimension one singular fibers 
$D_i^{\mathfrak{g}_\nu}$. In fact, all the curves $S_1 \cdot D_i^{\mathfrak{g}_\nu}$ for $\nu=u,v$ are already fully wrapped inside $S_1$. 
In general, the resulting Calabi--Yau threefold is still singular, and requires further small resolutions of the type 
\be
\{u_i, \delta_j , \delta_{j+1} \} \qquad \hbox{or} \qquad \{v_i, \delta_j , \delta_{j+1} \} \,,
\ee
which result in $r$ surface components $S_i$, whose union $\mathcal{S}$ contains all codimension one rational curves. 
The CFD associated to this blow-up up is precisely the marginal CFD. 

From this discussion, it is obvious that this marginal CFD will always contain the affine Dynkin diagram of $\mathfrak{g}_u$ and $\mathfrak{g}_v$ as subgraphs, as these are by construction always fully wrapped, and their vertices are marked (colored). 
The marginal CFD contains in addition  curves with $n_i \geq -1$, which connect the two marked subgraphs. 

For the rank one 5d SCFTs, the relevant starting point is the rank one E-string, and the blow-up for the marginal CFD already appeared in (\ref{BaseBU}) combined with (\ref{gdP9BU}). 
For the rank two 5d SCFTs, the marginal CFDs are derived in appendices \ref{app:TopCFDsRank2E} and \ref{app:TopCFDsD10}. In the following we will use these resolutions to determine all 5d SCFTs of rank one and two, and furthermore provide the marginal CFDs  higher rank conformal matter theories. 


\subsection{Rank one Classification from CFDs}
\label{sec:rank_one_CFDs}

For rank one 5d SCFTs, the marginal CFD corresponds precisely to the geometry  of curves inside the generalized del Pezzo surface gdP$_9$, shown in table \ref{tbl:rank1Res}. In addition there is a  $(-1)$-curve, which is also contained in the gdP$_9$ and part of the generators of the Mori cone. In summary we find the marginal CFD for the rank one theories
\be
\includegraphics*[width=6cm]{CFD-E8-Rank1-Top.pdf}
\ee
as proposed already in our earlier paper \cite{Apruzzi:2019vpe}.
There is precisely one CFD-transition that can be applied to the vertex with $n= -1$. The complete rank one CFD-tree is shown in figure \ref{fig:Rank1CFDTree}. 
Each descendant CFD is accompanied by the information about the superconformal flavor symmetry that is read off from the marked sub-graph, and the geometry of the non-flat fiber --- as determined by direct resolutions in section \ref{sec:EstringGeo}.
Needless to say, this tree agrees with the Seiberg theories and their  mass deformations \cite{Seiberg:1996bd, Morrison:1996xf} and the flavor symmetries predicted therein. Note that the CFD includes the theory that does not have a weakly-coupled gauge theory description (associated to the geometry $\mathbb{P}^2$).

It is useful to compare the CFD-tree with the complete fiber diagrams in section \ref{sec:EstringGeo}. The CFDs only capture the information that is relevant for the 5d SCFT --- in particular, information about the additional curves that are unaffected by the singular limit is not retained. This makes these graphs particularly efficient.

\subsection{Rank two Classification from CFDs}
\label{sec:rank_two_CFDs}

To illustrate the power of the CFD approach, we next consider the rank two  5d SCFTs. 
The key is to find the marginal CFD for each of the starting points in 6d listed in appendix \ref{sec:origin}. 
This contains the codimension one $(-2)$-curves $F_k^{\mathfrak{g}_u}$ and $F_l^{\mathfrak{g}_v}$ in the fibers, but has to be determined from a resolution of the singular elliptic fibration, where the two surface components of the resolved geometry contain all these curves. We determine these resolutions in the appendices \ref{app:TopCFDsRank2E} and \ref{app:TopCFDsD10} for the rank two marginal theories. 

There are seven starting points in total, two of which, $(E_8,SU(2))$ and $(D_{10},I_1)$, have a direct conformal matter origin.
We infer the marginal CFDs for two further models, which arises as quotients from 6d. Starting with these marginal CFDs we generate the complete CFD-tree by transitions and confirm the tree-structure of SCFTs, that was obtained by independent methods in 
\cite{Jefferson:2018irk,Hayashi:2018lyv}. Furthermore, we determine the superconformal flavor symmetries in all cases. 

\begin{sidewaysfigure}
\centering
\includegraphics[width=\textwidth]{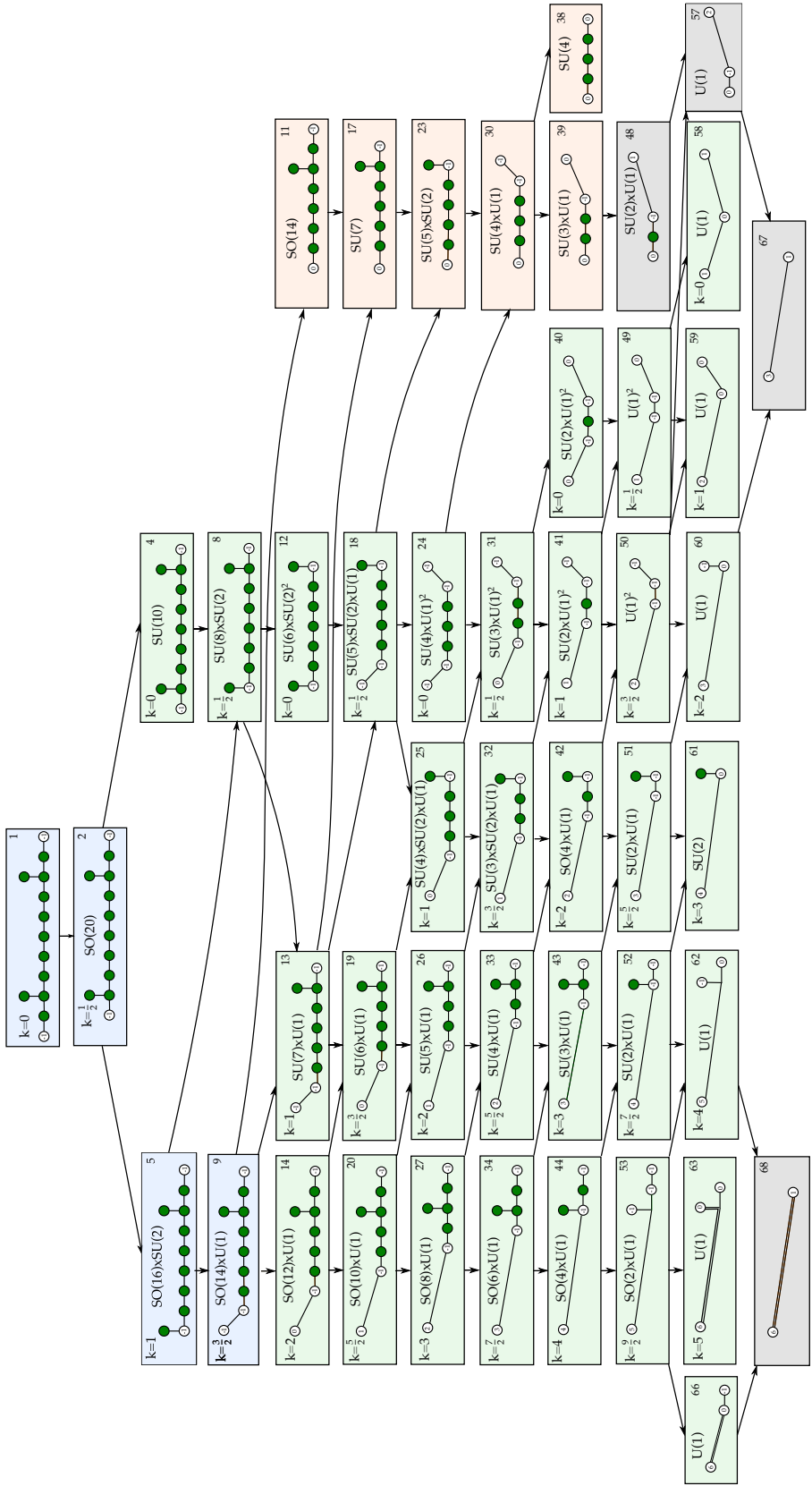}
\caption{$(D_{10},I_1)$ CFD-tree. For caption see text.
\label{fig:D10FibsAll}}
\end{sidewaysfigure}

\begin{sidewaysfigure}
\centering
\includegraphics[width=\textwidth]{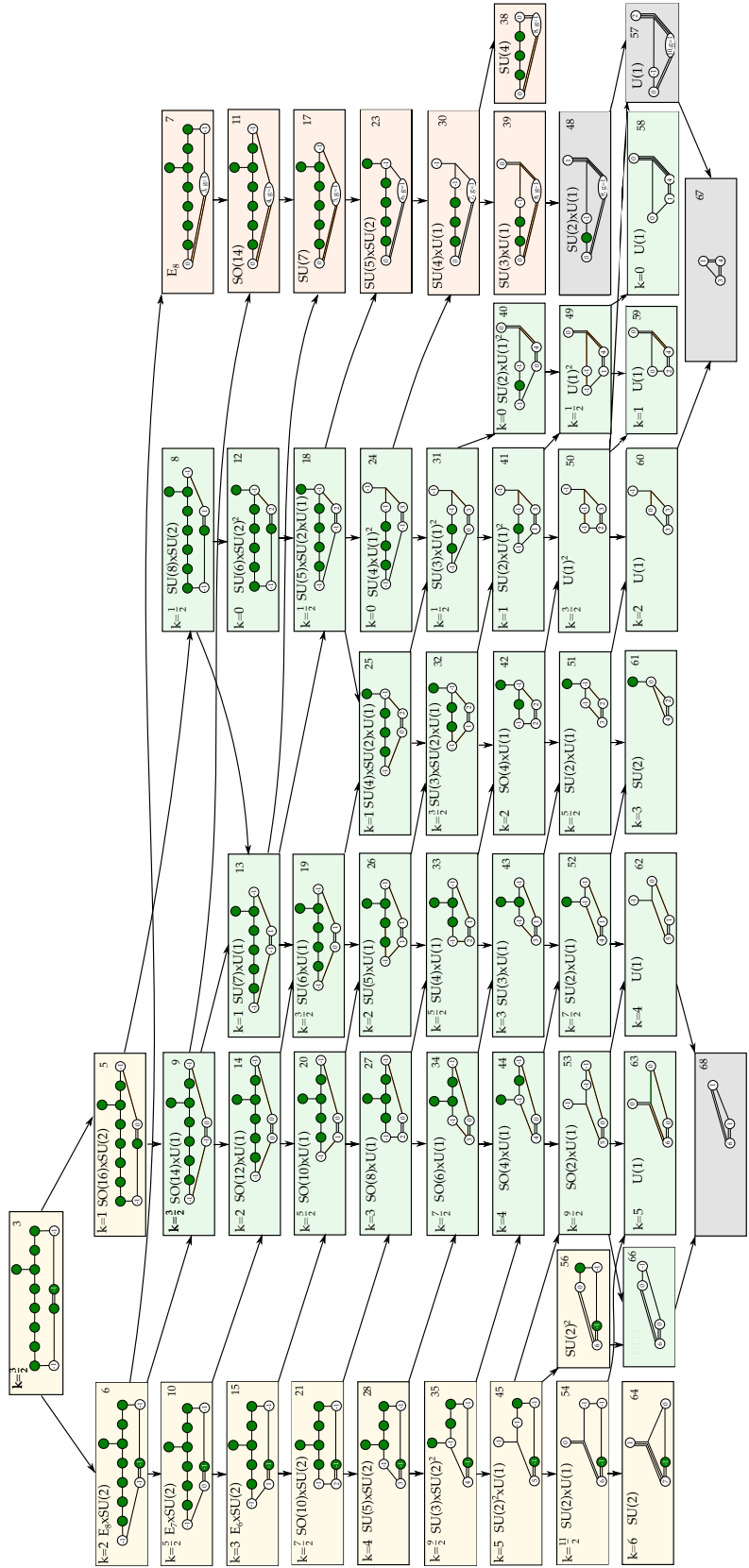}
\caption{$(E_8,SU(2))$ CFD-tree. For caption see text. \label{fig:E8FibsAll}}
\end{sidewaysfigure}

\begin{enumerate}
\item \underline{$(D_{10}, I_1)$}: \\
The marginal CFD for the $(D_5, D_5)$, or equivalently $(D_{10}, I_1)$ conformal matter theory is 
\be
\includegraphics*[scale =0.24]{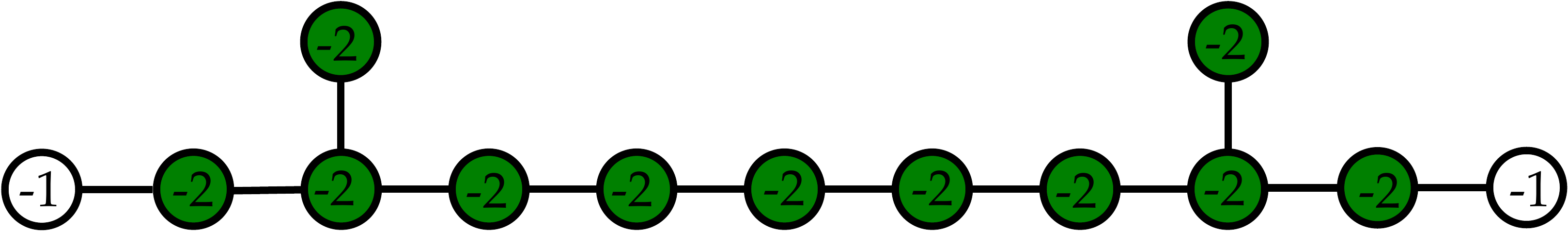}\label{D10-top-CFD}
\ee
The complete fiber geometry was determined in figure \ref{fig:D10topFib} and the geometry of the surfaces in this non-flat resolution is 
\be
S_1 \cup S_2 = \mb{F}_1\cup \hbox{Bl}_{10}\mb{F}_6\,.
\ee
Note that this is related after a number of flops to the surfaces in  \cite{Jefferson:2018irk}, which are Bl$_6\mathbb{F}_1\cup \text{dP}_6$. As emphasized repeatedly, however, our description is somewhat better suited 
 for the purposes of reading off vital physical information of the SCFTs.
This marginal CFD is extracted from figure~\ref{fig:D10topFib}, by first of all noting that the entire affine Dynkin diagram of $D_{10}$ is contained in the surface component $S_2$. These are all $(-2)$-curves and therefore marked vertices in the CFD. There is also a $(-1)$-curve $e_1$ on $S_1$, which connects the $(-2)$-node $F_0$ (labeled by $U$) and a $(-1)$-curve $e_{11}$ on $S_2$ that connects the node $F_{10}$ (labeled by $u_9$). These are unmarked $(n, g)= (-1, 0)$ vertices, which will be key in determining the CFD-transition tree.
There is no other Mori cone generator for $S_1\cup S_2$, hence the marginal CFD for the $(D_{10},I_1)$ geometry  is exactly the one in (\ref{D10-top-CFD}).

The complete CFD-tree that descends from this is shown in figure
    \ref{fig:D10FibsAll}. We note down the (generally enhanced) superconformal
    flavor symmetry $G_\text{F}$. The number of $U(1)$ flavor symmetry factors
    is given by the number of white nodes minus two, and the generators of the
    $U(1)$s are chosen to be the white curves that do not intersect any of the
    green curves. Whenever an $SU(3)$ gauge theory description exists we also
    specify the CS-level $k$. Theories with no gauge theory description are
    shaded in grey.

\noindent
{\bf Caption for figure \ref{fig:D10FibsAll}:}\\
 Inside the box: strongly coupled flavor symmetry, including the non-abelian part, which is realized in terms of the marked (green) vertices. 
Box colors: yellow: theories that are only realized as descendants of this marginal CFD;  $k$ the CS level for the $SU(3)$ gauge theory description; green: CFDs that have an alternative realization as descendants of $(E_8,SU(2))$; 
orange: $SU(2)\times SU(2)$ gauge theory description; grey: no gauge theory description. Further details for each theory are listed in the tables in appendix \ref{sec:origin}.

\item \underline{Rank two E-string $(E_8, SU(2))$}: \\
The marginal CFD for the rank two  E-string is computed to be
\be
\includegraphics*[scale =.25]{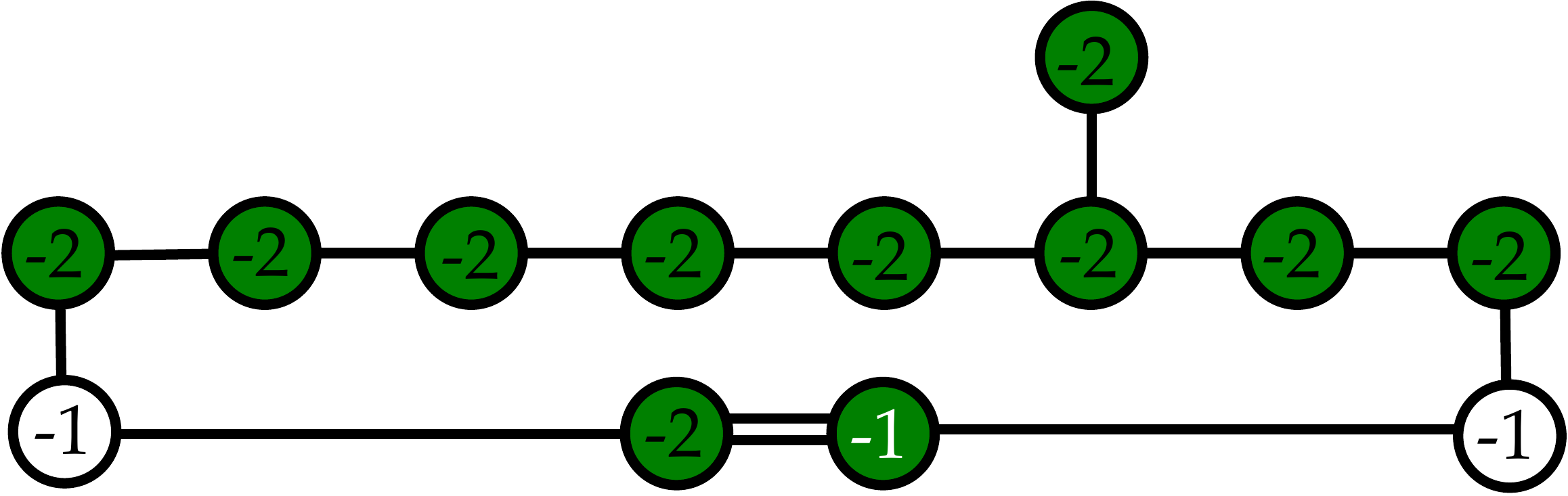}\label{E8SU(2)-top-CFD}
\ee
This was obtained from a non-flat fiber resolution, whose geometry is shown in figure \ref{f:E8SU(2)10_7}. The two non-flat surface components are identified to be 
\be
S_1 \cup S_2 =  \hbox{gdP}_3(A_1+A_1)\cup \hbox{gdP}_8(E_7+A_1)\,,
\ee
where in addition to the gdP we also specify the Lie algebra type that the $(-2)$ curves realize, inside brackets; see appendix \ref{app:gdP} for a discussion of the generalized del Pezzo surfaces.

This geometry is related to that of  \cite{Jefferson:2018irk} (where this is referred to as Bl$_9\mathbb{F}_4 \cup \mathbb{F}_0$) by a number of flops between the two non-flat fiber components $S_1$ and $S_2$. 

From this geometry, it is clear that the fibral curves of the Cartan divisors of $E_8$ and $SU(2)$ are all fully contained in $S_1\cup S_2$, and form a marked subgraph of  affine Dynkin diagrams in the marginal CFD. In addition there is a $(-1)$-curve $e_2$ on $S_1$, which connects the two affine nodes $F_0^{E_8}$ and  $F_0^{SU(2)}$. The other $(-1)$-curve $h-e_1-e_8$ on $S_2$ connects the two nodes $F_7^{E_8}$ and $F_1^{SU(2)}$. There are no other curves that generate the Mori cone of $S_1\cup S_2$. Hence we conclude that the topology of the marginal CFD is the one shown in (\ref{E8SU(2)-top-CFD}).

A subtlety here is that the rightmost marked vertex of the affine $SU(2)$ in (\ref{E8SU(2)-top-CFD}) should be interpreted as a $(-1)$-curve with multiplicity 2, see Appendix \ref{app:TopCFDsRank2E}. 
In particular, applying a CFD-transition to blow down the $(-1)$-curve to the right of this marked curve, this multiplicity 2 $(-1)$-curves maps to a multiplicity two $0$-curve, rather than a single $(-1)$-curve.
Because of this fact, the rightmost vertex of the affine $SU(2)$ can never be blown down. 

The complete CFD-tree obtained by applying CFD-transitions to this marginal CFD is shown in figure \ref{fig:E8FibsAll}. The number of $U(1)$ flavor symmetry factors is given by the number of unmarked vertices minus three.\footnote{Note that we need to subtract one more than the $(D_{10},I_1)$ case, because unlike the $I_1$ fiber there, the $I_2$ comes with an additional linear relation of type \eqref{eq:homology_relation_vert_div}.}, and the generators of the $U(1)$s  are chosen to be the white curves that do not intersect any of the green curves.

\noindent
{\bf Caption for figure \ref{fig:E8FibsAll}:}\\
 Inside the box: strongly coupled flavor symmetry, including the non-abelian part, which is realized in terms of the marked (green) vertices. 
Box colors: yellow: realized only in terms of descendants of the rank two E-string; green: models realized from $(D_{10}, I_1)$ as well; 
$k$ denotes the CS-level when an $SU(3)$ gauge theory description exists; 
orange: $SU(2)\times SU(2)$ gauge theory description; grey: no gauge theory description. Further details for each theory are listed in the tables in appendix \ref{sec:origin}.

We expect the  rank $Q>2$ E-string theories to have the same 6d flavor symmetry and the same marginal CFD as \eqref{E8SU(2)-top-CFD}. In particular, the tree-structure of descendants does not change --- only the rank distinguishes  the actual SCFTs. For instance, even though two theories have the same flavor symmetry, having a different rank means that their Coulomb branches are different, and hence must be distinguished. This is consistent with the fact that for every $Q$, there is a subset of descendants of the rank $Q$ E-string, whose Higgs branches are the $Q$-instanton of $E_{n}$ \cite{Intriligator:1997pq}. That is, for every $Q$ there is a theory with superconformal flavor symmetry $E_n \times SU(2)$, which nevertheless has a different Higgs branch depending on $Q$.
More generally, this applies to any case with different rank but same flavor symmetry of the marginal theory, where the 6d original theories are of conformal matter type. The CFD will be the same, but the rank distinguishes the Coulomb branch of these theories.

\end{enumerate}


\subsubsection{Marginal CFDs from Automorphisms}
\label{sec:CFD-automorphism}

The first task in our program for the determination of  all the descendant
SCFTs from a given 6d SCFT is to construct the marginal theory and its
associated marginal geometry. This is not always as straightforward as it may
seem; there can be ``outer automorphism twists'' which act on the tensor
branch geometry after the circle compactification, and this changes the
geometry of the singular Calabi--Yau threefold realizing the marginal
geometry. We discuss two rank two marginal theories that are obtained from 6d
in this way. 

In this paper we will not construct explicit resolutions of these singular geometries
to obtain the marginal theory, but instead we will directly propose the
marginal CFD associated to these geometries. These two theories arise as
$\mathbb{Z}_2$-twisted circle compactifications of the 6d theory that has a
one dimensional tensor branch, an $SU(3)$ gauge group, and, respectively,
twelve and six fundamental hypermultipets. The CFDs are quotients of the
CFD of the untwisted theory, and they have the structure of twisted affine
Dynkin diagrams\footnote{See \cite{Tachikawa:2011ch} for a nice summary of
twisted affine algebras.}, which is expected as the strongly coupled flavor
symmetries predicted in \cite{Jefferson:2017ahm} are $\widehat A^{\rm
tw}_{11}$ and $\widehat A^{\rm tw}_{5}$.

This proposal is supported by consistency with the following known facts
about these two marginal theories:
\begin{itemize}
  \item The reducible surface $\mathcal{S}$ and the curves inside agree with
    the explicit collections of rational surfaces given for these theories in
    \cite{Jefferson:2018irk}.
  \item The structure of the trees of the descendants of these marginal CFDs,
    which involves a detailed and involved structure of mass deformations,
    matches exactly the known trees for the descendant theories as determined in
    \cite{Jefferson:2018irk,Hayashi:2018lyv}. 
\end{itemize}

We hope to turn to the systematic construction of the marginal CFDs, which
involve outer automorphism twists and often involve geometries with a singular base, 
in future work.

\begin{sidewaysfigure}
\centering
\includegraphics*[width=\textwidth]{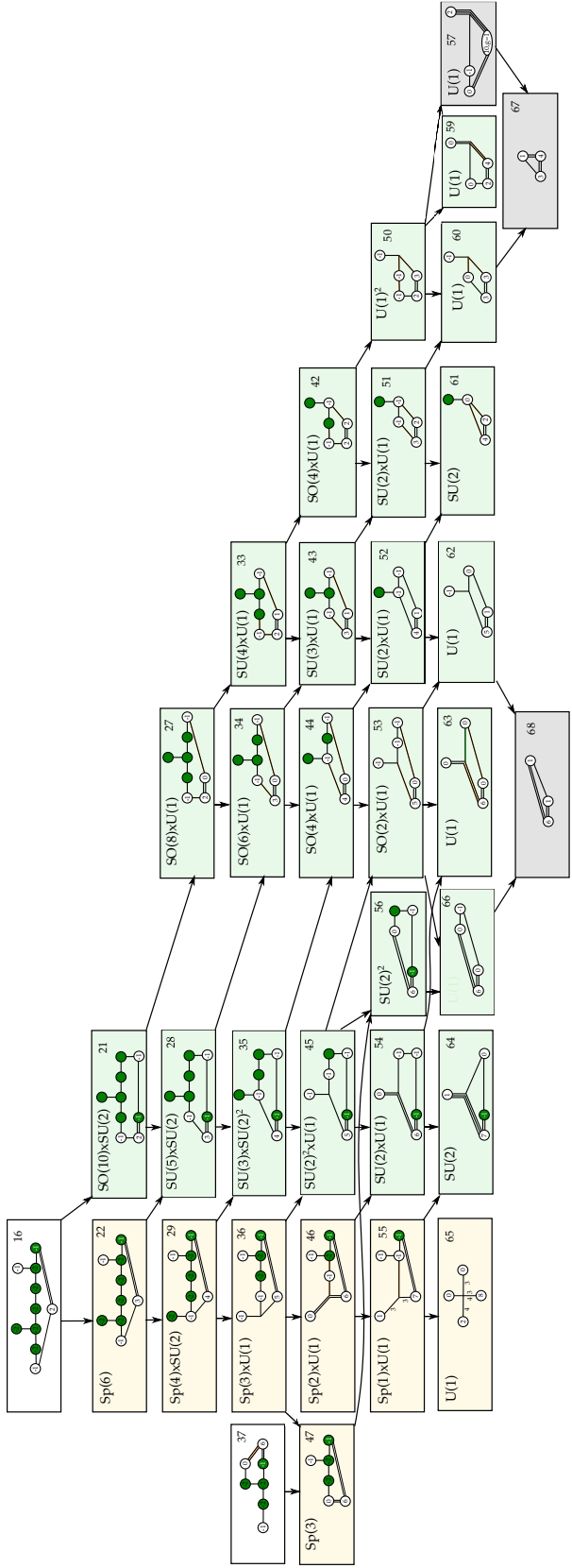}
\caption{Combined CFD tree for $SU(3)$ on $(-1)$-curve with $12$ hypers and
  $SU(3)$ on $(-2)$-curve with $6$ hypers (the marginal CFDs are shown in white boxes).
Box colors: yellow uniquely have a realization as descendants from these two marginal CFDs. The green/grey shaded boxes show models that have alternative realization in terms of the $(D_{10}, I_1)$ and $(E_8, SU(2))$ CFD-trees already; grey: no gauge theory description.  \label{fig:Model3FibsAll}}
\end{sidewaysfigure}



\begin{figure}
\centering
\subfloat[]{\includegraphics*[width= 4.5cm]{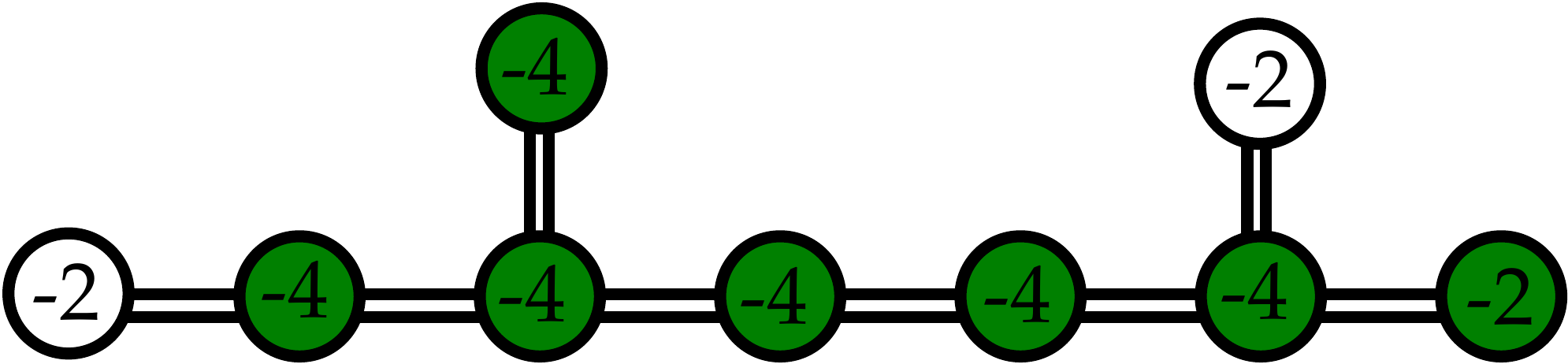}}
\\
  \subfloat[]{{\includegraphics[width=8cm]{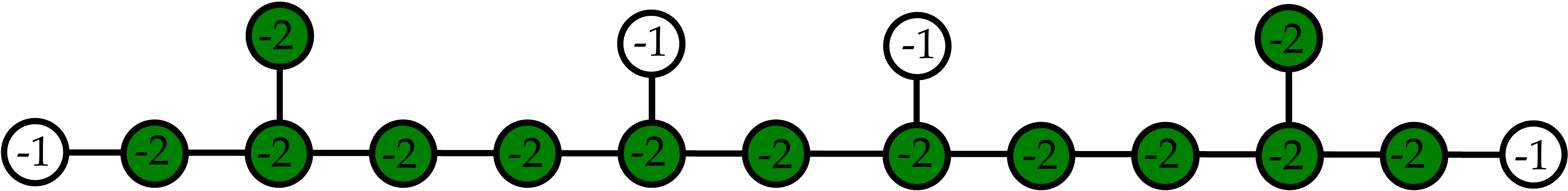}}}\\
\caption{(a) The `doubled' CFD of the marginal theory for $SU(3)$ on a
  $(-1)$-curve with $12$ hypermultiplets, which contains a twisted affine Dynkin diagram $\widehat A^{\rm tw}_{11}$. The labels here are doubled as they are obtained by folding of (b). The unfolded version of figure (a), which contains an affine $SO(24)$ Dynkin diagram.\label{fig:Model4-double}}
\end{figure}


\begin{enumerate}
	\setcounter{enumi}{2}

\item \underline{$SU(3)$ on a $(-1)$-curve with $12$ hypermultiplets}:\\
A third starting point for rank two theories is obtained by an
outer-automorphism reduction of a  6d $(1,0)$ SCFT, whose tensor branch is
described by a $(-1)$-curve with an $SU(3)$ gauge group and $12$ fundamental hypermultiplets. We propose the marginal CFD to be 
\be\label{Model3topCFD}
\includegraphics*[scale= 0.25]{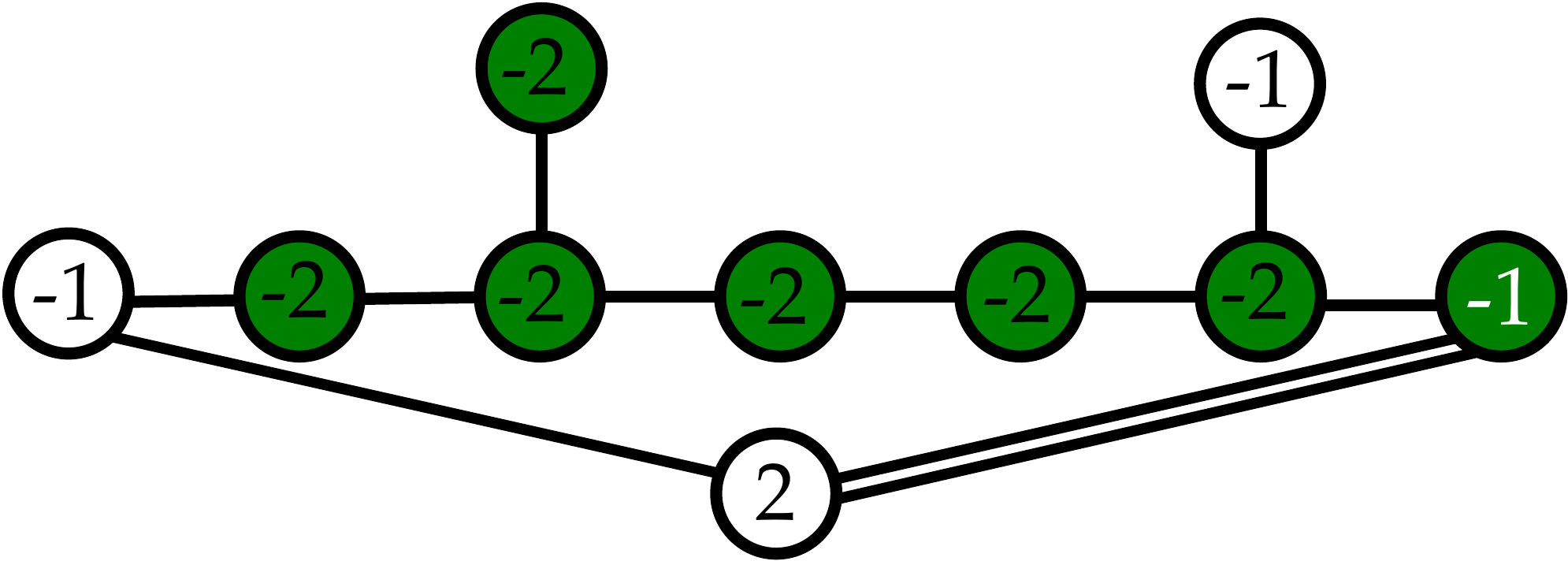} \,.
\ee
This model can be thought to arise from folding the diagram in figure \ref{fig:Model4-double} (b), which results in the 
 twisted affine Dynkin diagram of $\widehat A^{\rm tw}_{11}$\footnote{Further support for obtaining this CFD from the folding of a  $\widehat D$-type diagram is provided by a  IIA construction of this 6d theory. 
This involves a half NS5-brane stuck on top of  O8$^{-}+$ 11 D8-branes. The 8-brane stack is usually responsible for the flavor symmetry, which naively would be an $SO$-type group. However, the flavor symmetry is $SU(12)$, because of the presence of a stuck NS5. The string theory origin of this phenomenon is not yet clear \cite{Bhardwaj:2018jgp}.}. The additional $+2$ curve is obtained by consistency with the CFD descendants from other starting points and the resolution geometryThis is consistent with $\mathbb{F}_2 \cup \text{dP}_7$ in \cite{Jefferson:2018irk}. 
All the descendant CFDs are shown in figure \ref{fig:Model3FibsAll}, where we
also depict the one coming from the fourth marginal theory. At last, the
number of $U(1)$ flavor symmetry factors is given by the number of unmarked
vertices minus two.
\item \underline{$SU(3)$ on a $(-2)$-curve with $6$ hypermultiplets}: \\
This marginal theory is obtained by an outer-automorphism reduction of a 6d $(1,0)$ SCFT, which is given by a $(-2)$-curve with an $SU(3)$ gauge
group and 6 {fundamental} hypermultiplets. The marginal CFD is
\be \label{eq:CFD4}
\includegraphics*[scale = 0.25]{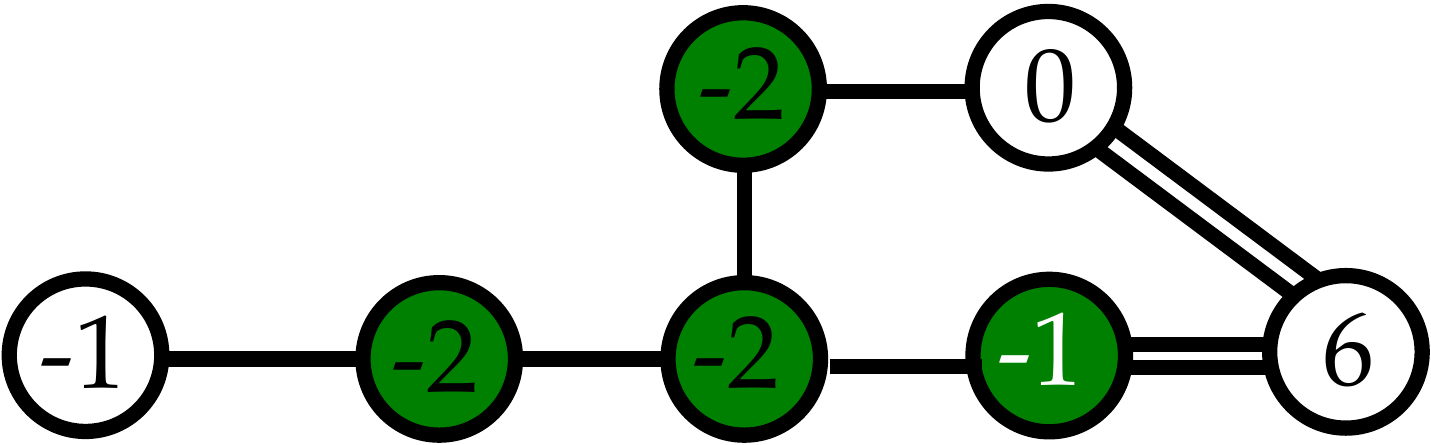}\,.
\ee
This is the geometry $\mathbb{F}_6\cup \text{dP}_4$ in \cite{Jefferson:2018irk}.
This case is similar to the
previous one, where \eqref{eq:CFD4} is the $\mathbb Z_2$ quotient of a
graph with doubled connections and self-intersection numbers, which would
correspond to the fiber in the geometric realization of $\widehat A^{\rm tw}_{5}$. 
Another consistency check is that the first descendant CFD which has $Sp(3)$ flavor symmetry has to 
match the descendant CFD from \eqref{Model3topCFD},
which is model $\# 47$ in figure \ref{fig:Model3FibsAll}, see also \cite{Jefferson:2018irk, Hayashi:2018lyv}. 
All the descendant CFDs of \eqref{eq:CFD4} are included in figure
\ref{fig:Model3FibsAll}. The number of $U(1)$ flavor factors is given
by the number of unmarked vertices minus two.
\end{enumerate}
There are three more starting points (see appendix \ref{sec:origin}), which in
total give rise to a single descendant 5d SCFT according to
\cite{Jefferson:2018irk, Hayashi:2018lyv} not contained in our combined
CFD-trees. This theory however requires $O7^+$ planes in the dual brane-web description, 
and therefore is not intrinsically geometric. 


\subsubsection{New Flavor Symmetry Predictions for Rank Two Theories}
\label{sec:new-flavor}
Our methods allow us to predict all the strongly coupled flavor symmetries of the rank two descendant 5d SCFTs, whenever they have or not have an effective low-energy gauge theory description. The gauge theory descriptions for the marginal theories can be found in appendix \ref{sec:origin}. The descendant gauge theories are obtained by successively decoupling flavor hypermultiplets. This procedure will form a subtree of our CFD-tree and is the topic of  \cite{Apruzzi:2019enx}.  

We present and summarize here some predictions of the so far unknown strongly coupled flavor symmetries for the following rank two gauge theories --- i.e. either $SU(3)$ or $Sp(2)$  gauge groups: 
\begin{equation}
\begin{array}{|c|c|c|}
\hline
\text{\# } & {\rm Gauge \; Theory} & \text{Superconformal Flavor Symmetry }G_\text{F}\cr \hline 
22 &SU(3)_{\frac{9}{2}} + 5 \mathbf F ; \ Sp(2) +  2 \mathbf{AS} + 3 \mathbf F & Sp(6)\cr
29&SU(3)_{5} + 4 \mathbf F ; \  Sp(2) + 2 \mathbf{AS} + 2 \mathbf F & Sp(4)\times SU(2) \cr 
36&SU(3)_{\frac{11}{2}} + 3 \mathbf F ; \  Sp(2) + \, 2 \mathbf{AS} +1 \mathbf F & Sp(3) \times U(1)\cr
46&SU(3)_{6} + 2 \mathbf F ; \  Sp(2)_{\pi} + 2 \mathbf{AS}  & Sp(2)\times U(1)\cr
55&SU(3)_{\frac{13}{2}} + 1 \mathbf F  & Sp(1) \times U(1)\cr
47&Sp(2)_{0} + 2 \mathbf{AS}  & Sp(3)\cr
65& SU(3)_{7} & U(1)\cr \hline
\end{array}\label{new-flavor-Sp}
\end{equation}
The $\#$ refers to the numbering of the SCFTs in the tables in appendix \ref{sec:origin}.
Moreover, we can predict the flavor symmetries of some descendant 5d SCFTs, which only have an $SU(2)\times SU(2)$ quiver gauge theory description:
\begin{equation}
\begin{array}{|c|c|c|}
\hline
\text { \# } & {\rm Gauge \; Theory} & \text{Superconformal Flavor Symmetry } G_\text{F}\cr \hline 
7& SU(2)_0 \times [SU(2) + 5 \mathbf F]  & E_8\cr
11&SU(2)_0 \times [SU(2) + 4 \mathbf F]  & SO(14)\cr
17&SU(2)_0 \times [SU(2) + 3 \mathbf F]  & SU(7)\cr
23&SU(2)_0 \times [SU(2) + 2 \mathbf F]  & SU(5)\times SU(2)\cr
30&SU(2)_0 \times [SU(2) + 1 \mathbf F]  & SU(4)\times U(1)\cr
38&SU(2)_0 \times SU(2)_0  &SU(4)\cr
39&SU(2)_0 \times SU(2)_{\pi}  &SU(3)\times U(1)\cr\hline
\end{array}\label{new-flavor-SU2SU2}
\end{equation}
Some of these were already anticipated using CFDs in \cite{Apruzzi:2019vpe}. 

Another interesting case is theory $\#$ 48 in figures \ref{fig:D10FibsAll} and \ref{fig:E8FibsAll} (see also the table in appendix \ref{sec:origin}), for which no effective gauge theory description is known, and its SCFT flavor symmetry is $SU(2) \times U(1)$.

This concludes the rank two theories. 
Two marginal CFDs in this case have a concrete geometric realizations, which we determined using non-flat resolutions.
Furthermore, the fact that the two conjectured marginal CFDs lead to descendants which perfectly fit known results including flavor symmetries lends very strong support to this approach. 
In particular, we find it encouraging that the tree structure generated from CFD-transitions agrees fully with the known RG-flow trees or flop trees in alternative constructions of these theories. 
We now turn to applying our method to the realm of the unexplored, namely to higher rank conformal matter theories and their descendant 5d SCFTs.

\subsection[\texorpdfstring{$(E_n,E_n)$}{(En,En)} Minimal Conformal Matter Theories]{\boldmath{$(E_n, E_n)$} Minimal Conformal Matter Theories}
\label{sec:EnEn}

A particularly interesting class of conformal matter theories are the exceptional $(E_n, E_n)$ minimal conformal matter theories for $n=6, 7, 8$. For $n=5$ this is the case of the rank two $(D_5, D_5)$ conformal matter that we have discussed already at length. For $n=6$ we proposed the marginal CFD and determined its descendant theories in \cite{Apruzzi:2019vpe}, which is a rank $5$ theory. The theories with $n=7$ have rank $10$, and $n=8$ has rank $21$. 

Here we will determine the marginal theories for all $n=6,7,8$ from a non-flat resolution sequence. We furthermore identify all descendant SCFTs and their flavor symmetries, and match those that have a quiver gauge theory description with the associated weakly-coupled description. 

The marginal CFDs for the three theories are shown in figure \ref{fig:EnEnMarginals} and are derived from non-flat resolutions in appendix  \ref{app:HigherRankBU}. 
The descendants are obtained in the supplementary material \cite{CFDMMA}. We find 
\be
\ba
(E_6, E_6): & \qquad 93 \text{ descendant SCFTs} \cr 
(E_7,E_7): & \qquad 56\text{ descendant SCFTs} \cr 
(E_8, E_8):& \qquad 127 \text{ descendant SCFTs} \,.
\ea
\ee

\begin{figure}
\centering
\includegraphics*[width=\textwidth]{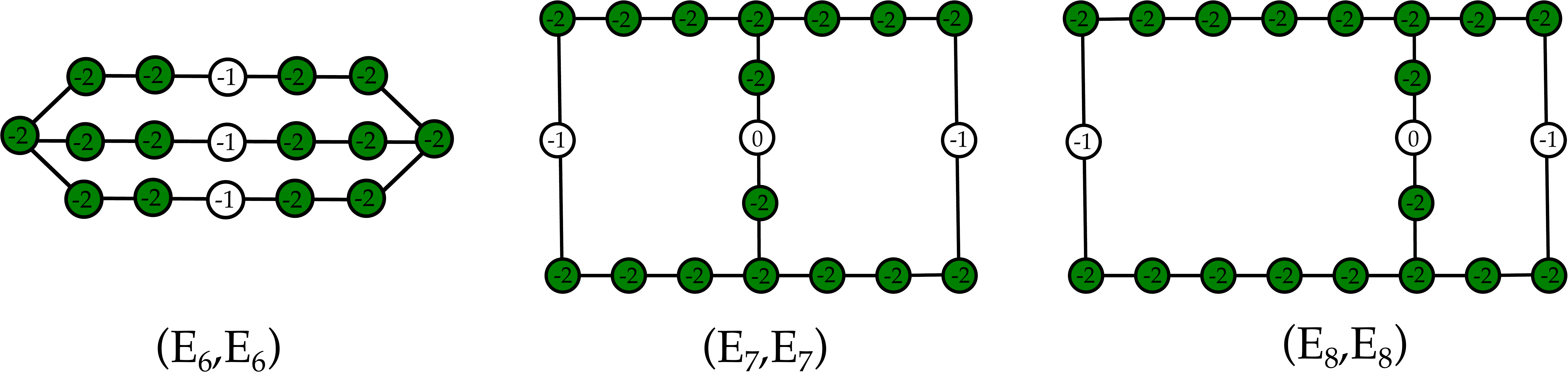}
\caption{The marginal CFDs for the $(E_n, E_n)$, $n=5,6,7$ minimal conformal matter theories.\label{fig:EnEnMarginals}}
\end{figure}

Some of these theories have a quiver description which we now discuss. However as will be clear, most of the 5d SCFTs that we found here lie outside of known weakly-coupled descriptions and it would be interesting to explore alternative descriptions to confirm our findings. 

\begin{figure}
\centering
{\includegraphics*[width=.8\textwidth]{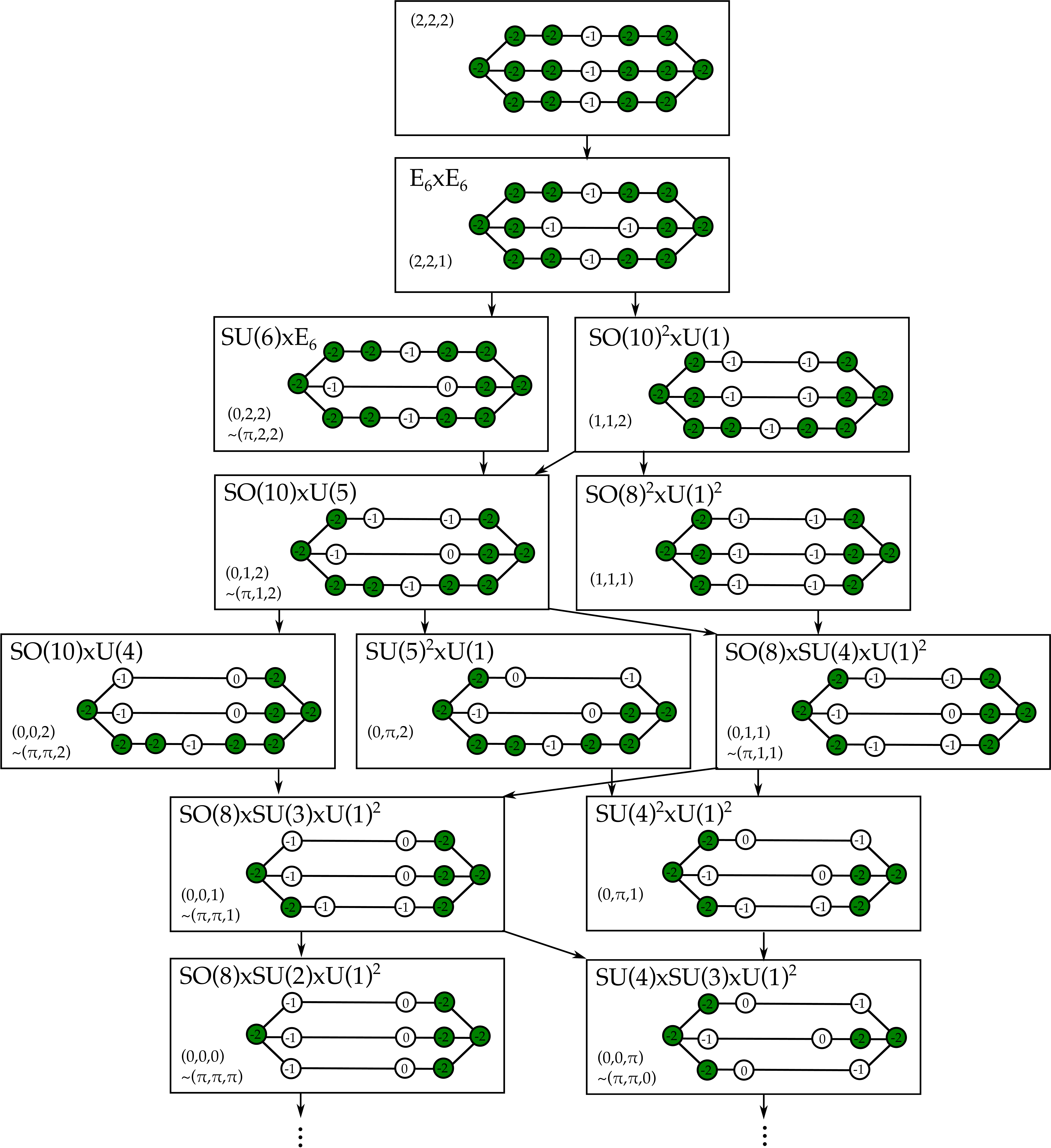}}
\caption{Subtree of the $(E_6, E_6)$ CFD-tree. The full CFD-tree can be seen in \cite{CFDMMA} and has 93 descendants.
The theories that have a known quiver description, and comprise a small subset, are shown here. 
 \label{fig:E6E6SubTrees}}
\end{figure}

The gauge theory description for the 5d marginal theory of the minimal $(E_7,E_7)$ conformal matter theory was conjectured in \cite{DelZotto:2014hpa} to be  
\begin{equation}
 [2]-SU(2)-SU(3)_{k=0}-\overset{%
\begin{array}
[c]{c}%
SU(2)_{\theta=0}\\
|
\end{array}
}{SU(4)_{k=0}}-SU(3)_{k=0}-SU(2)-[2] \,.
\end{equation}
By decoupling one-by one the flavor hypermultiplets, we argue that in total there are 9 different gauge theories, which fill a subtree of the descendant of the CFD. We label the descendant theories by the number of flavor on the left and right $SU(2)$ or their $\theta$-angle, $(q_1,q_2)$, and we also list their superconformal flavor symmetry:
\be
\begin{array}{|c|c|}
\hline
(q_1, q_2) & G_\text{F}\cr \hline 
(2,1) & E_7 \times E_7 \cr 
(2,0)\sim (2,\pi)& SO(12) \times E_7 \cr 
(1,1)  & E_6 \times E_6\cr 
(1,0) \sim  (1,\pi) & SO(10) \times E_6 \cr 
(0,0) \sim  (\pi,\pi) &SO(8) \times E_6 \cr 
(0,\pi) & SO(10) \times SO(10) \cr \hline
\end{array}
\ee
The gauge theory description for the 5d marginal theory of minimal $(E_8,E_8)$ conformal matter, \cite{DelZotto:2014hpa}, is 
\begin{equation}
 [2]-SU(2)-SU(3)_{k=0}-SU(4)_{k=0}-SU(5)_{k=0}-\overset{%
\begin{array}
[c]{c}%
SU(3)_{k=0}\\
|
\end{array}
}{SU(6)_{k=0}}-SU(4)_{k=0}-SU(2)_{\theta=0} \,.
\end{equation}
In total there are only 3 different gauge theories descendant from this effective description of the marginal theory, and form a subtree of the full descendent CFD-tree. We again label the descendant theories by the number of flavor on the left $SU(2)$ or its $\theta$-angle, $(q)$, and we also list their strongly coupled flavor symmetry:
\be
\begin{array}{|c|c|}
\hline
q & G_\text{F}\cr \hline 
1 & E_8 \times E_8 \cr 
0\sim \pi &E_7 \times E_8\cr \hline
\end{array}
\ee
Our approach allows us to predict some UV dualities between theories with different theta angles on the $SU(2)$ gauge nodes. This effect was already observed in simpler cases of $SU$-type quivers by studying the instanton operator spectrum or the superconformal index in \cite{Yonekura:2015ksa, Zafrir:2014ywa}. 
Clearly it would be highly desirable to determine alternative quiver or gauge theory descriptions for the remaining descendants and to confirm the flavor symmetry enhancement that we see in the SCFTs. 
To exemplify we show the CFDs for the theories that have a quiver description in figures \ref{fig:E6E6SubTrees}, \ref{fig:E7E7SubTrees} and \ref{fig:E8E8SubTrees}. 
The complete list of descendants are available in the supplementary material \cite{CFDMMA}.

\begin{figure}
\centering
{\includegraphics*[width=.8\textwidth]{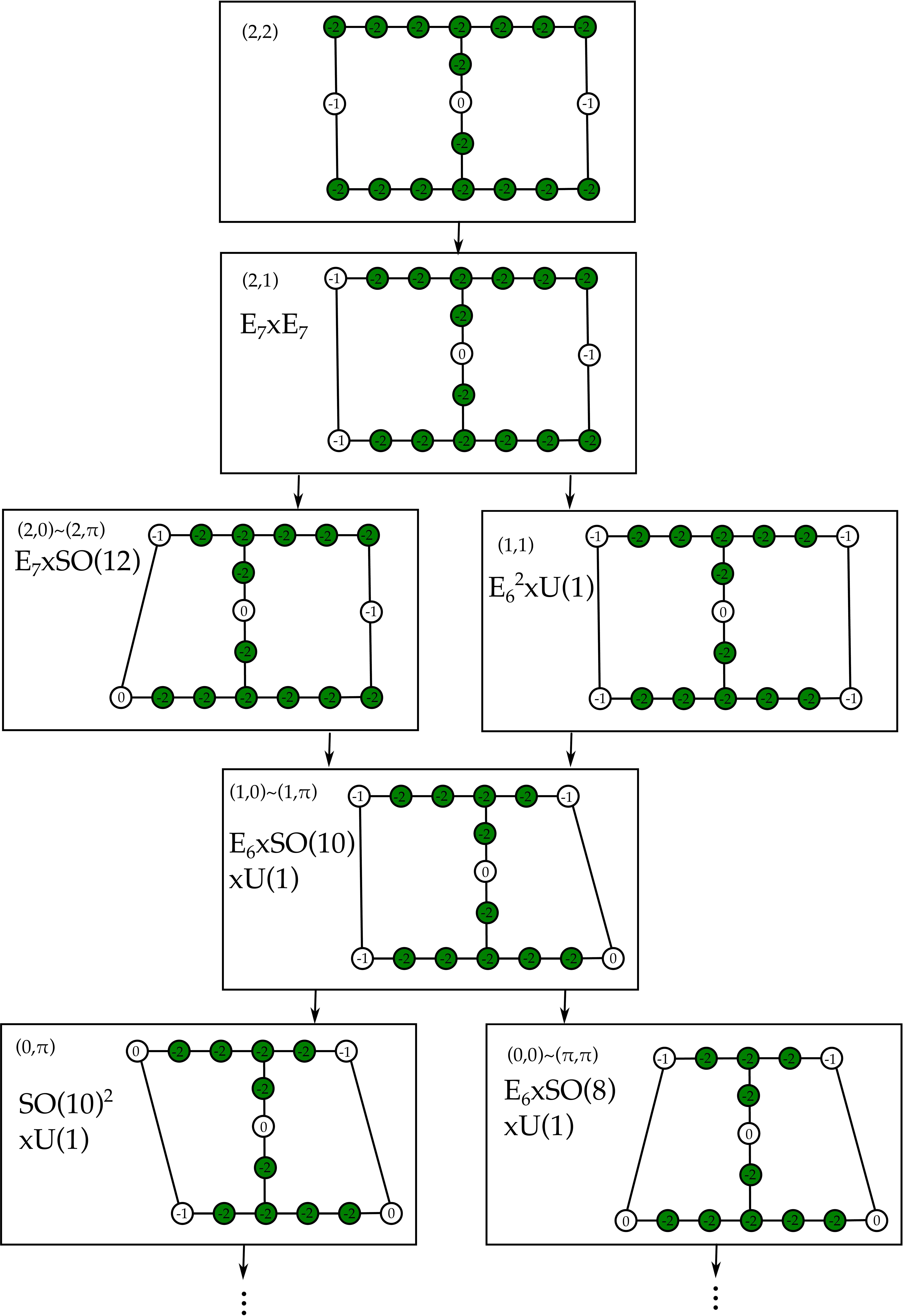}}
\caption{Subtree of the $(E_7, E_7)$ CFD-tree of theories that have a quiver description.
The full CFD-tree can be seen in \cite{CFDMMA} and has 56 descendants. 
The theories that have a known quiver description, and comprise a small subset, are shown here. 
 \label{fig:E7E7SubTrees}}
\end{figure}

\begin{figure}
\centering
{\includegraphics*[width=.4\textwidth]{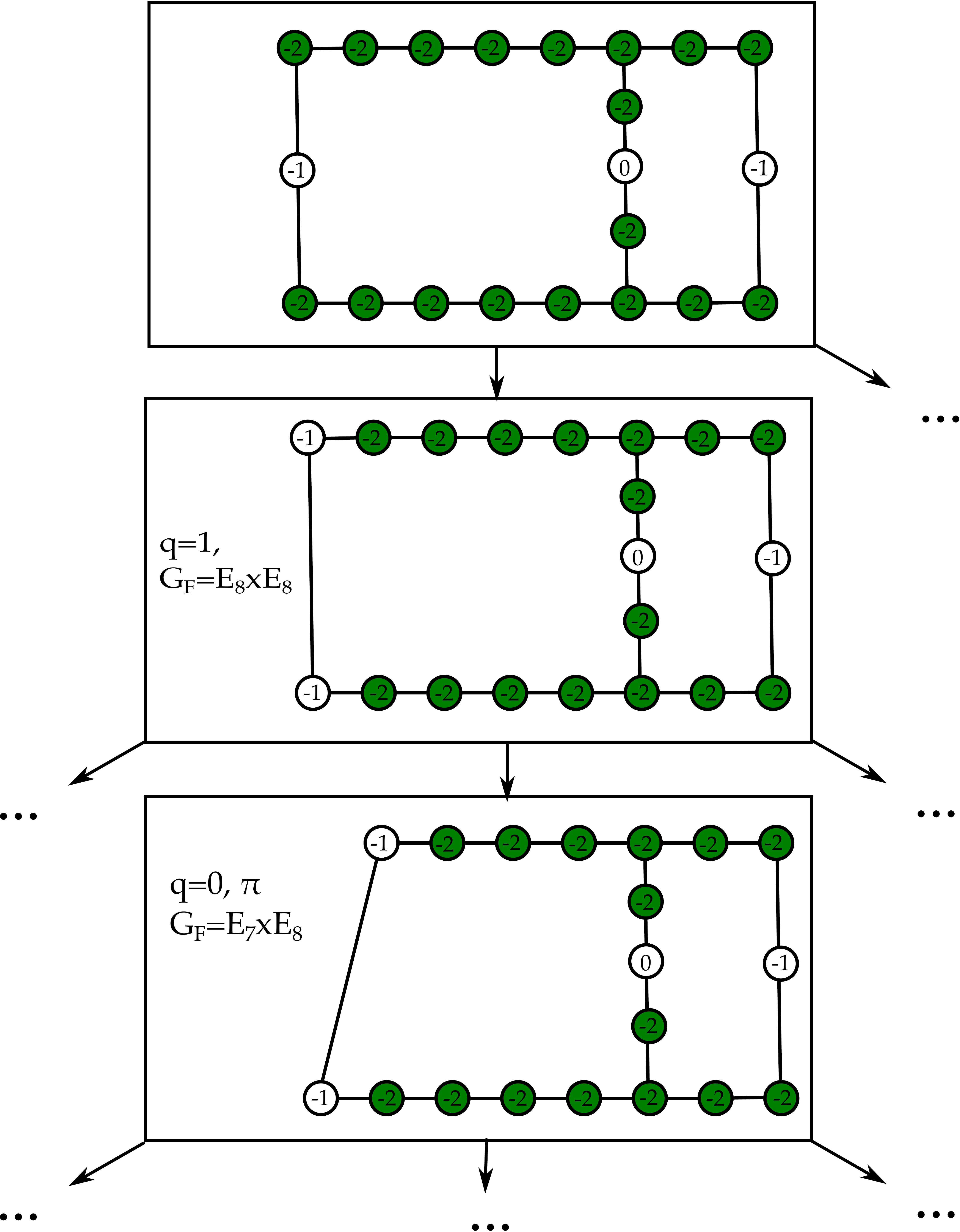}}
\caption{Subtrees of the $(E_8, E_8)$ CFD-tree of theories that have a quiver description, including the theta angles.
The full CFD-tree can be seen in \cite{CFDMMA} and has 127 descendants. 
The theories that have a known quiver description, and comprise a small subset, are shown here. 
 \label{fig:E8E8SubTrees}}
\end{figure}

\subsection[Higher Rank: \texorpdfstring{$(E_8,SU(n))$}{(E8,SU(n))}]{Higher Rank: \boldmath{$(E_8, SU(n))$}}
\label{sec:higher-rank}

In \cite{Apruzzi:2019vpe}, we obtained an infinite sequence of 5d SCFTs which descend from the $(D_k,D_k)$ minimal conformal matter using the CFD approach (a complete CFD-tree for $k=9$ can be seen in \cite{CFDMMA}). An in depth analysis of the higher rank 5d SCFTs will appear in a future work \cite{Apruzzi:2019kgb}.
 
Another class of 5d SCFTs descend from $(E_8, SU(n))$ conformal matter. These have marginal CFDs that depend on whether $n$ is even or odd, and are shown in figure \ref{fig:E8SUnMarginal}. The derivation of the marginal CFDs from a non-flat resolution is provided in appendix \ref{app:HigherRankBU}. 
 
Here we present one example that we discussed already in section \ref{sec:HigherRankBU}: the $(E_8, SU(3))$ conformal matter. This is a rank 4 theory, and the marginal CFD can be computed from the resolution to be 
\be\label{CFDE8Rank4Top}
\includegraphics[scale=.25]{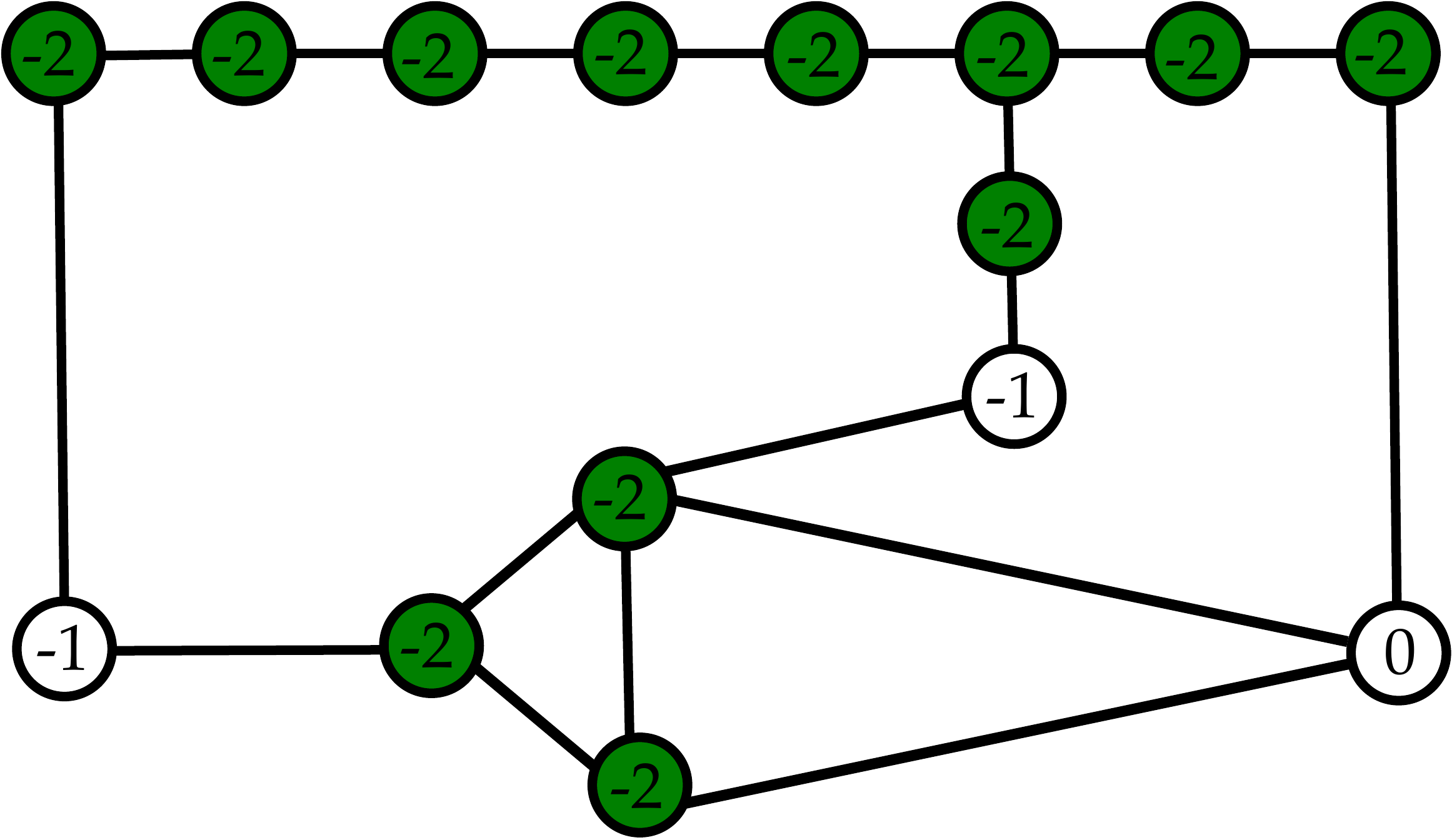} 
\ee
There are 102 descendant CFDs/SCFTs.
The example resolution that was constructed in section \ref{sec:HigherRankBU} corresponds to the CFD
\be
\includegraphics[scale= .25]{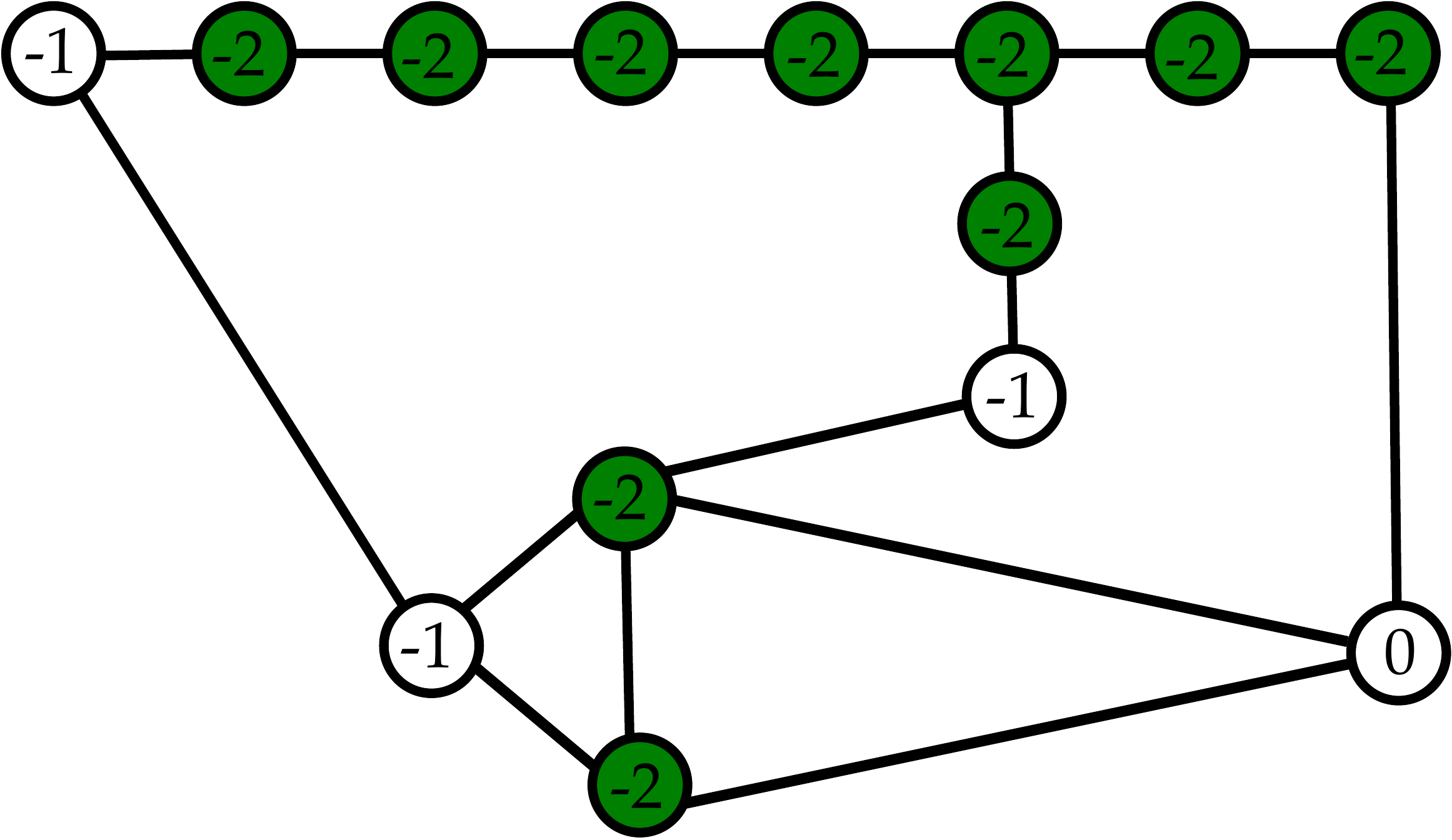}
\ee
which is a descndent obtained by transitioning with the lower left $(-1)$-vertex in (\ref{CFDE8Rank4Top}). As promised, the CFD manifestly shows the $E_7\times SU(3)$ strongly coupled flavor symmetry of the corresponding 5d SCFT.

\begin{figure}
\centering
\includegraphics*[width=6cm]{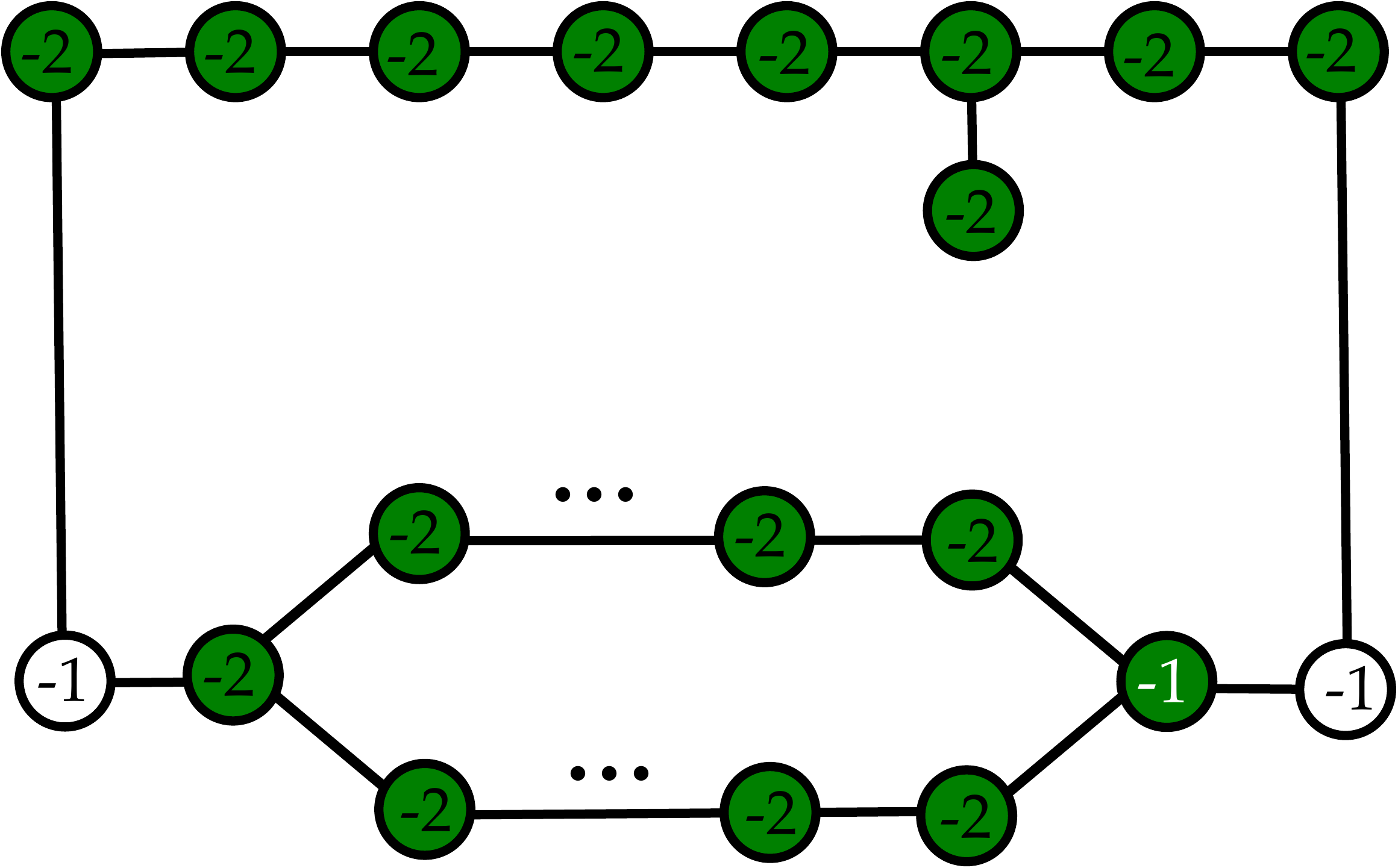}\qquad 
\includegraphics*[width=6cm]{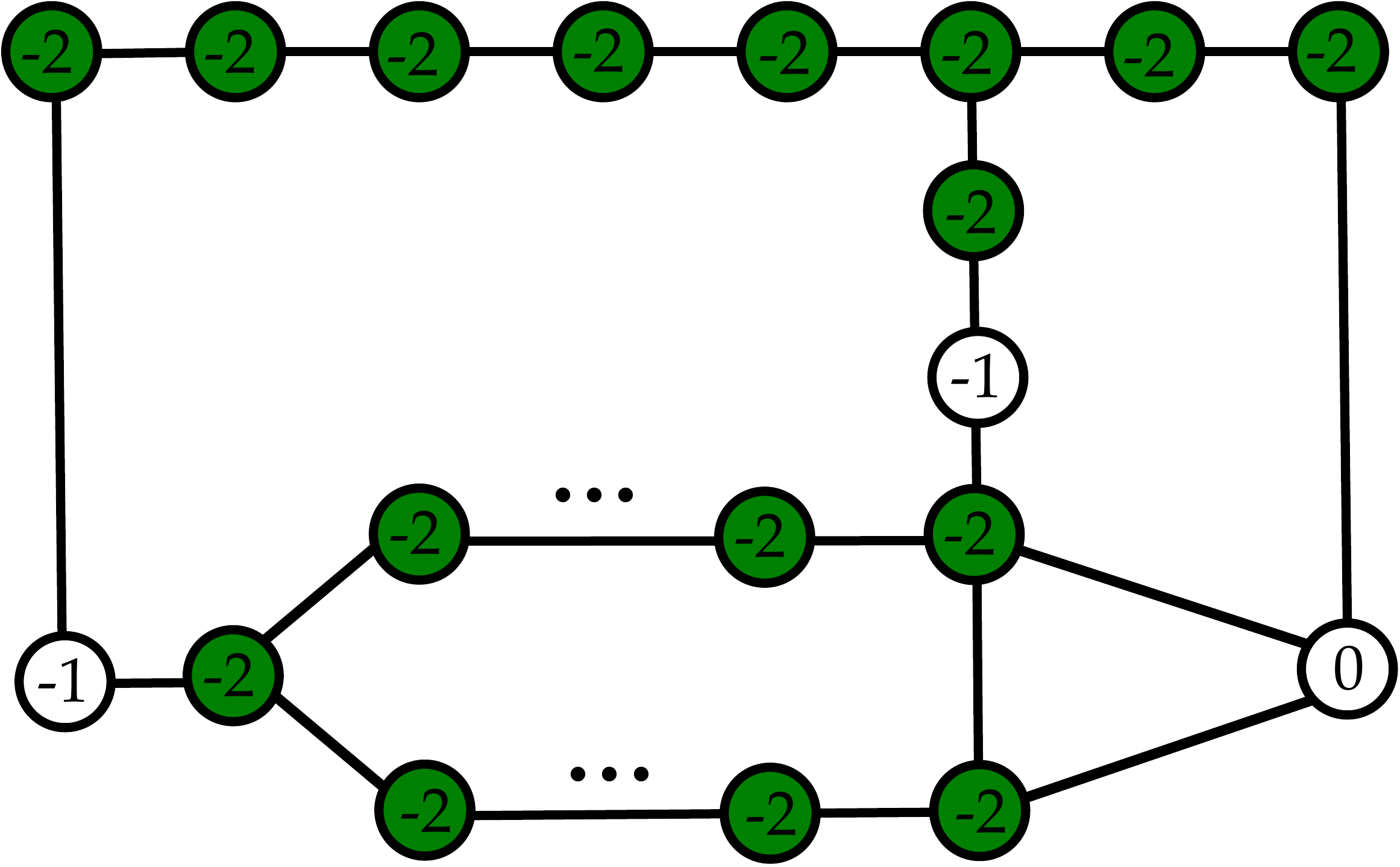}
\caption{The marginal CFDs for $(E_8, SU(2k))$ and $(E_8, SU(2k+1))$ minimal conformal matter. \label{fig:E8SUnMarginal}}
\end{figure}

Starting with the $(E_8,G_2)$ the marginal CFD is given by 
\be \label{eq:CFDE8G2}
\includegraphics[scale =.25]{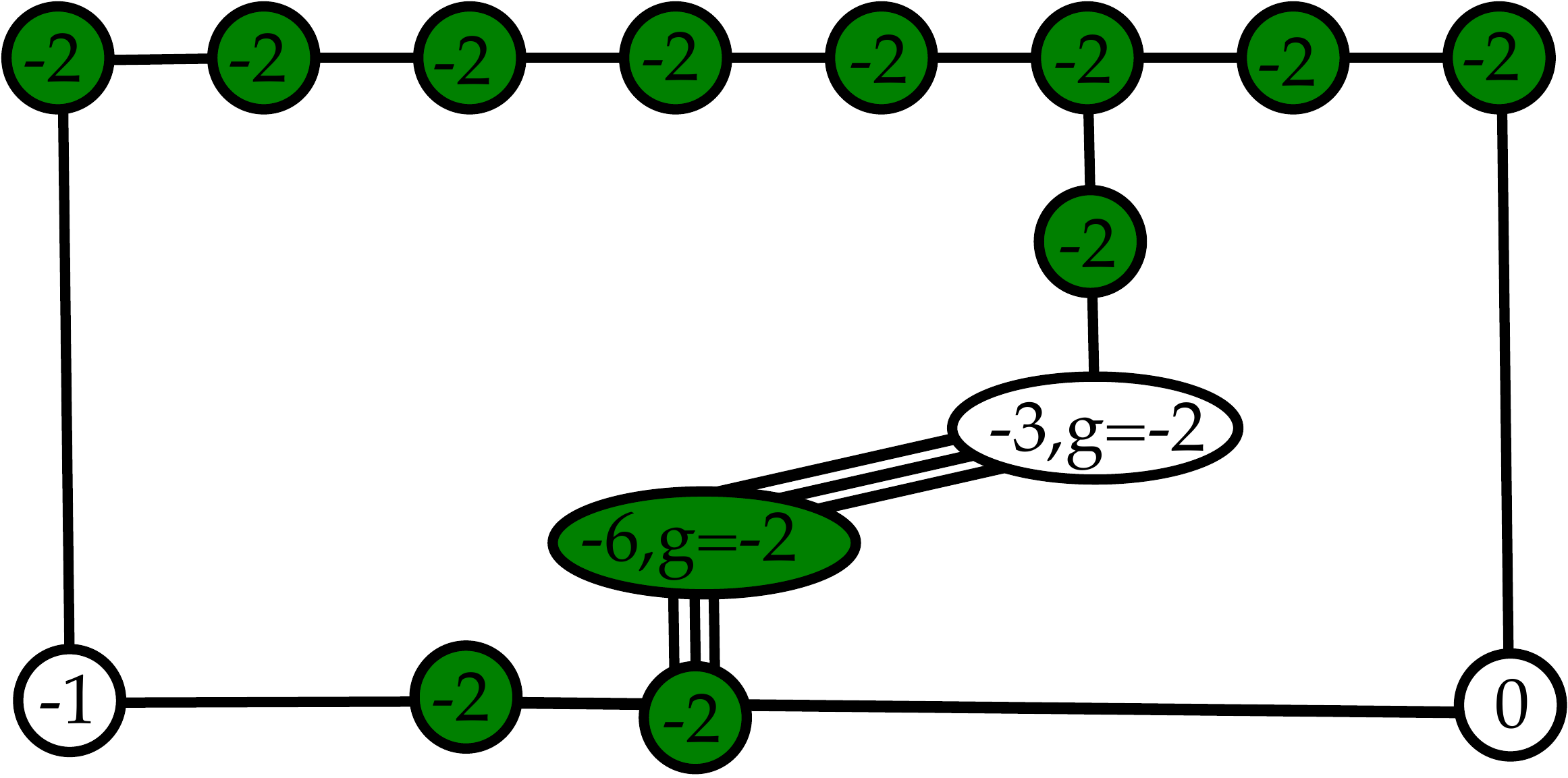} 
\ee
Note that the curves $(n, g)=(-6, -2)$ and $(-3, -2)$ are in fact multiplicity 3 curves of the type $(-2, 0)$ and $(-1, 0)$, respectively. The latter can be transitioned on, but requires removing all three curves simultaneously.

We propose a gauge theory description for this marginal theory in terms of  
\begin{equation} \label{eq:gtmarE8G2}
SU(2)_{\theta=0} - SU(4)_{k=0} - [8]\,.
\end{equation}
We will explain in \cite{Apruzzi:2019enx} how this gauge theory description can be derived from a ruling of the surface components in the resolved geometry. 
This gauge was realized as a collision in \cite{Esole:2017hlw}. 
Note also that the quiver $SU(2)_{\theta=0} - SU(4)_{k=0} - [6]$ is an effective model, which describes one of the descendants\footnote{I.e., $M=11-2$, where $M=11$ is the number total number of possible mass deformation of the marginal theory for  $(E_8,G_2)$ conformal matter.}. 
Its flavor symmetry is $SU(6) \times SU(3) \times SU(3)$ \cite{Yonekura:2015ksa, Apruzzi:2018nre}, and we can see that it comes from two consecutive transitions on the $(-1)$ curves connecting one $SU(3)$ node and the spcial $E_8$ node in \eqref{CFDE8Rank4Top}. At last, the total number of descendants for \eqref{eq:gtmarE8G2} is 24, which can be obtained by successively decoupling fundamental hypermultiplets, i.e., $m_f\rightarrow \pm \infty$. This shifts the Chern--Simons level of $SU(4)$, $k$, by $\pm \frac{1}{2}$. All the obtained descendant gauge theories fill a subtree of the CFDs/SCFTs tree arising from \eqref{CFDE8Rank4Top} and \eqref{eq:CFDE8G2}.


\section{BPS States}\label{sec:BPS}

In a 5d $\mathcal{N} = 1$ theory engineered through a compactification of
M-theory on a smooth Calabi--Yau threefold, which is the setup considered
throughout this paper, each compact two-cycle and four-cycle contributes a BPS
state.  M2-branes wrapping a holomorphic 2-cycle, in our case
$C\subset\bigcup_i S_i=\mathcal S$, with the topology of a genus $g$ Riemann
surface give rise to electrically charged particle states. On the other hand
M5-branes wrapping four-cycles, in the case studied herein the non-flat
surfaces, $S_i$, give rise to magnetically charged BPS string states.  At the
singular/SCFT limit the area of the two- and four-cycles goes to zero, and,
therefore, the BPS particles become massless, and the magnetically charged
string become tensionless. In what follows we will focus on the electrically
charged states coming from M2-branes wrapping $C\subset \mathcal S$, using the
general results  of \cite{Gopakumar:1998jq, Witten:1996qb, Kachru:2018nck}.

\subsection{BPS States from wrapped M2-branes}

The 8 real supercharges of the 5d $\mc{N}=1$ super-Poincar\'{e} algebra
transform as $2\cdot (1/2,0)\oplus 2\cdot (0,1/2)$ of the 5d little group
$SO(4)=SU(2)_L\times SU(2)_R$. On the M2-brane worldvolume two complex
supercharges $2\cdot (0,1/2)$ are broken. These two supercharges do not
annihilate the ground state anymore, and they consist of 4 real fermionic
degrees of freedom. In particular, by acting on a fermionic ground states,
these degrees of freedom reorganize into a  half-hypermultiplet
\begin{equation}
  H_0=\left(0,\frac{1}{2}\right)\oplus 2\left(0,0\right)\, ,
\end{equation}

More precisely, the BPS states are zero-modes associated to the M2-brane
worldvolume theory reduced on the genus $g$ curve $C\subset \mathcal S$. In
particular the theory has $8$ fermions and $8$ scalars corresponding to the
directions perpendicular to the brane. The fermionic zero-modes pair with the
bosonic ones, whose quantum numbers are specified by representations of the
little group $SO(4)=SU(2)_L\times SU(2)_R$. If we assume that the Riemann
surface is not degenerate, the BPS states consist of two different
contributions. The first one is given by the zero-modes coming from the 0-form
and the 1-form on $C$ \cite{Gopakumar:1998jq},
\begin{equation}
  \left[\left(0,\frac{1}{2}\right)\oplus 2\left(0,0\right)\right]^{g_C+1}\, .
\end{equation}
The second contribution comes from the modes associated to the deformation of
$C$ inside the Calabi--Yau threefold, or, in other words, the moduli space
$\mathcal M_C$. In terms of representations of the little group these states
are given by 
\begin{equation}
  \sum_{j=0}^{\frac{n}{2}} a_j [(j,0)]\,,
\end{equation}
where $n\equiv {\rm dim_{\mathbb C}}(\mathcal M_C)$, and the $a_j$ are
gradings associated to the $SU(2)_L$ (Lefschetz) decomposition of the
cohomology groups of $\mc{M}_C$.  Summing up, the total contribution for the
electrically charged BPS states coming from an M2-brane wrapping  curve,
$C\subset\bigcup_i S_i$, of genus $g_C$ is \cite{Gopakumar:1998jq}:
\begin{equation} \label{eq:BPSsttot}
  \left[\left(0,\frac{1}{2}\right)\oplus 2\left(0,0\right)\right]^{g_C+1} \otimes \left(\sum_{j=0}^{\frac{n}{2}} a_j [\left( j, 0  \right) ]\right).
\end{equation}
In order to compute the $a_j$, we need to look at the four unbroken
supercharges transforming in the representation $2\cdot (1/2,0)$ of $SO(4)$,
which act as the four differential operators
$\partial,\bar{\partial},\partial^*$ and $\bar{\partial}^*$ on the cohomology
of $\mc{M}_C$, $H^{p,q}(\mc{M}_C)$, \cite{Witten:1996qb}. For example, the
generator $J_3$ of $SU(2)_L$ acts on a differential form $\omega$ as 
\be \label{eq:SL2act}
J_3\omega=\frac{(p+q-\mathrm{dim}_{\mb{C}}(\mc{M}_C))}{2}\omega\ ,\ \omega\in H^{p,q}(\mc{M}_C)\,.
\ee
The differential form in $H^{p,q}(\mc{M}_C)$ will reorganize according to
representation of $SU(2)_L$, labelled by $j$.\footnote{For each $j$ we will
  have $(2j+1)$ values of
  $-j\leq\frac{(p+q-\mathrm{dim}_{\mb{C}}(\mc{M}_C))}{2}\leq j$.} The
  multiplicity at each $j$ is given by the number of zero-modes, which are in
  a full representation of helicity $j$, i.e.
\begin{equation} \label{eq:aj}
a_j=h^{p,q}(\mc{M}_C)\, .
\end{equation}

We will be interested in two cases in this paper, which are the BPS states
that arise from M2-branes wrapping genus zero curves with normal bundle
$\mathcal{O}(-1) \oplus \mathcal{O}(-1)$ or $\mathcal{O}(0) \oplus
\mathcal{O}(-2)$ inside of the Calabi--Yau threefold. These states have been
analyzed by Witten \cite{Witten:1995zh} who showed that they are,
respectively, spin $0$ and spin $1$ states in the sense of (\ref{eq:BPSsttot}).
In the smooth Calabi--Yau geometry, before we take the SCFT limit where the
volume of the surfaces is taken to zero, the former states are
hypermultiplets, whilst the latter are vector multiplets.

In order to determine the BPS states that we are interested in directly from
the CFD data we will assume that each vertex, or curve, in the CFD can be
written as a complete intersection between two divisors in the Calabi--Yau
threefold. One of these divisors we will require to be the reducible surface,
$\mathcal{S}$. As we will be interested in genus zero curves for the
determination of the BPS states it is necessary to first determine the genus
of the linear combination of vertices. This can be computed recursively
through the formula
\begin{equation}
  g(C + C^\prime) = g(C) + g(C^\prime) + C \cdot_\mathcal{S} C^\prime - 1 \,.
\end{equation}

Using the Calabi--Yau condition together with the complete intersection
property one can see that the normal bundle to any genus zero curve, $C$, that
is a non-negative linear combination of vertices in the CFD, can be written as
\begin{equation}
  N_{C/Y} = \mathcal{O}(C \cdot_\mathcal{S} C) \oplus \mathcal{O}(-2 - C
  \cdot_\mathcal{S} C) \,.
\end{equation}
Thus we can see that spin $0$ states, which come from M2-branes wrapping
curves that have normal bundle $\mathcal{O}(-1) \oplus \mathcal{O}(-1)$ inside
the Calabi--Yau threefold come from $C$s such that $C \cdot_\mathcal{S} C =
-1$, such $C$ will be referred to as $(-1)$-curves. The spin $1$ states,
which come from curves with normal bundle $\mathcal{O}(0) \oplus
\mathcal{O}(-2)$ have two origins, either $C \cdot_\mathcal{S} C = -2$ or $C
\cdot_\mathcal{S} C = 0$. We will not consider the former as they give rise to
``decoupled states'' \cite{Taki:2014pba,Bhardwaj:2018vuu} that decouple in the
SCFT limit, and moreover, the partition function including such states do not
preserve the $G_\text{F}$ flavor symmetry \cite{Taki:2014pba}. The latter, which we
refer to as $(0)$-curves give rise to the spin $1$ BPS states of the SCFT.

For the representation of such states under the superconformal flavor symmetry
$G_\text{F}$, one merely needs to compute the intersection numbers $C\cdot D_i$,
where $D_i$s are various divisors that generate the non-abelian and abelian
parts of $G_\text{F}$. In the language of CFD, these numbers can be computed from the
intersection relation of the nodes that correspond to $C$ and $D_i$.

To get the highest weight state for a representation of the non-abelian part,
we require that such a curve $C$ does not intersect negatively with any $D_i$
that generates the non-abelian flavor symmetry (equivalently, the fully
wrapped Cartan nodes, or flavor curves, $F_i$ in the CFD). We will list all the genus zero curves
generating spin $0$ and spin $1$ BPS states for all the 5d SCFTs considered in
this paper.

In conclusion, the strategy for determining a part\footnote{Throughout this
  paper we will generally state that we are ``determining the BPS states'' of
  a particular SCFT. By this we will refer only to those BPS states that are
  spin $0$ or spin $1$ and arise from M2-branes wrapping genus zero curves
  with self-intersection either $0$ or $-1$, as described in this paragraph.}
  of the BPS spectrum of the SCFT is as follows. We consider a curve $C$
  formed as a non-negative linear combination of vertices of the CFD
  (associated to the SCFT of interest). There are then two cases in which we
  are interested
\begin{equation}
  \begin{aligned}
    g(C) = 0 &\text{ and } C \cdot_\mathcal{S} C = -1 & \quad \Rightarrow \quad
    \text{Spin $0$ BPS state} \,, \cr
    g(C) = 0 &\text{ and } C \cdot_\mathcal{S} C = 0 & \quad \Rightarrow \quad
    \text{Spin $1$ BPS state} \,.
  \end{aligned}
\end{equation}
The former are referred to as $(-1)$-curves, and the latter as $(0)$-curves.
Each such curve $C$ is associated to a weight of the superconformal flavor
symmetry by considering the intersection numbers $C \cdot_\mathcal{S} F_i$,
where, again, the $F_i$ are the marked vertices in the CFD. The collection of
all such curves, $C$, which give rise of BPS states of the same spin, and with
their corresponding weights form themselves into full representations of the
superconformal flavor symmetry. To determine the appearing representations it
is enough to determine the curves, satisfying the above conditions, that are
associated to weights within the fundamental Weyl chamber of the flavor
algebra; in short, we need to obtain the $C$ associated to spin $0$ and spin
$1$, as above, which further satisfy 
\begin{equation}
  C \cdot_\mathcal{S} F_i \geq 0 \,.
\end{equation}

\subsection{BPS States of Rank One SCFTs}
\label{BPS:rank-1}

We begin by determining the BPS states associated to the ten rank one
interacting 5d $\mathcal{N} = 1$ SCFTs, for which the CFDs were written down
in figure \ref{fig:Rank1CFDTree}. The BPS states are summarized in table
\ref{t:Rank1BPS}. The Lie group representations follow the conventions of
\cite{Yamatsu:2015npn}. The $U(1)$ charges are written in the subscripts
unless the flavor symmetry group $G_\text{F}$ is entirely abelian. The spin
$0$ and $1$ spectrum from the genus $0$ curves matches the results in
\cite{Huang:2013yta}, which are derived using
orthogonal methods. 

As a representative example we will consider the rank one theory with
superconformal flavor symmetry group $E_8$, which is associated to the CFD
\begin{equation}
  \includegraphics[scale=0.2]{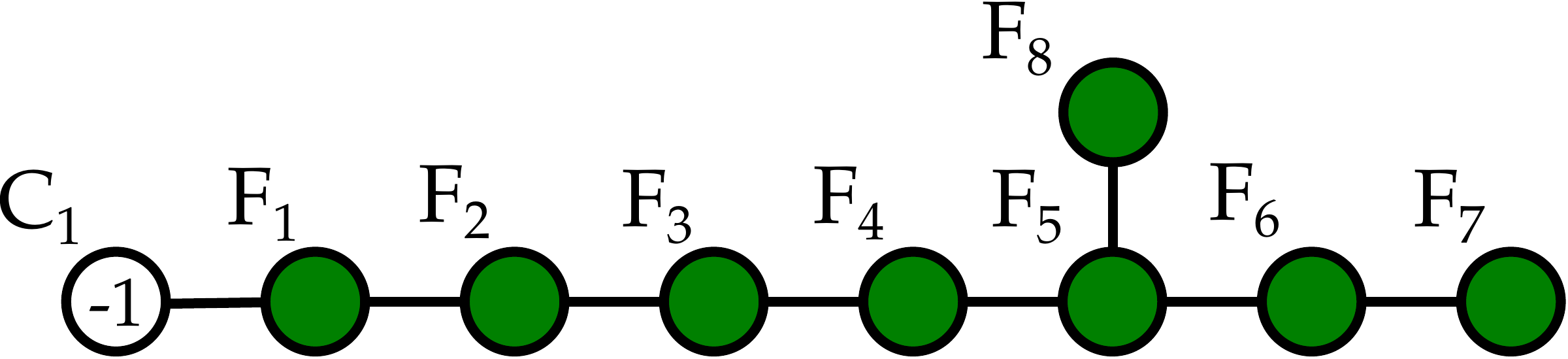} \,, 
\end{equation}
where we have introduced the labels $C_1$ and $F_i$ for the, respectively
unmarked and marked, vertices. There is a single $(-1)$-curve inside of the
fundamental Weyl chamber
\begin{equation}
  C^\text{spin $0$} = C_1 \,,
\end{equation}
which is associated to the weight $(1,0,0,0,0,0,0,0)$ of the $E_8$ flavor
group. This is the highest weight of the $\bm{248}$ representation of $E_8$. 

Similarly, the only $(0)$-curve that corresponds to a highest weight vector is
\begin{equation}
  C^\text{spin $1$} = 2C_1 + 2F_1 + 2F_2+2F_3+2F_4+2F_5+F_6+F_8  \,,
\end{equation}
which corresponds to the weight vector $(0,0,0,0,0,0,1,0)$ of $E_8$. The
M2-brane wrapping mode over such curve gives rise to a spin $1$ state in the
representation $\mbf{3875}$.

We summarize in table \ref{t:Rank1BPS} our results for all rank one theories.
In particular we present the CFDs, the weakly-coupled gauge theory
description, the strongly coupled flavor symmetry and the spin $0$ and $1$ BPS
states coming from $g_C=0$ holomorphic curves.


\subsection{BPS States of Rank Two SCFTs}
\label{sec:classrank2}

In this section we consider some representative examples of the application of
the above given method for the determination of the BPS spectra in the case of
the rank two 5d SCFTs. We list all of the rank two SCFTs with their CFD,
flavor symmetry, gauge theory descriptions and BPS states in the tables in appendix
\ref{app:Rank2Classification}.

Consider first the following illustrative example, where we analyze the BPS
states from the CFD (\ref{CFDBU3}). We label the nodes as below:
\be
\includegraphics[height=2.5cm]{CFD-E8-15-labels.pdf}\label{CFDBU3label}
\ee
The rational $(-2)$-curves $F_1,\cdots,F_5$ generate the $SU(6)$ flavor
symmetry and the $(0)$-curve $C_1$ generates the $U(1)$ flavor symmetry.  The
rational $(-1)$-curves in (\ref{CFDBU3label}) are $C_2$ and $C_3$, which
correspond to the highest weight vector of the representation $\mbf{6}$ and
$\overline{\mbf{15}}$ of $SU(6)$ respectively. Their $U(1)$ charges can be
read off from
\be
C_2\cdot C_1=1\ ,\ C_3\cdot C_1=0\,.
\ee
Hence the M2-brane wrapping modes give rise to 5d hypermultiplets in the
representations $\mbf{6}_1$ and $\overline{\mbf{15}}_0$.  The rational
$(0)$-curves are $C_1$, $C_2+F_1+F_2+F_3+F_4+C_3$ and $2C_3+2F_4+F_3+F_5$,
which are in the representations $\mbf{1}_0$, $\bar{\mbf{6}}_1$ and
$\mbf{15}_0$ of $SU(6)\times U(1)$ respectively, which gives rise to the
spin $1$ BPS spectrum. 

We also analyze one of the more complicated examples here, which is the 5D
SCFT with flavor symmetry $G_\text{F}=E_8\times SU(2)$. We label the nodes in the CFD
as follows:
\be
\includegraphics[height=2.5cm]{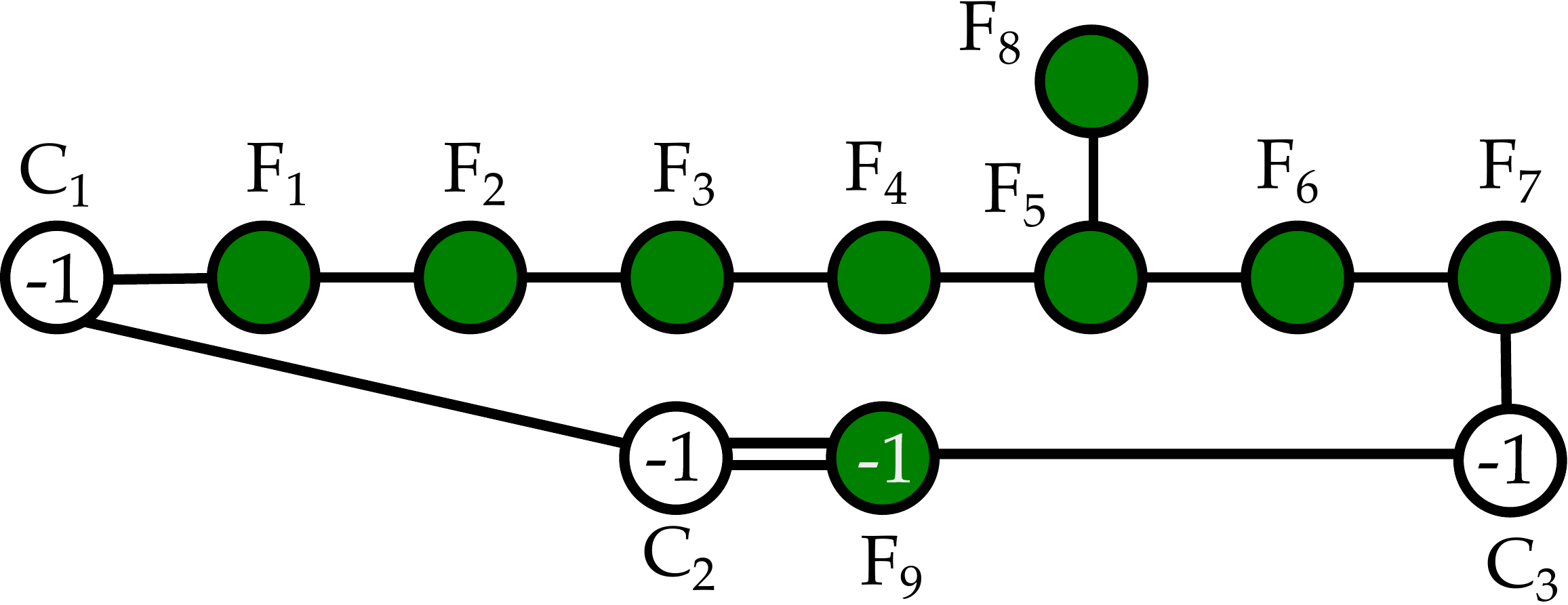}\label{CFDE8SU2label}
\ee
Notice that the Cartan node $F_9$ is actually a $(-1)$-curve with weight 2,
which is then effectively fully wrapped and contributes to the non-abelian
flavor symmetry. We list the genus-0 curves contributing to the spin-0/1 BPS
spectrum in table~\ref{t:E8SU2spin0} and table~\ref{t:E8SU2spin1}
respectively, with their genus, self-intersection number, intersection numbers
with the Cartan nodes $(F_1,\dots,F_9)$ and the representation under
$G_\text{F}=E_8\times SU(2)$.

\begin{table}
\centering
\begin{tabular}{|c|c|c|c|c|}
\hline
\hline
curve $C$ &  {\small $(C\cdot F_i)\ ,(i=1,\dots,9)$} & Rep. \\ \hline
$C_1$ & $(1,0,0,0,0,0,0,0,0)$ & $(\mbf{248},\mbf{1})$\\
$C_2$ & $(0,0,0,0,0,0,0,0,2)$ & $(\mbf{1},\mbf{3})$\\
$C_3$ & $(0,0,0,0,0,0,1,0,1)$ & $(\mbf{3875},\mbf{2})$\\
\hline
\hline
\end{tabular}
\caption{The rational $(-1)$-curves giving rise to spin $0$ BPS states on the CFD of the SCFT with $G_\text{F}=E_8\times SU(2)$.}\label{t:E8SU2spin0}
\end{table}

\begin{table}
\centering
\begin{tabular}{|c|c|c|c|c|}
\hline
\hline
curve $C$ &  {\small $(C\cdot F_i)\ ,(i=1,\dots,9)$} & Rep. \\ \hline
{\small $C_1+C_2$} & $(1,0,0,0,0,0,0,0,2)$ & $(\mbf{248},\mbf{3})$\\
{\small $2F_1+2F_2+2F_3+2F_4+2F_5+2F_6+F_8+2C_1$} & $(0,0,0,0,0,0,1,0,0)$ & $(\mbf{3875},\mbf{1})$\\
{\small $F_2+2F_3+3F_4+4F_5+3F_6+2F_7+2F_8+C_1+C_3$} & $(2,0,0,0,0,0,0,0,1)$ & $(\mbf{27000},\mbf{2})$\\
{\small $F_1+F_2+F_3+F_4+F_5+F_6+F_7+C_1+C_3$} & $(0,0,0,0,0,0,0,1,1)$ & $(\mbf{147250},\mbf{2})$\\
{\small $F_4+2F_5+2F_6+2F_7+2F_8+2C_3$} & $(0,0,1,0,0,0,0,0,2)$ & $(\mbf{2450240},\mbf{3})$\\
\hline
\hline
\end{tabular}
\caption{The rational $(0)$-curves giving rise to spin $1$ BPS states on the CFD of the SCFT with $G_\text{F}=E_8\times SU(2)$.}\label{t:E8SU2spin1}
\end{table}

As the final example, we discuss the case of the 5D SCFT with flavor symmetry $G_\text{F}=Sp(6)$, with the following CFD:
\be
\includegraphics[height=2cm]{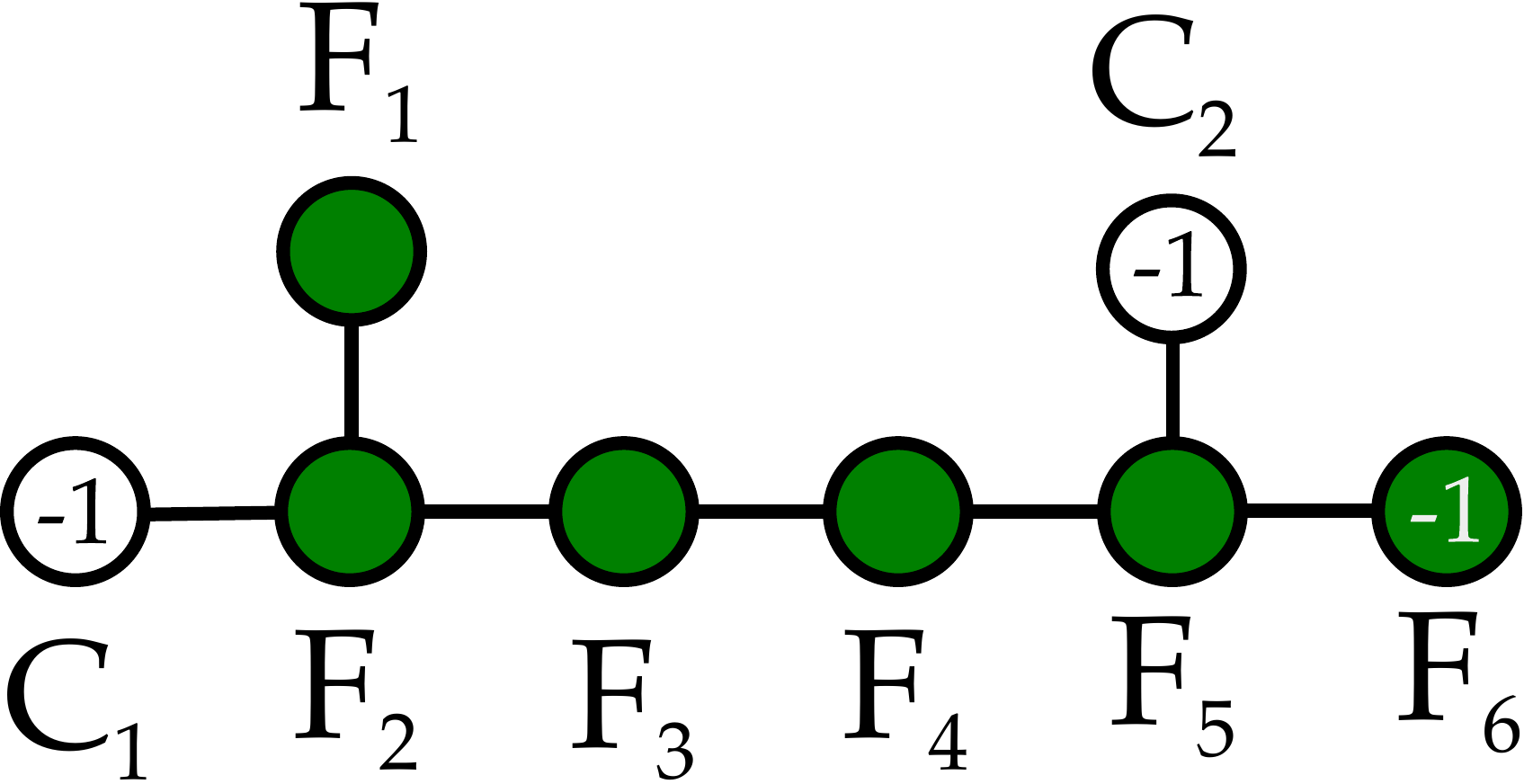}\label{CFDModel3label}
\ee
Note that the right most Cartan node (long node) of the $Sp(6)$ is in fact a
$(-1)$-curve with multiplicity $2$, and the intersection number between $F_6$ and
$F_5$ should be $1$ if $F_6$ is interpreted to be a $(-1)$-curve. Only when this
is true, can we  match the CFDs from the starting point $(E_8,SU(2))$ and the
Model 3, see the 5D SCFT with $G_\text{F}=SO(10)\times SU(2)$ and
$SU(3)_{7/2}+5\mbf{F}$ gauge theory description for example.

Then the curves giving rise to spin $0$ BPS states are the rational curves
$C_1$ and $C_2$, which are the highest weight vector of the representations
$\mbf{65}$ and $\mbf{572}$ respectively. The curves giving rise to spin-1 BPS
states are the rational curves $F_1+2F_2+F_3+2C_1$ and
$C_1+F_2+F_3+F_4+F_5+F_6+C_2$, which give rise to $\mbf{429}$ and $\mbf{4576}$
representations respectively.


\section{Conclusions and Outlook}
\label{sec:Conclusions}

In this work we have put forward a description of 5d ${\cal N}=1$ SCFTs in terms of graphs that encodes relevant mass deformations, superconformal flavor symmetries and certain BPS states.
The underlying structures are founded upon the realization of 5d SCFTs in M-theory on elliptic Calabi--Yau threefolds $Y$ with non-minimal singularities.
Associated to an elliptic fibration, there is a 6d ${\cal N} = (1,0)$ SCFT (which is obtained by compactifying F-theory on $Y$), whose circle reduction including holonomies in the flavor symmetry yields 5d SCFTs on their Coulomb branch.
Geometrically, the latter is described by topologically distinct configurations of compact surfaces ${\cal S} = \bigcup_j S_j$ and ruled non-compact divisors $F_i \hookrightarrow D_i$ that resolve the non-minimal singularity.
The limit where ${\cal S}$ collapses to a point --- which by construction exists as a partial resolution of the singularity --- corresponds to the origin of the Coulomb branch, where the strongly coupled SCFT lives.

While there can be in general many birationally equivalent geometries that realize the same 5d SCFT in this fashion, the key aspect of our approach is to manifestly keep track of the flavor symmetries. This is achieved by tracking those fibral curves $F_i$ of the non-compact divisors $D_i$ that are contained within $\mathcal{S}$. In a given resolution geometry, these so-called flavor curves intersect in the Dynkin diagram of the 5d superconformal flavor group $G_\text{F}$. 
It is these flavor curves, together with additional curves inside $\mathcal{S}$ describing possible mass deformations, that are encoded in the CFDs and provide a succinct characterization of each 5d SCFT. 
In section \ref{sec:CFD}, we have associated to every (equivalence class of) resolution(s) a CFD. 
Linear combinations of the vertices inside a CFD correspond to a chain of holomorphic curves, whose mutual intersections are indicated by edges of the graph.
This information furthermore allows us to efficiently determine spin 0 and 1 BPS states from M2-branes wrapping rational curves in $\cal S$, see section \ref{sec:BPS}.
Moreover, any such graph depicts the flavor curves specifying the strongly coupled flavor symmetry $G_\text{F}$, as well as possible mass deformations of the corresponding 5d SCFT.
Such deformations correspond to transitions between CFDs, which are encoded in simple combinatorial rules. 

This sets up an elegant way to classify all 5d SCFTs originating from a given 6d SCFT via circle reduction.
Starting from the CFD of the unique 5d marginal theory associated with the 6d SCFT, the graph transitions generate a tree of descendant 5d SCFTs including information about their superconformal flavor symmetry and spin 0 and 1 BPS states.
At rank one and two, this graph-based classification agrees perfectly with known results \cite{Jefferson:2018irk,Hayashi:2018lyv} and furthermore adds the flavor symmetry and BPS states for each theory. The full list of rank one and rank two theories are in the tables in appendix \ref{app:SUMMARY}, and the CFD-trees showing the transitions are in figures \ref{fig:Rank1CFDTree} for rank one, and in figures \ref{fig:D10FibsAll},
\ref{fig:E8FibsAll}, and \ref{fig:Model3FibsAll} for rank two.

A major advantage of this approach is that it easily generalizes to higher rank, once the CFD for the marginal theory is computed. We determine the marginal CFDs and their descendants of $(E_n, E_n)$, $(D_k,D_k)$, and $(E_8, SU(n))$ conformal matter theories, providing predictions for previously unknown superconformal field theories and flavor enhancements.

At the technical level, the intuition for our proposal is largely based on so-called non-flat resolutions of non-minimal elliptic singularities.
In these blow-ups, the smooth space is still elliptically fibered over the same base $B$, but has the compact surfaces $S_j$ inserted into a special fiber in codimension two, resulting in a non-flat fibration. 
However, as emphasized by the rank two examples in section \ref{sec:rank_two_CFDs}, our graph-based approach is in no way limited to non-flat resolutions, and only requires tracking the flavor curves that are contained in the compact divisors in the geometry. 
Indeed, the description in terms of CFDs is applicable both to non-flat resolutions and the tensor branch geometries. Resolutions that are a combination of base and fiber resolutions are also useful in determining the marginal theories for certain higher rank cases. 
All that our proposal requires is a thorough understanding of the marginal geometry, i.e., a resolution where \emph{all} codimension one fibers $F_i$ are contained in $S_j$.

The complementary approach in the companion paper Part II \cite{Apruzzi:2019enx} is to determine the SCFT-relevant information from the effective gauge theory descriptions of the marginal theory.
There, we derive the CFDs entirely from the analysis of the Coulomb branch of 5d SCFTs, including the gauge theory descriptions of the descendant theories --- whenever such descriptions exist. 
This gives another, independent check of our proposal. 

A complete classification program will require determining the marginal CFDs for all 6d SCFTs and we return to this in Part III \cite{Apruzzi:2019kgb}. An obvious generalization includes combining CFDs and formulating rules for gluing marginal CFDs by gauging common subgroups of the flavor symmetry. These are gluing operations on CFDs for instance of the type
\be
\includegraphics*[width= 12cm]{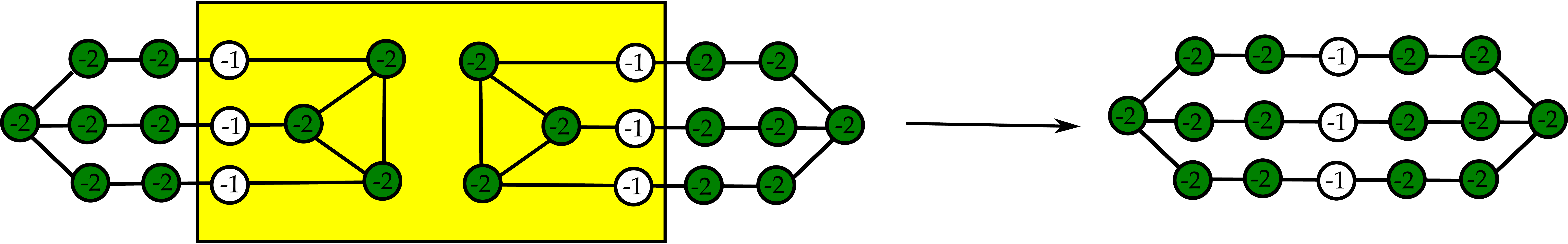} \,.
\ee
This combines two $(E_6, SU(3))$ rank one conformal matter theories by gauging a common $SU(3)$ flavor symmetry. Geometrically, this corresponds to compactifying the surface components, where the $SU(3)$s are realized. The $(-1)$-curve in the $(E_6,E_6)$ marginal CFD can be interpreted as a combination of three curves $(-1)/(-3)/(-1)$ from the tensor branch resolution of $(E_6,E_6)$ non-minimal Weierstrass model, and there are three of them due to the $S_3$ symmetry of the affine $E_6$ and $A_2$ Dynkin diagram. After this gauging the resulting CFD is that of the $(E_6, E_6)$ conformal matter theory.

The main hallmark of our approach is that the superconformal flavor symmetries are manifest in the description of 5d SCFTs. 
Equally, we have seen that certain BPS states associated to genus zero curves can be easily read off from the CFD. Developing the enumeration of more general BPS states, using this description in terms of CFDs, is something we believe is worthwhile exploring. It may be useful to connect this to the approach in \cite{Cecotti:2013mba, Banerjee:2018syt}.\footnote{We thank D. Gaiotto and P. Longhi for bringing this to our attention.}

\subsection*{Acknowledgements}

We thank H.~Hayashi, J.~J.~Heckman, N.~Mekareeya, D.~Morrison, W.~Taylor and M.~Weidner for discussions and A.~Bourget, C.~Closset and J.~Eckhard for pointing out typos in an earlier version. 
The work of FA, SSN, YNW is supported by the ERC Consolidator Grant number 682608 ``Higgs bundles: Supersymmetric Gauge Theories and Geometry (HIGGSBNDL)''. CL is supported by NSF CAREER grant PHY-1756996. 
LL is supported by DOE Award DE-SC0013528Y. YNW thanks the Aspen Center for Physics, and support in part by the Simons Foundation. CL and FA acknowledge the Pollica Workshop 2019 for hospitality.
We thank the String-Phenomenology and String-Math 2019 conferences, where this work was presented. 

\appendix

\section{Summary for Rank one and Two 5d SCFTs}
\label{sec:origin}

This appendix serves two purposes: we first list all rank one and two marginal 5d theories and their geometric, CFD, and gauge theoretic realization. 
In appendix \ref{app:SUMMARY} we summarize all descendant 5d SCFTs in rank one and two. These are a detailed description of the theories already shown in the CFD-trees in section \ref{sec:CFD}.

\subsection{Marginal Theories in 5d}

Marginal theories are 5d theories obtained from 6d SCFTs by circle-reduction. They do not have  UV completion in 5d, but flow to a fixed point in 6d. 
In terms of classifications of 5d SCFTs, they are the starting points of our CFD-trees. In this section, we summarize the marginal theories of rank one and two, combining all the data that goes into the CFD-descendant computation. In addition we also supplement the gauge theoretic description for the marginal theories, which are the topic of the companion paper \cite{Apruzzi:2019enx}.

To characterize the 6d theories relevant for rank one and two, it is useful to list their tensor branch geometries, which are essentially the resolved base of the F-theory realization. We use standard notation conventions:
non-minimal points in the base of the F-theory elliptic Calabi--Yau threefold are blowun up by inserting a chain of $\mathbb{P}^1$s, until there are no more non-minimal singularities.  On a curve $\Sigma$ with $\Sigma^2 =-n$, the elliptic fiber can still have  a singularity, of type $\mathfrak{g}$, and we denote this by 
\begin{equation}
\overset{\mathfrak{g}}n \,.
\end{equation}
Over non-compact curves the fiber can also be singular, which corresponds to  flavor symmetries of the tensor branch theory, and we distinguish this case by denoting these in square brackets, as usual
\begin{equation}
[\mathfrak{g}]\,.
\end{equation}

\subsubsection{Marginal Theories for Rank one SCFTs}

The marginal theory, from which the rank one 5d SCFTs descend, is the 6d rank one E-string theory, whose tensor branch is 
\begin{equation} \label{eq:TES}
[\mathfrak{e}_8]- 1 \,,
\end{equation}
i.e., this has an $E_8$ flavor symmety and one compact self-intersection $-1$ curve in the base. 
The summary table for the rank one case is: 
\begin{equation}
\centering
\begin{array}{|c|c|c|}\hline 
\text{Marginal CFD} & \text{Gauge Theory} & \text{Box Graph} \cr \hline 
\multirow{4}{*}{
\includegraphics*[width= 6cm]{CFD-E8-Rank1-Top.pdf}
}  
&\multirow{5}{*}{$SU(2)+ 8 \mathbf{F}$} & \\
&&\includegraphics*[width=4cm]{BG-Rank1EString-KK.pdf} \\
\hline 
\end{array}
\end{equation} 
This table describes all the data for the marginal theory in 5d: the CFD (from which we determine all descendant 5d SCFTs by CFD-transitions), the weakly-coupled gauge theory description in terms of an $SU(2)$ gauge theory with 8 fundamental flavors, and the box graph, which gives a simple graphical characterization of the Coulomb branch phase, and will be part of the compantion paper \cite{Apruzzi:2019enx}. 


\subsubsection{Marginal Theories for Rank two SCFTs}
\label{app:Rank2Classification}

There are several 6d SCFTs, which upon circle compactification and, in some cases, outer automorphism twists, give rise to rank two  5d theories \cite{Jefferson:2017ahm, Jefferson:2018irk}. We list these here, as well as the data of the CFDs and box graphs that are key to our classification approach:
\begin{enumerate}
\item  \underline{Rank two E-string:} \\
The tensor branch for this theory is 
\begin{equation} \label{eq:TBEstr}
[\mathfrak{e}_8]- 1 - 2 - [\mathfrak{su}_2] \,.
\end{equation}
The summary table for the marginal theory is as follows:
\begin{equation}
\centering
\begin{array}{|c|c|c|}\hline 
{\rm Marginal \ CFD} & {\rm Gauge\ Theory} & {\rm Box\  Graphs} \cr \hline 
\multirow{12}{*}{
\includegraphics*[width= 4.8cm]{CFD-E8-Rank2-Top.pdf}
}  
&\multirow{4}{*}{$SU(3)_{\frac{3}{2}} + 9\mathbf{F}$} & \\
&&\includegraphics*[width= 3cm]{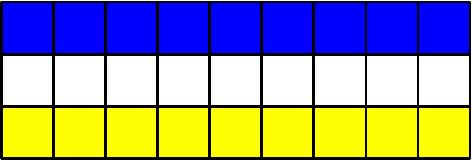} \\\cline{2-3}
&\multirow{8}{*}{$Sp(2) + 8\mathbf F + 1\mathbf{AS}$} & \\
& &\ \includegraphics*[width= 3.5cm]{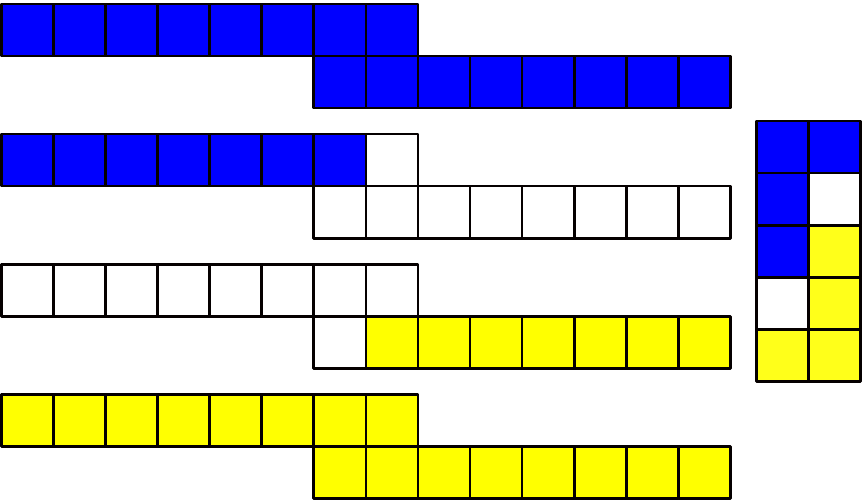}\ \\
&&\\\cline{2-3}
&\multirow{4}{*}{$ [2\mathbf F + SU(2)] \times [SU(2) + 5\mathbf F]$} & \\
&& \includegraphics*[width= 3cm]{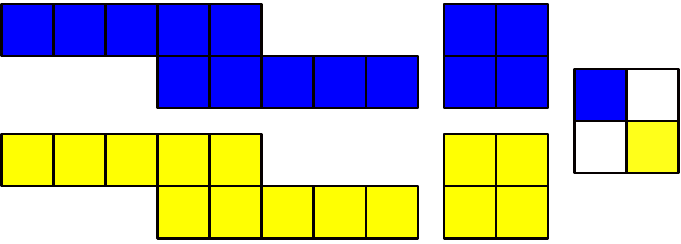} \\
\hline 
\end{array}
\end{equation}
Again the box graph description for each of the weakly-coupled gauge theory descriptions is added for ease of comparison of this paper and Part II \cite{Apruzzi:2019enx}. 
The above table lists the Chern-Simons level for the $SU$ groups. 
\item \underline{$(D_5, D_5)$ minimal conformal matter theory:} 
This 6d SCFT, or alternatively $(D_{10}, I1)$, has flavor symmetry $SO(20)$. The tensor branch geometry is 
\begin{equation} \label{eq:TBD5CM}
[\mathfrak{so}_{10}]- \overset{\mathfrak{sp}_1} 1  - [\mathfrak{so}_{10}] \,.
\end{equation}
The summary table for the marginal theory is as follows:
\begin{equation}
\centering
\begin{array}{|c|c|c|}\hline 
{\rm Marginal\  CFD} & {\rm Gauge\  Theory} & {\rm Box\  Graphs} \cr \hline 
\multirow{12}{*}{
\includegraphics*[width= 4.8cm]{CFD-D10-Top.pdf}
}  
&\multirow{4}{*}{$SU(3)_0 + 10\mathbf F$} & \\
&&\includegraphics*[width= 3cm]{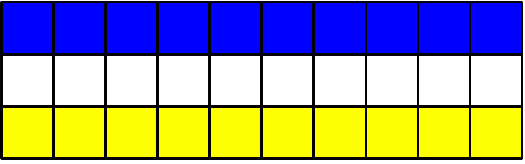} \\\cline{2-3}
&\multirow{8}{*}{$Sp(2) + 10\mathbf F$} & \\
& &\includegraphics*[width= 4cm]{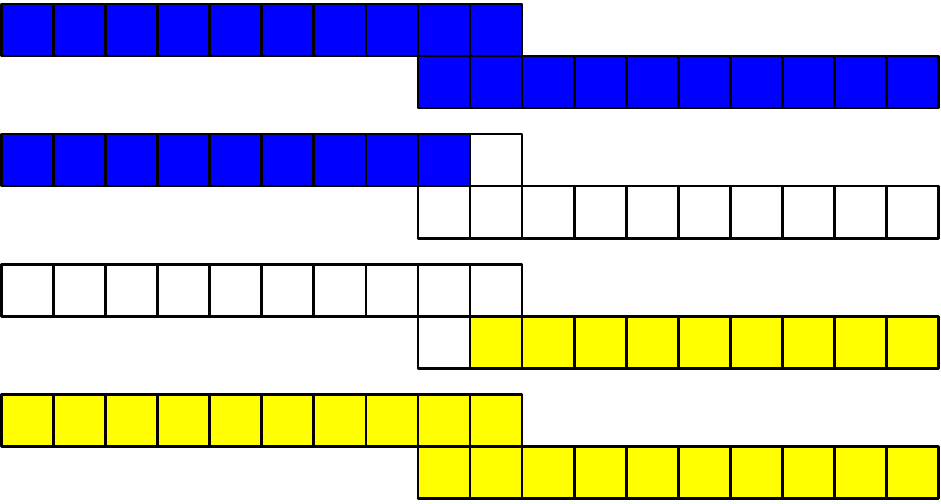}\\
&&\\\cline{2-3}
&\multirow{4}{*}{$[4\mathbf F + SU(2)] \times [SU(2) + 4\mathbf F] $} & \\
&& \includegraphics*[width= 4cm]{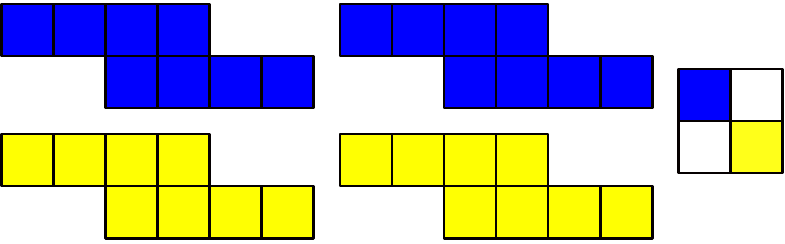} \\
\hline 
\end{array}
\end{equation}
\item \underline{$SU(3)$ on a $(-1)$ curve with 12 hypers:}\\
This theory has in fact one hypermultiplet in the antisymmetric representation of $SU(3)$ and $11$ hypers in the fundamental representation. The flavor symmetry at the superconformal point is $SU(12)$, and the tensor branch geometry is
\begin{equation}\label{eq:TBSU3-1}
[1 \mathbf{AS}]\; // \; \overset{\mathfrak{su}_3} 1  - [11]\,,
\end{equation}
where $//$ indicates that a non-compact curve intersect tangentially the $-1$ curve, and 1\textbf{AS} refers to one hypermultiplet in the antisymmetric representation of the gauge group. Upon $S^1$ compactification with a $\mathbb{Z}_2$ automorphism twist 
we obtain a rank two theory that is the marginal theory with the following data: 
\begin{equation}\label{Model3Sum}
\centering
\begin{array}{|c|c|c|}\hline 
{\rm Marginal \ CFD} & {\rm Gauge\ Theory} & {\rm Box\ Graphs} \cr \hline 
\multirow{14}{*}{
\includegraphics*[width= 4cm]{CFD-Model3-Rank2-Top.pdf}
}  
&\multirow{4}{*}{$SU(3)_4 + 6 \mathbf F$} & \\
&&\includegraphics*[width= 2.5cm]{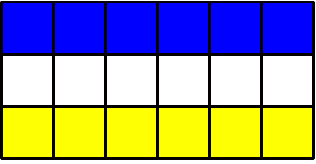} \\\cline{2-3}
&\multirow{8}{*}{$Sp(2) +2 \mathbf{AS} + 4 \mathbf F$} & \\
&&\includegraphics*[width= 3cm]{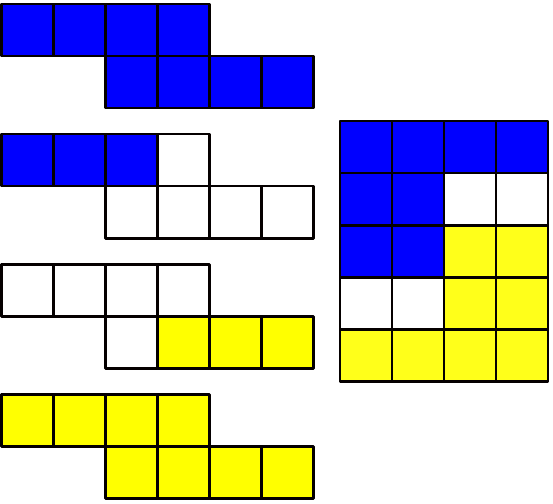}  \\
&&  \\
\hline 
\end{array}
\end{equation}
There is also a $G_2 + 6\mathbf F $ gauge theory description and we will expand on this in \cite{Apruzzi:2019enx}.

\item \underline{$SU(3)$ theory on a $(-2)$-curve with 6 hypers:}\\
This $SU(3)$ theory with 6 fundamental hypers has 6d  superconformal flavor symmetry $SU(6)$. 
Its tensor branch is
\begin{equation} \label{eq:TBsu36fl}
[\mathfrak{su}_{3}]- \overset{\mathfrak{su}_3} 2  - [\mathfrak{su}_{3}] \,.
\end{equation}
Applying a $\mathbb{Z}_2$-automorphism reduces this to a 5d marginal theory of rank two. The summary data is as follows:
\begin{equation}
\centering
\begin{array}{|c|c|c|}\hline 
{\rm Marginal \ CFD} & {\rm Gauge\  Theory} & {\rm Box\  Graphs} \cr \hline 
\multirow{6}{*}{
\includegraphics*[width= 3cm]{CFD-Model4-Rank2-Top.pdf}
}  
&\multirow{6}{*}{$Sp(2)_0+ 3 \mathbf{AS}$} & \\
&& \includegraphics*[width= 3cm]{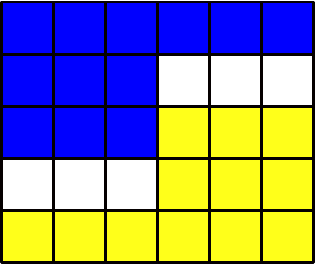}  \\
\hline 
\end{array}
\end{equation}
\end{enumerate}

In addition to these four marginal theories, there are two, which 
which have gauge theoretic descriptions, but a less clear geometric description in all the marginal theory phases \cite{Jefferson:2018irk}.  In terms of the rank two  classification \cite{Jefferson:2018irk, Hayashi:2018lyv}, they give rise to one single extra theory that are not obtained otherwise as descendants of the above four marginal theories (a theory descending from a model with an $O7^+$-plane, which is not purely geometric).
\begin{enumerate}
\item[5.] The $A_4$ $(2,0)$ 6d theory has a tensor branch of the form
\begin{equation} \label{eq:TBA420}
 2  \; - \;2 \;-\; 2  \,.
\end{equation}
The marginal geometry is given by three surfaces, whose geometry is $T^2 \times \mathbb P^1$, and which are glued along the $T^2$.
The spectrum of the marginal theory apart from the vector multiplet has an additional adjoint scalar due to the $g=1$ curve as base of the ruling \cite{Intriligator:1997pq}. This  leads to the 5d $SU(4)$ $\mathcal N=2$  gauge theory, which consists of a $\mathcal N=1$ vector multiplet and a real scalar in the adjoint of $SU(4)$.  If we turn on a $\mathbb{Z}_2$-automorphism twist the 5d marginal theory is specified by
\begin{equation}
\centering
\begin{array}{|c|}\hline 
 {\rm Gauge\; Theory}  \cr \hline 
Sp(2)_{\pi}+ 1 \mathbf{Adj}   \\
\hline 
SU(3)_{\frac{3}{2}}+ 1 \mathbf{Sym}   \\
\hline 
\end{array}
\end{equation}
where 1\textbf{Sym} means one hypermultiplet in the symmetric representation of the gauge group.

The duality between the $SU(3)$ gauge theory description and the $Sp(2)$  can be understood in terms of Hanany-Witten moves from a point of view of the $(p,q)$ 5-branes web \cite{Hayashi:2018lyv}. In the geometric M-theory description this correspond to some flop transitions together with some complex structure deformations, and it should work similarly to the duality between the $SU$ and $Sp$ descriptions of the marginal theory for the rank two  E-string, which we will discuss in more detail in section \ref{subsec:non-flat_res_E8SU2}. 
\item[6.] Finally, there is the 6d theory, with tensor branch
\begin{equation}  \label{eq:TB2su2su22}
[\mathfrak{su}_{2}]- \overset{\mathfrak{su}_2} 2- \overset{\mathfrak{su}_2} 2  - [\mathfrak{su}_{2}] \,.
\end{equation}
Upon circle compactification and $\mathbb{Z}_2$-automorphism twist the 5d marginal theory data are given by
\begin{equation}
\centering
\begin{array}{|c|}\hline 
 {\rm Gauge\; Theory} \cr \hline 
SU(3)_{0}+ 1 \mathbf{Sym}+ 1 \mathbf{F} \\
\hline 
\end{array}
\end{equation}
\end{enumerate}

There are two more theories, which are listed in \cite{Jefferson:2017ahm}, that could potentially give new 5d rank two theories: 
 $SU(3)$ theory on a $-3$ curve, whose 5d marginal theory is $SU(3)_9$, however this does not give rise to any new 5d descendant SCFTs.

The tensor branch theories, described by the resolved base geometries \eqref{eq:TBEstr}, \eqref{eq:TBD5CM}, and \eqref{eq:TBSU3-1}, blow down to a smooth base, whereas \eqref{eq:TBsu36fl}, \eqref{eq:TBA420} and \eqref{eq:TB2su2su22} all blow down to a singular base. In the former case, the non-minimal singularity can be directly resolved into a non-flat fibrations, without any base blow-ups, where in the latter the base needs to be resolved as well.

\begin{table}
\begin{tabular}{|r|c|c|c|c|}\hline
CFD for SCFT & Flavor & Gauge Theory & BPS Spin 0 & BPS Spin 1 \cr \hline\hline 
\includegraphics[height=.8cm]{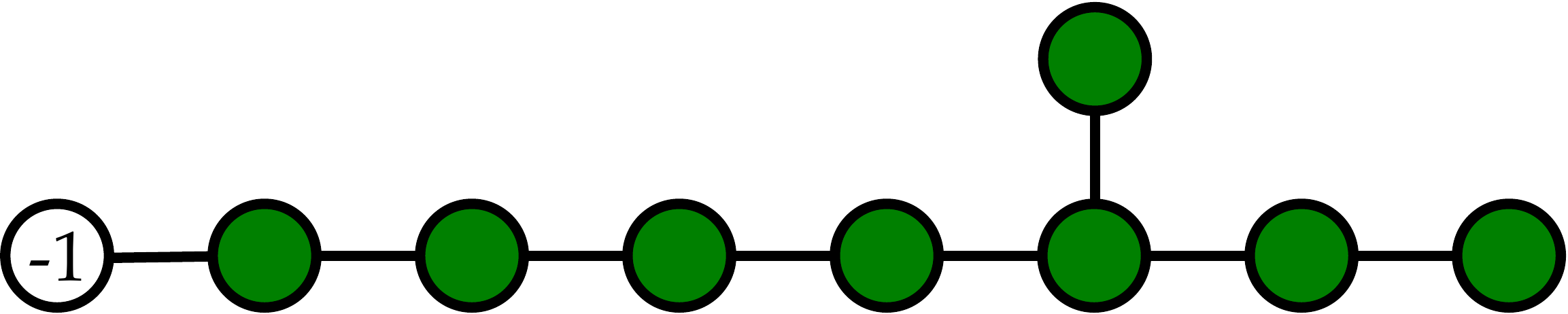} & $E_8$ & $SU(2)+7\mbf{F}$ & $\mbf{248}$ & $\mbf{3875}$\cr\hline
\includegraphics[height=.8cm]{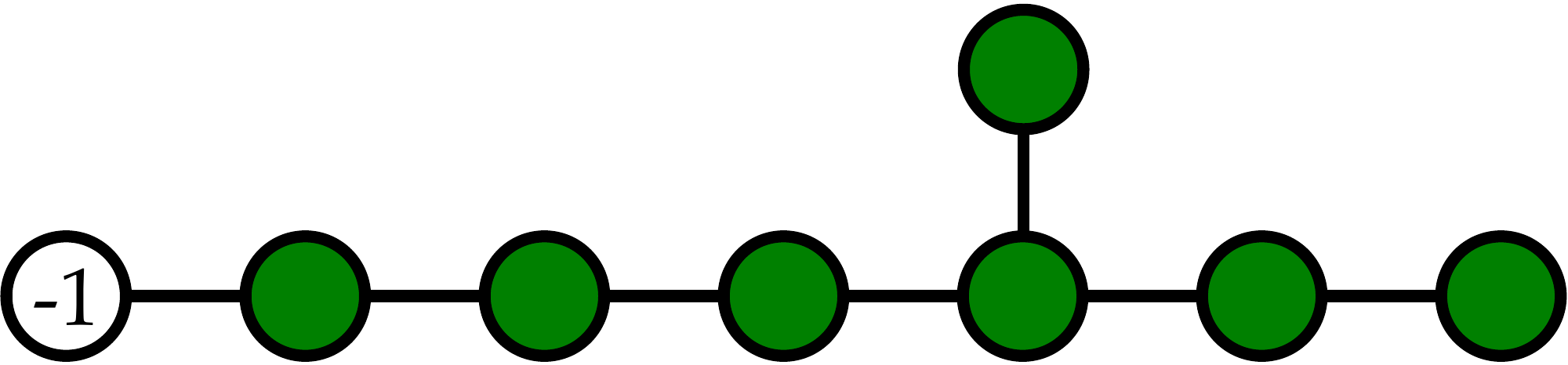} & $E_7$ & $SU(2)+6\mbf{F}$ & $\mbf{56}$ & $\mbf{133}$\cr\hline
\includegraphics[height=.8cm]{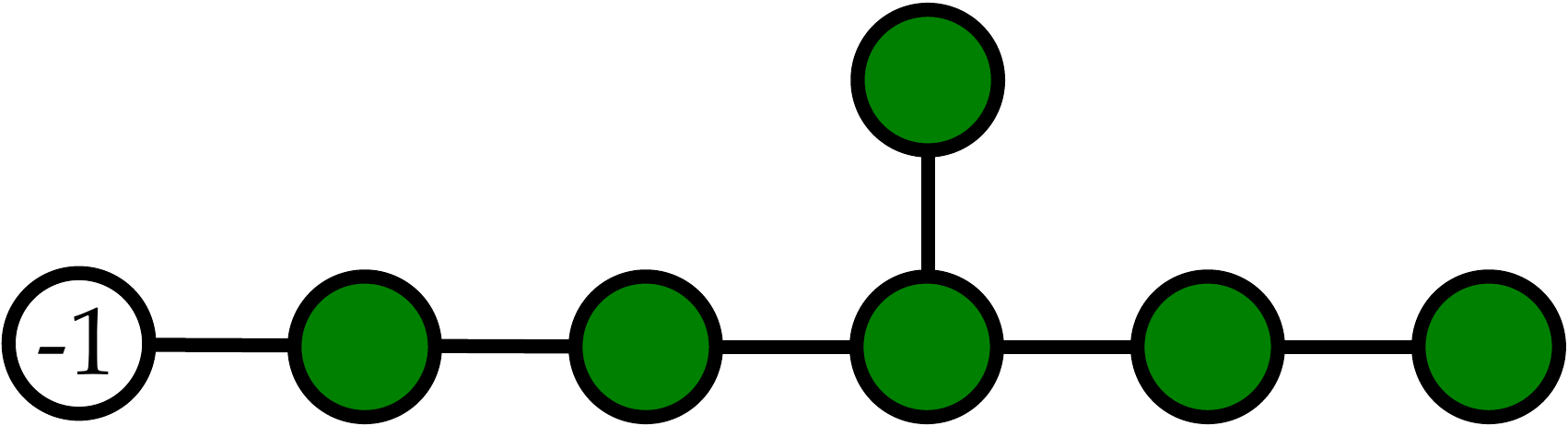} & $E_6$ & $SU(2)+5\mbf{F}$ & $\mbf{27}$ & $\overline{\mbf{27}}$\cr\hline
\includegraphics[height=.8cm]{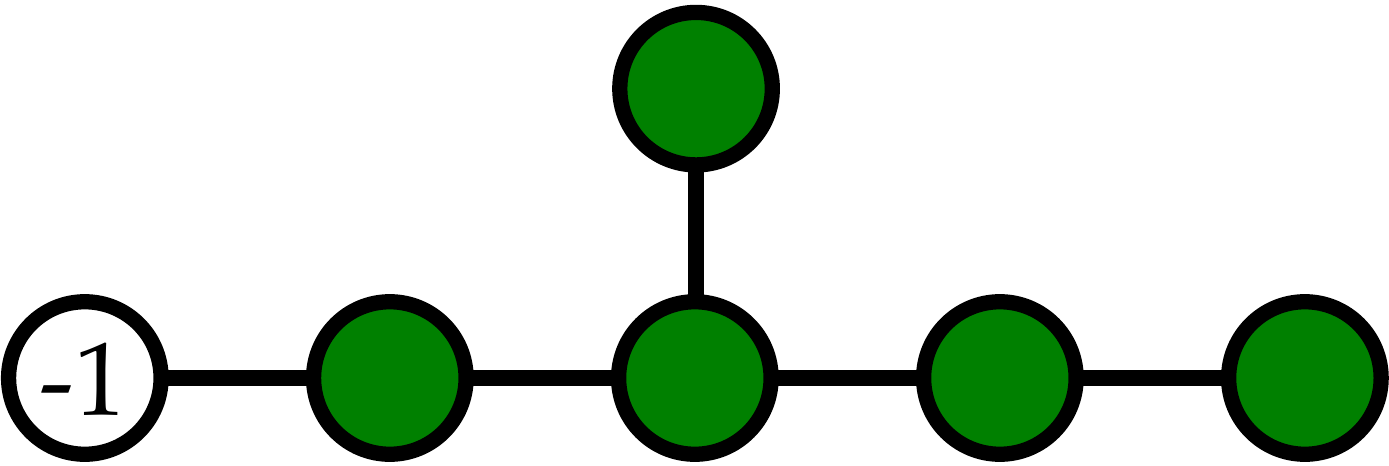} & $SO(10)$ & $SU(2)+4\mbf{F}$ & $\mbf{16}$ & $\mbf{10}$\cr\hline
\includegraphics[height=.8cm]{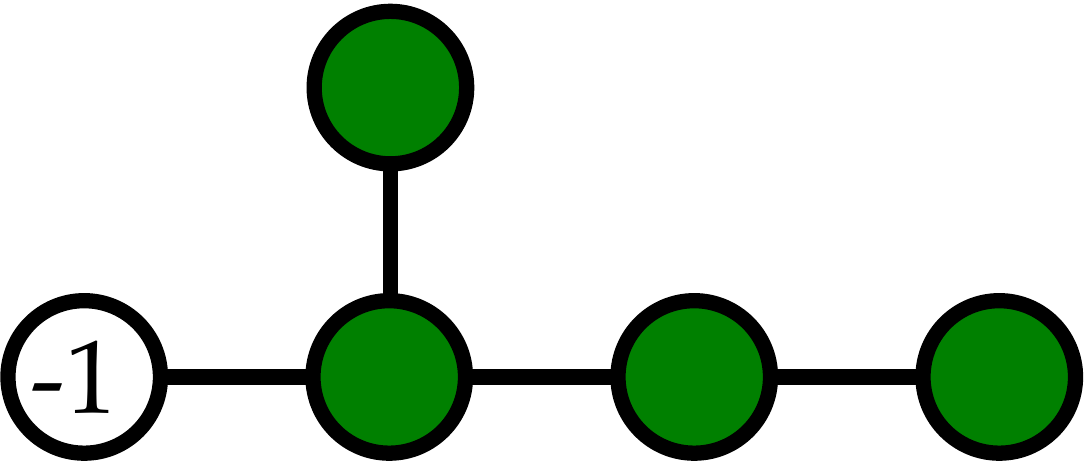} & $SU(5)$ & $SU(2)+3\mbf{F}$ & $\mbf{10}$ & $\bar{\mbf{5}}$\cr\hline
\includegraphics[height=.8cm]{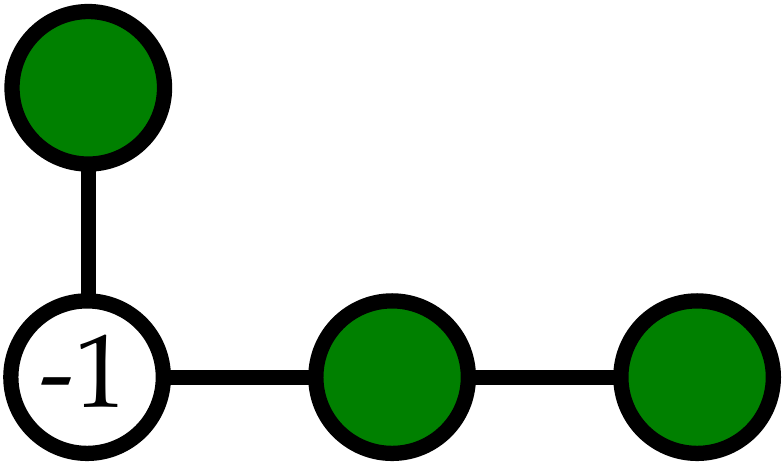} & $SU(3)\times SU(2)$ & $SU(2)+2\mbf{F}$ & $(\mbf{3},\mbf{2})$ & $(\bar{\mbf{3}},\mbf{1})$ \cr\hline
\includegraphics[height=.8cm]{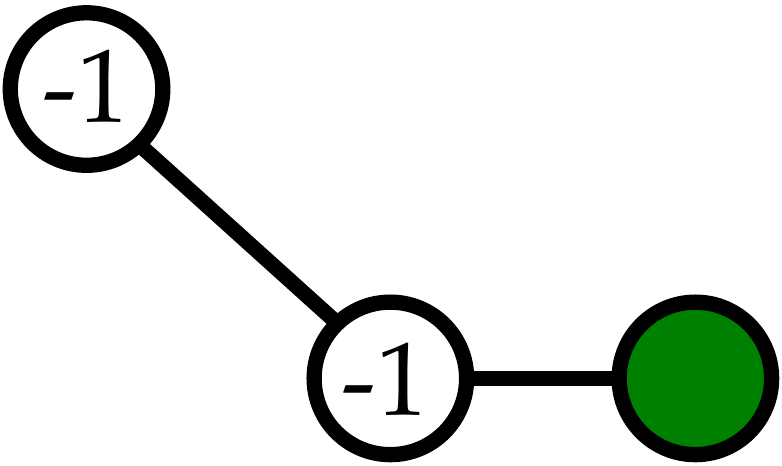} & $SU(2)\times U(1)$ & $SU(2)+1\mbf{F}$ & $\mbf{1}_{-1},\mbf{2}_1$ & $\mbf{2}_0$\cr\hline
\includegraphics[height=.3cm]{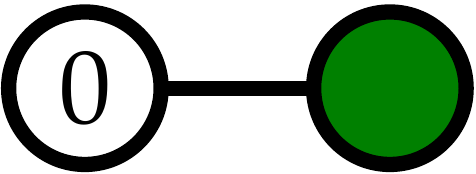} & $SU(2)$ & $SU(2)_0$ &  & $\mbf{2}$\cr\hline
\includegraphics[height=.8cm]{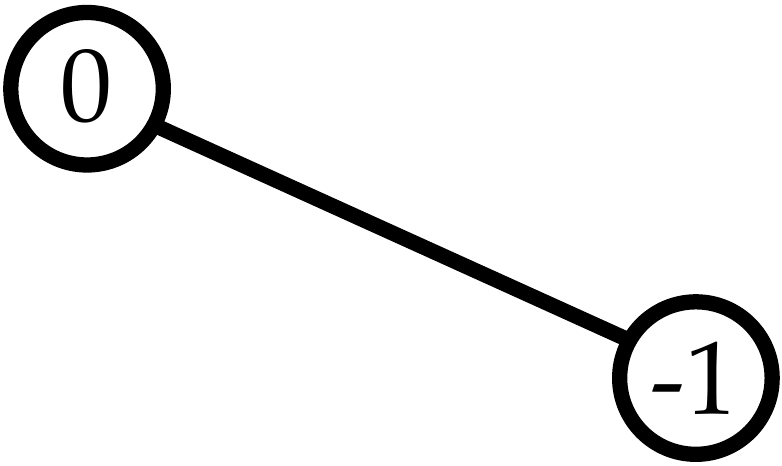} & $U(1)$ & $SU(2)_\pi$ & 1 & $0$\cr\hline
\includegraphics[height=.3cm]{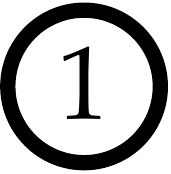} & - & - & & \cr\hline\hline
\end{tabular}
\caption{All 5d rank one SCFTs and the lowest spin BPS states from M2-brane wrapping modes over genus zero curves. We also list their  flavor symmetry, gauge theory description.}\label{t:Rank1BPS}
\end{table}


\subsection{Summary Tables for Rank one and Rank two 5d SCFTs}
\label{app:SUMMARY}

In this appendix, we summarize our findings in rank one and two,  by tabulating all 5d SCFTs organized by $M$, the number of mass deformations. 
The tables contain the CFDs (in cases when there are different realizations we give all CFDs), their weakly-coupled gauge theory descriptions, the strongly coupled flavor symmetry as read off from the CFD, and the spin 0 and 1 BPS states. Model 3/4 refer to the marginal theories (\ref{Model3Sum}) and (\ref{eq:TBsu36fl}), respectively.

The CFDs in the following tables are connected by CFD-transitions, and these are shown in figures are shown in figure \ref{fig:Rank1CFDTree} for rank one, and in figures \ref{fig:D10FibsAll}, \ref{fig:E8FibsAll}, \ref{fig:Model3FibsAll} for rank two.

\begin{sidewaystable}
\begin{tabular}{|c|c|c|c|c|c|c|c|}\hline
No. & $M$ & $(D_{10}, I_1)$ CFD & $(E_8,SU(2))$ CFD&  Flavor $G_\text{F}$ & Gauge Theory  & BPS Spin 0 & BPS Spin 1 \cr \hline\hline 
\multirow{3}{*}{1} & \multirow{3}{*}{11} & \multirow{3}{*}{\includegraphics[height=.6cm]{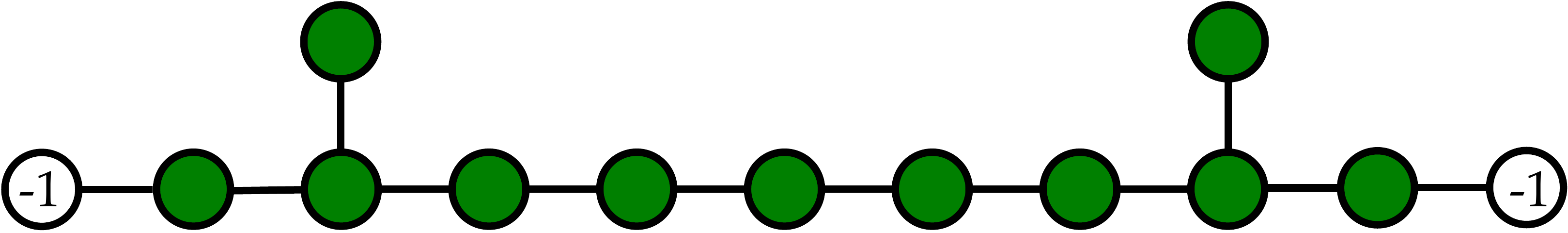}} & &  \multirow{3}{*}{$-$} & 
{\small $SU(3)_0+10\mbf{F}$} && \cr 
 & &&&& {\small $Sp(2)+10\mbf{F}$} & &\cr 
& & &&  & \tiny{$[4]-SU(2)-SU(2)-[4]$} & &\cr\hline
\multirow{3}{*}{2} & \multirow{3}{*}{10} & \multirow{3}{*}{\includegraphics[height=.7cm]{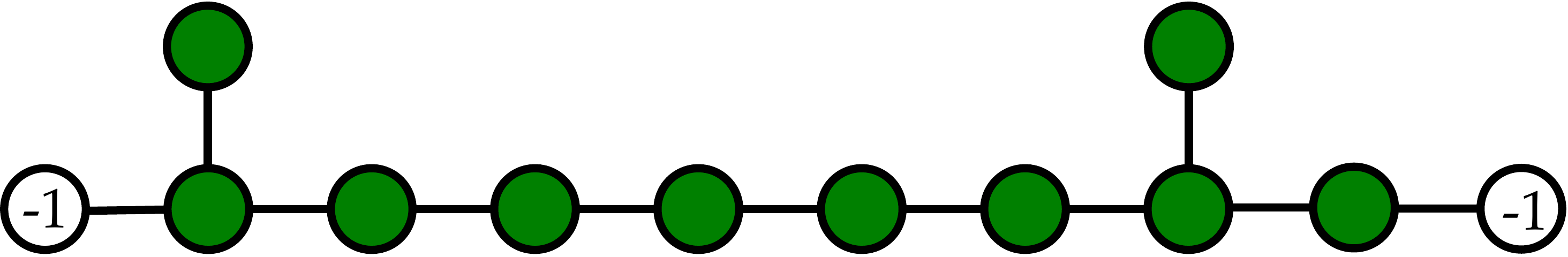}} & 
&\multirow{3}{*}{$SO(20)$} & {\small $SU(3)_{1/2}+9\mbf{F}$} & \multirow{3}{*}{$\mbf{190}$,$\mbf{512}$} & $\mbf{209}$,$\cdot\mbf{4845}$\cr 
& & & &  & {\small $Sp(2)+9\mbf{F}$}   &  & $\mbf{9728}$ \cr
& & & &  & \tiny{$[3]-SU(2)-SU(2)-[4]$} & & $\mbf{38760}$\cr\hline
\multirow{3}{*}{3} & \multirow{3}{*}{10} & & \multirow{3}{*}{\includegraphics[height=1cm]{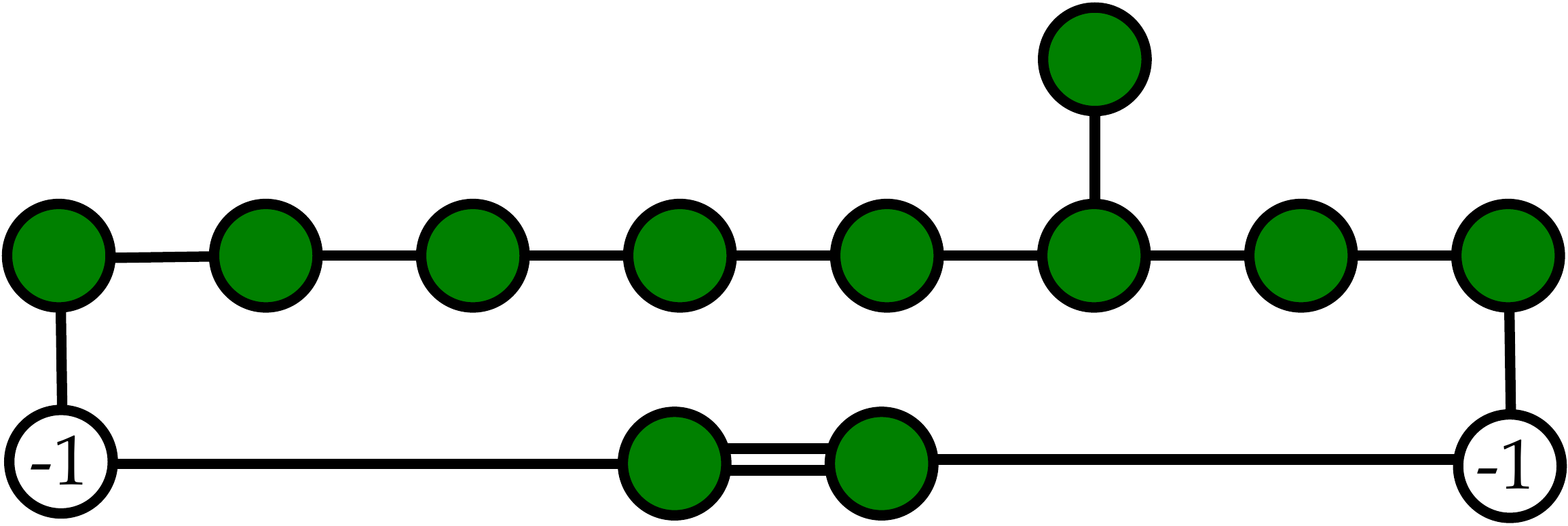}} &  \multirow{3}{*}{$-$} & {\small $SU(3)_{3/2}+9\mbf{F}$} & & \cr 
& & & & &  {\small $Sp(2)+1\mbf{AS}+8\mbf{F}$} & & \cr 
& & & & & \tiny{$[5]-SU(2)-SU(2)-[2]$} & &\cr\hline  

\multirow{2}{*}{4} & \multirow{2}{*}{9} & \multirow{2}{*}{\includegraphics[height=.7cm]{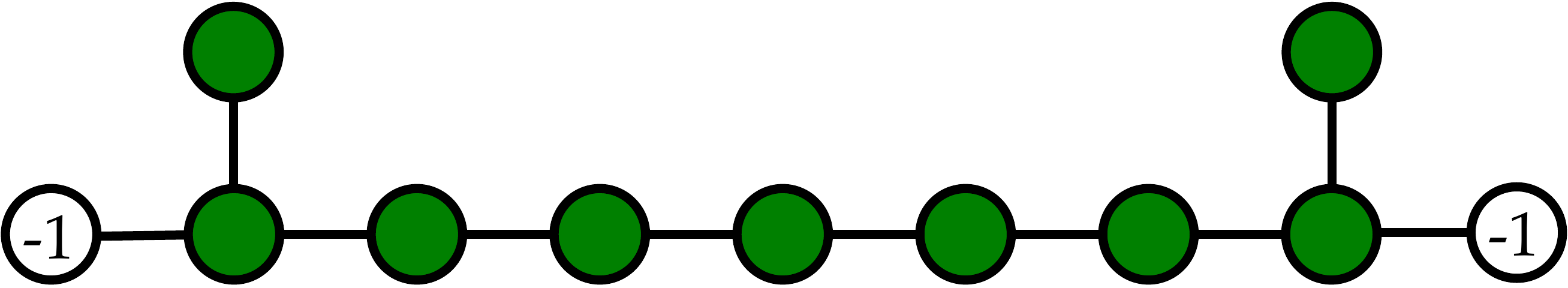}} & &  \multirow{2}{*}{$SU(10)$} & {\small $SU(3)_0+8\mbf{F}$} & \multirow{2}{*}{$\mbf{45},\overline{\mbf{45}}$}& $\mbf{99},\mbf{210}$ \cr 
& & & & &  \tiny{$[3]-SU(2)-SU(2)-[3]$} & & $\overline{\mbf{210}}$ \cr\hline
  
\multirow{4}{*}{5} &\multirow{4}{*}{9} & \multirow{4}{*}{\includegraphics[height=.8cm]{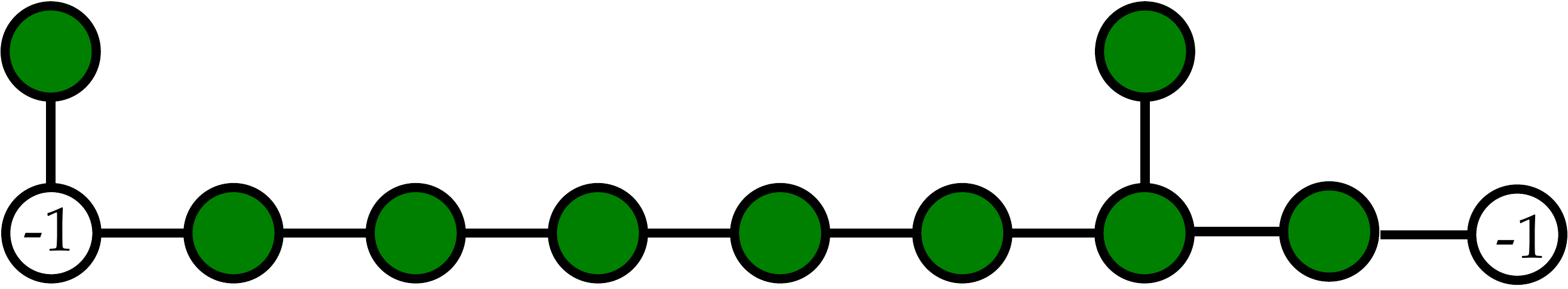}} &
\multirow{4}{*}{\includegraphics[height=.9cm]{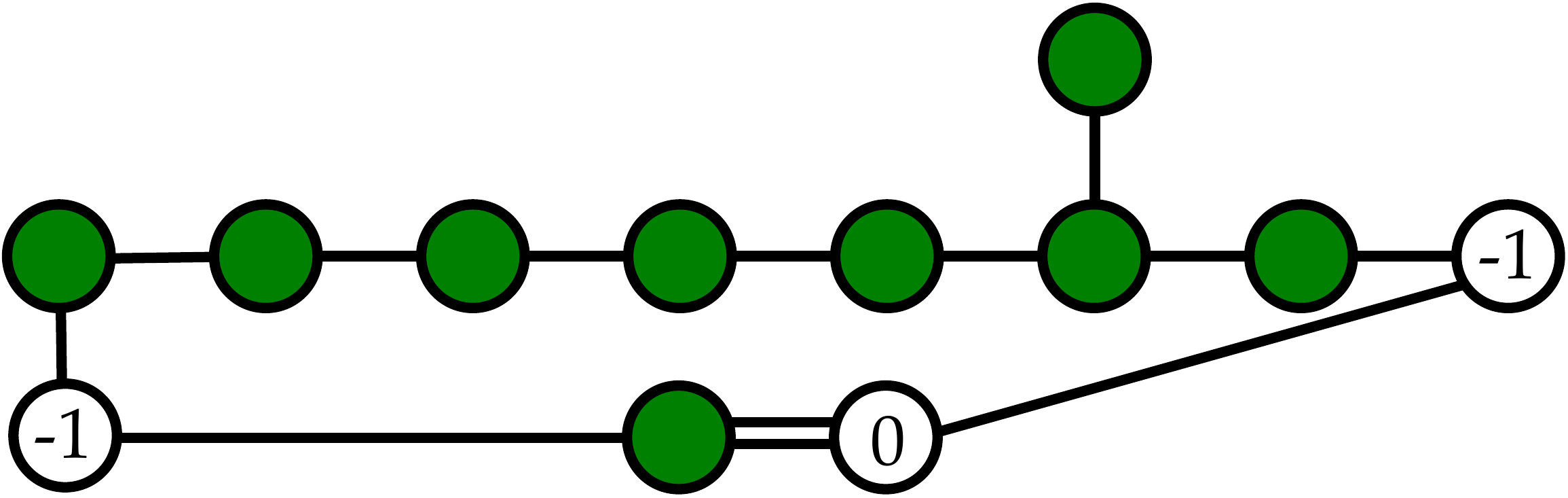}}
&  \multirow{4}{*}{$SO(16) \times SU(2)$} & {\small $SU(3)_{1}+8\mbf{F}$} &  & $(\mbf{1},\mbf{3})$ \cr 
& & & &  & {\small $Sp(2)+8\mbf{F}$} & $(\mbf{16},\mbf{2})$ & $(\mbf{120},\mbf{1})$ \cr
& & & &  & {\tiny{$[2]-SU(2)-SU(2)-[4]$}} & $(\mbf{128},\mbf{1})$ & $(\mbf{128}',\mbf{2})$ \cr
& & &  & & & & $(\mbf{1820},\mbf{1})$\cr\hline

\multirow{5}{*}{6} & \multirow{5}{*}{9} & & \multirow{5}{*}{\includegraphics[height=1cm]{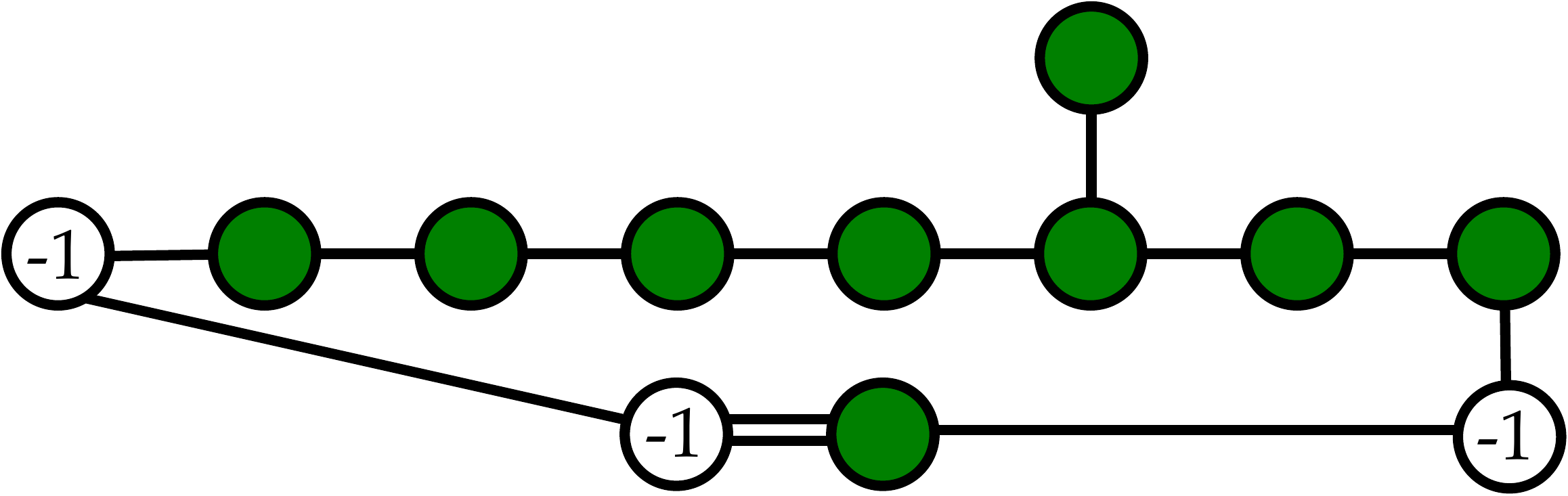}} &  \multirow{5}{*}{$E_8\times SU(2)$} & & & $(\mbf{248},\mbf{3})$ \cr 
& && & & {\small $SU(3)_2+8\mbf{F}$} & $(\mbf{1},\mbf{3})$ & $(\mbf{3875},\mbf{1})$\cr
& &&& & {\small $Sp(2)+1\mbf{AS}+7\mbf{F}$}  & $(\mbf{248},\mbf{1})$ & $(\mbf{27000},\mbf{2})$\cr
&& & &   & \tiny{$[5]-SU(2)-SU(2)-[1]$} & $(\mbf{3875},\mbf{2})$ & $(\mbf{147250},\mbf{2})$\cr
&&& & &  & & $(\mbf{2450240},\mbf{3})$\cr\hline

\end{tabular}

\caption{Table summarizing all Rank Two SCFTs: $M$ is the number of mass deformations; $(D_{10}, I_1)$, $(E_8,SU(2))$, Model 3/4 CFD show for each theory the CFD obtained as a decendant from one of these marginal theories. ``Flavor" lists the superconformal flavor symmetry $G_\text{F}$; ``Gauge Theory" provides the weakly coupled gauge theory descriptions. Finally, we list the BPS states with spin 0 and 1, as representations of $G_\text{F}$.}
\end{sidewaystable}

\begin{sidewaystable}
\begin{tabular}{|c|c|c|c|c|c|c|c|}\hline
No. & $M$ & $(D_{10}, I_1)$ CFD & $(E_8,SU(2))$ CFD &  Flavor $G_\text{F}$& Gauge Theory  & BPS Spin 0 & BPS Spin 1 \cr \hline\hline

\multirow{3}{*}{7} & \multirow{3}{*}{8} & &  \multirow{3}{*}{\includegraphics[height=1cm]{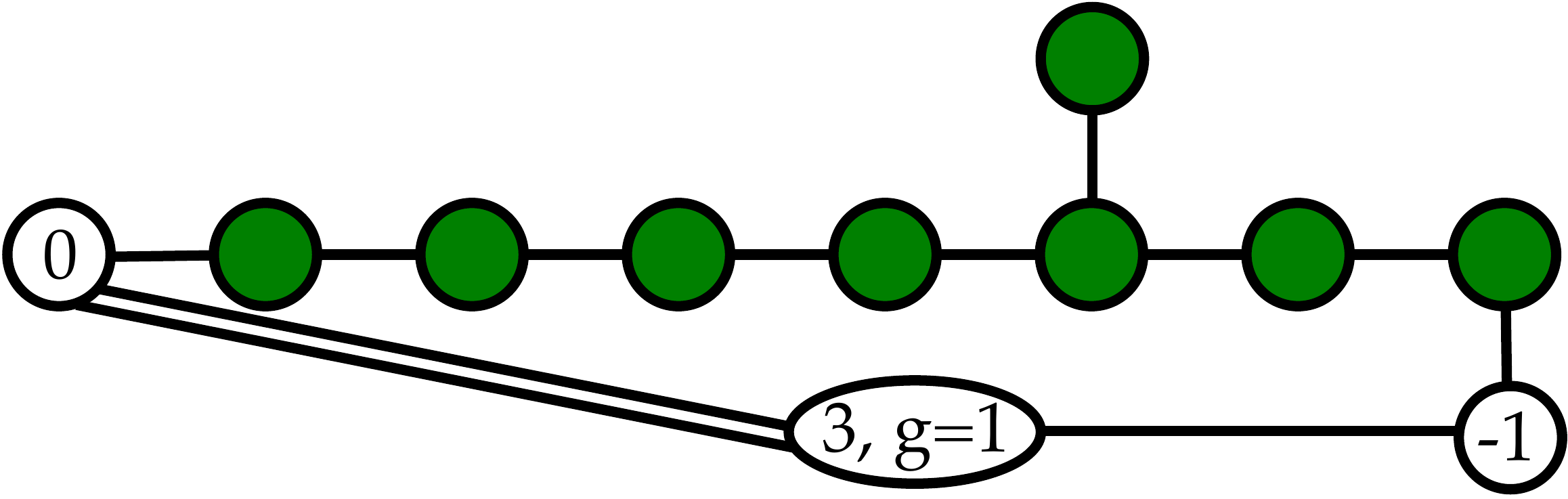}} & \multirow{3}{*}{$E_8$} & \multirow{3}{*}{{\small $[5]-SU(2)-SU(2)_0$}} & \multirow{3}{*}{$\mbf{3875}$} & $\mbf{248}$\cr
& & & & & & & $\mbf{2450240}$\cr
& & & & & & &\cr 
\hline

\multirow{3}{*}{8} & \multirow{3}{*}{8} & \multirow{3}{*}{\includegraphics[height=.8cm]{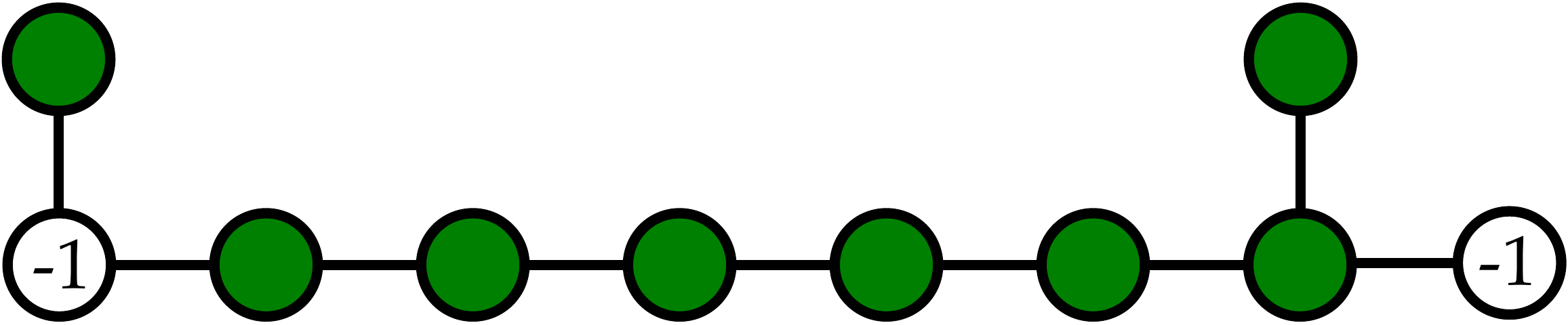}} & \multirow{2}{*}{\includegraphics[height=1cm]{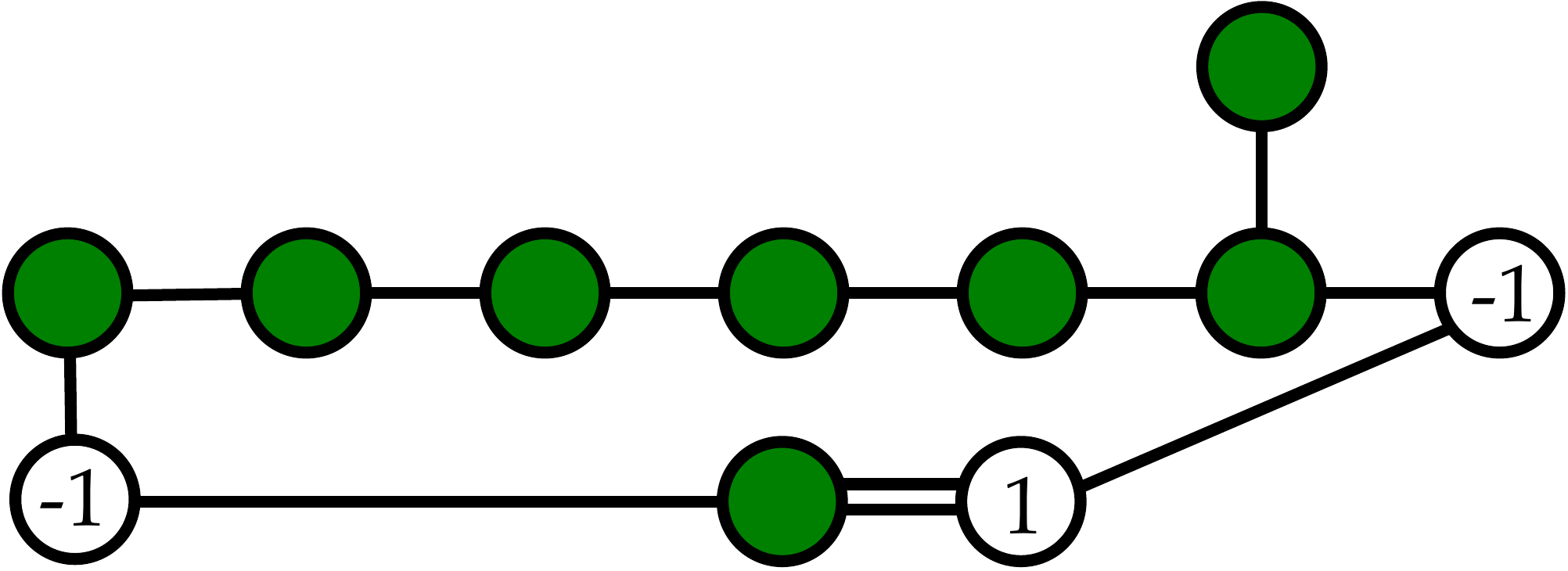}} & \multirow{3}{*}{$SU(8)\times SU(2)$} & {$SU(3)_{1/2}+7\mbf{F}$}  & $(\mbf{8},\mbf{2})$ & $(\bar{\mbf{8}},\mbf{2})$ \cr
 & & & & & {\small $[2]-SU(2)-SU(2)-[3]$} & $(\overline{\mbf{28}},\mbf{1})$ & $(\mbf{28},\mbf{1})$ \cr
  & & & & & &  & $(\mbf{70},\mbf{1})$\cr\hline

\multirow{3}{*}{9} & \multirow{3}{*}{8} & \multirow{3}{*}{\includegraphics[height=.8cm]{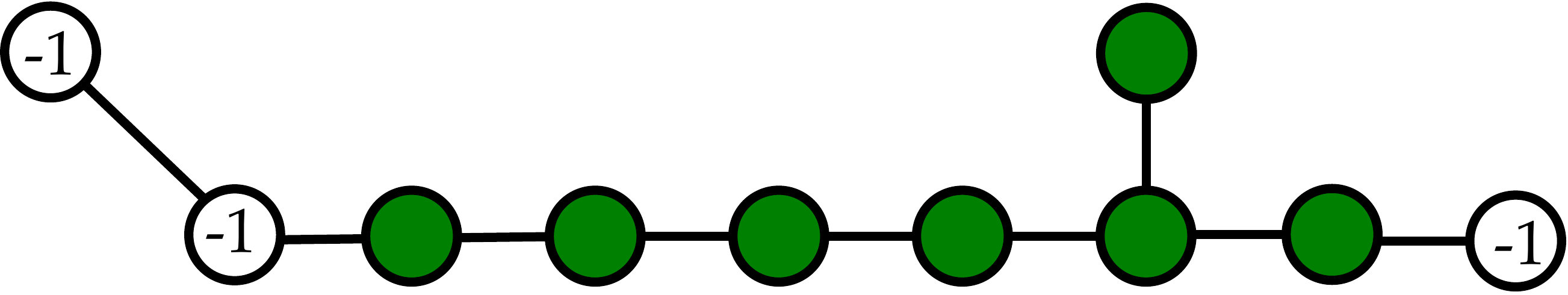}} & 
\multirow{3}{*}{\includegraphics[height=1cm]{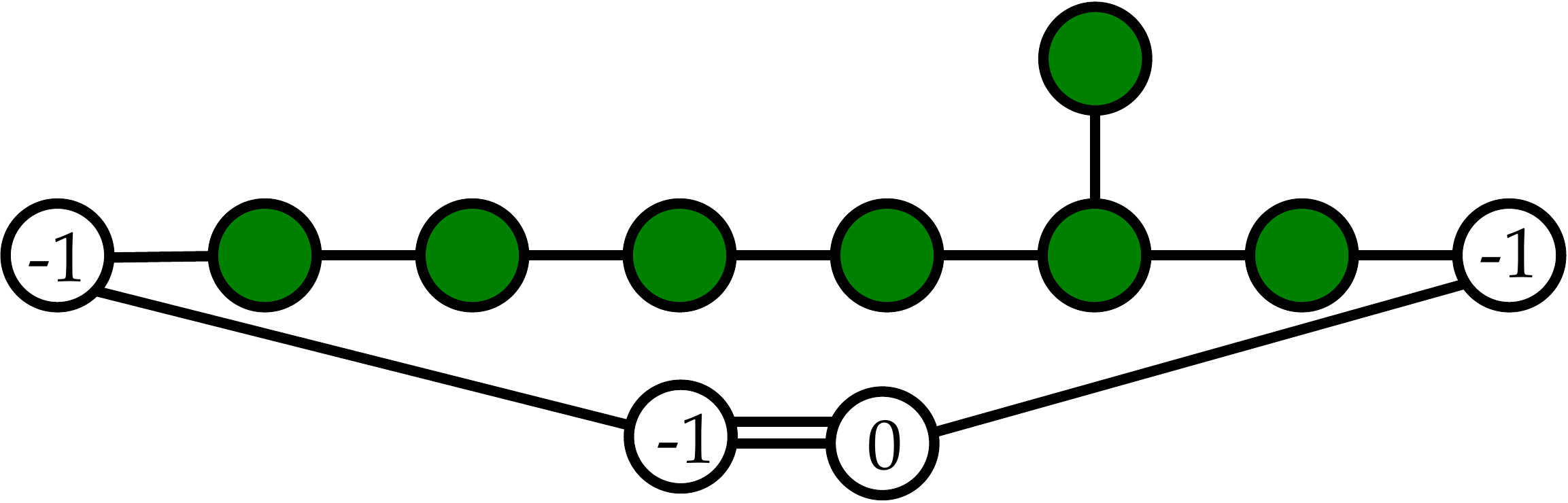}}
& \multirow{3}{*}{$SO(14) \times U(1)$} & {$SU(3)_{3/2}+7\mbf{F}$}  &  $\mbf{14}_1$ & $\mbf{1}_2, \mbf{14}_0$ \cr
&& & & & {$Sp(2)+7\mbf{F}$} & $\mbf{1}_{-1}$ & $\overline{\mbf{64}}_1$\cr
&& & & & {{\small $[1]-SU(2)-SU(2)-[4]$}}& $\mbf{64}_0$ & $\mbf{364}_0$\cr\hline

\multirow{3}{*}{10} &\multirow{3}{*}{8} & & \multirow{3}{*}{\includegraphics[height=1cm]{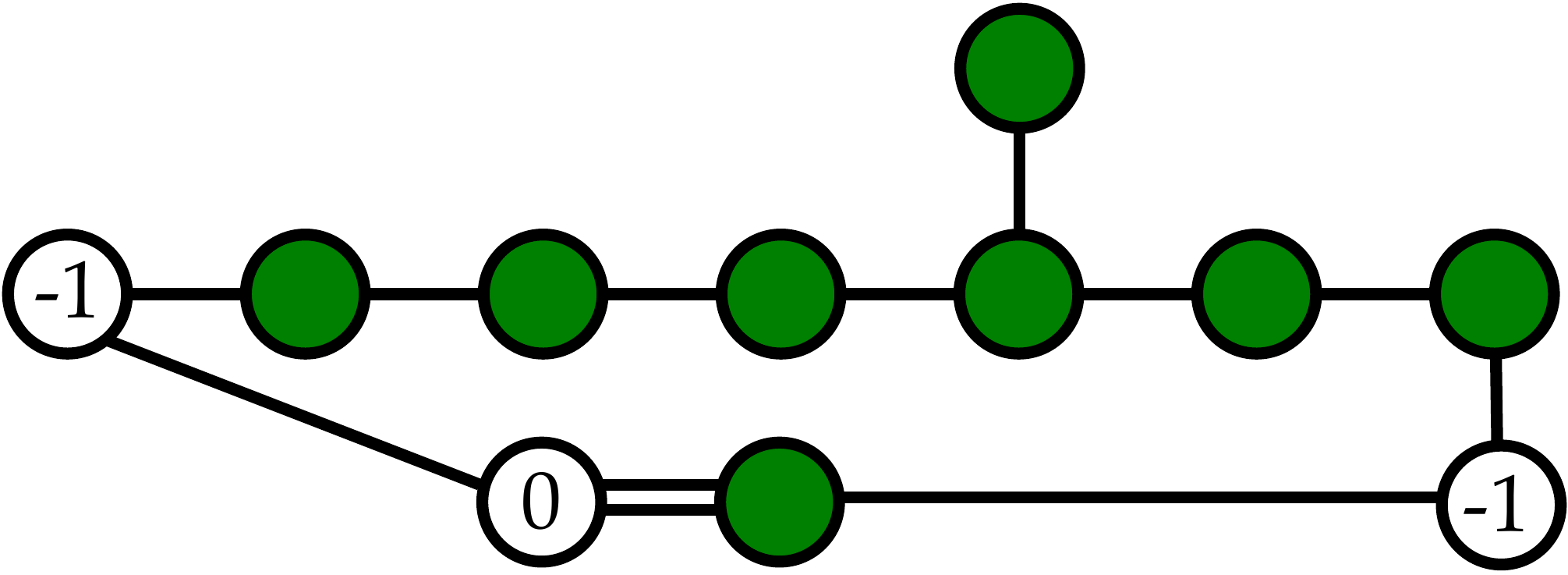}}  & \multirow{3}{*}{$E_7\times SU(2)$} &  {$SU(3)_{5/2}+7\mbf{F}$}  & $(\mbf{56},\mbf{1})$ & $(\mbf{1},\mbf{3})$ \cr
& & & & &  {$Sp(2)+1\mbf{AS}+6\mbf{F}$} &  $(\mbf{133},\mbf{2})$ & $(\mbf{912},\mbf{2})$\cr
& & & & &{\small $[5]-SU(2)-SU(2)_\pi$} &  & $(\mbf{1539},\mbf{3})$\cr\hline

11 & 7 & \includegraphics[height=.8cm]{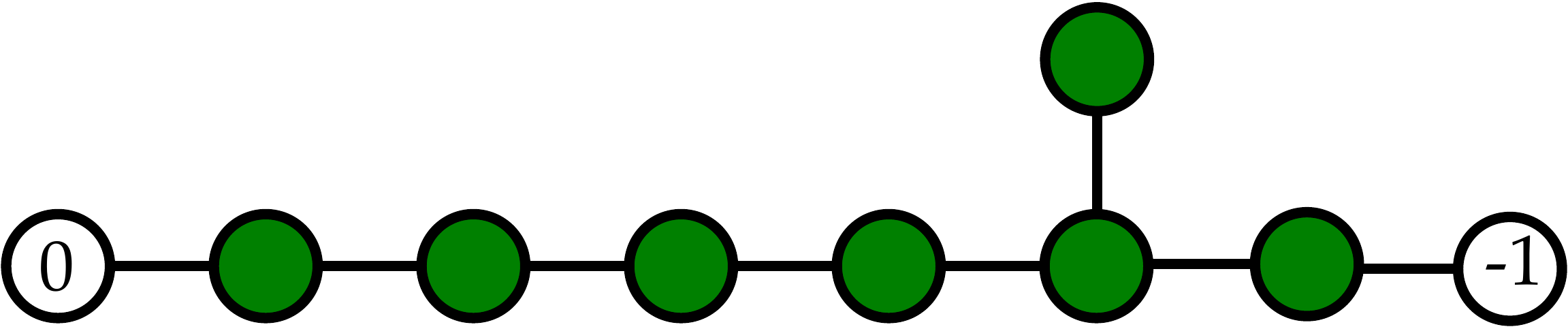} & \includegraphics[height=1cm]{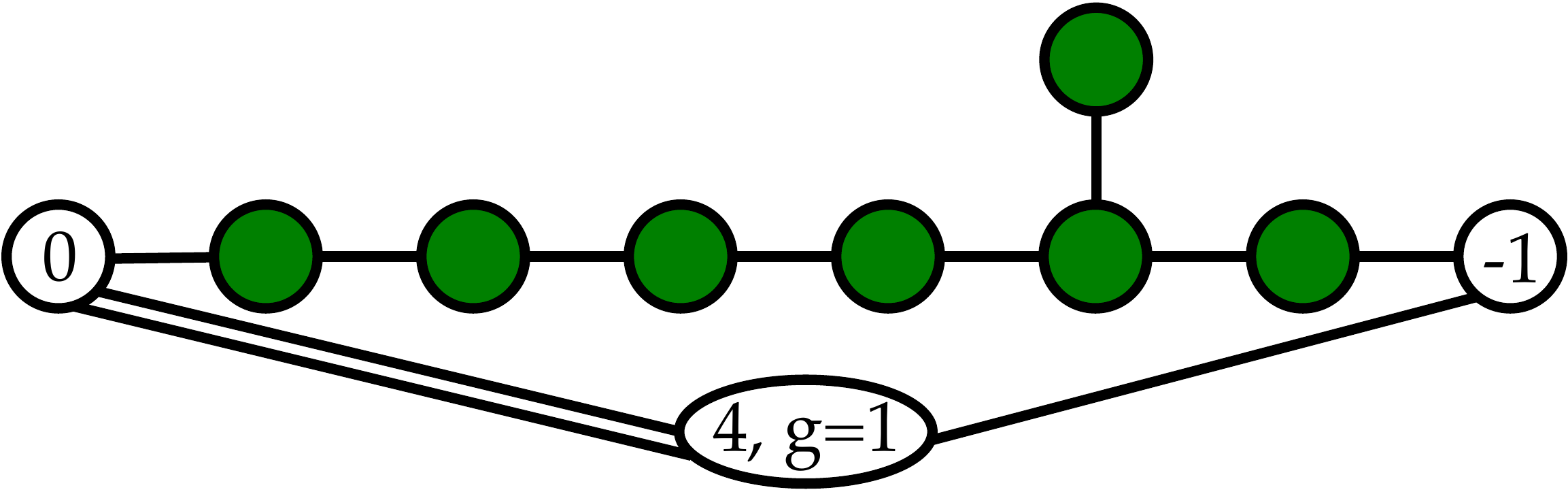} & $SO(14)$ & {\small $[4]-SU(2)-SU(2)_0$}& $\mbf{64}$ & $\mbf{14},\mbf{364}$ \cr \hline
 
\multirow{3}{*}{12} & \multirow{3}{*}{7} & \multirow{3}{*}{\includegraphics[height=.8cm]{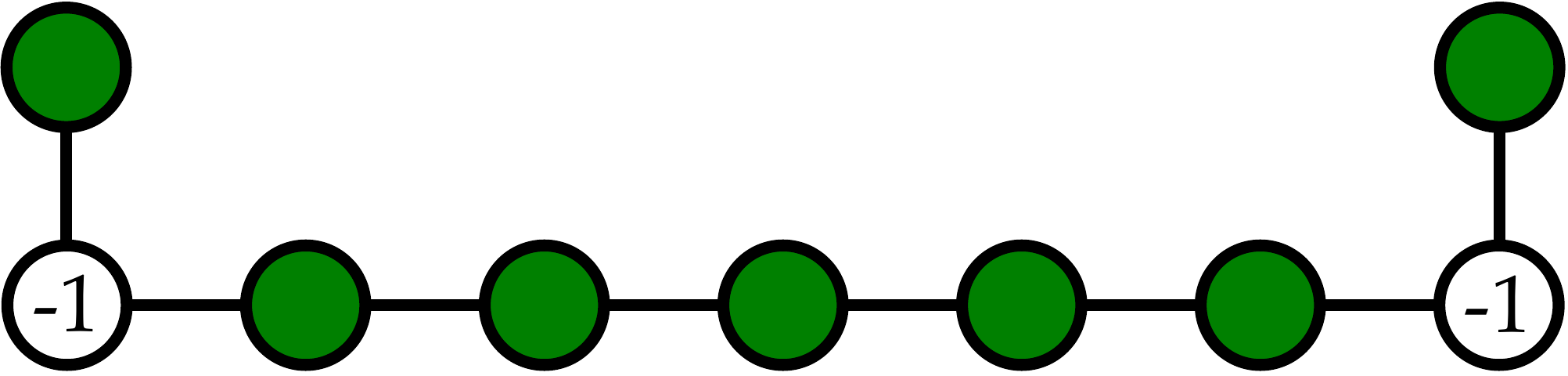}} & \multirow{3}{*}{\includegraphics[height=1cm]{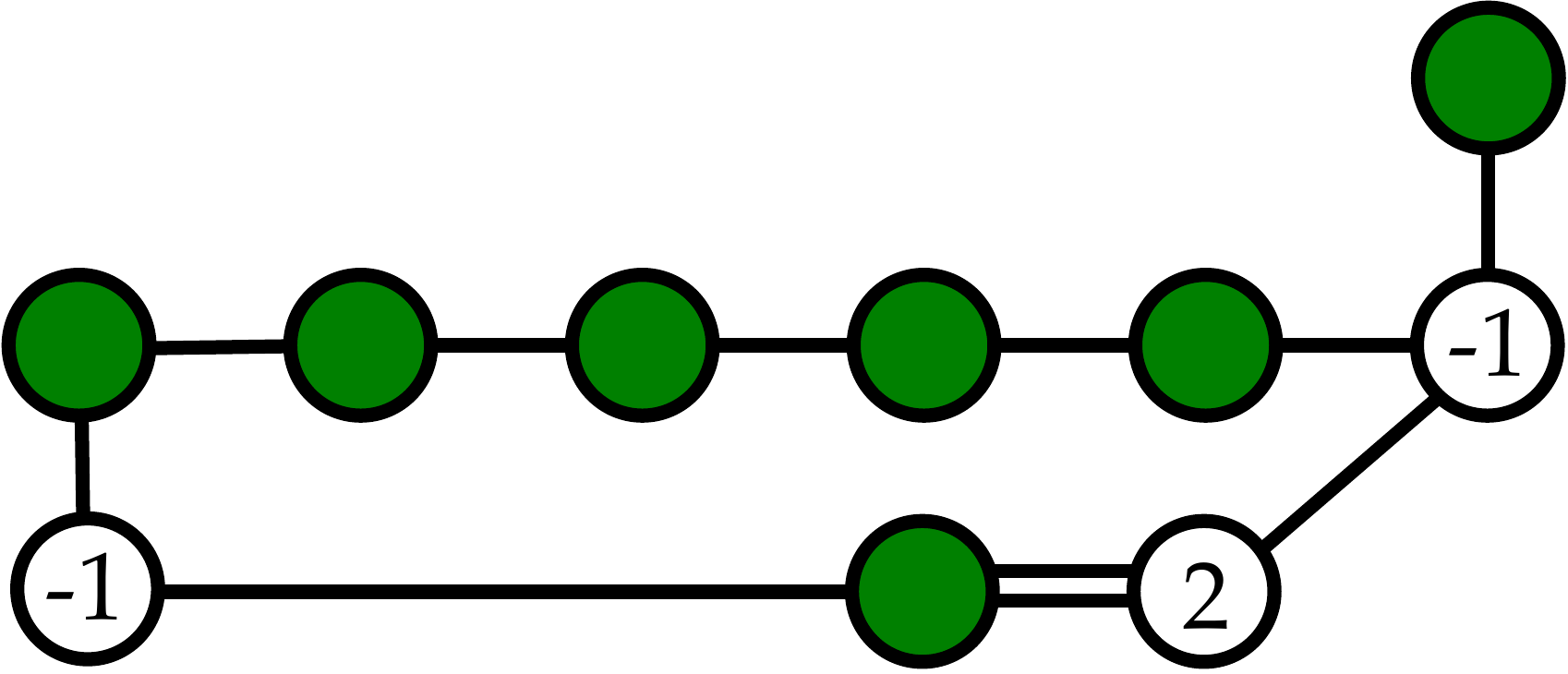}} & \multirow{3}{*}{$SU(6)\times SU(2)^2$} & {$SU(3)_{0}+6\mbf{F}$} &  $(\mbf{6},\mbf{2},\mbf{1})$& $(\mbf{15},\mbf{1},\mbf{1})$\cr 
& & & & & {\small $[2]-SU(2)-SU(2)-[2]$}  & $(\bar{\mbf{6}},\mbf{1},\mbf{2})$ & $(\overline{\mbf{15}},\mbf{1},\mbf{1})$\cr
& & & & & & & $(\mbf{1},\mbf{2},\mbf{2})$ \cr\hline

\multirow{3}{*}{13} &\multirow{3}{*}{7} & \multirow{3}{*}{\includegraphics[height=1cm]{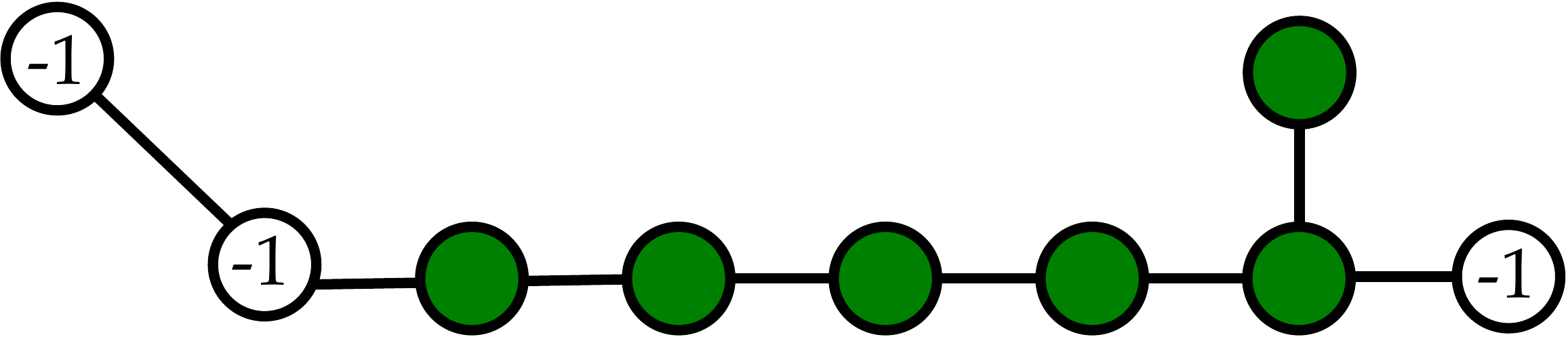}} & \multirow{3}{*}{\includegraphics[height=1cm]{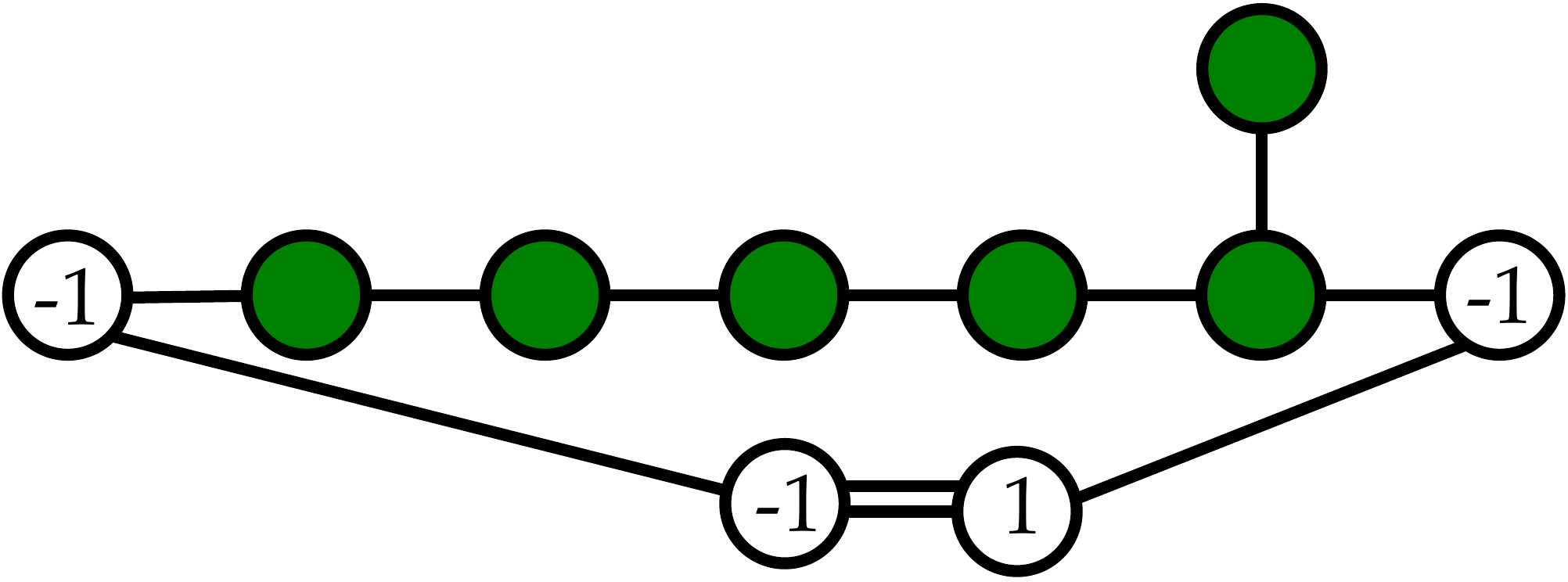}} &  \multirow{3}{*}{$SU(7)\times U(1)$} & {$SU(3)_1+6\mbf{F}$}& $\mbf{1}_{-1},\mbf{7}_1$& $\mbf{7}_0,\bar{\mbf{7}}_1$ \cr 
& & & & & {{\small $[1]-SU(2)-SU(2)-[3]$}} & $\overline{\mbf{21}}_0$ & $\mbf{35}_0$ \cr
& & & & & & & \cr\hline

\multirow{3}{*}{14} &\multirow{3}{*}{7} & \multirow{3}{*}{\includegraphics[height=.8cm]{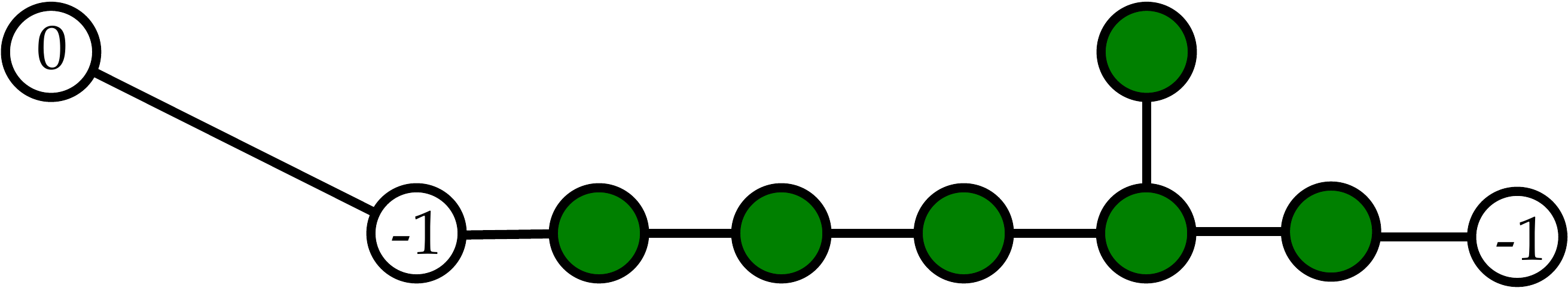}} &
\multirow{3}{*}{\includegraphics[height=1cm]{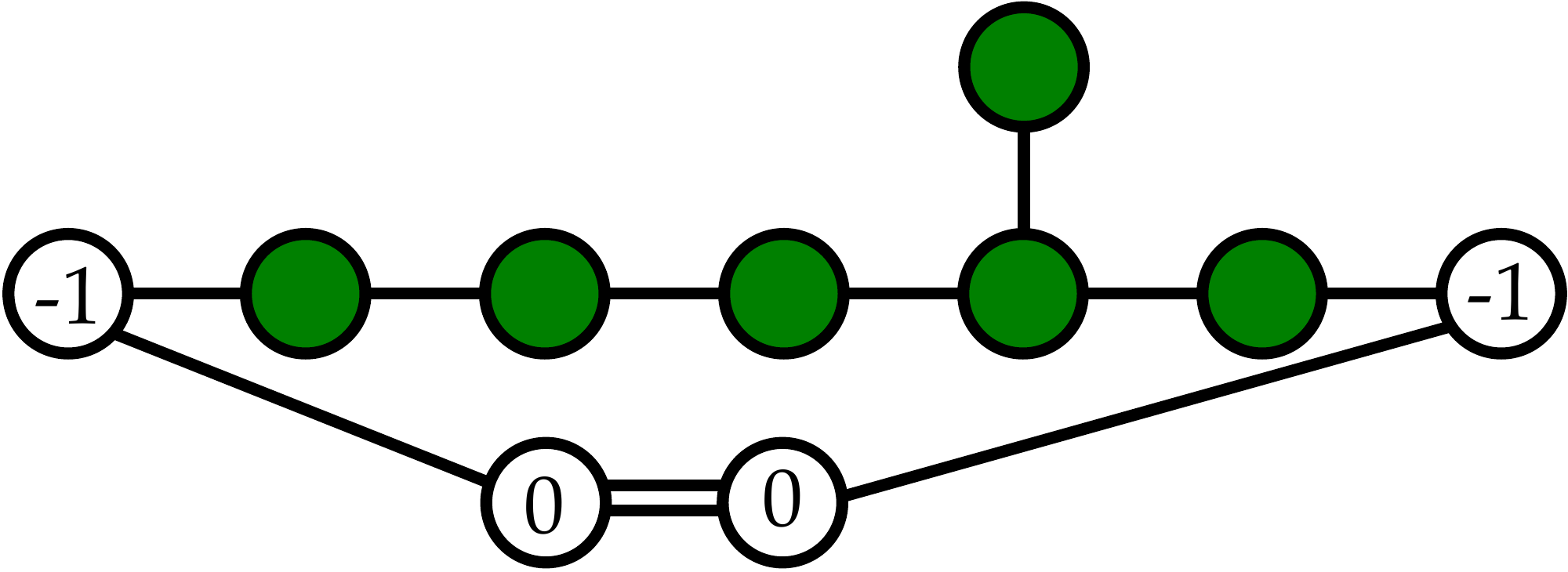}}
& \multirow{3}{*}{$SO(12) \times U(1)$} & {{$SU(3)_{2}+6\mbf{F}$}} & \multirow{2}{*}{$\mbf{12}_1,\mbf{32}_0$} & $\mbf{1}_0,\mbf{1}_2$ \cr
&& & & & {$Sp(2)+6\mbf{F}$} &  & $\mbf{32}'_1,\mbf{66}_0$\cr
& & & & & {{\small $[4]-SU(2)-SU(2)_\pi$}} & &\cr\hline

\multirow{3}{*}{15} &\multirow{3}{*}{7} & & \multirow{3}{*}{\includegraphics[height=1cm]{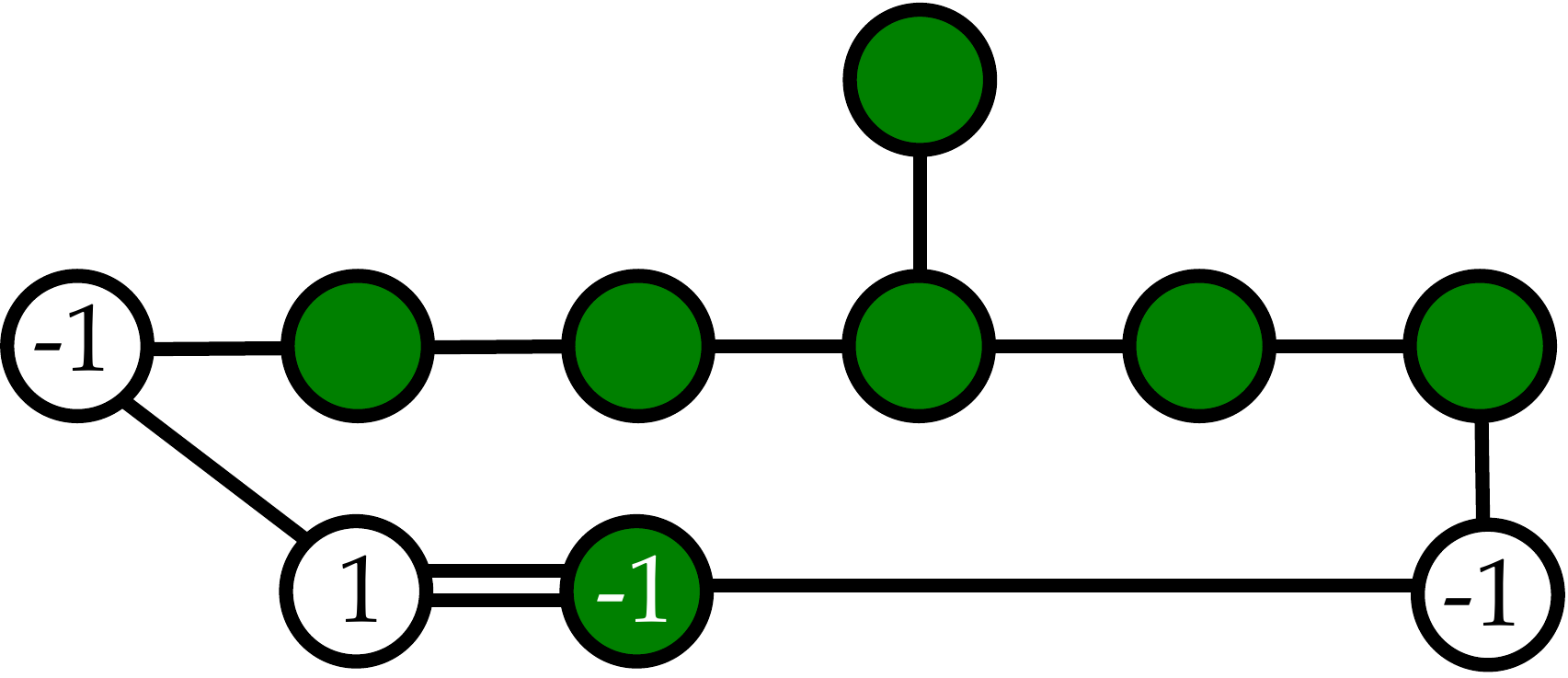}} &  \multirow{3}{*}{$E_6\times SU(2)$} & {$SU(3)_3+6\mbf{F}$} & $(\mbf{27},\mbf{1})$ & $(\overline{\mbf{27}},\mbf{1})$\cr
& & & & & {$Sp(2)+1\mbf{AS}+5\mbf{F}$}  & $(\overline{\mbf{27}},\mbf{2})$  & $(\mbf{27},\mbf{3})$\cr
& & & & & &  & $(\mbf{78},\mbf{2})$\cr\hline
\end{tabular}
\caption{Table summarizing all Rank Two SCFTs (Continued).}
\end{sidewaystable}

\begin{sidewaystable}
\begin{tabular}{|c|c|c|c|c|c|c|c|c|}\hline
No. & $M$ & $(D_{10}, I_1)$ CFD & $(E_8,SU(2))$ CFD &  Model 3 CFD &  Flavor  $G_\text{F}$& Gauge Theory  & BPS Spin 0 &  Spin 1 \cr \hline\hline 
\multirow{3}{*}{16} &\multirow{3}{*}{7} & & & \multirow{3}{*}{\includegraphics[height=1cm]{CFD-Model3-Rank2-Top.pdf}} & \multirow{3}{*}{$-$}& {\small $SU(3)_4+6\mbf{F}$} & & \cr
& & & & & & {\small $Sp(2)+2\mbf{AS}+4\mbf{F}$} & & \cr
& & &  & & & {\small $G_2+6\mbf{F}$} & & \cr\hline

17 & 6 & \includegraphics[height=.8cm]{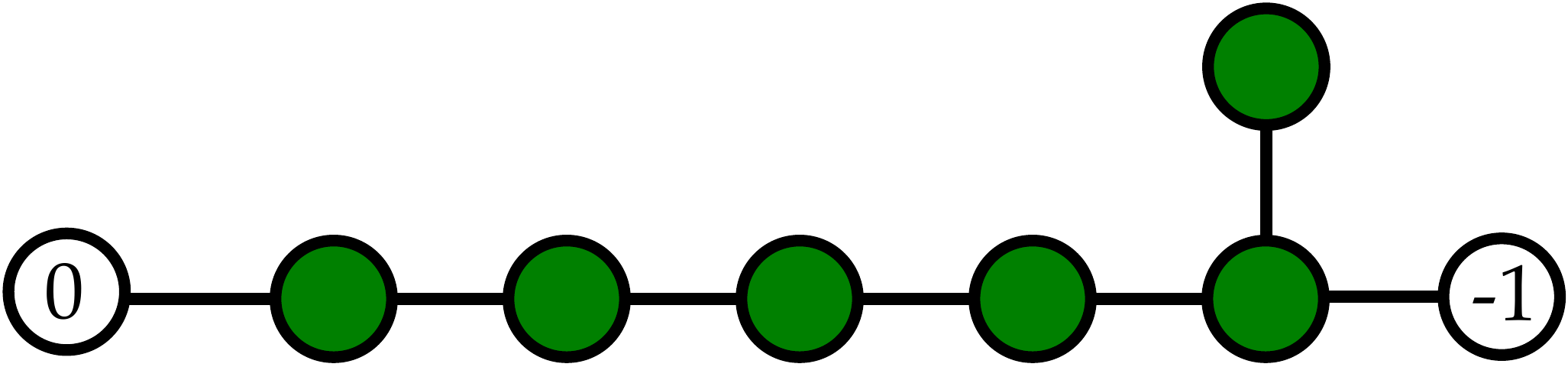} & \includegraphics[height=1cm]{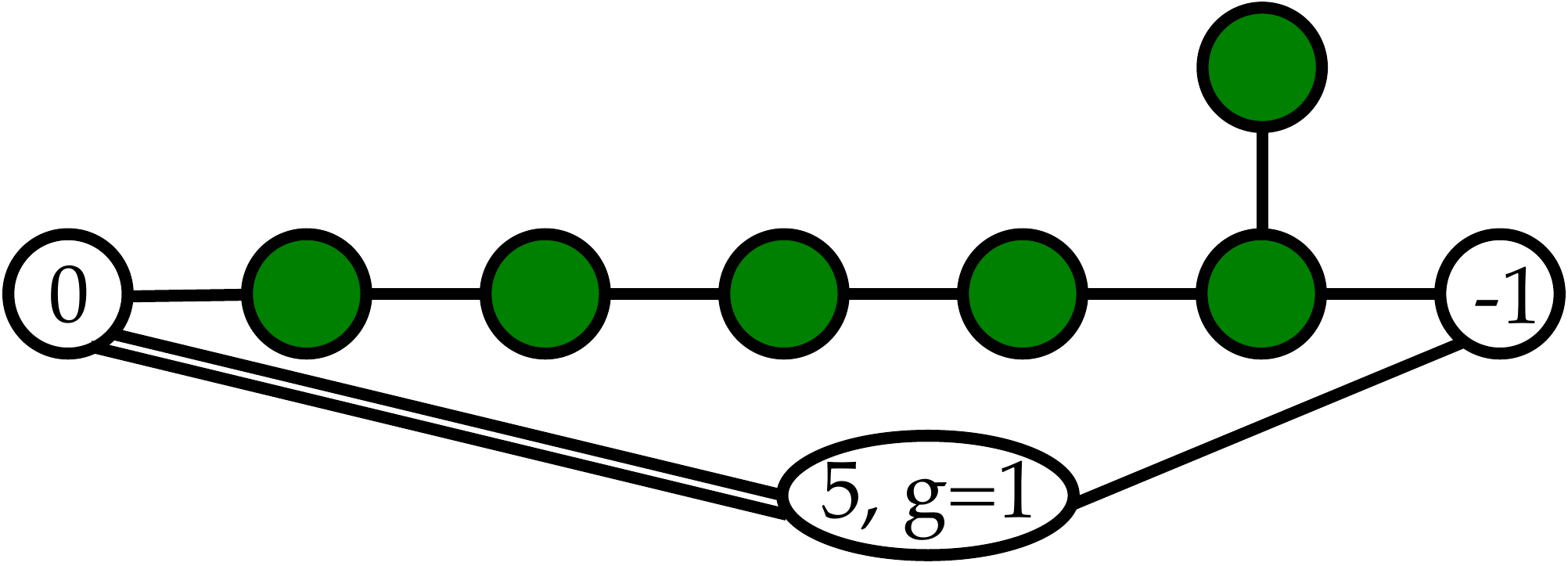} & & $SU(7)$ & 
{\small $[3]-SU(2)-SU(2)_0$} & $\overline{\mbf{21}}$ & $\mbf{7},\mbf{35}$ \cr\hline

\multirow{3}{*}{18} &\multirow{3}{*}{6} & \multirow{3}{*}{\includegraphics[height=.8cm]{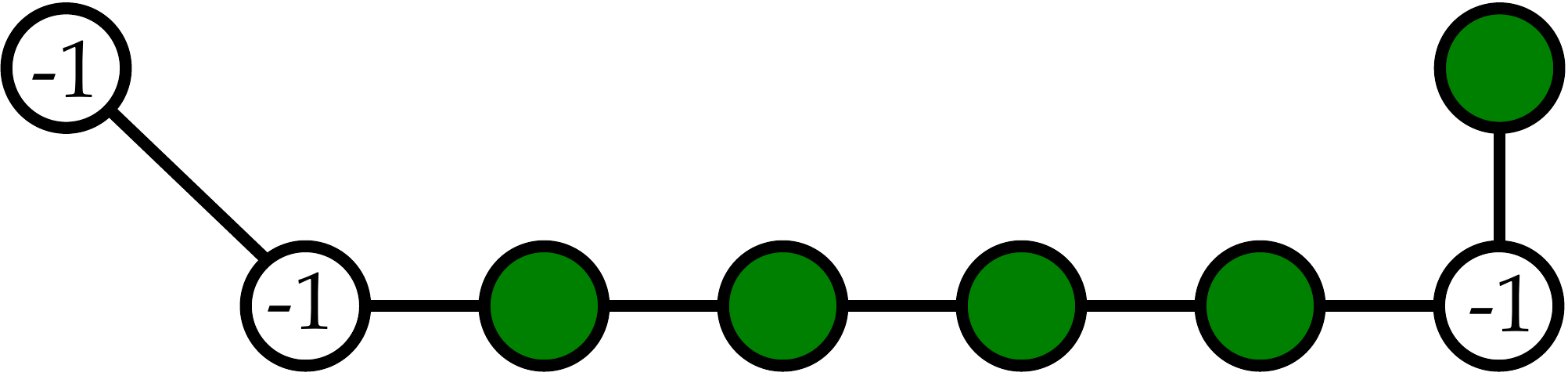}} & \multirow{3}{*}{\includegraphics[height=1cm]{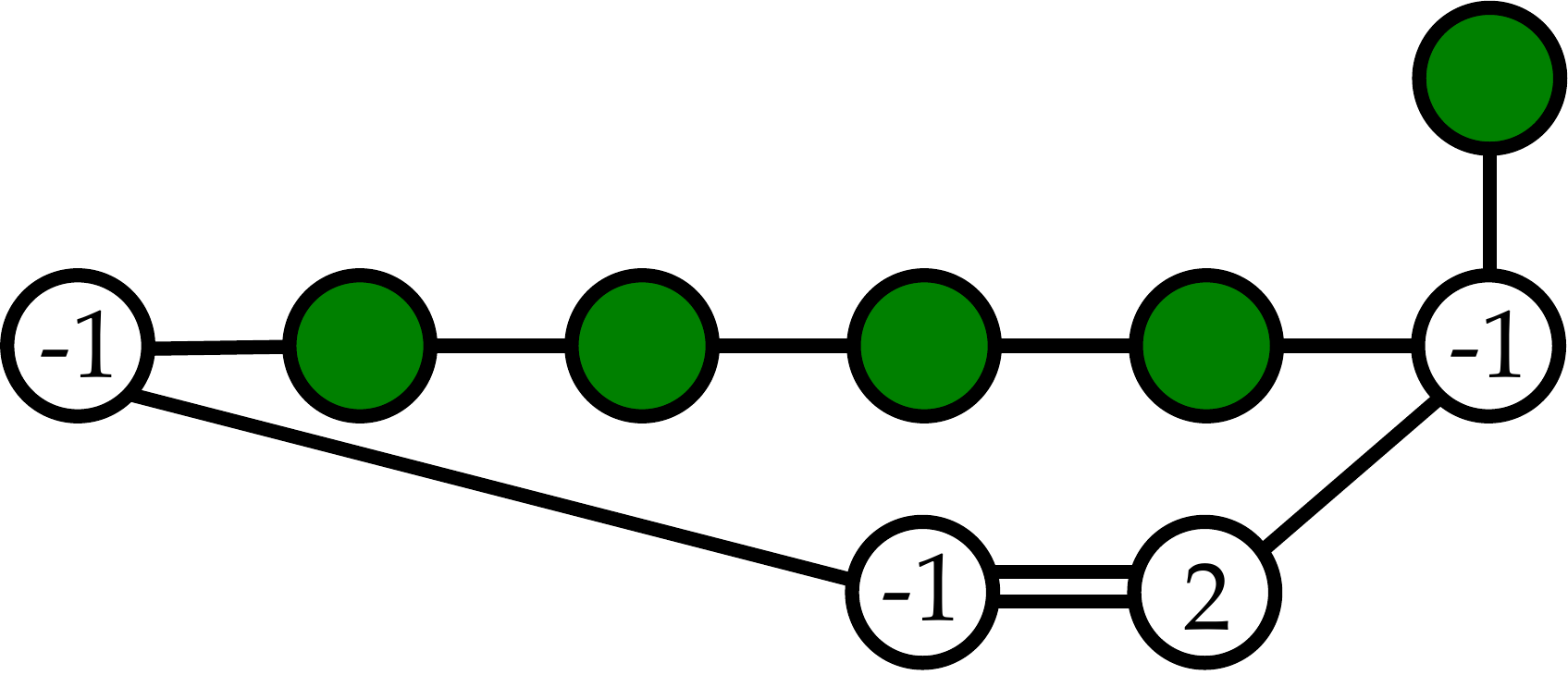}}  & & \multirow{3}{*}{\footnotesize{$SU(5)\times SU(2) \times U(1)$}} &  {\small $SU(3)_{1/2}+5\mbf{F}$}  & $(\mbf{1},\mbf{1})_{-1}$ & $(\mbf{1},\mbf{2})_1$ \cr
& & & & & & \tiny{$[1]-SU(2)-SU(2)-[2]$}  & $(\mbf{5},\mbf{1})_1$ & $(\mbf{5},\mbf{1})_0$ \cr
& & & & & &  & $(\bar{\mbf{5}},\mbf{2})_0$ & $(\overline{\mbf{10}},\mbf{1})_0$\cr\hline

\multirow{3}{*}{19} &\multirow{2}{*}{6} & \multirow{3}{*}{\includegraphics[height=.7cm]{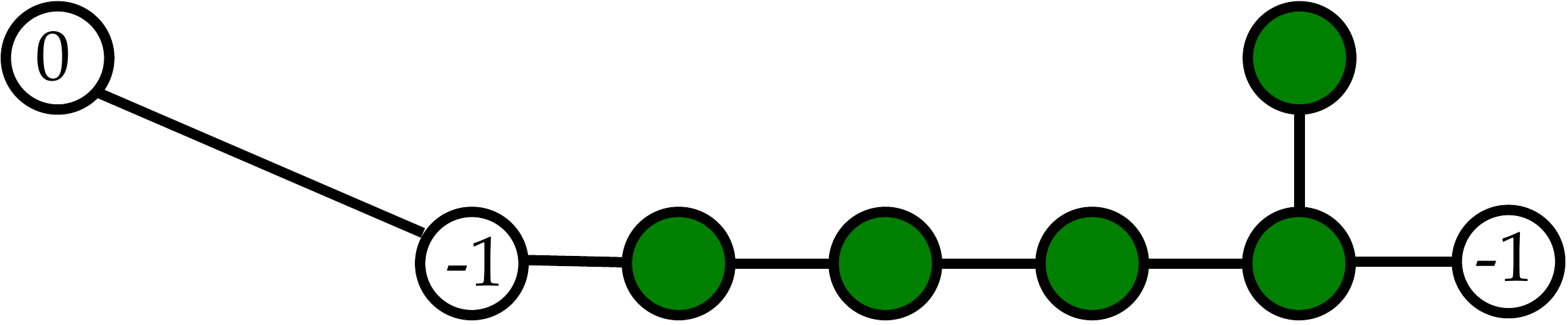}} &  \multirow{3}{*}{\includegraphics[height=1cm]{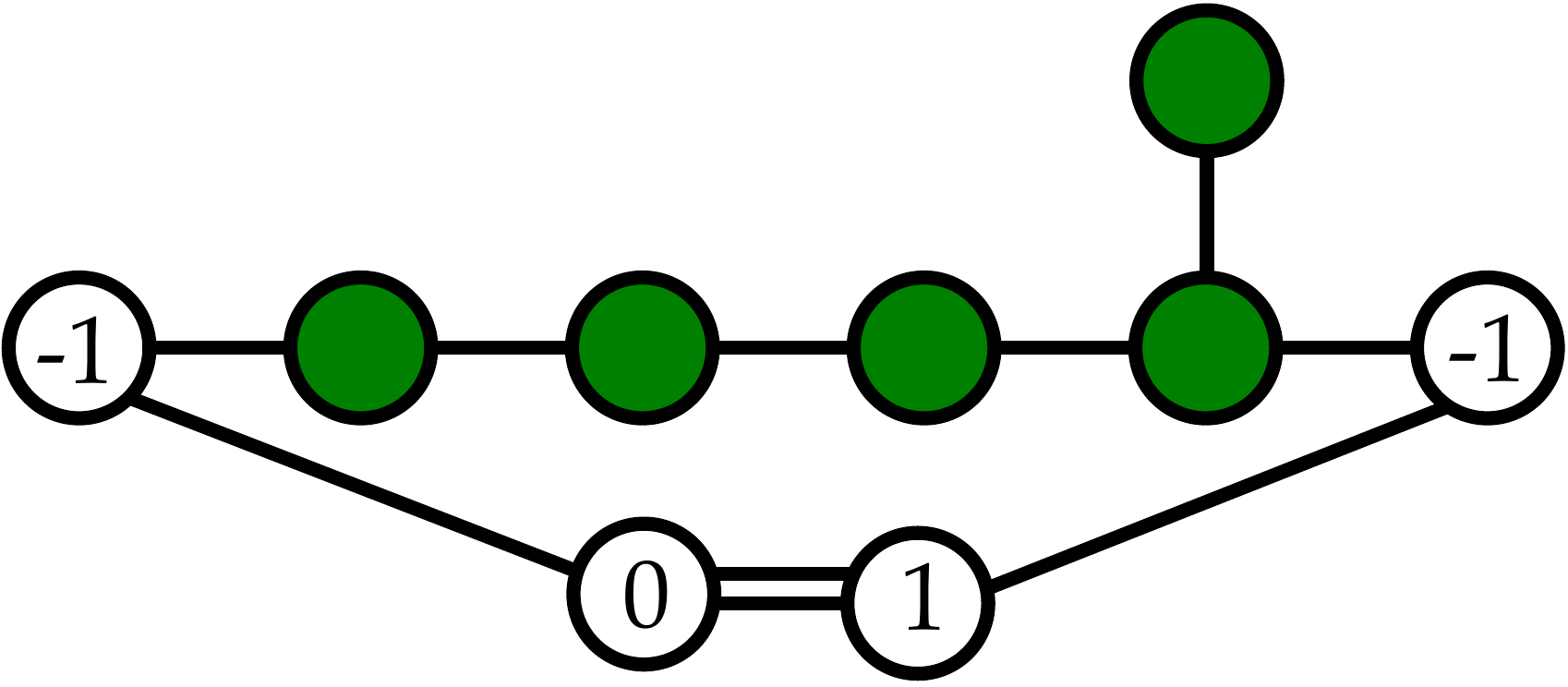}} & &  \multirow{3}{*}{$SU(6)\times U(1)$} & {\small $SU(3)_{3/2}+5\mbf{F}$} & \multirow{3}{*}{$\mbf{6}_1,\overline{\mbf{15}}_0$} & $\mbf{1}_0,\overline{\mbf{6}}_1$\cr 
 & & & & & & {{\small $[3]-SU(2)-SU(2)_\pi$}} & & $\mbf{15}_0$\cr
 &&&&&&&&\cr 
 \hline
 
\multirow{3}{*}{20} & \multirow{3}{*}{6} & \multirow{3}{*}{\includegraphics[height=.6cm]{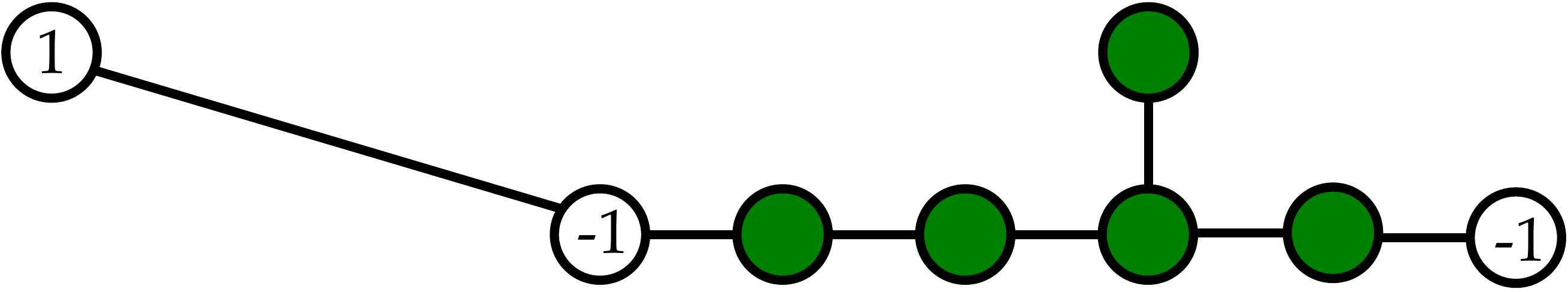}} & 
\multirow{3}{*}{\includegraphics[height=1cm]{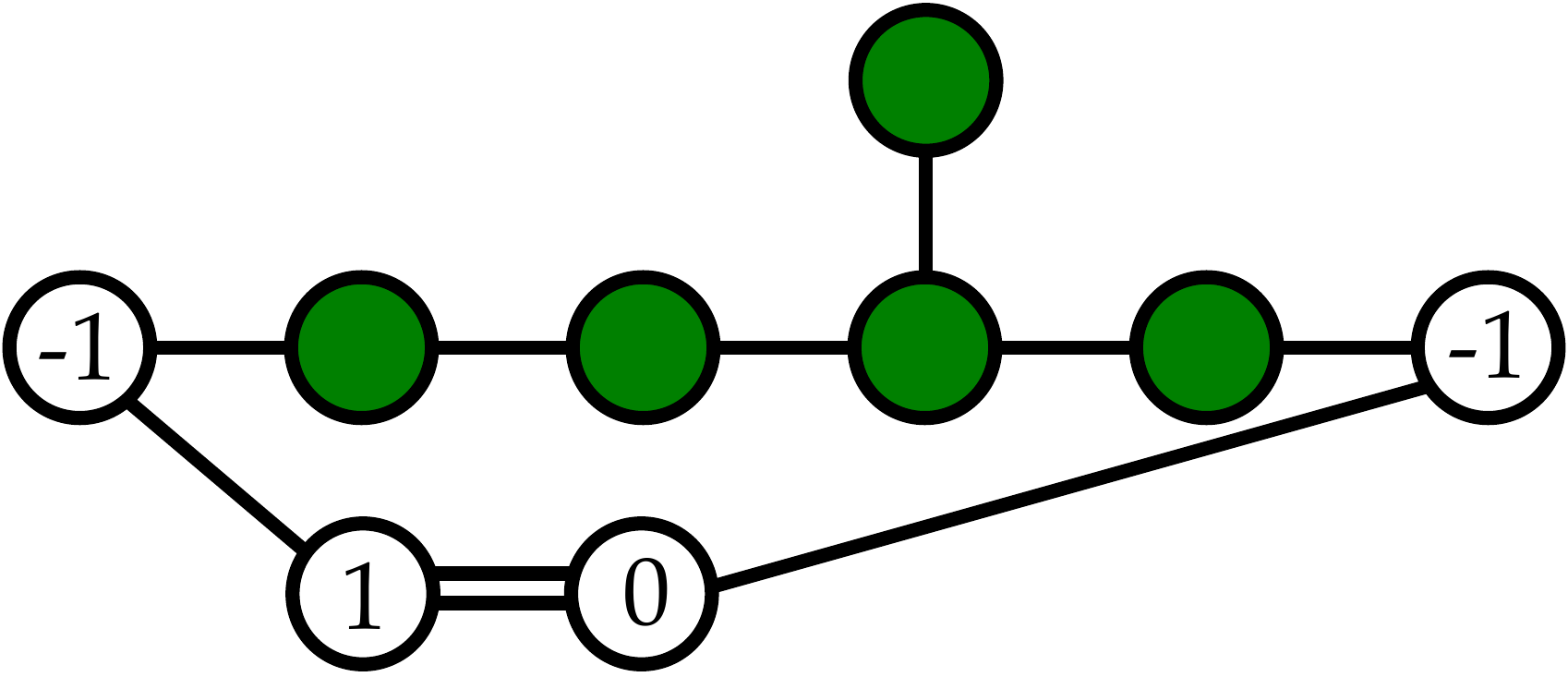}}
& & \multirow{2}{*}{$SO(10) \times U(1)$} & {\small $SU(3)_{5/2}+5\mbf{F}$} & \multirow{3}{*}{$\mbf{10}_1,\mbf{16}_0$} & $\mbf{1}_2,\mbf{10}_0$ \cr 
&& & & & & {\small $Sp(2)+5\mbf{F}$} & & $\overline{\mbf{16}}_1$\cr
 &&&&&&&&\cr \hline

\multirow{3}{*}{21} & \multirow{3}{*}{6} & & \multirow{3}{*}{\includegraphics[height=1cm]{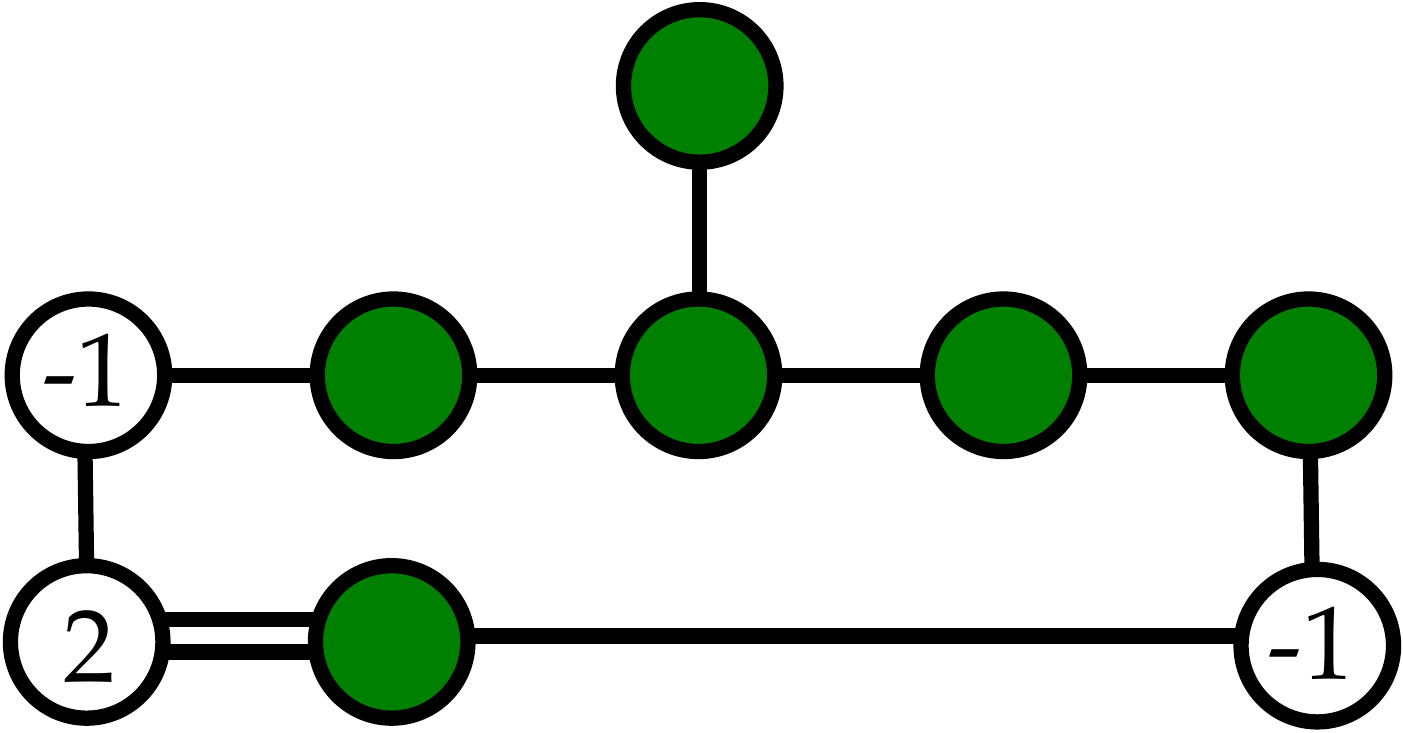}} & \multirow{2}{*}{\includegraphics[height=1cm]{CFD-E8-13.pdf}} &  \multirow{3}{*}{$SO(10)\times SU(2)$} & {\small $SU(3)_{7/2}+5\mbf{F}$} & $(\mbf{16},\mbf{1})$  & $(\mbf{10},\mbf{1})$\cr
& & & & & & {\small $Sp(2)+1\mbf{AS}+4\mbf{F}$} & $(\mbf{10},\mbf{2})$ & $(\overline{\mbf{16}},\mbf{2})$\cr
& & & & & & &&\cr 
\hline

\multirow{3}{*}{22} & \multirow{3}{*}{6} & & & \multirow{3}{*}{\includegraphics[height=1cm]{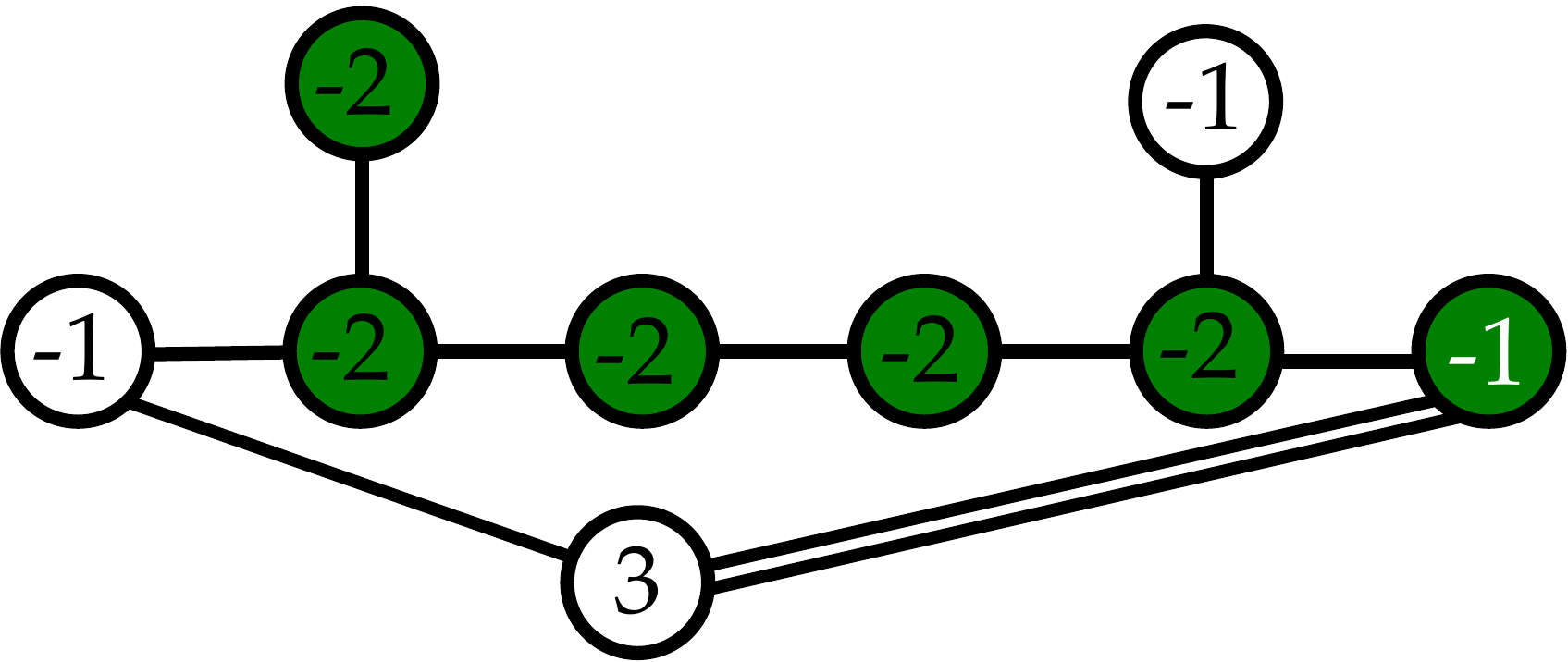}} & \multirow{3}{*}{$Sp(6)$} & {\small $SU(3)_{9/2}+5\mbf{F}$} & $\mbf{65}$ & $\mbf{429}$ \cr
& & & & & & {\small $Sp(2)+2\mbf{AS}+3\mbf{F}$} & $\mbf{572}$ & $\mbf{4576}$ \cr
& & & & & & {\small $G_2+5\mbf{F}$} & & \cr\hline

\multirow{3}{*}{23} & \multirow{3}{*}{5} & \multirow{3}{*}{ \includegraphics[height=.8cm]{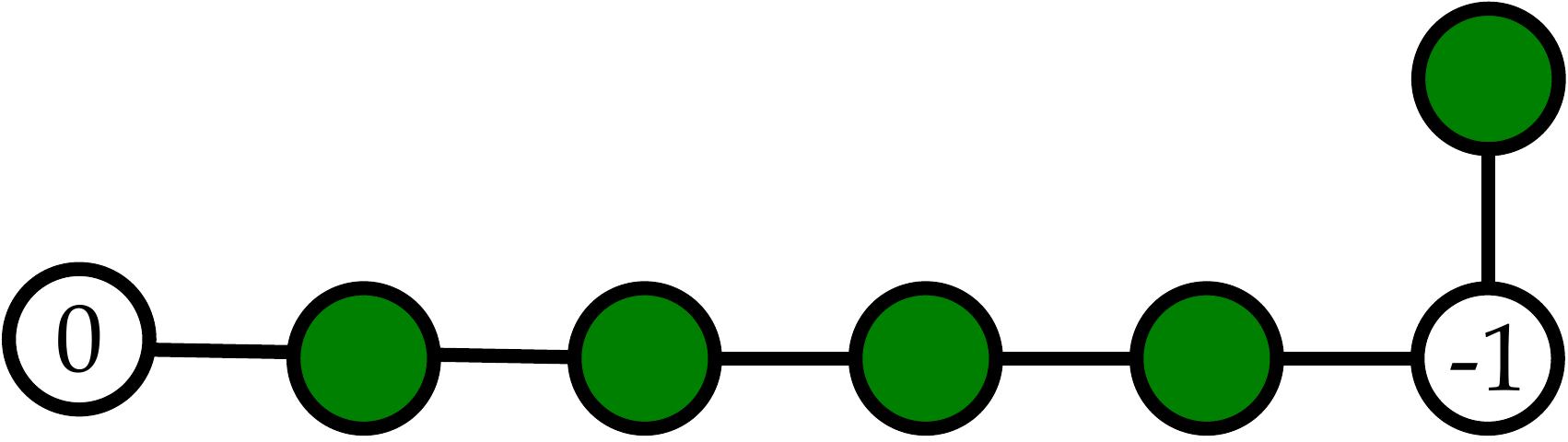}} & \multirow{3}{*}{\includegraphics[height=1cm]{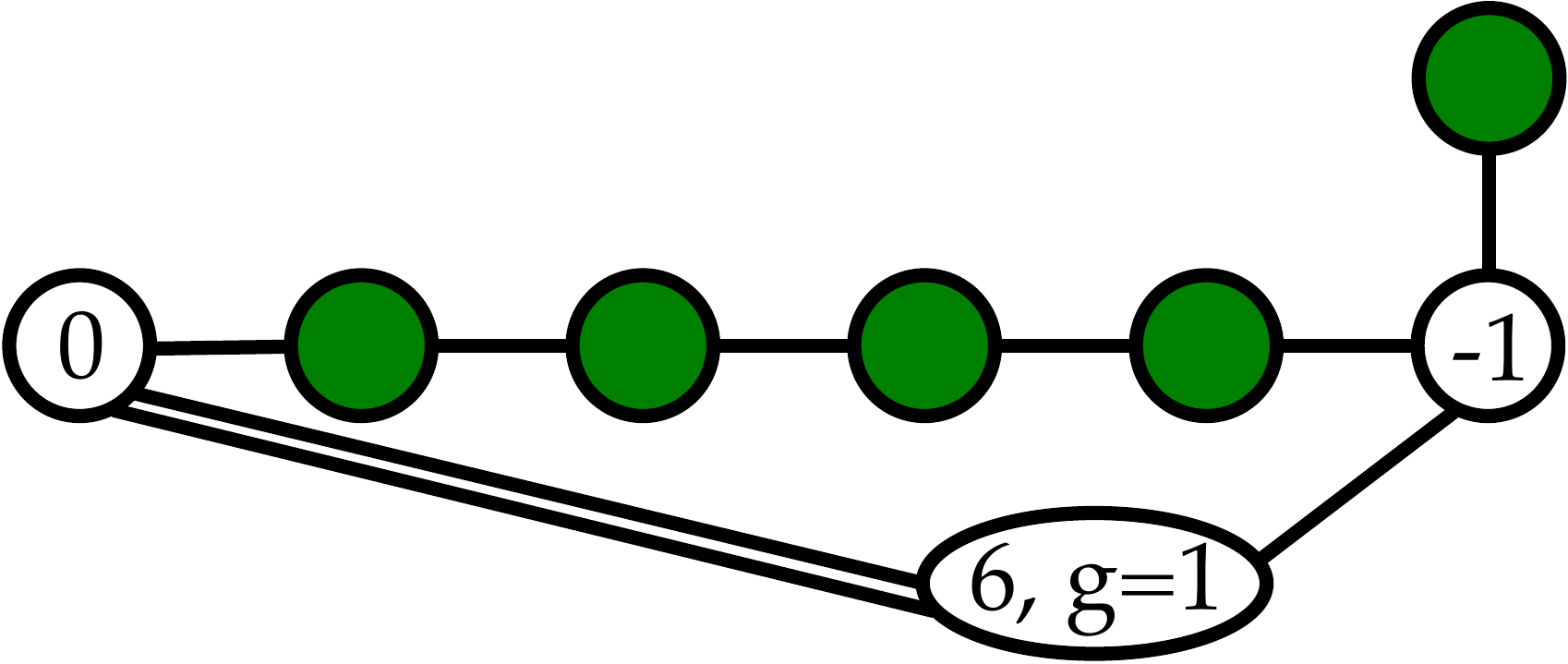}} & & \multirow{3}{*}{$SU(5)\times SU(2)$} &\multirow{3}{*}{{\small $[2]-SU(2)-SU(2)_0$}} & \multirow{2}{*}{$\bar{\mbf{5}},\mbf{2}$} & $(\mbf{5},\mbf{1})$ \cr 
 & & & & & & & &$(\overline{\mbf{10}},\mbf{1})$\cr
 & & & & & & &&\cr \hline
 
\multirow{4}{*}{24} & \multirow{4}{*}{5} & \multirow{4}{*}{\includegraphics[height=.8cm]{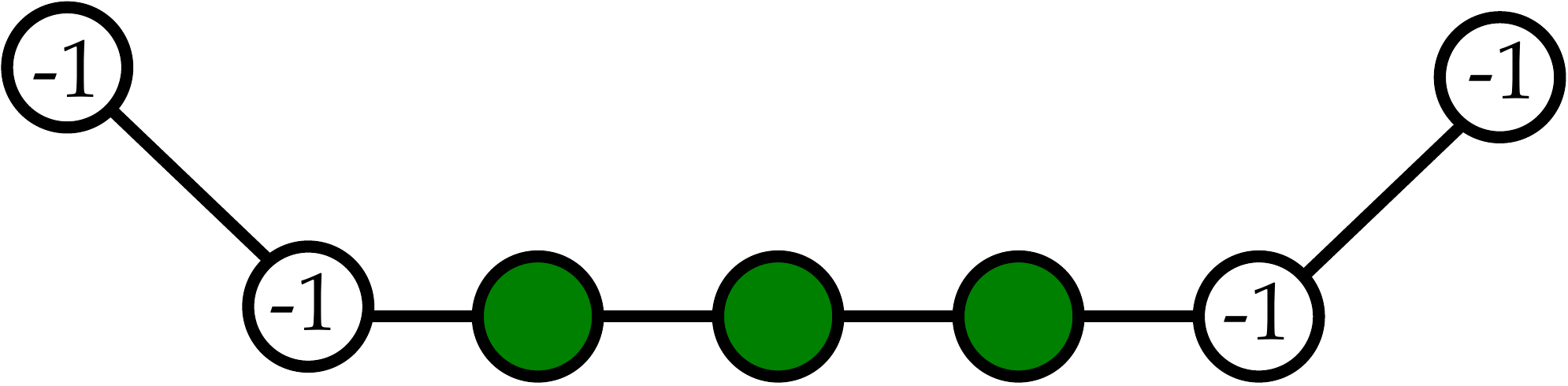}} & \multirow{4}{*}{\includegraphics[height=1cm]{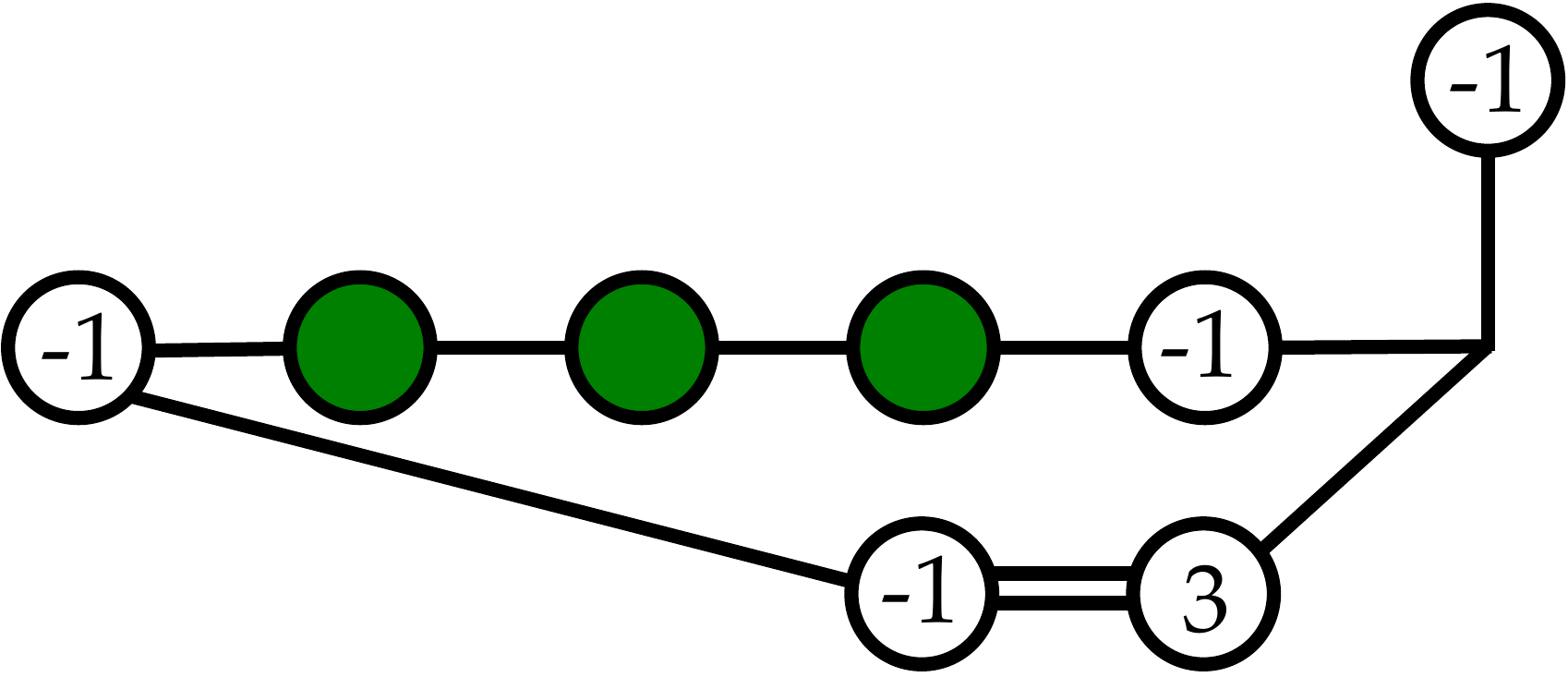}} & & \multirow{4}{*}{$SU(4)\times  U(1)^2$} & & $\mbf{1}_{(-1,0)}$ & $\mbf{4}_{(0,0)}$ \cr  
& & & & & & {\small $SU(3)_0+4\mbf{F}$} & $\mbf{1}_{(0,-1)}$ & $\bar{\mbf{4}}_{(0,0)}$\cr
& & & & & & \tiny{$[1]-SU(2)-SU(2)-[1]$} & $\mbf{4}_{(1,0)}$ & $\mbf{1}_{(1,1)}$\cr
& & & & & & & $\bar{\mbf{4}}_{(0,1)}$ & \cr\hline

\multirow{3}{*}{25} & \multirow{3}{*}{5} & \multirow{3}{*}{\includegraphics[height=.7cm]{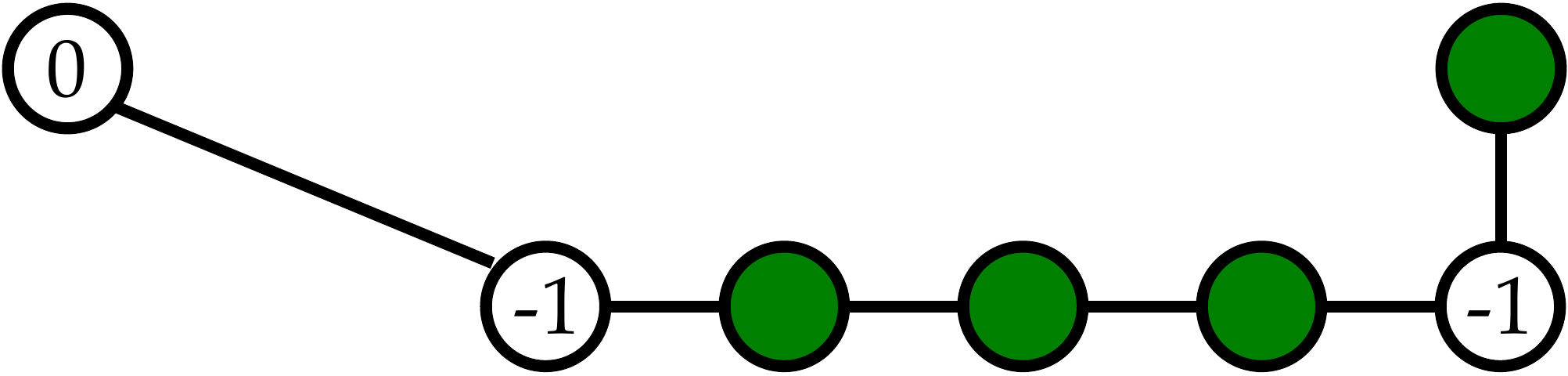}} &  \multirow{3}{*}{\includegraphics[height=1cm]{CFD-E8-21.pdf}} & & \multirow{3}{*}{\footnotesize{$SU(4)\times SU(2) \times U(1)$}} & {\small $SU(3)_1+4\mbf{F}$}& $(\mbf{4},\mbf{1})_1$ & $(\mbf{1},\mbf{1})_0$ \cr
& & & & & & {\small $[2]-SU(2)-SU(2)_\pi$} & $(\bar{\mbf{4}},\mbf{2})_0$ & $(\mbf{1},\mbf{2})_1$\cr
& & & & & & & & $(\mbf{6},\mbf{1})_0$\cr\hline
\end{tabular}
\caption{Table summarizing all Rank Two SCFTs (Continued).}
\end{sidewaystable}

\begin{sidewaystable}
\begin{tabular}{|c|c|c|c|c|c|c|c|c|}\hline
No. & $M$ & $(D_{10}, I_1)$ CFD & $(E_8,SU(2))$ CFD &  Model 3 CFD &  Flavor  $G_\text{F}$& Gauge Theory  & BPS Spin 0 &  Spin 1 \cr \hline\hline

26 & 5 & \includegraphics[height=.6cm]{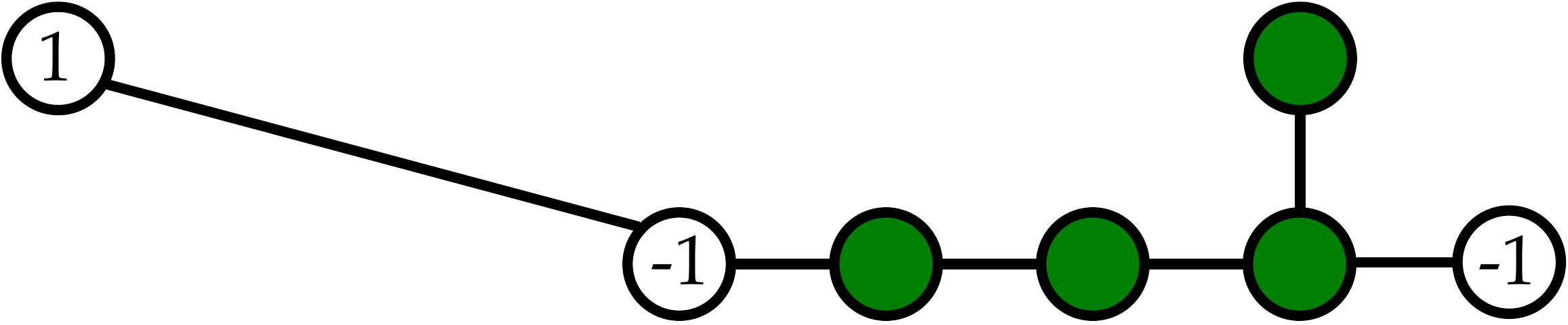} & \includegraphics[height=1cm]{CFD-E8-20.pdf} & & $SU(5)\times U(1)$ & {\small $SU(3)_2+4\mbf{F}$}& $\mbf{5}_1,\overline{\mbf{10}}_0$ & $\mbf{5}_0,\bar{\mbf{5}}_1$ \cr \hline

\multirow{3}{*}{27} &\multirow{3}{*}{5} & \multirow{3}{*}{\includegraphics[height=.6cm]{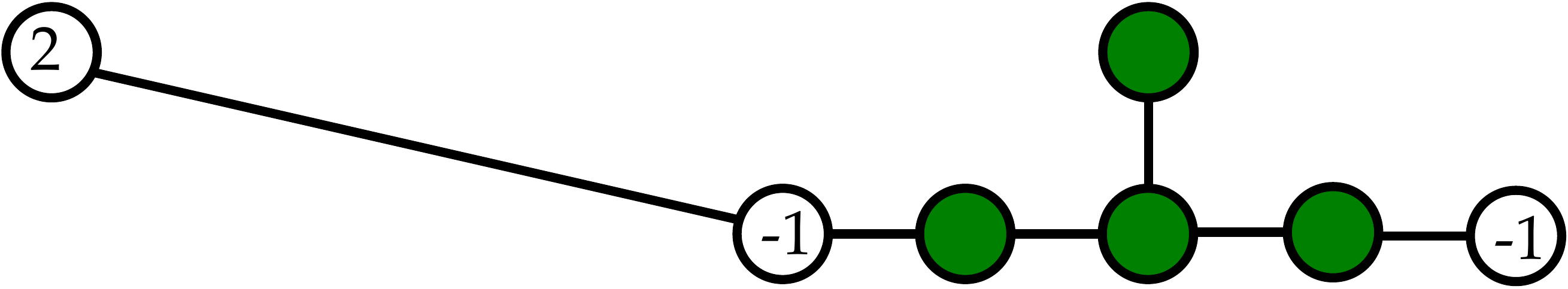}} &
\multirow{3}{*}{\includegraphics[height=1cm]{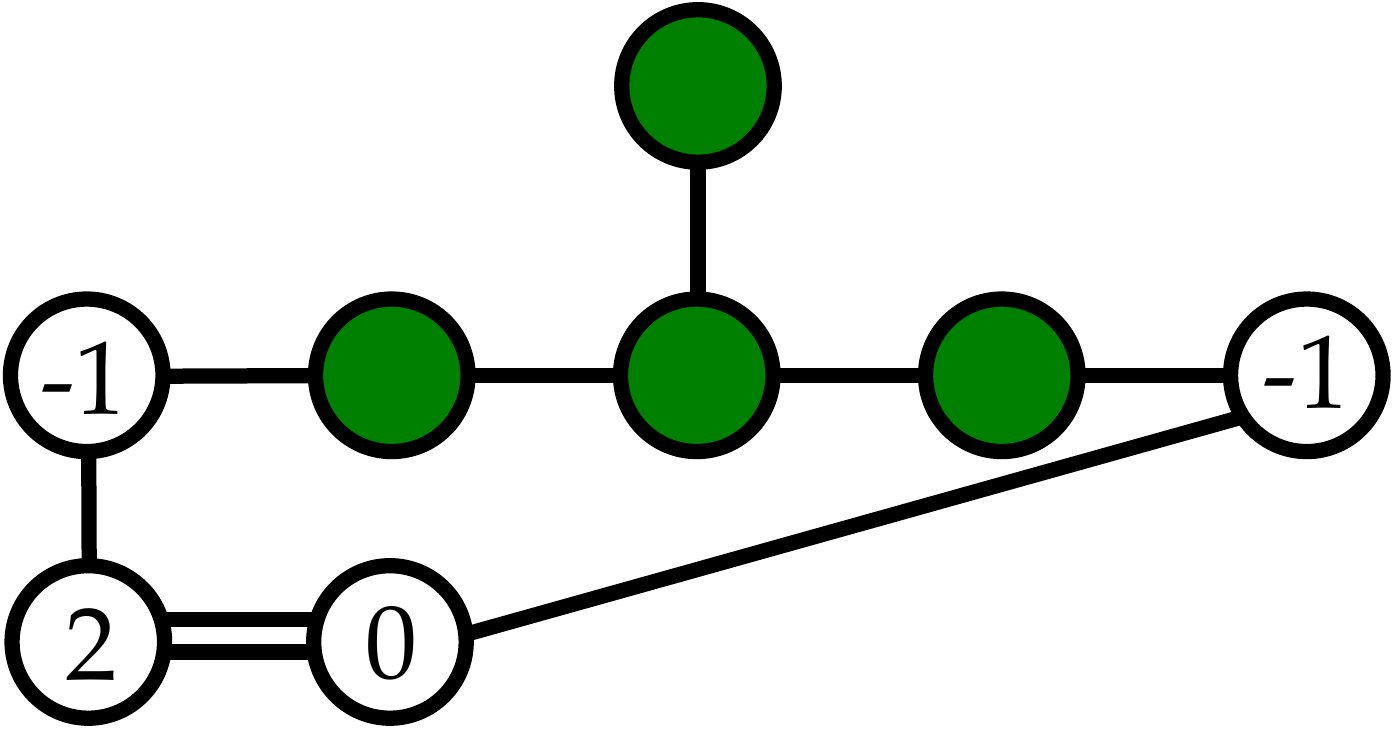}}
& \multirow{3}{*}{\includegraphics[height=1cm]{CFD-E8-19.pdf}} & \multirow{2}{*}{$SO(8) \times U(1)$} & {\small $SU(3)_{3}+4\mbf{F}$} & 
\multirow{3}{*}{$\mbf{8}_{v,1},\mbf{8}_{s,0}$}& $\mbf{1}_0,\mbf{1}_2$\cr
& & & & & & {\small $Sp(2)+4\mbf{F}$} & & $\mbf{8}_{c,1}$\cr
& & & & & & &&\cr 
\hline

\multirow{3}{*}{28} & \multirow{2}{*}{5} & & \multirow{3}{*}{\includegraphics[height=1cm]{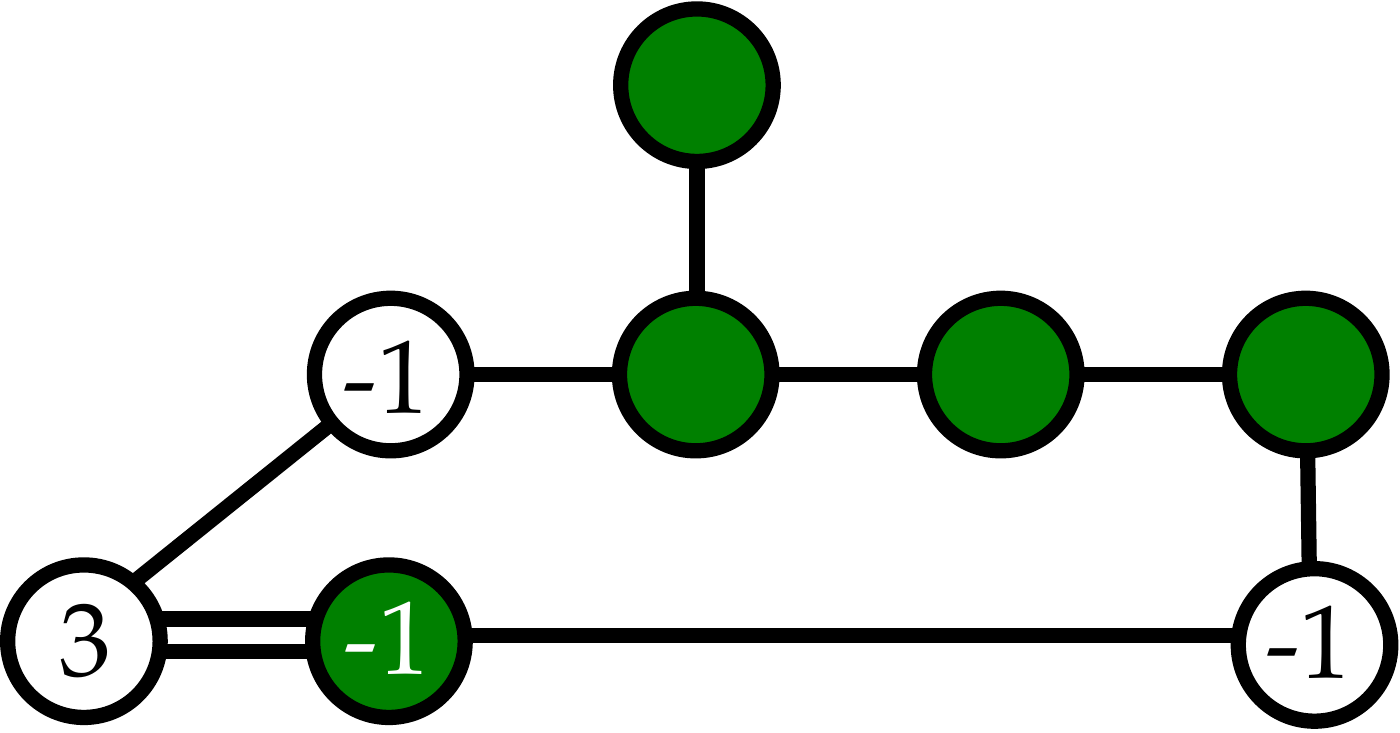}} & \multirow{3}{*}{\includegraphics[height=1cm]{CFD-E8-18.pdf}} &  \multirow{3}{*}{$SU(5)\times SU(2)$} & {\small $SU(3)_4+4\mbf{F}$} & $(\mbf{10},\mbf{1})$ & $(\mbf{5},\mbf{2}),(\bar{\mbf{5}},\mbf{1})$\cr
& & & & & &  {\small $Sp(2)+1\mbf{AS}+3\mbf{F}$} & $(\bar{\mbf{5}},\mbf{2})$ & $(\overline{\mbf{10}},\mbf{1})$\cr
& & & & & & &&\cr \hline

\multirow{3}{*}{29} & \multirow{3}{*}{5} & & & \multirow{3}{*}{\includegraphics[height=1cm]{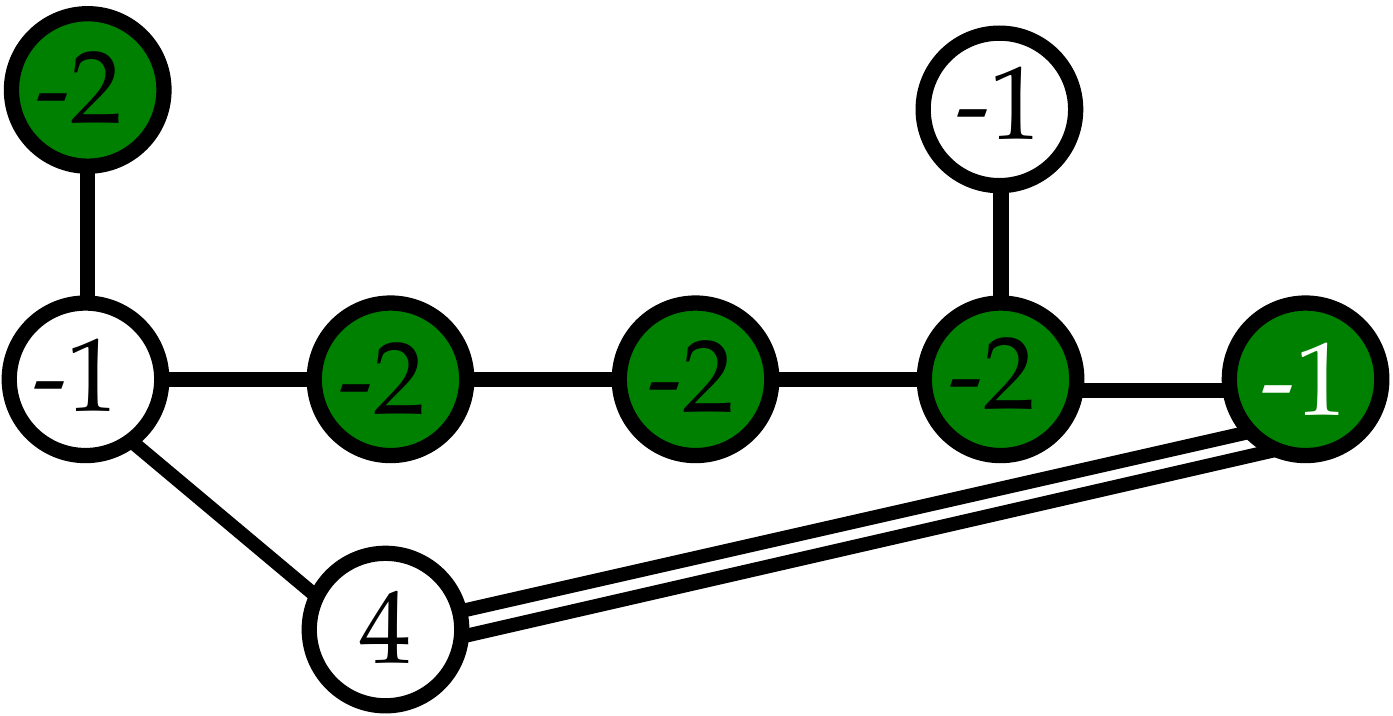}} & 
\multirow{3}{*}{$Sp(4)\times SU(2)$} & {\small $SU(3)_5+4\mbf{F}$} & $(\mbf{8},\mbf{2})$ & $(\mbf{27},\mbf{1})$\cr
& & & & & & {\small $Sp(2)+2\mbf{AS}+2\mbf{F}$} & $(\mbf{48},\mbf{1})$ & $(\mbf{42},\mbf{2})$\cr
& & & & & & {\small $G_2+4\mbf{F}$} & & \cr
 \hline

30 & 4 &\includegraphics[height=.8cm]{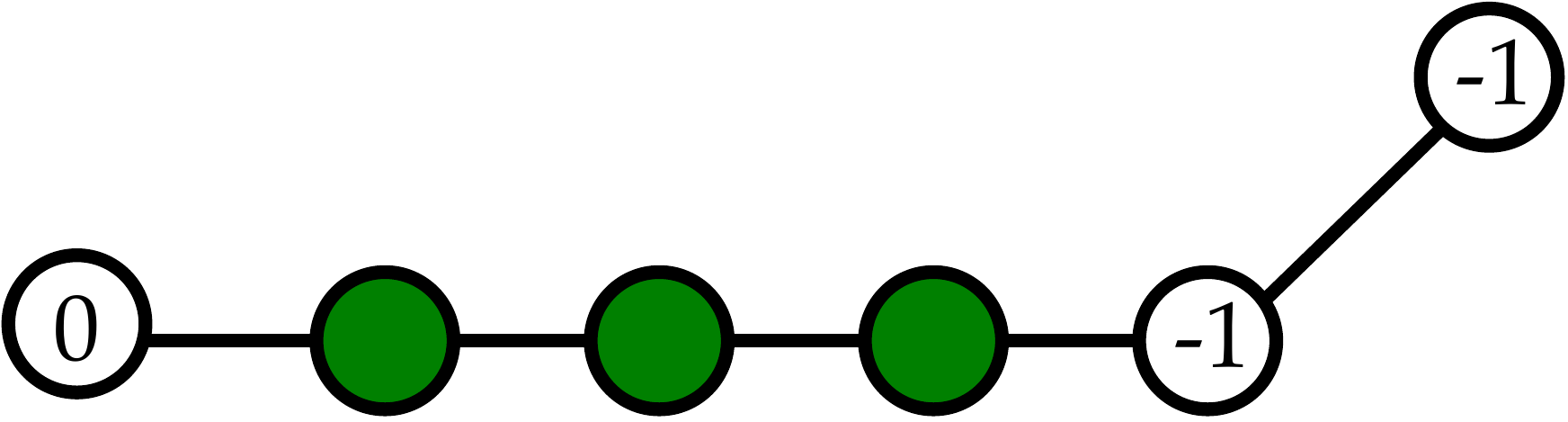} & \includegraphics[height=1cm]{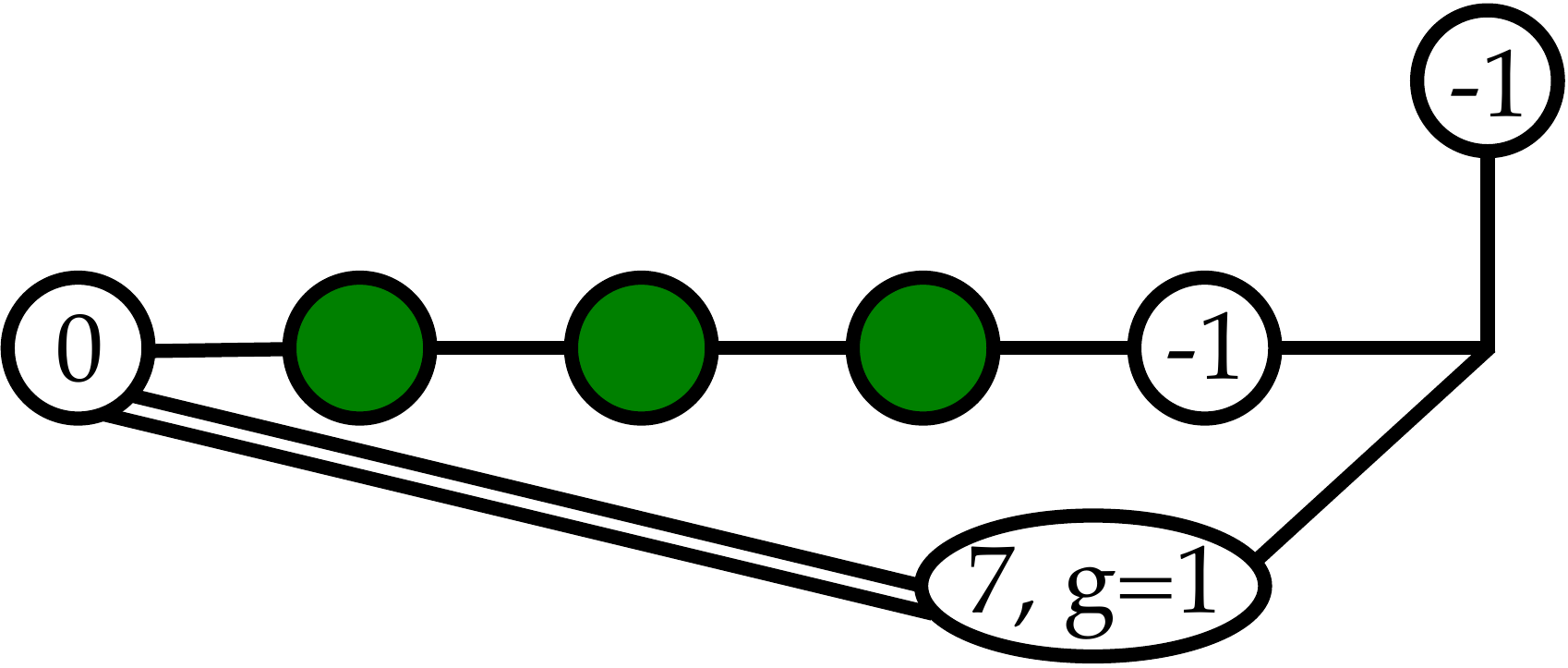} & &
$SU(4)\times U(1)$ & \tiny{$[1]-SU(2)-SU(2)_0$} & $\mbf{1}_{-1},\bar{\mbf{4}}_1$ & $\mbf{4}_0,\bar{\mbf{4}}_0$\cr\hline

\multirow{3}{*}{31} & \multirow{3}{*}{4} & \multirow{2}{*}{\includegraphics[height=.8cm]{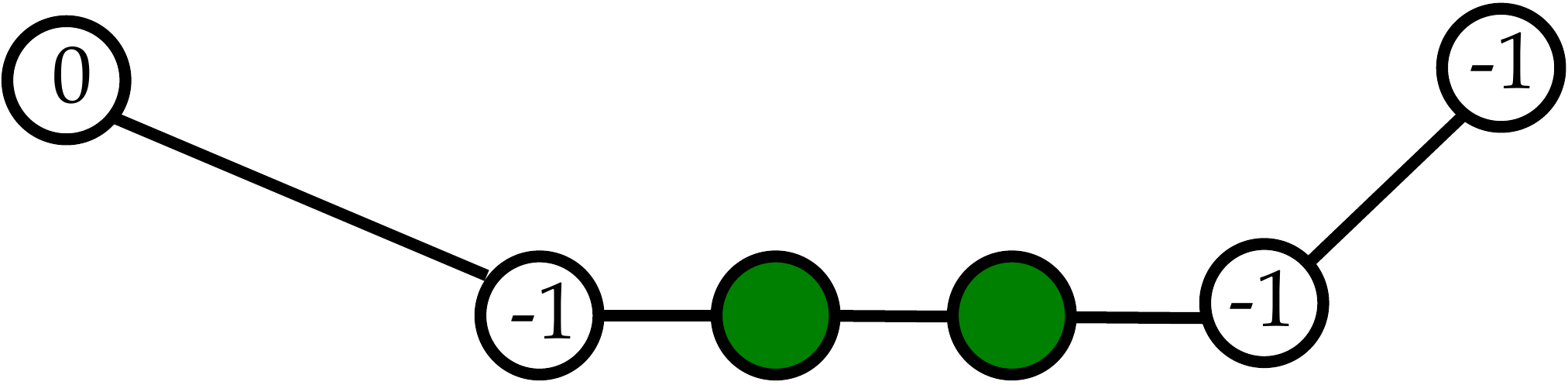}} & \multirow{3}{*}{\includegraphics[height=1cm]{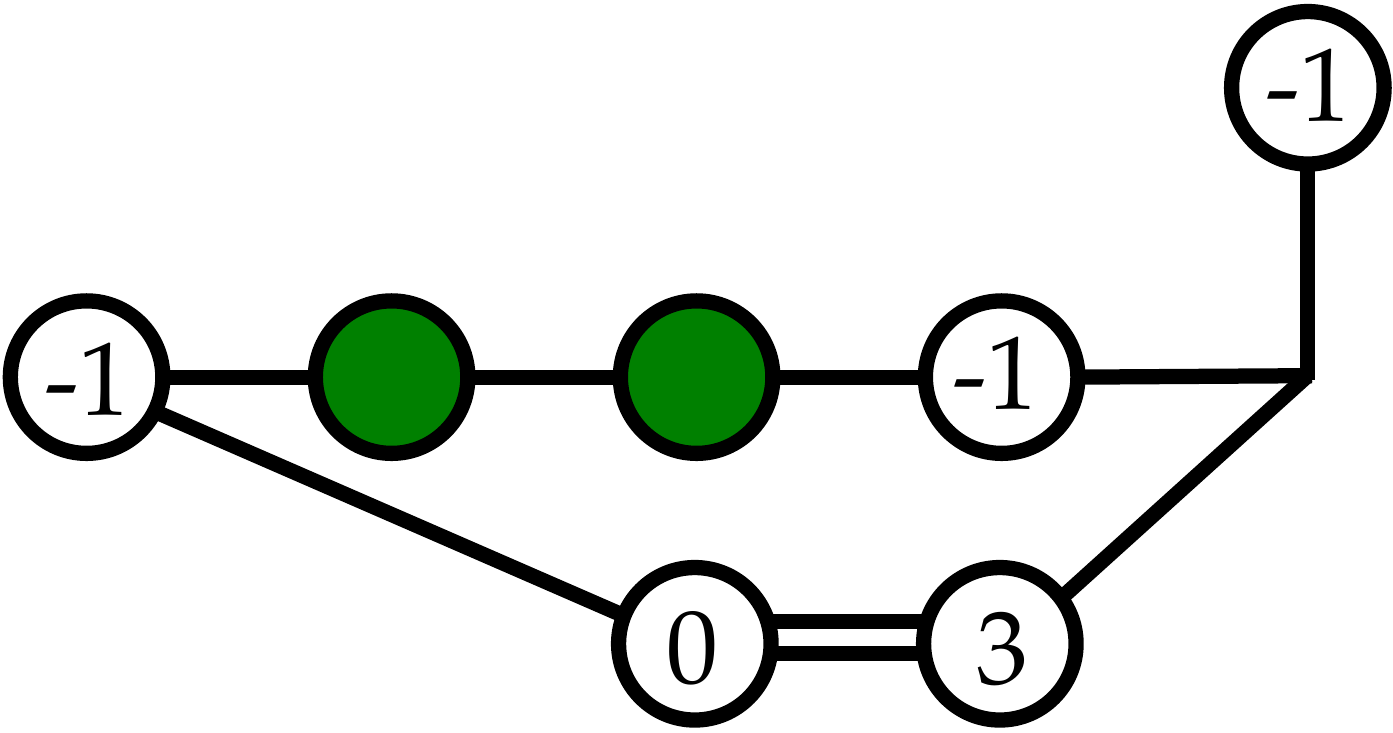}} & & \multirow{3}{*}{$SU(3)\times  U(1)^2$} &{\small $SU(3)_{1/2}+3\mbf{F}$} & $\mbf{1}_{(0,-1)},\mbf{3}_{(1,0)}$ & $\mbf{1}_{(0,0)},\bar{\mbf{3}}_{(0,0)}$ \cr 
& & & & & & {\small $[1]-SU(2)-SU(2)_\pi$} & $\mbf{3}_{(0,1)}$  & $\mbf{1}_{(1,1)}$\cr
& & & & & & &&\cr\hline

\multirow{3}{*}{32} & \multirow{3}{*}{4} &\multirow{2}{*}{\includegraphics[height=.7cm]{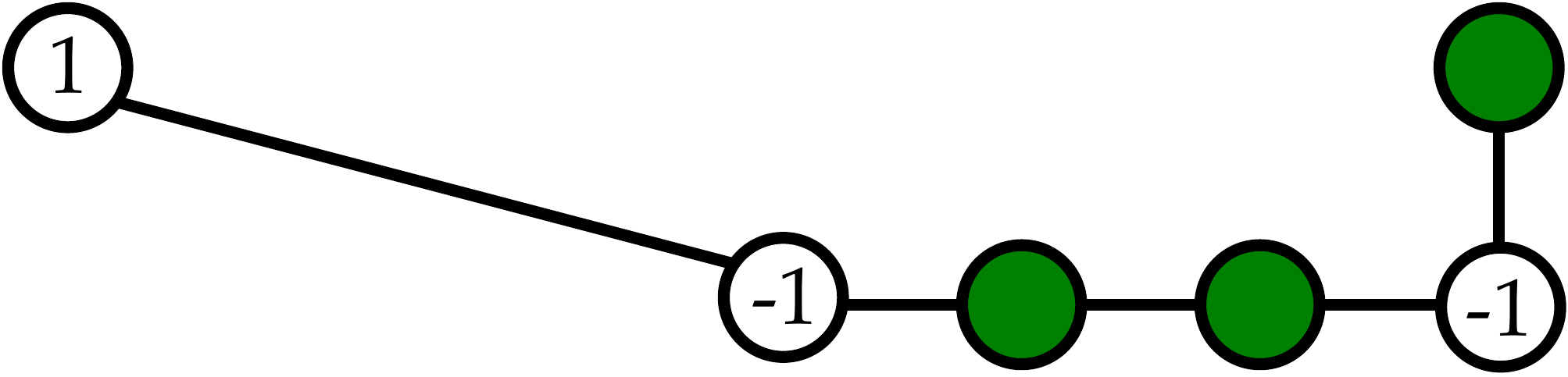}} & \multirow{3}{*}{\includegraphics[height=1cm]{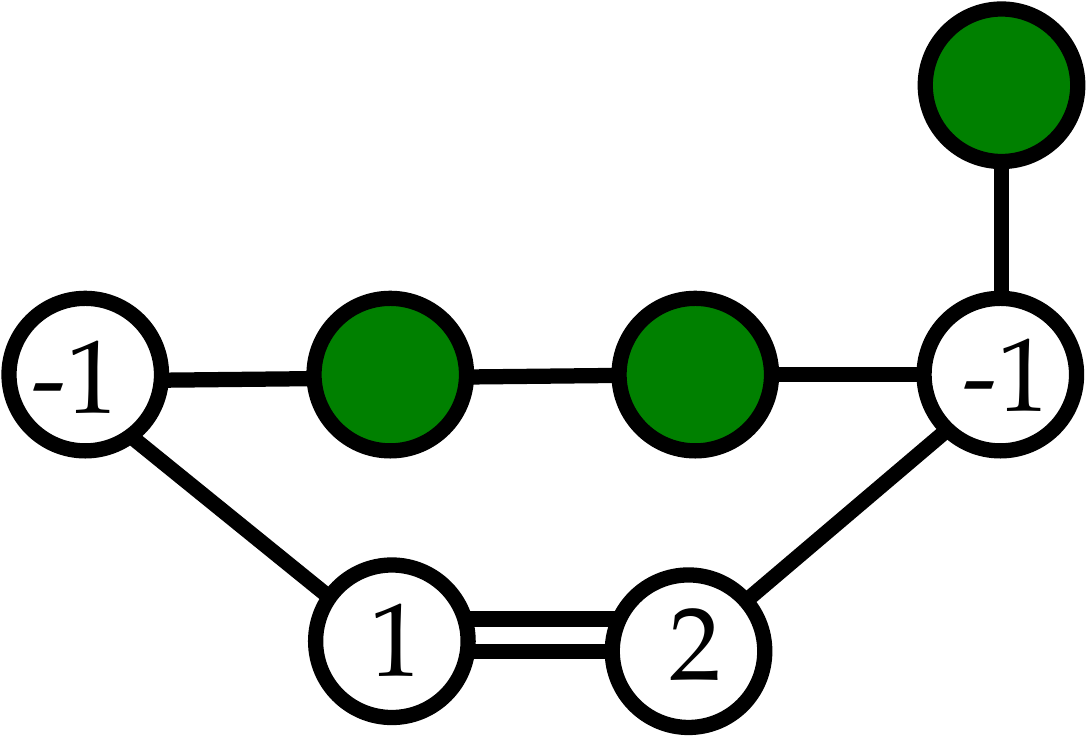}} & & \multirow{3}{*}{\footnotesize{$SU(3)\times U(2)$}} & \multirow{3}{*}{{\small $SU(3)_{3/2}+3\mbf{F}$}} & $(\mbf{3},\mbf{1})_1$ & $(\mbf{1},\mbf{2})_0$ \cr
& & & & & & & $(\bar{\mbf{3}},\mbf{2})_0$ & $(\bar{\mbf{3}},\mbf{1})_0$ \cr
& & & & & & &&\cr\hline

33 & 4 & \includegraphics[height=.6cm]{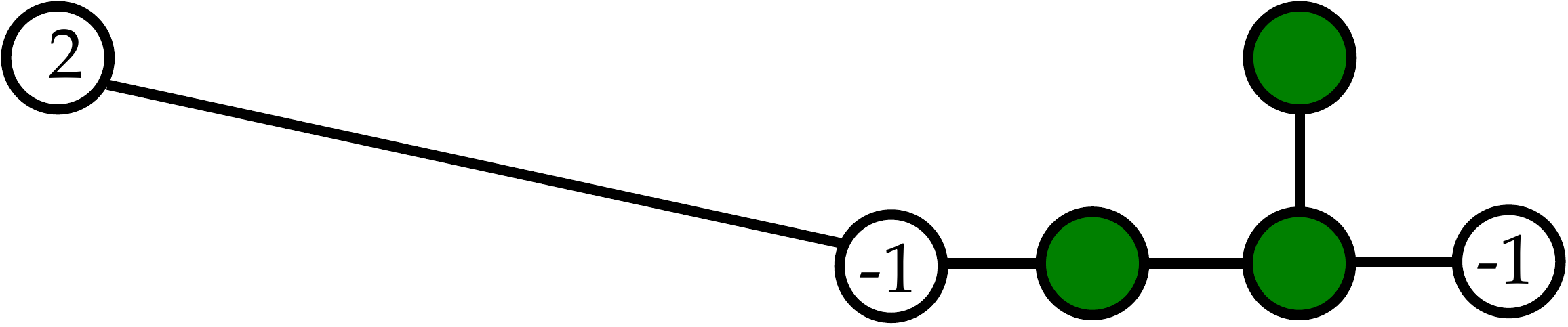} & \includegraphics[height=1cm]{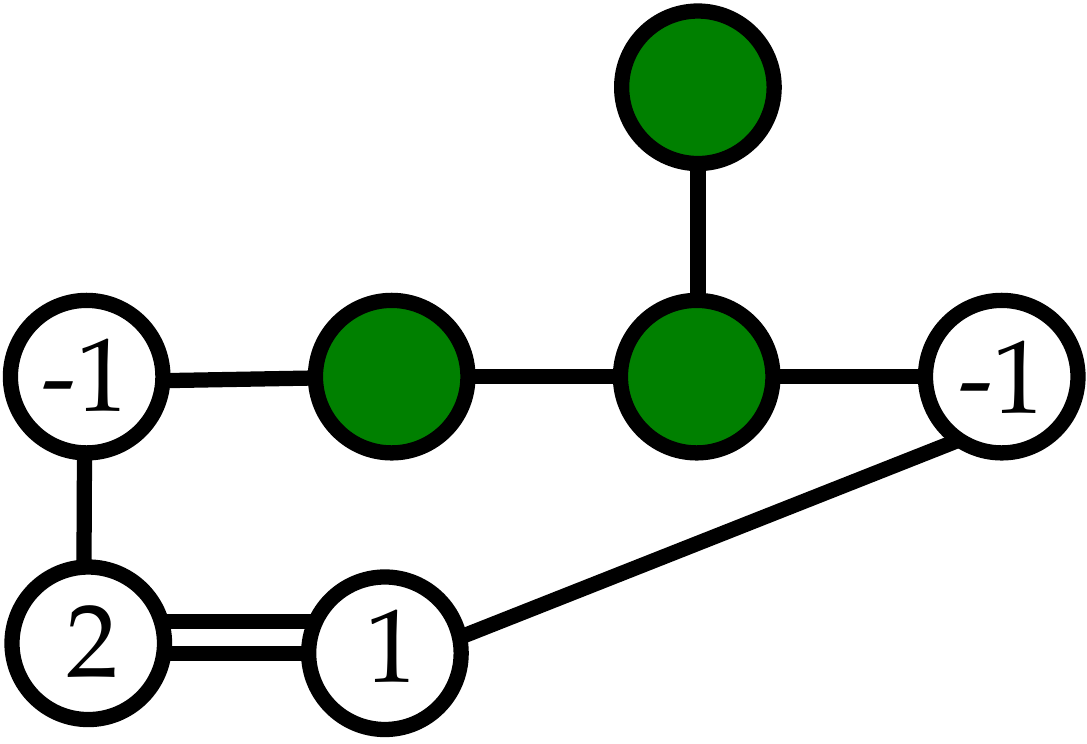}  & \includegraphics[height=1cm]{CFD-E8-26.pdf} & $SU(4)\times U(1)$ & {\small $SU(3)_{5/2}+3\mbf{F}$} & $\mbf{4}_1,\mbf{6}_0$ & $\mbf{1}_0,\bar{\mbf{4}}_1$\cr\hline

\multirow{3}{*}{34} & \multirow{3}{*}{4} &\multirow{3}{*}{\includegraphics[height=.6cm]{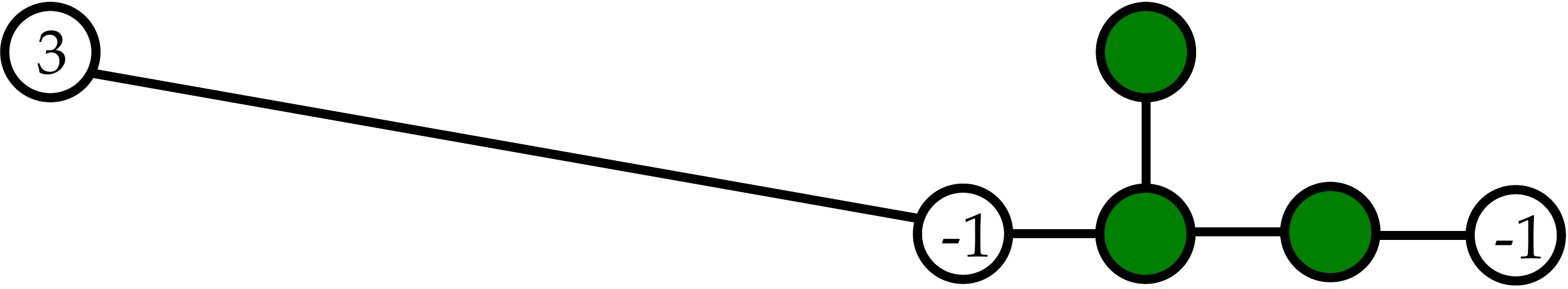}} &
\multirow{3}{*}{\includegraphics[height=1cm]{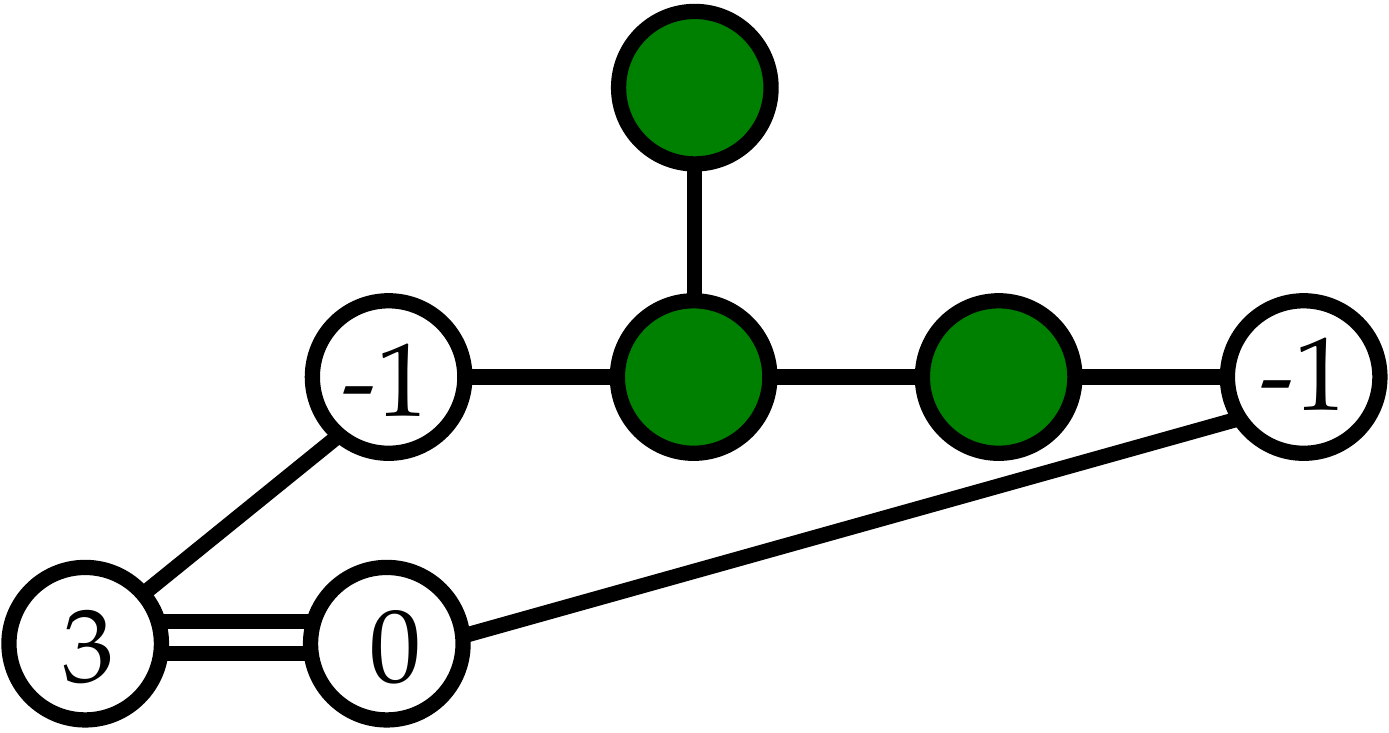}}
& \multirow{3}{*}{\includegraphics[height=1cm]{CFD-E8-25.pdf}} & \multirow{3}{*}{$SO(6) \times U(1)$} & {\small $SU(3)_{7/2}+3\mbf{F}$}  & \multirow{3}{*}{$\mbf{6}_1,\mbf{4}_0$} & \multirow{3}{*}{$\mbf{1}_2,\bar{\mbf{4}}_1$} \cr 
& & & & & & {\small $Sp(2)+3\mbf{F}$} & &  \cr
& & & & & & &&\cr\hline

\multirow{3}{*}{35} & \multirow{3}{*}{4} & & \multirow{3}{*}{\includegraphics[height=1cm]{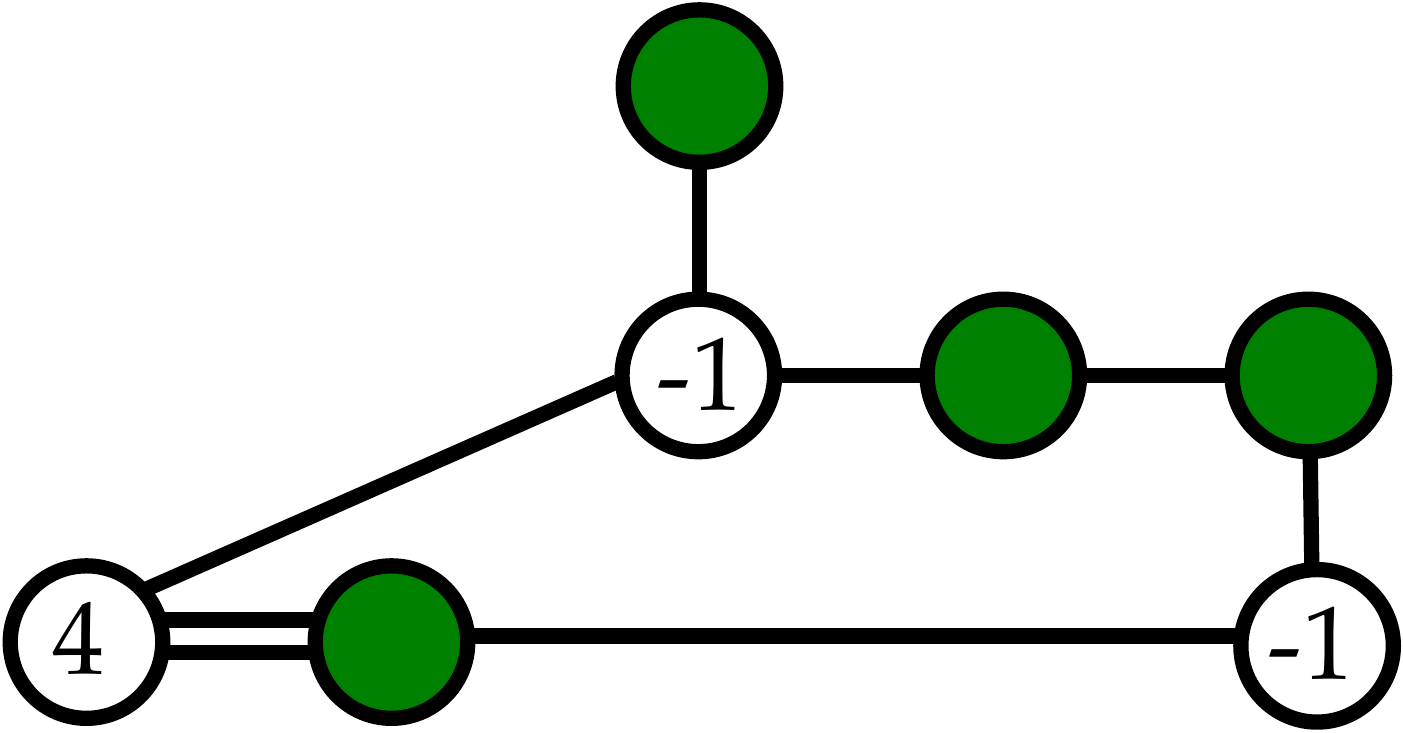}} & \multirow{3}{*}{\includegraphics[height=1cm]{CFD-E8-24.pdf}}& \multirow{3}{*}{\footnotesize{$SU(3)\times SU(2)^2$}} & {\small $SU(3)_{9/2}+3\mbf{F}$} & $(\mbf{3},\mbf{2},\mbf{1})$ & $(\mbf{3},\mbf{1},\mbf{1})$\cr
& & & & & & {\small $Sp(2)+1\mbf{AS}+2\mbf{F}$} & $(\bar{\mbf{3}},\mbf{1},\mbf{2})$ & $(\bar{\mbf{3}},\mbf{1},\mbf{1})$ \cr
& & & & & & & & $(\mbf{1},\mbf{2},\mbf{2})$\cr\hline

\multirow{3}{*}{36} & \multirow{3}{*}{4} & & & \multirow{3}{*}{\includegraphics[height=1cm]{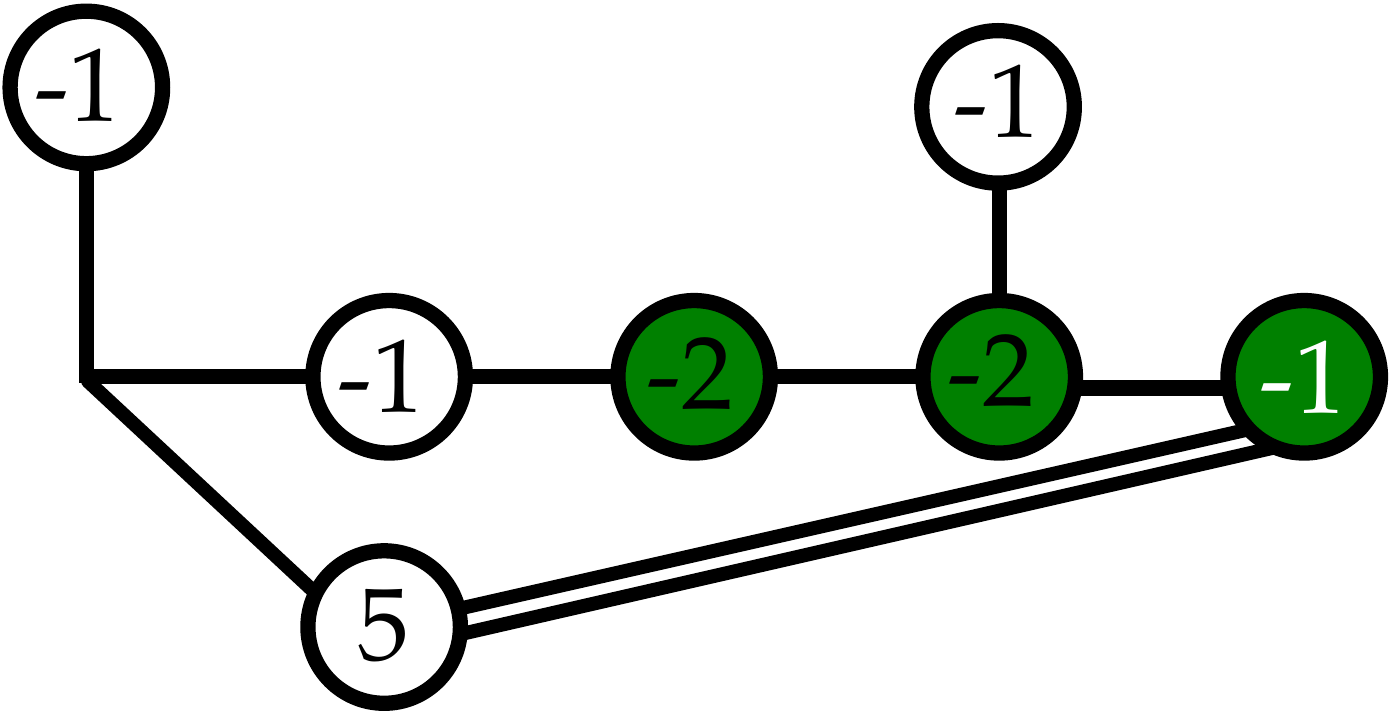}}  & \multirow{3}{*}{$Sp(3)\times U(1)$} & {\small $SU(3)_{11/2}+3\mbf{F}$} & $\mbf{1}_0$ & $\mbf{6}_0$\cr
& & & & & & {\small $Sp(2)+2\mbf{AS}+1\mbf{F}$} & $\mbf{6}_1$ & $\mbf{14}'_1$ \cr
& &  & & & & {\small $G_2+3\mbf{F}$} & $\mbf{14}_0$ & \cr\hline

\end{tabular}
\caption{Table summarizing all Rank Two SCFTs (Continued).}
\end{sidewaystable}

\begin{sidewaystable}
\begin{tabular}{|c|c|c|c|c|c|c|c|c|}\hline
No. & $M$ & $(D_{10}, I_1)$ CFD & $(E_8,SU(2))$ CFD &  Model 3\&4 CFD &  Flavor  $G_\text{F}$& Gauge Theory  & BPS Spin 0 &  Spin 1 \cr \hline\hline 

37 & 4 & & & \includegraphics[height=.8cm]{CFD-Model4-Rank2-Top.pdf} & $-$ & {\small $Sp(2)_0+3\mbf{AS}$} & & \cr\hline
38 & 3 & \includegraphics[height=.8cm]{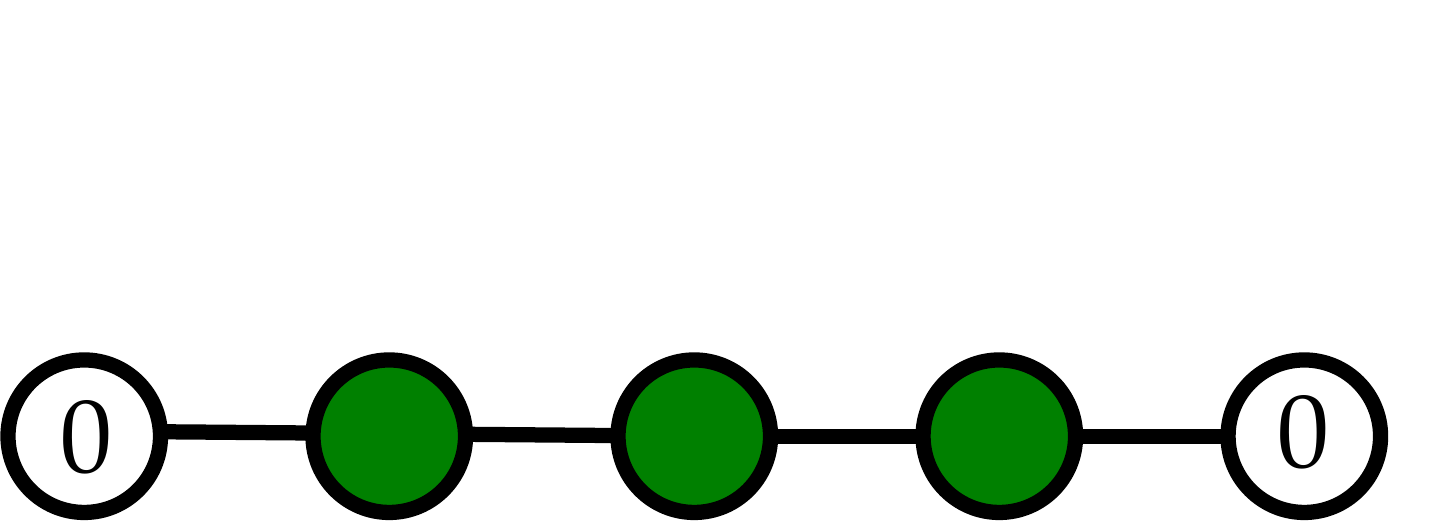} & \includegraphics[height=.8cm]{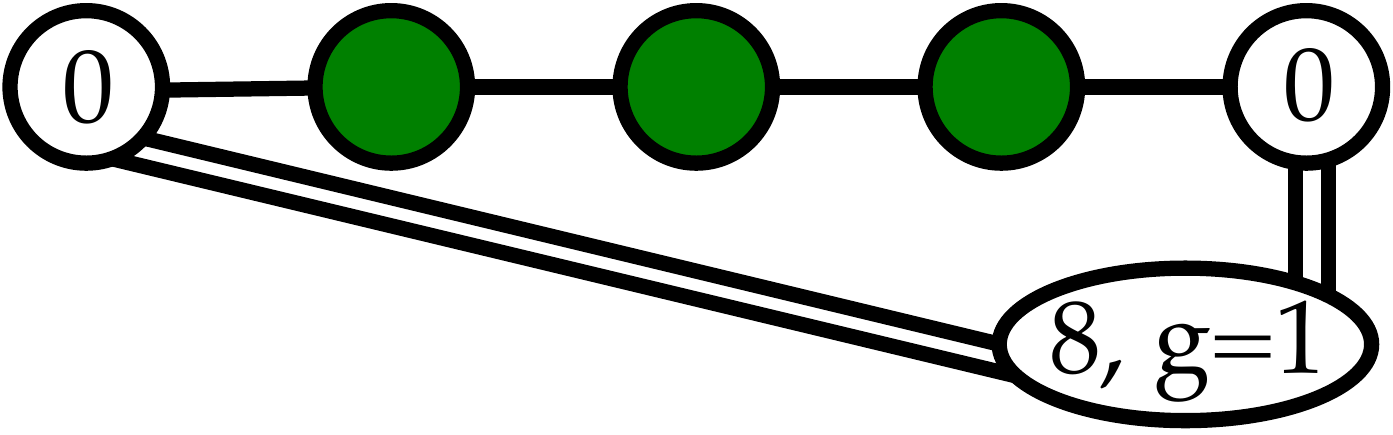}  & & $SU(4)$ & {\small $SU(2)_0-SU(2)_0$} & & $\mbf{4},\bar{\mbf{4}}$ \cr \hline
39 & 3 & \includegraphics[height=.8cm]{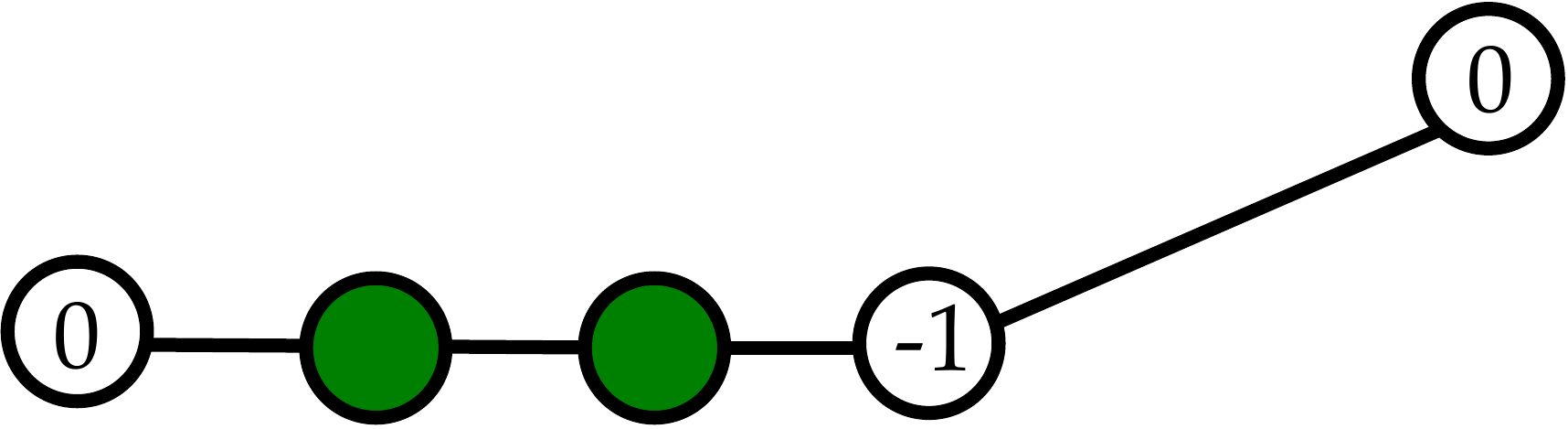} & \includegraphics[height=1cm]{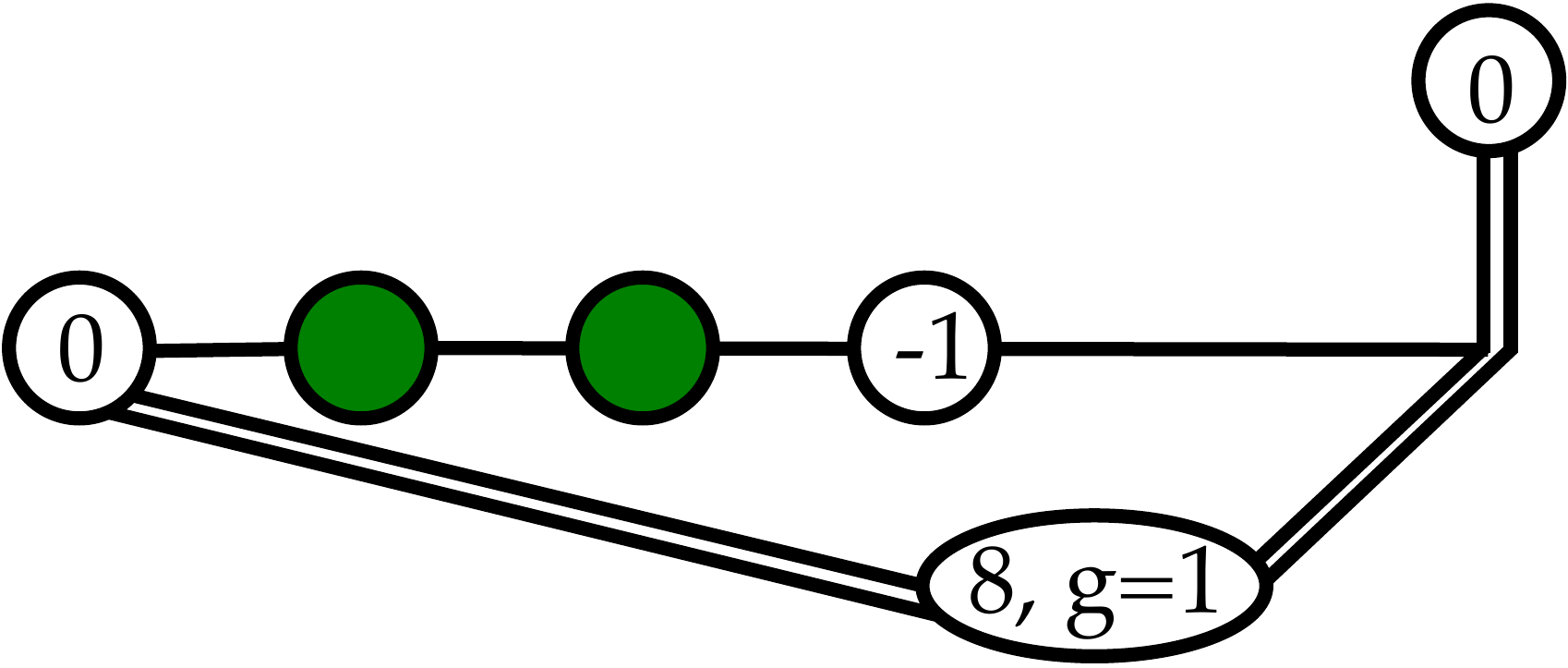} & & $SU(3)\times U(1)$ & {\small $SU(2)_\pi-SU(2)_0$} & $\bar{\mbf{3}}_1$ & $\mbf{1}_0,\mbf{3}_0$ \cr\hline

\multirow{2}{*}{40} & \multirow{2}{*}{3} & \multirow{2}{*}{\includegraphics[height=.8cm]{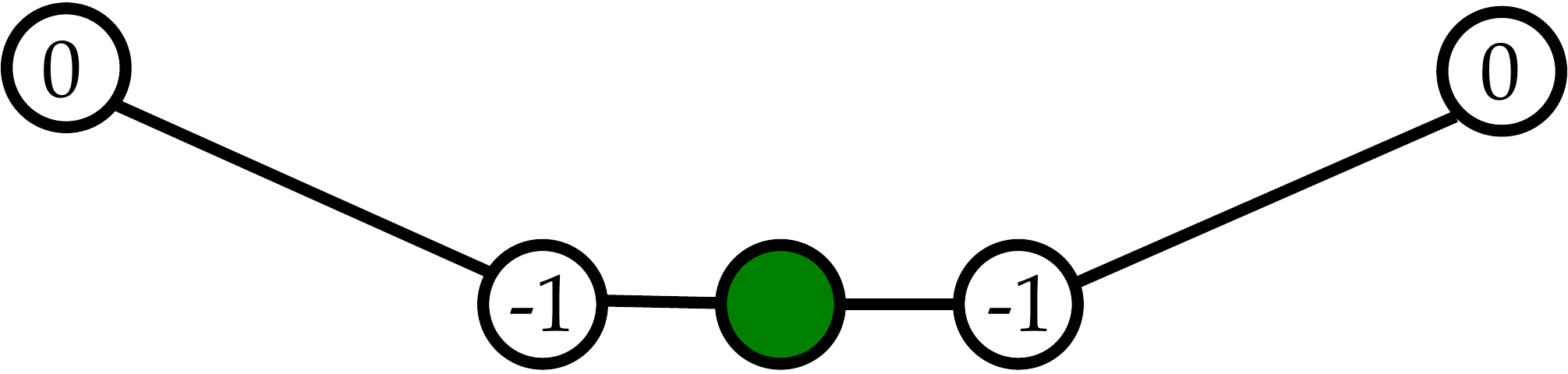}} & \multirow{2}{*}{\includegraphics[height=.8cm]{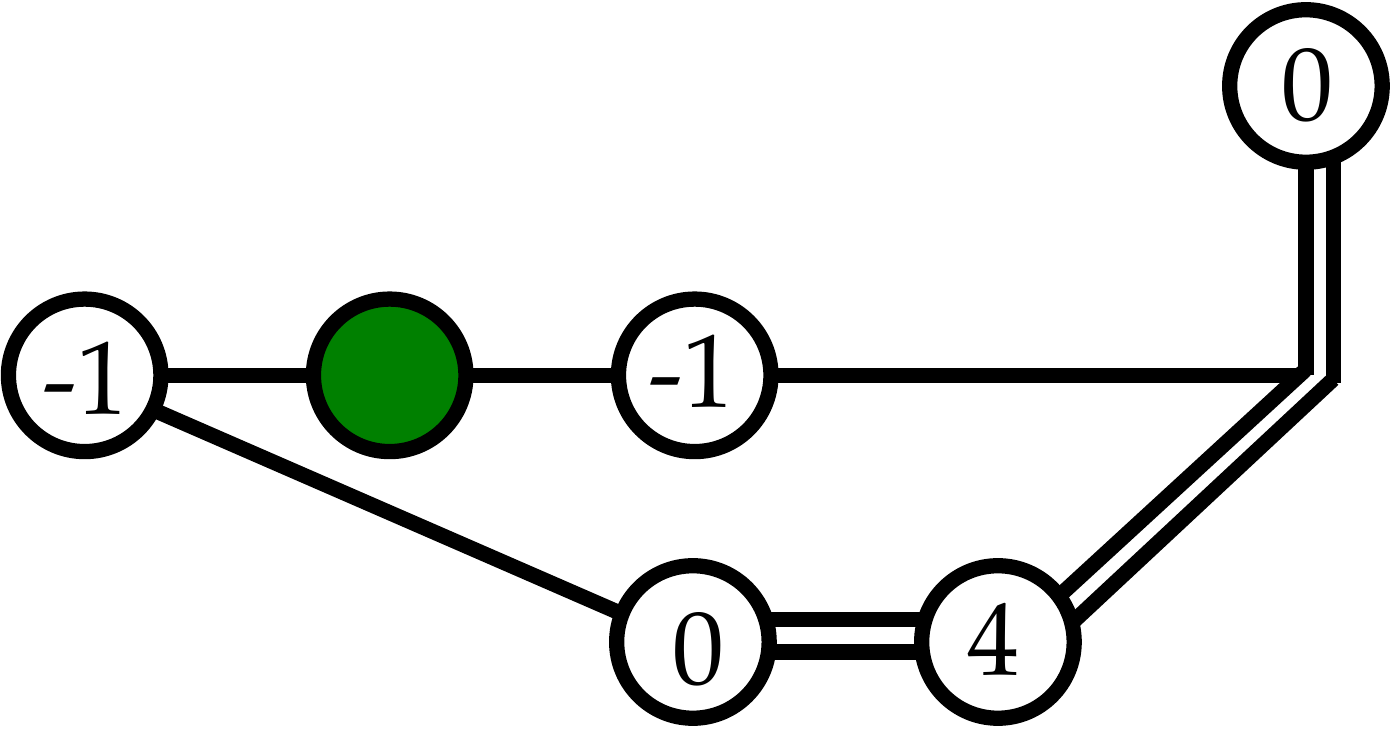}} & & \multirow{2}{*}{$SU(2)\times  U(1)^2$} & {\small $SU(3)_0+2\mbf{F}$} & $\mbf{2}_{(1,0)}$ & $2\cdot \mbf{1}_{(0,0)}$ \cr 
& & & & & & {\small $SU(2)_\pi-SU(2)_\pi$} & $\mbf{2}_{(0,1)}$ &  $\mbf{1}_{(1,1)}$\cr\hline

\multirow{3}{*}{41} & \multirow{3}{*}{3} & \multirow{3}{*}{\includegraphics[height=.8cm]{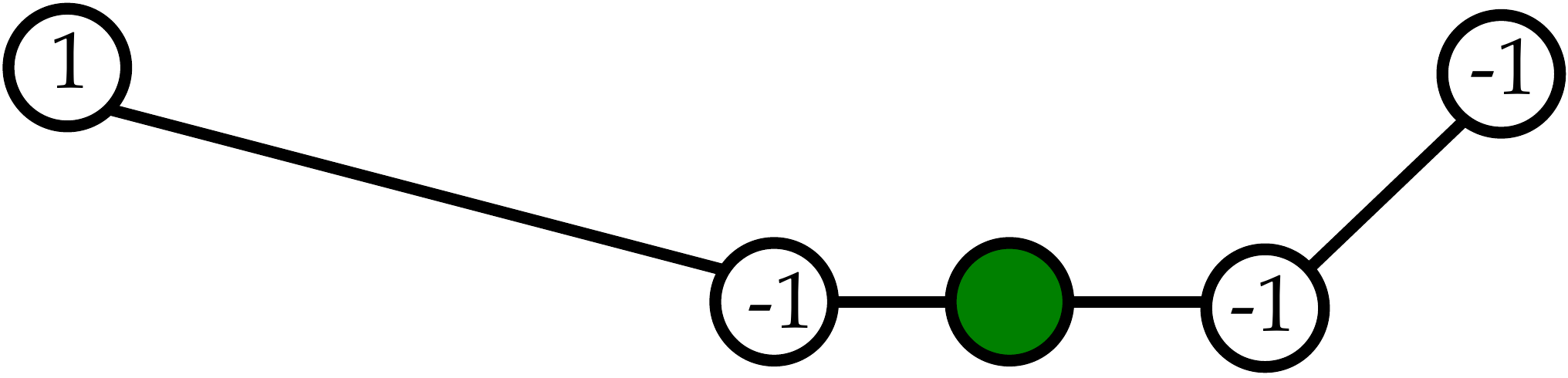}} & \multirow{3}{*}{\includegraphics[height=1cm]{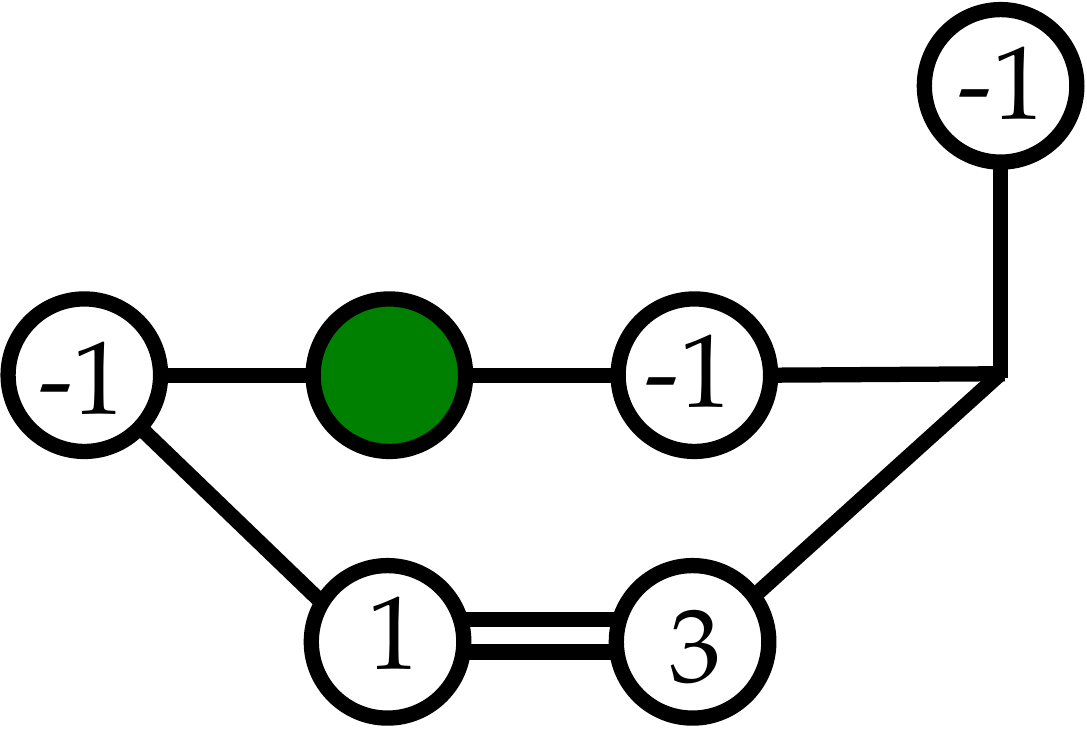}} & & \multirow{3}{*}{$SU(2)\times  U(1)^2$} & \multirow{3}{*}{{\small $SU(3)_{1}+2\mbf{F}$}} & $\mbf{1}_{(0,-1)}$ & \multirow{3}{*}{$\mbf{2}_{(0,0)}$} \cr 
 & & & & & & & $\mbf{2}_{(1,0)}$   &  \cr
 & & & & & & & $\mbf{2}_{(0,1)}$ & \cr\hline
 
\multirow{3}{*}{42} & \multirow{3}{*}{3} & \multirow{3}{*}{\includegraphics[height=.8cm]{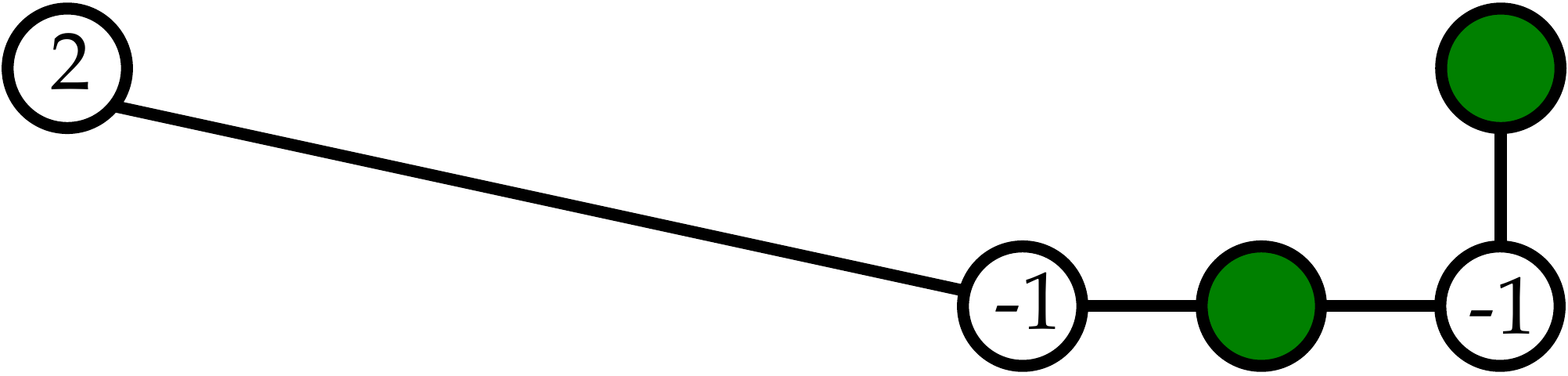}} & \multirow{3}{*}{\includegraphics[height=1cm]{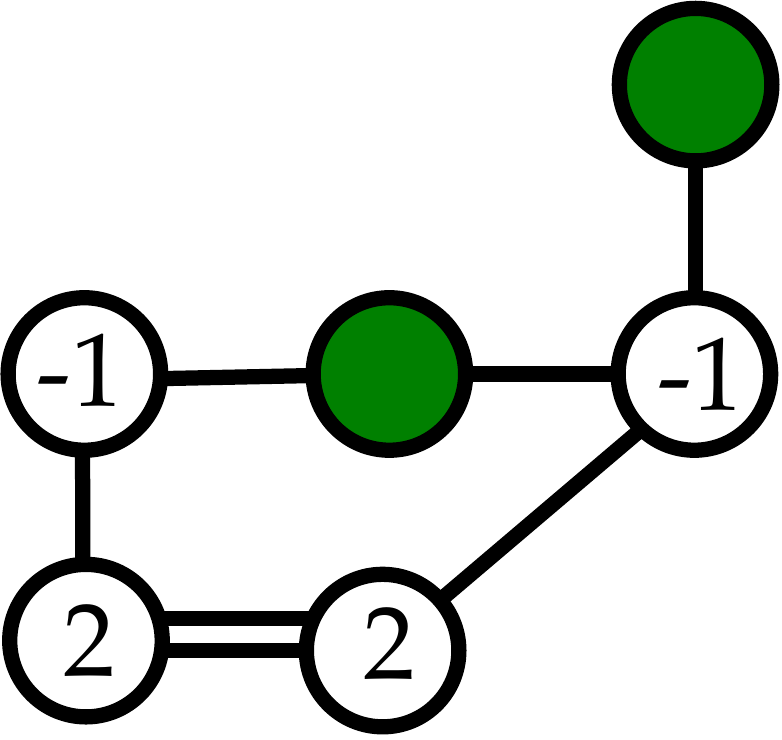}} & \multirow{3}{*}{\includegraphics[height=1cm]{CFD-E8-33.pdf}} & \multirow{3}{*}{$SO(4)\times  U(1)$} & \multirow{3}{*}{{\small $SU(3)_2+2\mbf{F}$}} & $(\mbf{2},\mbf{1})_1$ & $(\mbf{1},\mbf{1})_0$\cr
 & & & & & & & $(\mbf{2},\mbf{2})_0$ & $(\mbf{1},\mbf{2})_1$\cr
 & & & & & & &&\cr\hline
 
 43 & 3 & \includegraphics[height=.6cm]{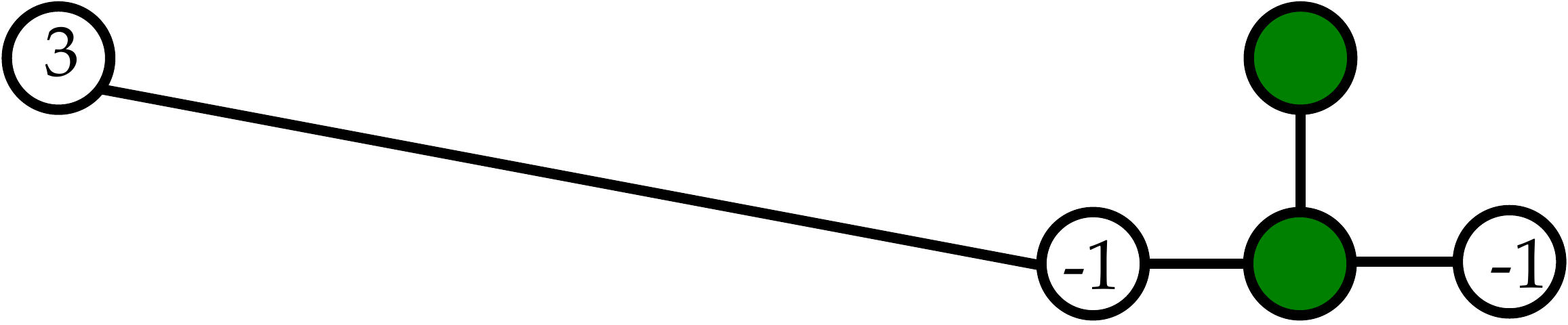} & \includegraphics[height=1cm]{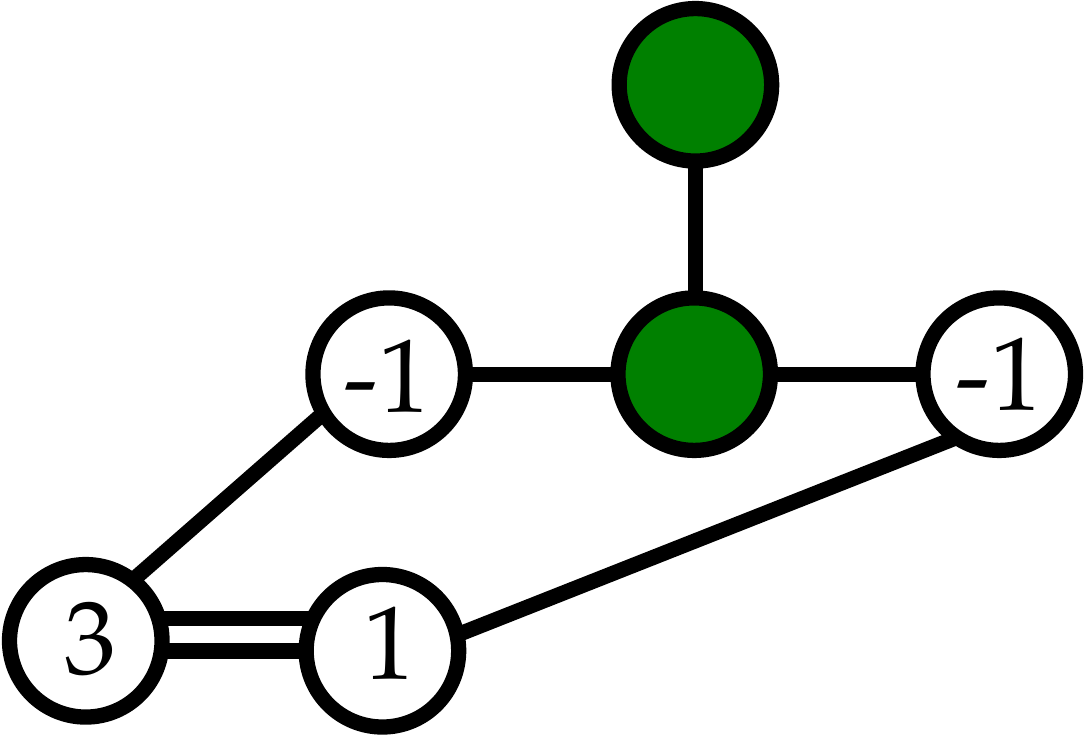} & \includegraphics[height=1cm]{CFD-E8-32.pdf} & $SU(3)\times U(1)$ & {\small $SU(3)_{3}+2\mbf{F}$} & $\mbf{3}_0,\mbf{3}_1$ & $\bar{\mbf{3}}_1$\cr \hline
 
\multirow{3}{*}{44} & \multirow{2}{*}{3} & \multirow{3}{*}{\includegraphics[height=.6cm]{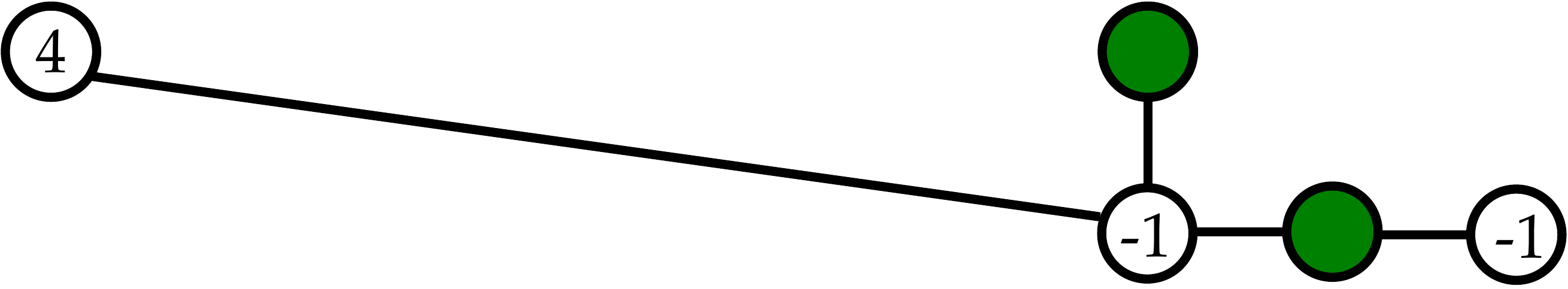}} &
\multirow{3}{*}{\includegraphics[height=1cm]{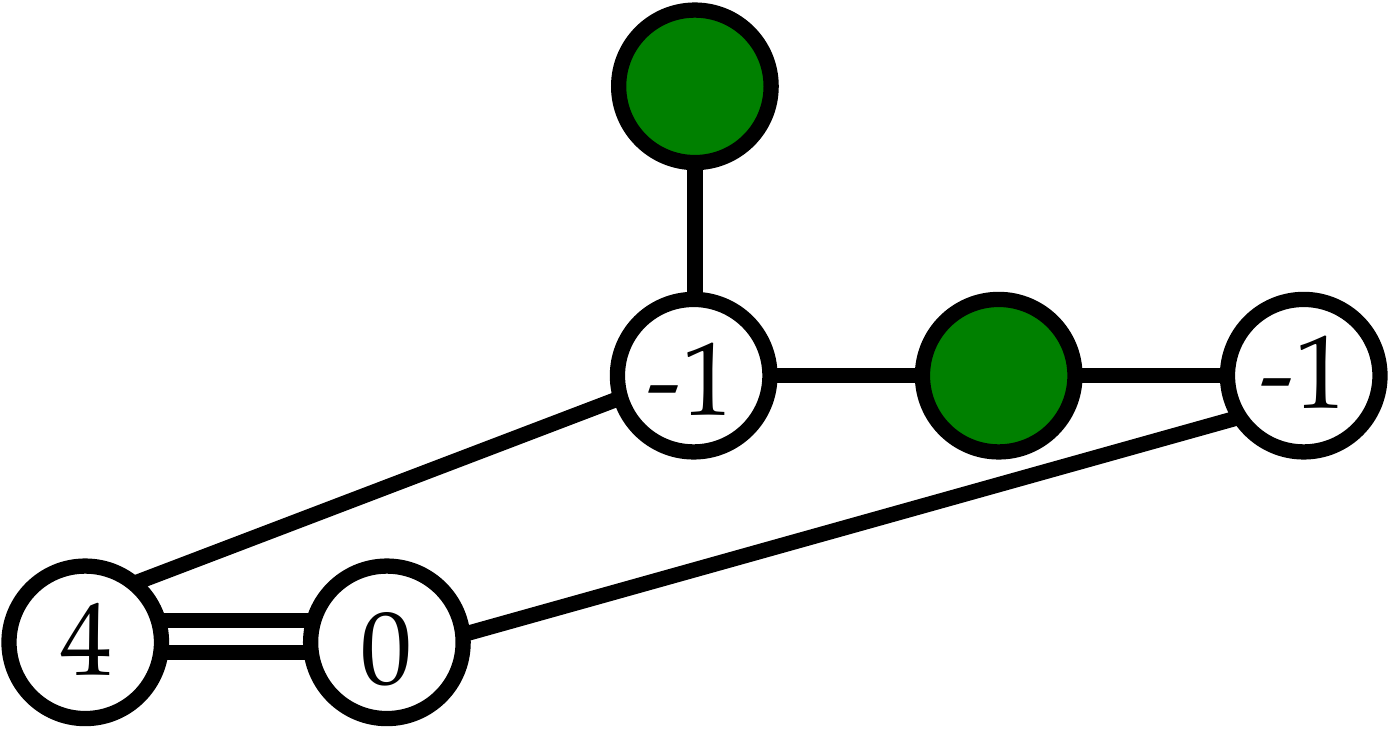}} 
& \multirow{3}{*}{\includegraphics[height=1cm]{CFD-E8-31.pdf}} & \multirow{3}{*}{$SO(4) \times U(1)$} & {\small $SU(3)_{4}+2\mbf{F}$} & $(\mbf{2},\mbf{2})_1$  & $(\mbf{1},\mbf{1})_2$ \cr
&& & & & & {\small $Sp(2)+2\mbf{F}$}  & $(\mbf{2},\mbf{1})_0$ & $(\mbf{1},\mbf{2})_1$ \cr
& & & & & & &&\cr\hline

\multirow{3}{*}{45} & \multirow{3}{*}{3} & &\multirow{3}{*}{\includegraphics[height=1cm]{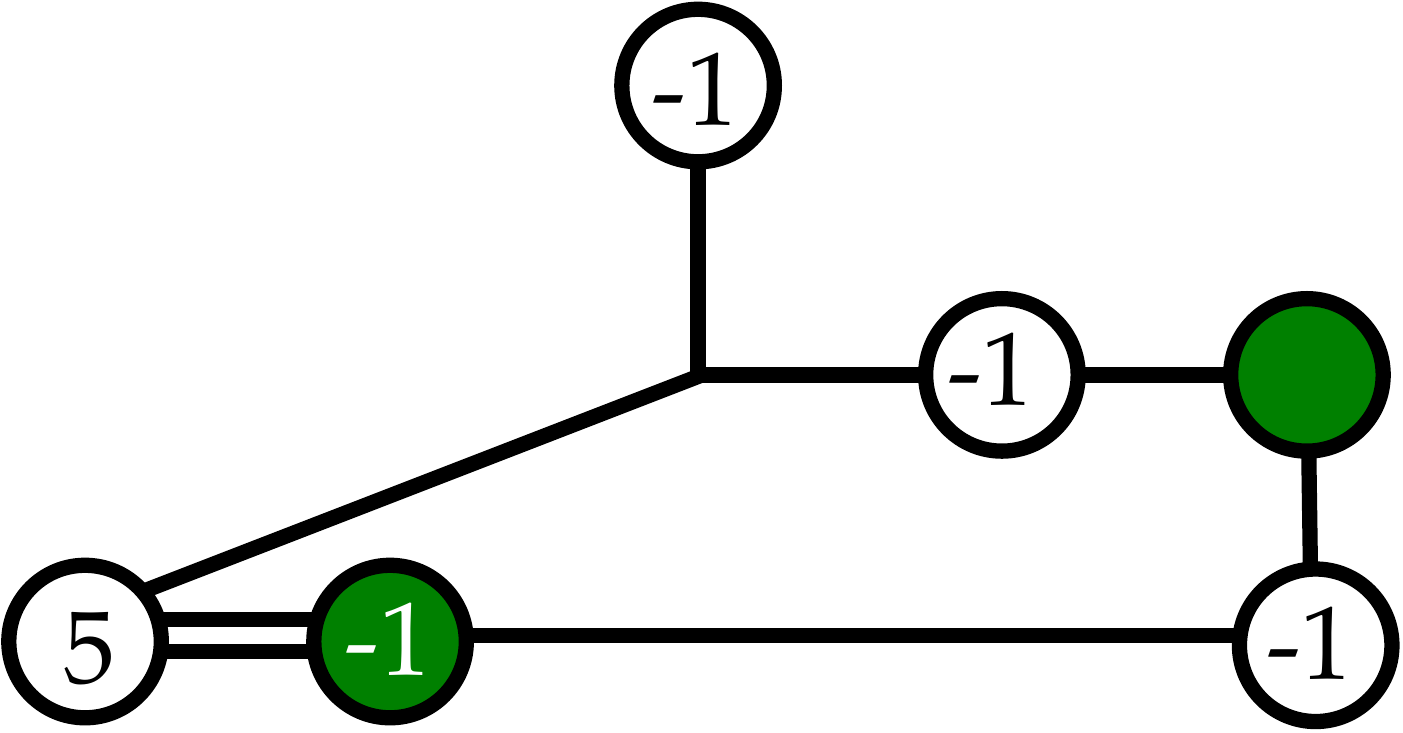}} & \multirow{3}{*}{\includegraphics[height=1cm]{CFD-E8-30.pdf}} & \multirow{3}{*}{\footnotesize{$SU(2)^2 \times U(1)$}} & {\small $SU(3)_{5}+2\mbf{F}$} & $(\mbf{1},\mbf{1})_{-1}$ & $(\mbf{1},\mbf{1})_0$\cr
& & &  & & & {\small $Sp(2)+1\mbf{AS}+1\mbf{F}$} &$(\mbf{2},\mbf{1})_1$  & $(\mbf{1},\mbf{2})_1$\cr
& & & & & & & $(\mbf{2},\mbf{2})_0$ & $(\mbf{2},\mbf{1})_0$\cr\hline

\multirow{3}{*}{46} & \multirow{3}{*}{3} & & & \multirow{3}{*}{\includegraphics[height=1cm]{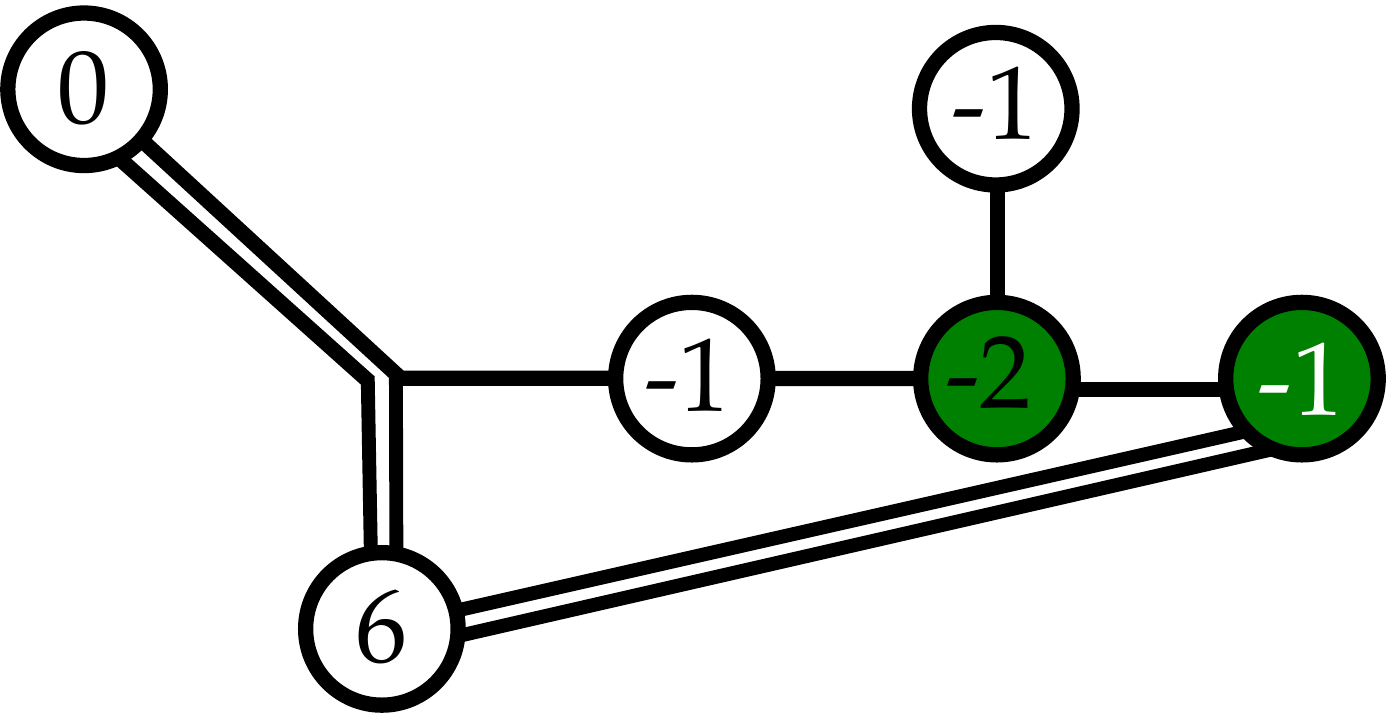}} & \multirow{3}{*}{$Sp(2)\times U(1)$} & {\small $SU(3)_6+2\mbf{F}$} & $\mbf{4}_0$ & $\mbf{1}_0$ \cr
& & & & & & {\small $Sp(2)_\pi+2\mbf{AS}$} & $\mbf{4}_1$ & $\mbf{5}_1$ \cr
& &  & & & & {\small $G_2+2\mbf{F}$} & & \cr\hline

47 & 3 & & & \includegraphics[height=1cm]{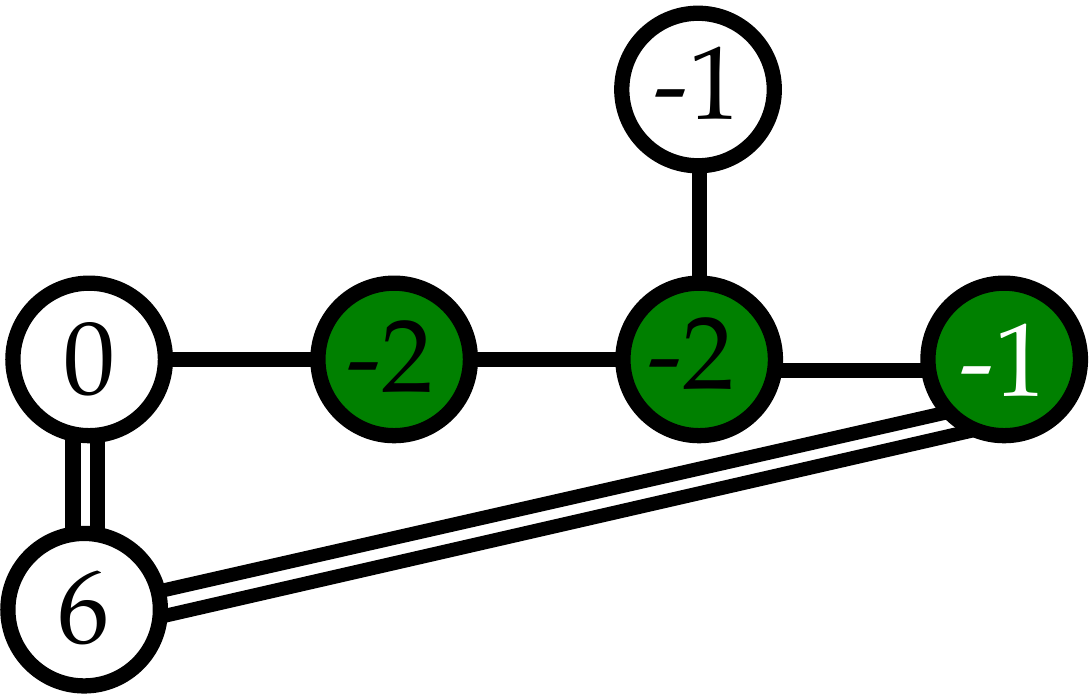} & $Sp(3)$ & {\small $Sp(2)_0+2\mbf{AS}$} & $\mbf{6}$ & $\mbf{4}$ \cr\hline

\end{tabular}
\caption{Table summarizing all Rank Two SCFTs (Continued).}
\end{sidewaystable}

\begin{sidewaystable}
\begin{tabular}{|c|c|c|c|c|c|c|c|c|}\hline
No. & $M$ & $(D_{10}, I_1)$ CFD & $(E_8,SU(2))$ CFD &  Model 3\&4 CFD &  Flavor  $G_\text{F}$& Gauge Theory  & BPS Spin 0 &  Spin 1 \cr \hline\hline 

48 & 2 & \includegraphics[height=.8cm]{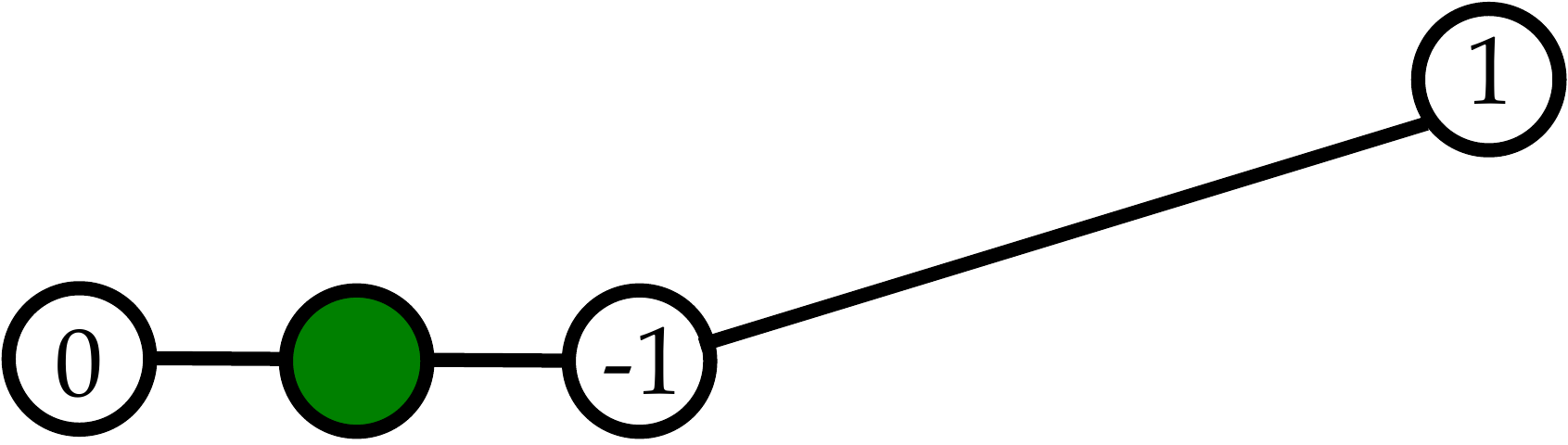} & \includegraphics[height=.8cm]{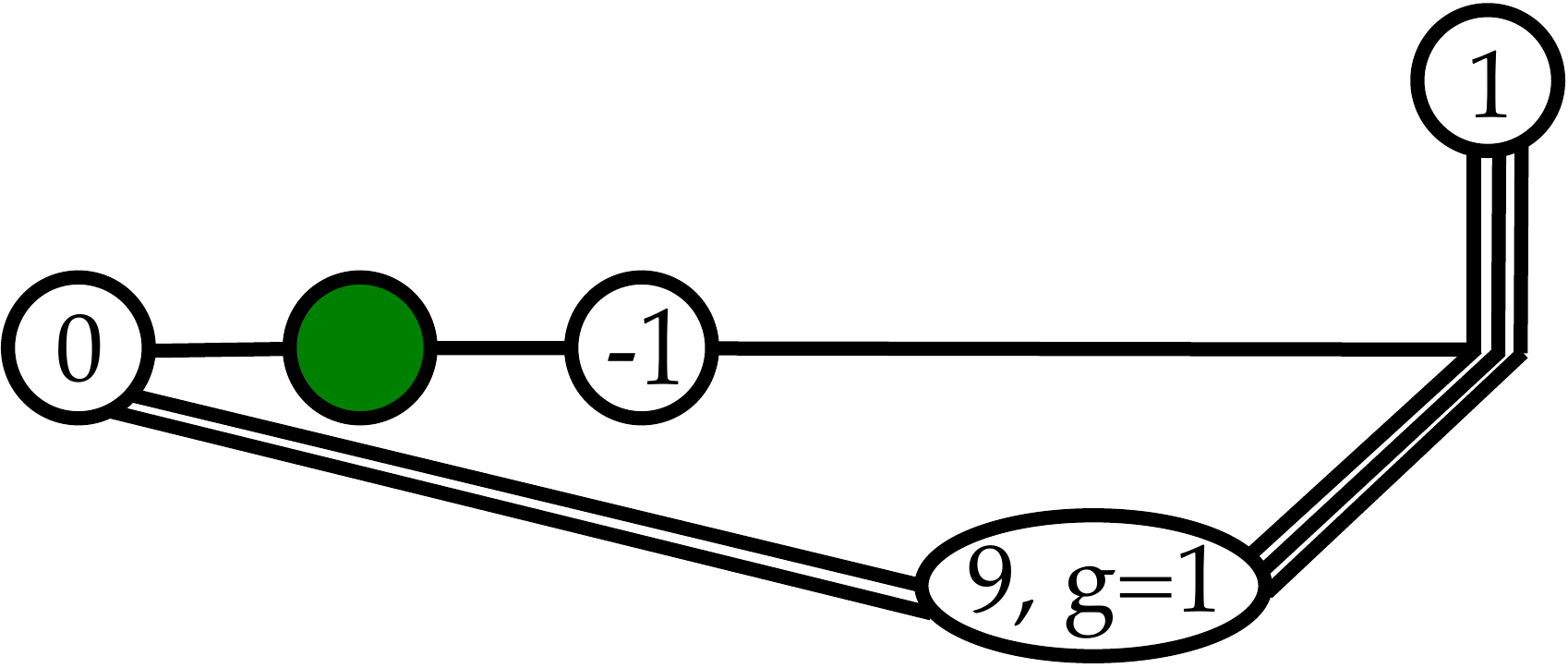} & & $SU(2)\times U(1)$ & $-$ & $\mbf{2}_1$ & $\mbf{2}_0$\cr\hline 
49 & 2 & \includegraphics[height=.8cm]{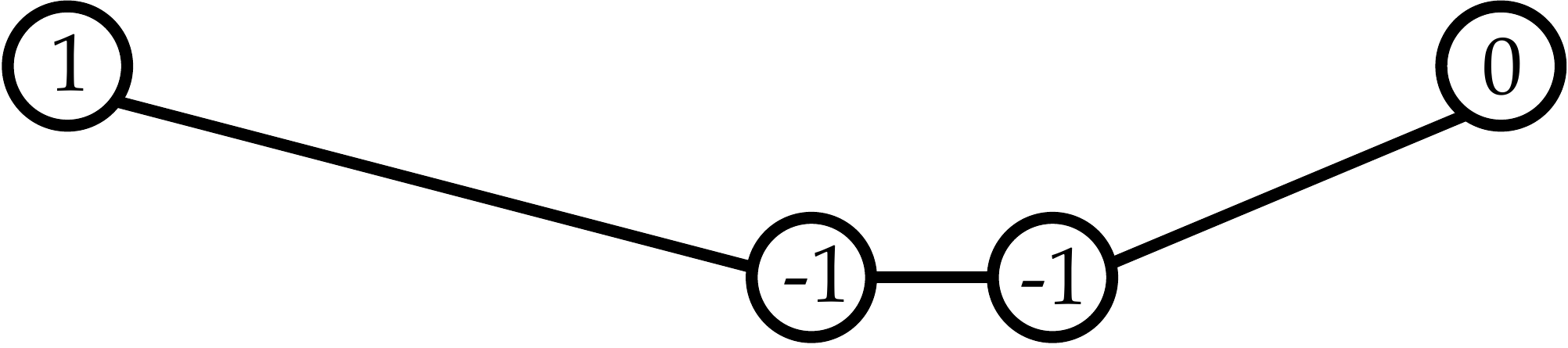} & \includegraphics[height=1cm]{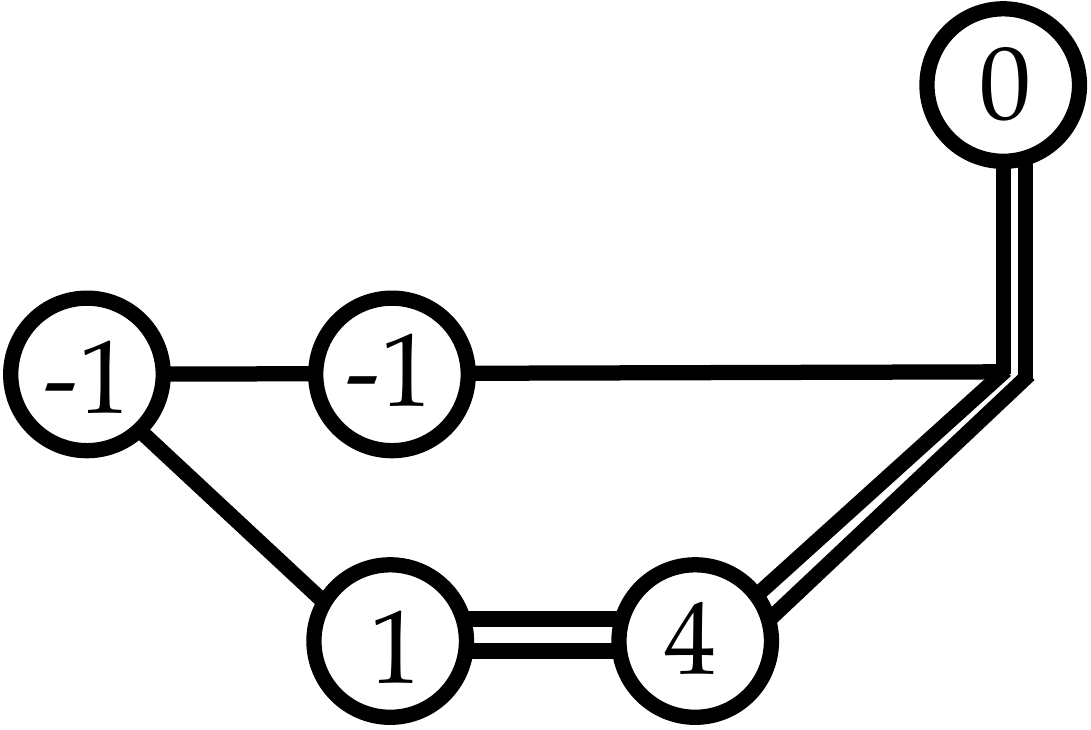} & & $ U(1)^2$ & {\small $SU(3)_{1/2}+1\mbf{F}$} & $(1,0),(0,1)$ & $(0,0),(1,1)$ \cr\hline

\multirow{3}{*}{50} & \multirow{3}{*}{2} & \multirow{3}{*}{\includegraphics[height=.8cm]{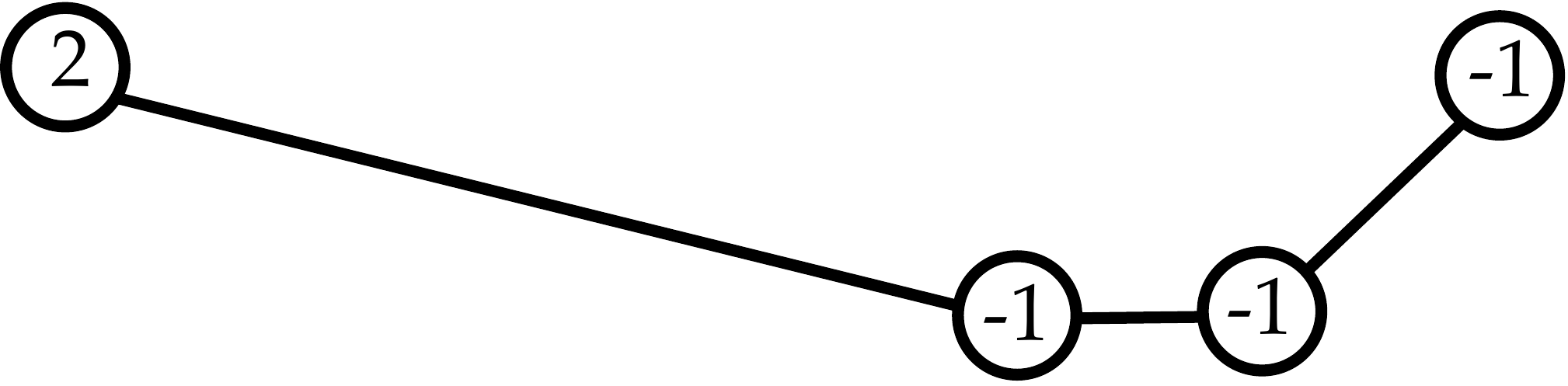}} & \multirow{3}{*}{\includegraphics[height=1cm]{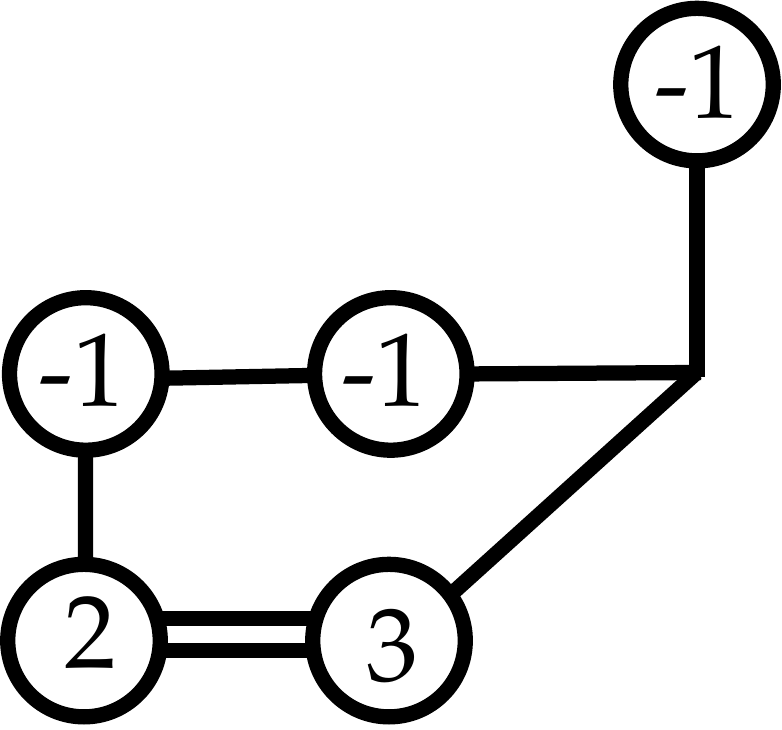}} & \multirow{3}{*}{\includegraphics[height=1cm]{CFD-E8-43.pdf}} & \multirow{3}{*}{ $ U(1)^2$} & \multirow{3}{*}{{\small $SU(3)_{3/2}+1\mbf{F}$}} & $(1,0),(0,1)$ & $(1,1)$ \cr 
& & & & & & & $(0,-1)$ & $(0,0)$\cr
& & & & & & &&\cr\hline

51 & 2 & \includegraphics[height=.8cm]{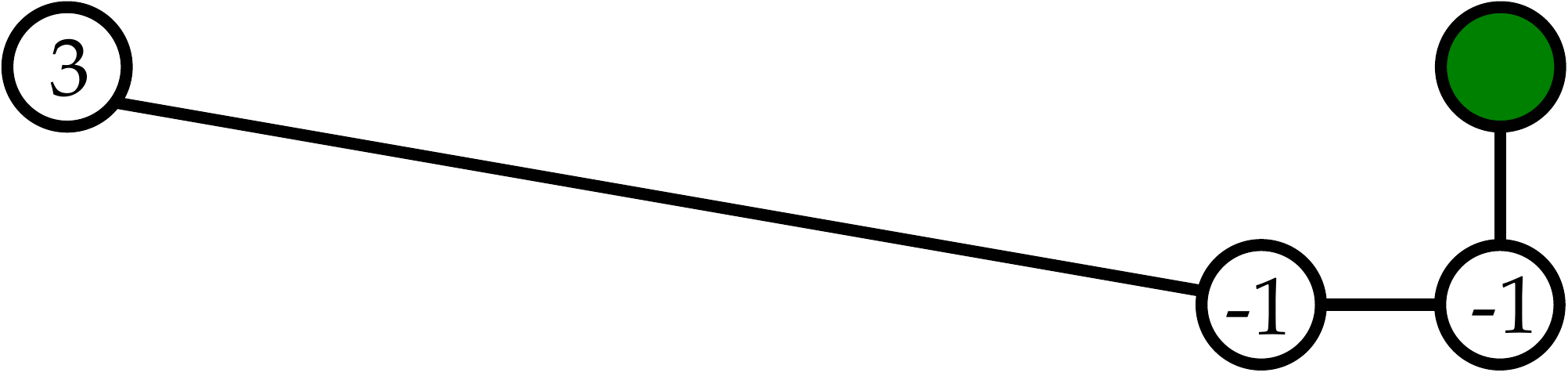} & \includegraphics[height=1cm]{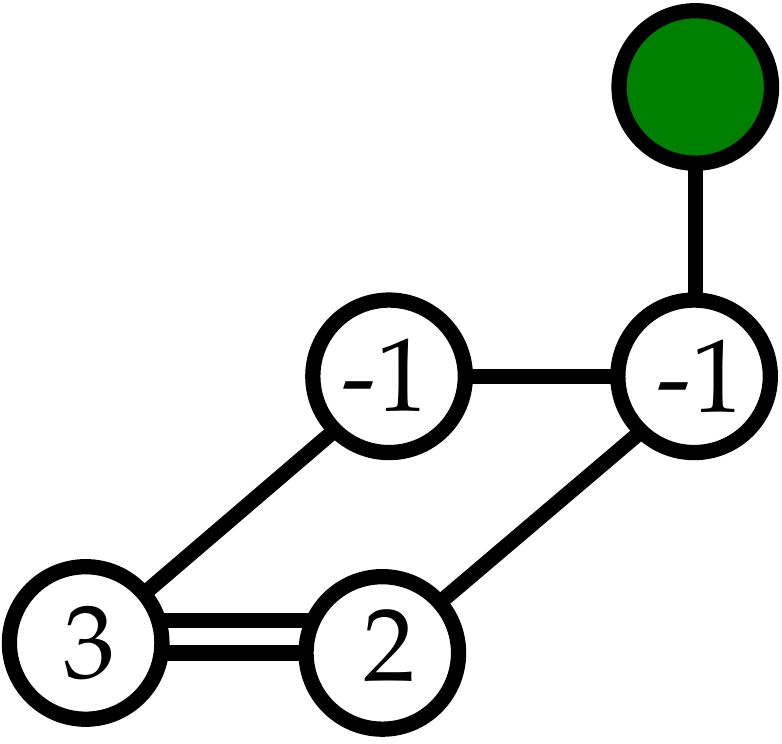} & \includegraphics[height=1cm]{CFD-E8-42.pdf} & $ SU(2)\times  U(1)$ & {\small $SU(3)_{5/2}+1\mbf{F}$} & $\mbf{1}_1,\mbf{2}_0$ & {$\mbf{2}_1$}\cr \hline

52 & 2 & \includegraphics[height=.8cm]{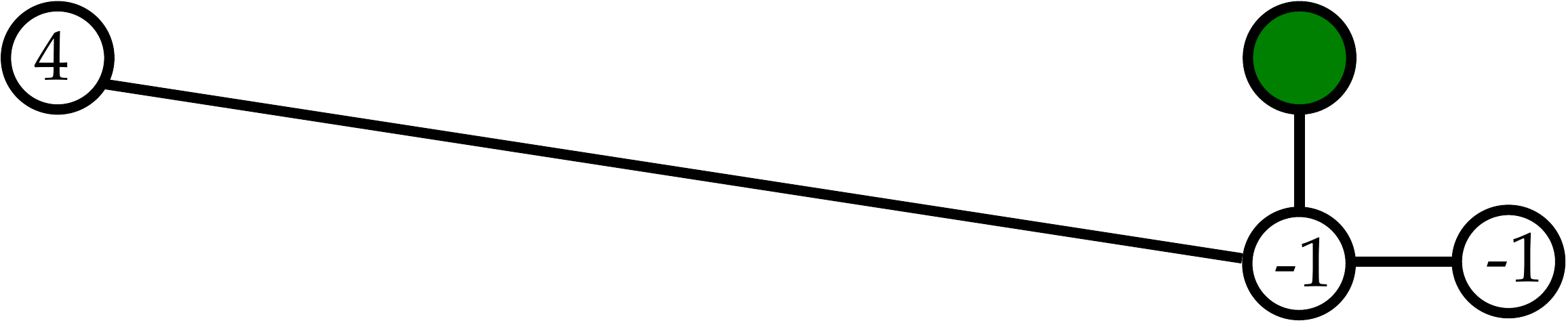} & \includegraphics[height=1cm]{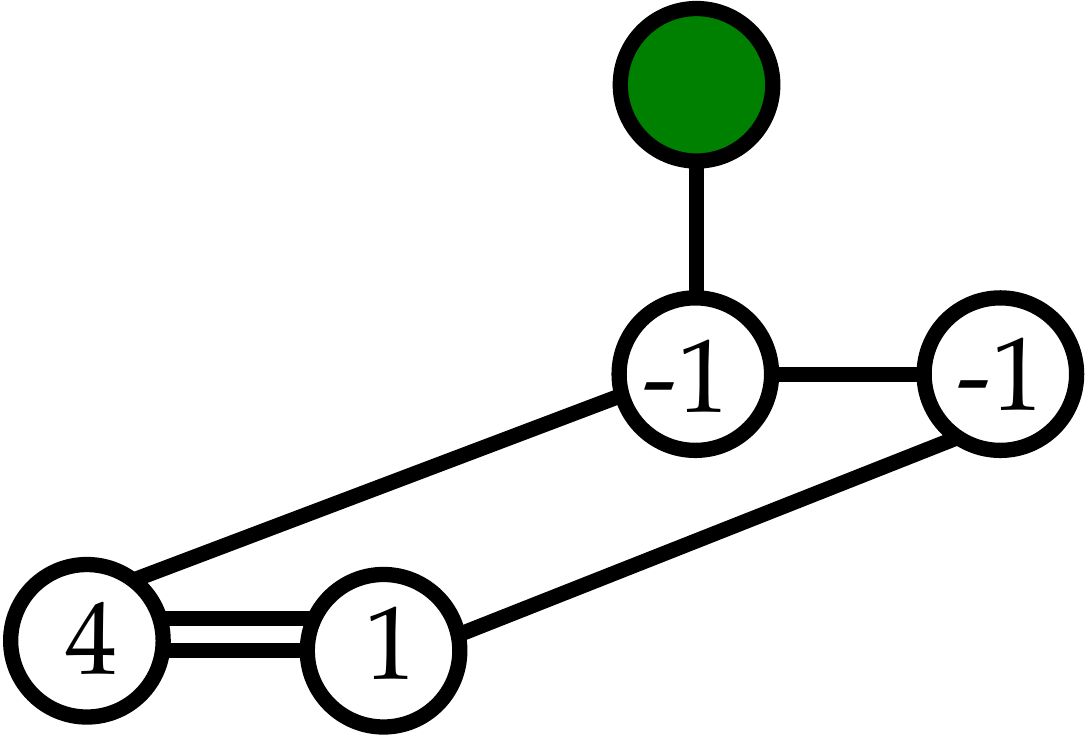} & \includegraphics[height=1cm]{CFD-E8-41.pdf} & $SU(2)\times U(1)$ & $SU(3)_{7/2}+1\mbf{F}$ & $\mbf{1}_0,\mbf{2}_1$ & $\mbf{2}_1$\cr \hline

\multirow{3}{*}{53} & \multirow{3}{*}{2} & \multirow{3}{*}{\includegraphics[height=.8cm]{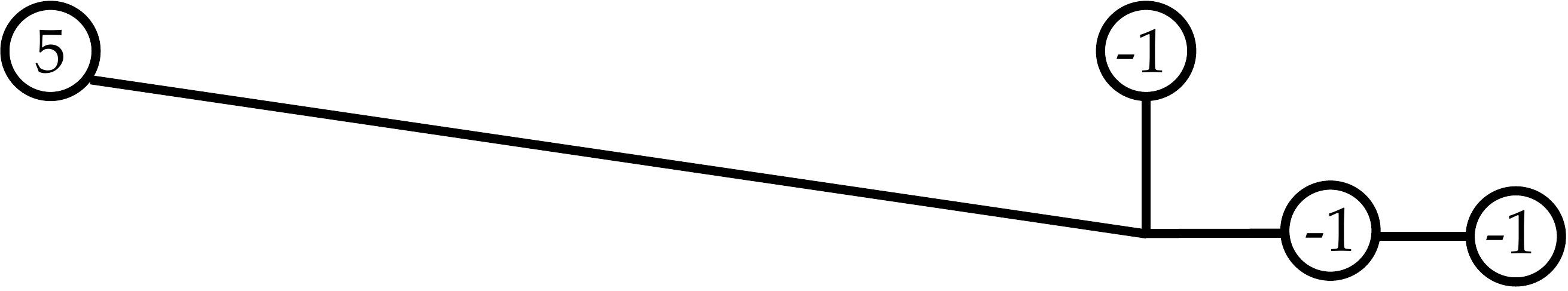}}&
 \multirow{3}{*}{  \includegraphics[height=1cm]{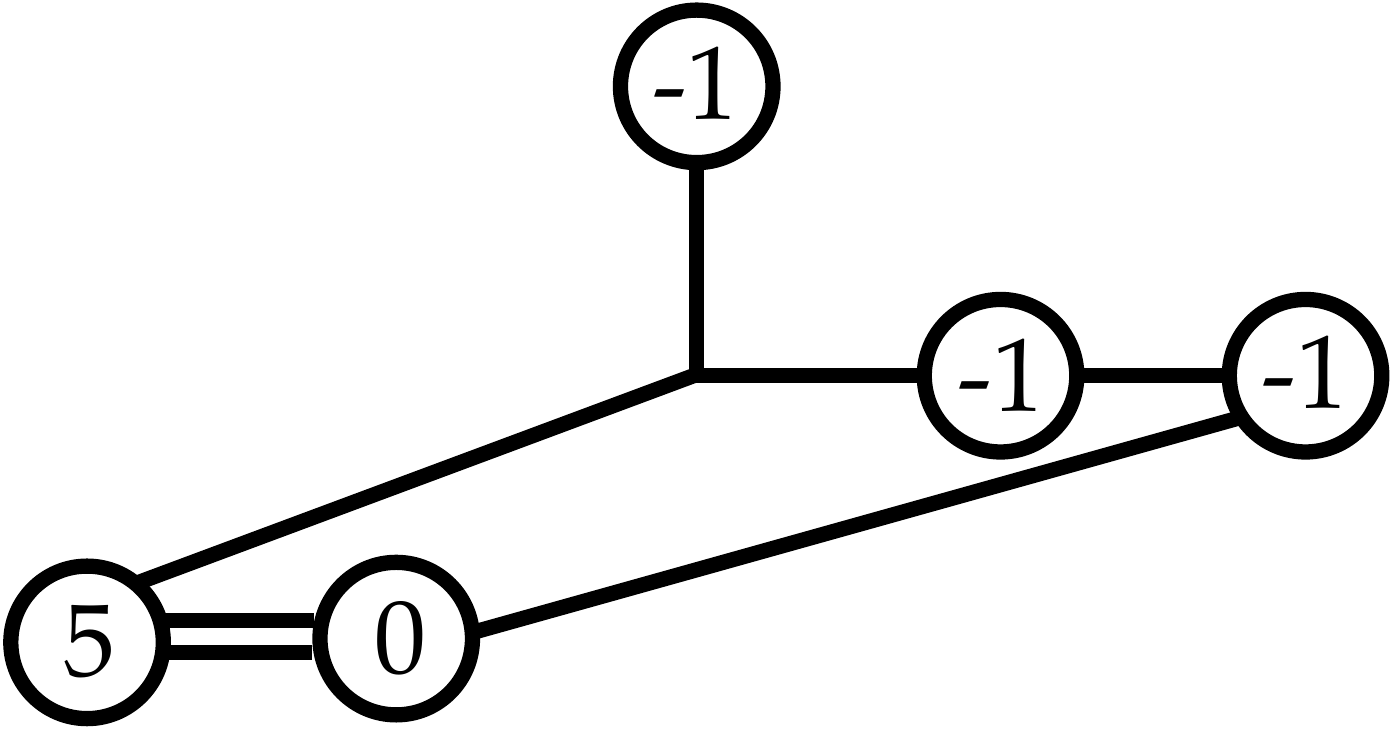} }
 & \multirow{3}{*}{  \includegraphics[height=1cm]{CFD-E8-40.pdf} } & \multirow{3}{*}{$SO(2)\times U(1)$} & {\small $SU(3)_{9/2}+1\mbf{F}$} & $(\mbf{1},\mbf{0})$&\multirow{3}{*}{$(\mbf{2},\mbf{0}),(\mbf{1},\mbf{1})$} \cr 
&& & & & & {\small $Sp(2)+1\mbf{F}$}& $(\mbf{0},\mbf{1})$ & \cr
& & & & & & &&\cr\hline

\multirow{3}{*}{54} & \multirow{3}{*}{2} & & \multirow{3}{*}{\includegraphics[height=1cm]{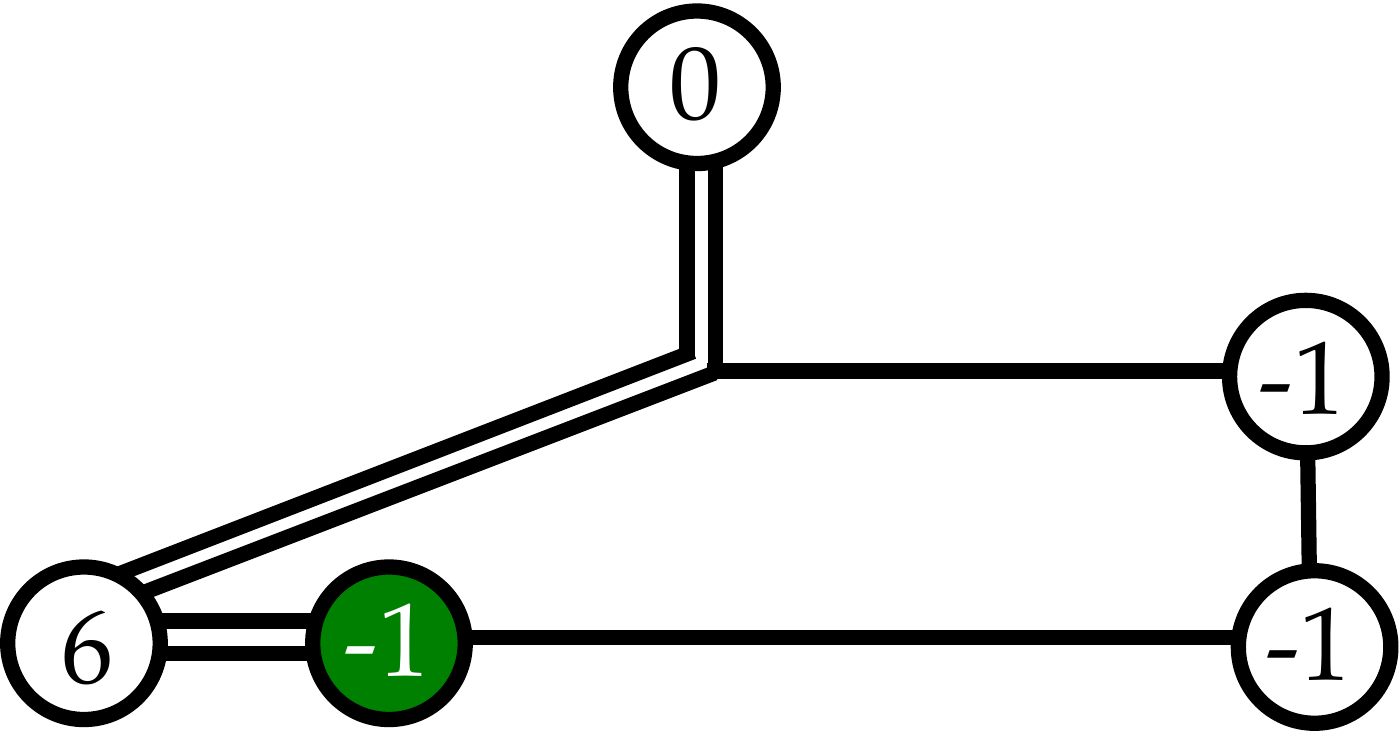}} & \multirow{3}{*}{\includegraphics[height=1cm]{CFD-E8-38.pdf}} & \multirow{3}{*}{$SU(2)\times U(1)$} & {\small $SU(3)_{11/2}+1\mbf{F}$} & \multirow{3}{*}{$\mbf{1}_1,\mbf{2}_0$} & \multirow{3}{*}{$\mbf{1}_0,\mbf{2}_1$}\cr
& & & & & & {\small $Sp(2)_\pi+1\mbf{F}$} & & \cr
& & & & & & &&\cr\hline

\multirow{3}{*}{55} & \multirow{3}{*}{2} & & & \multirow{3}{*}{\includegraphics[height=1cm]{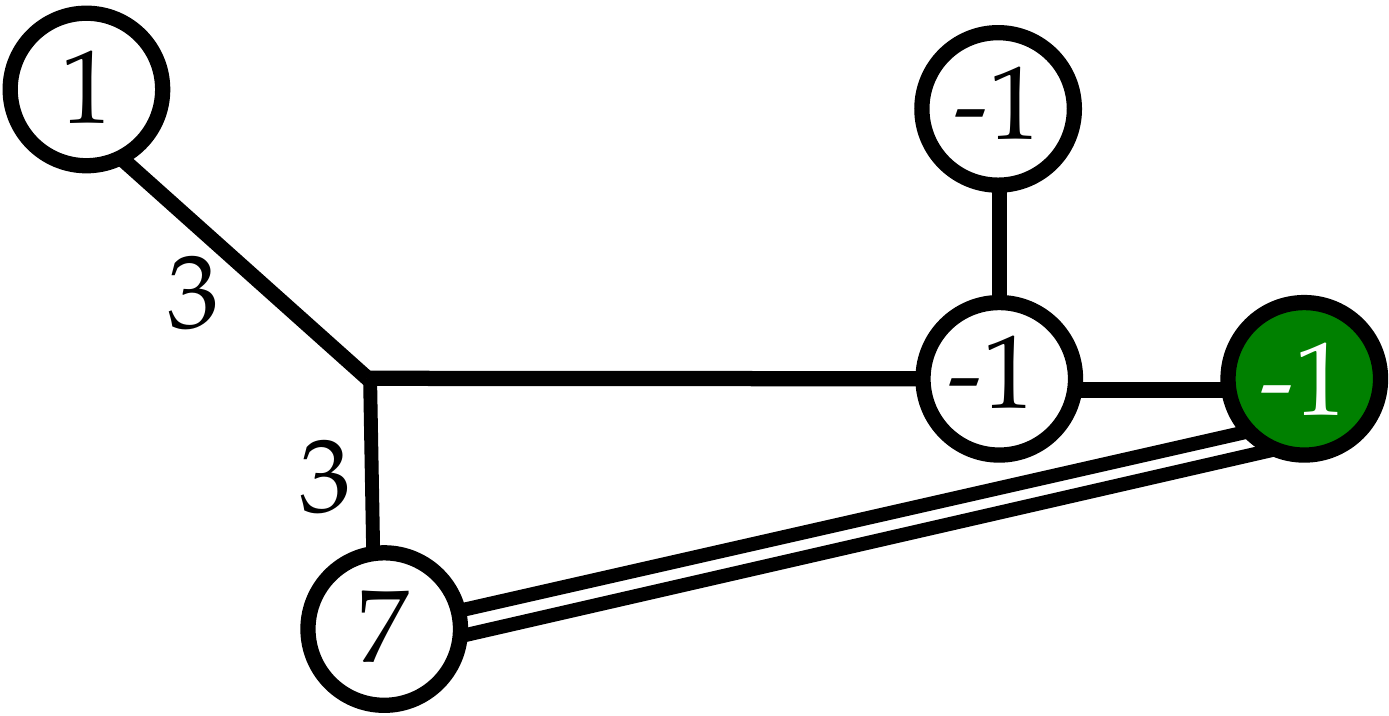}} & \multirow{3}{*}{$SU(2)\times U(1)$} & {\small $SU(3)_{13/2}+1\mbf{F}$} & $\mbf{1}_0$ & $\mbf{2}_1$ \cr
& &  & & & & {\small $G_2+1\mbf{F}$} & $\mbf{2}_1$ & \cr
& & & & & & &&\cr\hline

\multirow{2}{*}{56} & \multirow{2}{*}{2} & & \multirow{2}{*}{\includegraphics[height=.6cm]{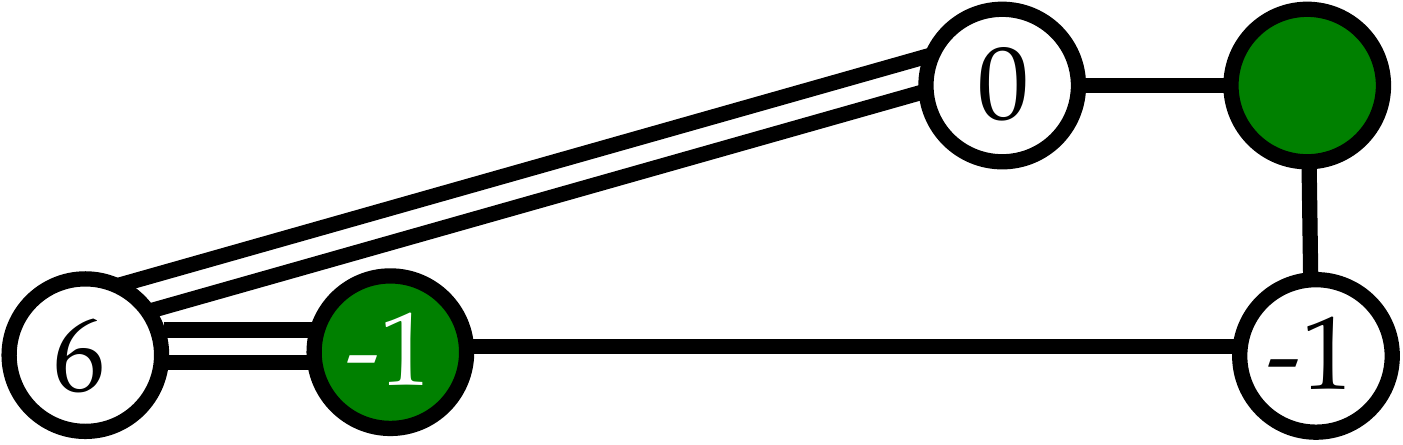}} & \multirow{2}{*}{\includegraphics[height=.6cm]{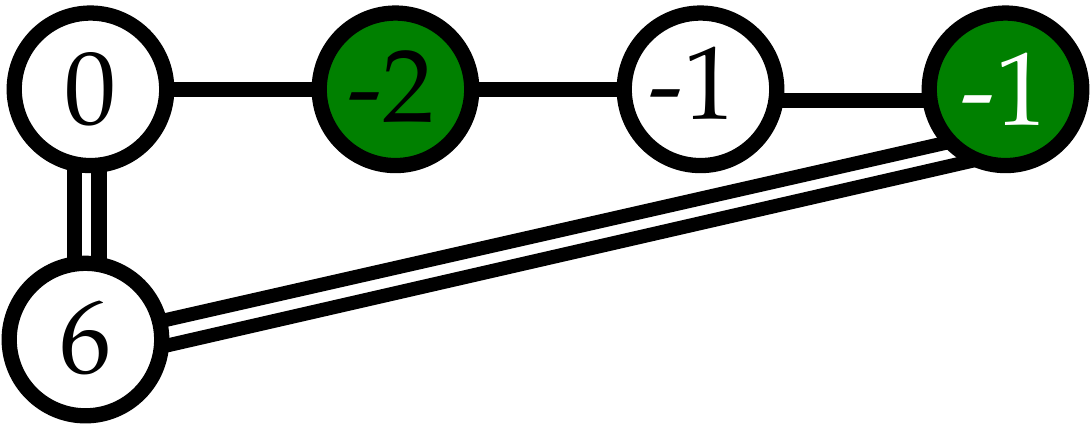}}  & \multirow{2}{*}{$SU(2)^2$} & \multirow{2}{*}{{\small $Sp(2)_0+1\mbf{AS}$}} & \multirow{2}{*}{$(\mbf{2},\mbf{2})$} & $(\mbf{1},\mbf{1})$\cr
& & & & & & & & $(\mbf{2},\mbf{1})$ \cr\hline

57 & 1 & \includegraphics[height=.8cm]{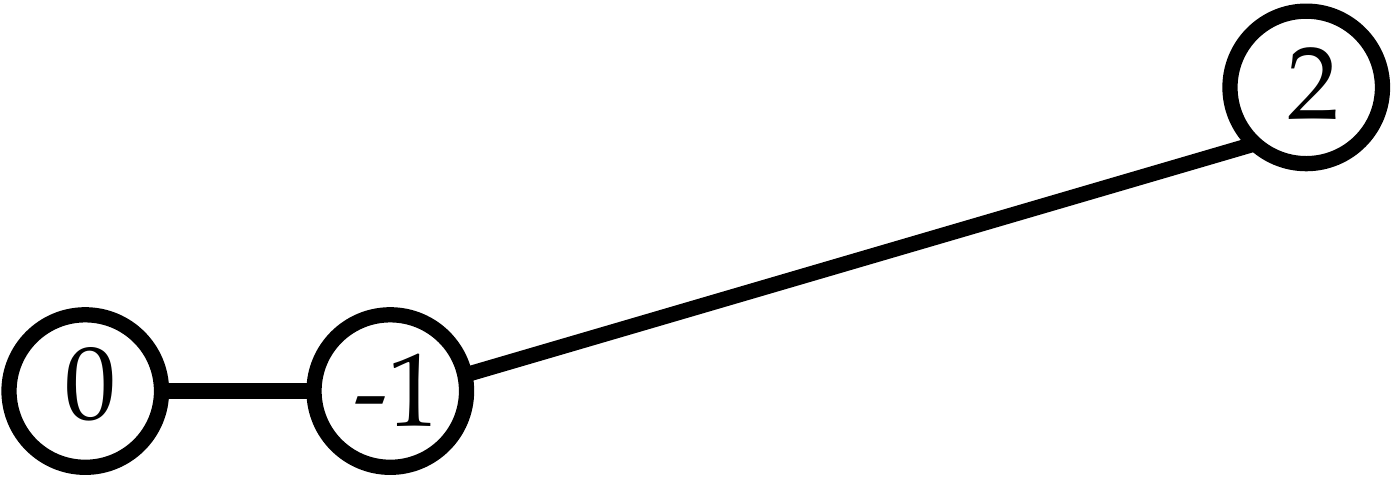} & \includegraphics[height=.8cm]{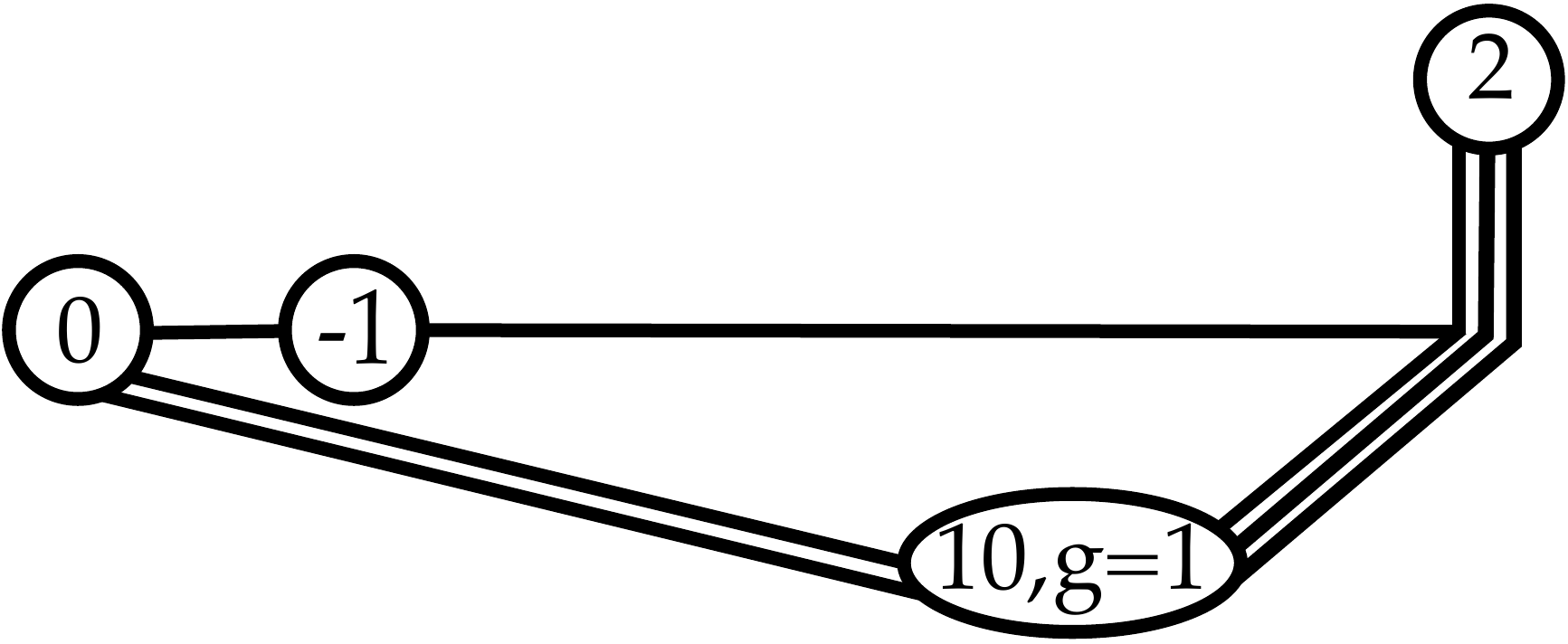}  &  \includegraphics[height=.8cm]{CFD-E8-54.pdf} & $U(1)$ & $-$ & 1 & 0\cr\hline

58 & 1 & \includegraphics[height=.8cm]{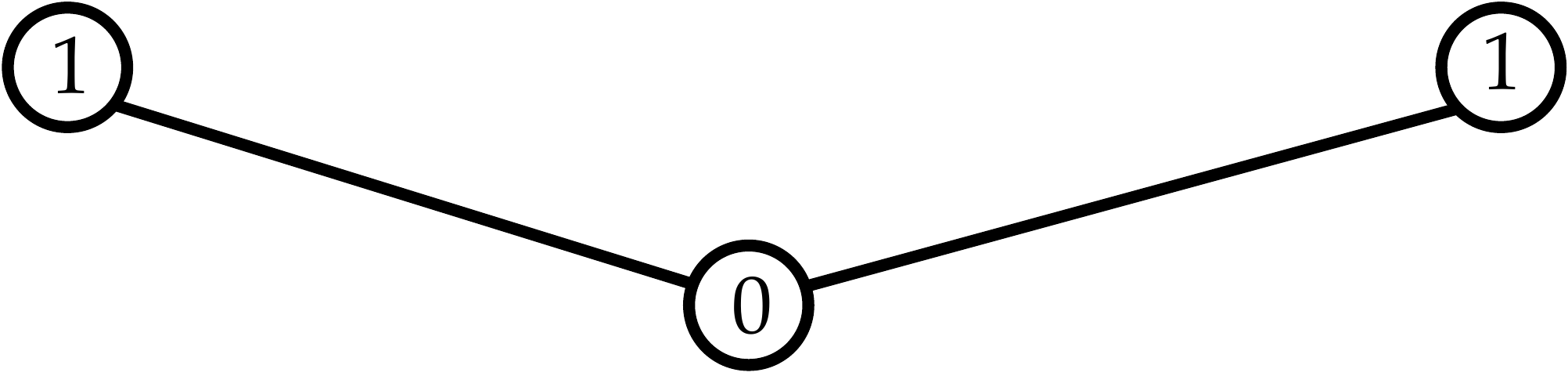} & \includegraphics[height=.8cm]{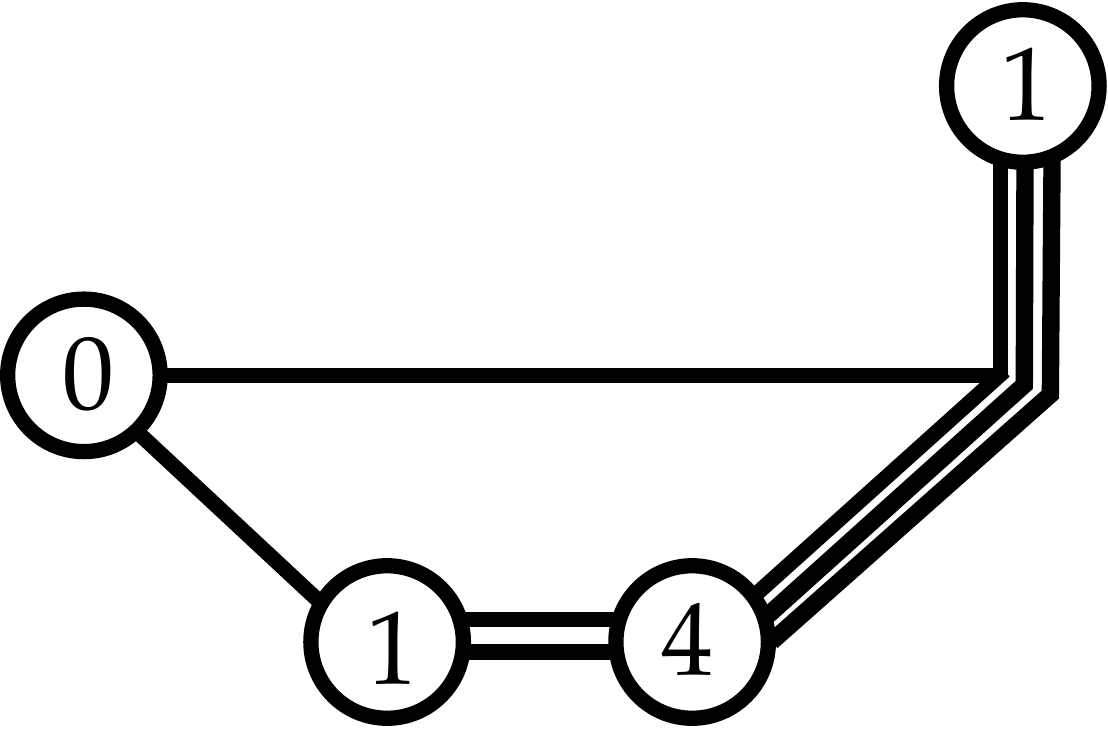} & & $U(1)$ & {\small $SU(3)_0$} & & 1\cr\hline

59 & 1 & \includegraphics[height=.8cm]{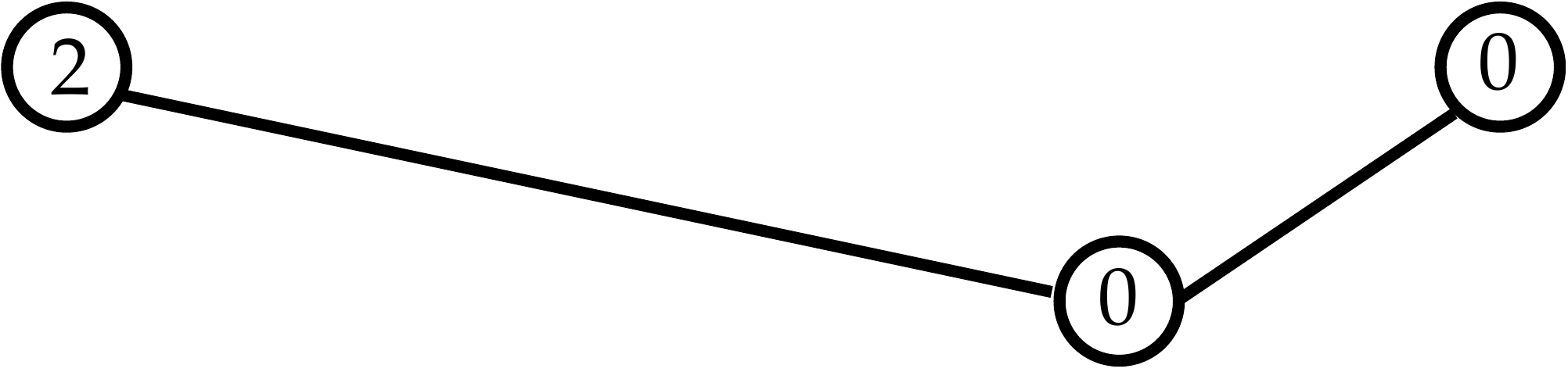} & \includegraphics[height=1cm]{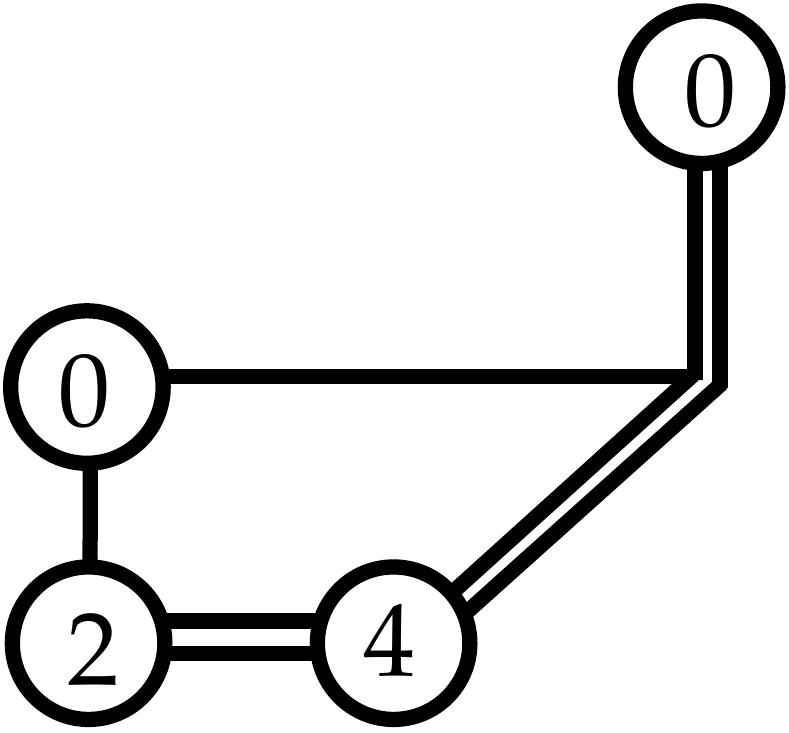} & \includegraphics[height=1cm]{CFD-E8-52.pdf} & $ U(1)$ & {\small $SU(3)_{1}$} &  & {0,1} \cr\hline

\end{tabular}\caption{Table summarizing all Rank Two SCFTs (Continued).}
\end{sidewaystable}

\begin{sidewaystable}
\begin{tabular}{|c|c|c|c|c|c|c|c|c|}\hline
No. & $M$ & $(D_{10}, I_1)$ CFD & $(E_8,SU(2))$ CFD &  Model 3\&4 CFD &  Flavor  $G_\text{F}$& Gauge Theory  & BPS Spin 0 &  Spin 1 \cr \hline\hline 
60 & 1 & \includegraphics[height=.8cm]{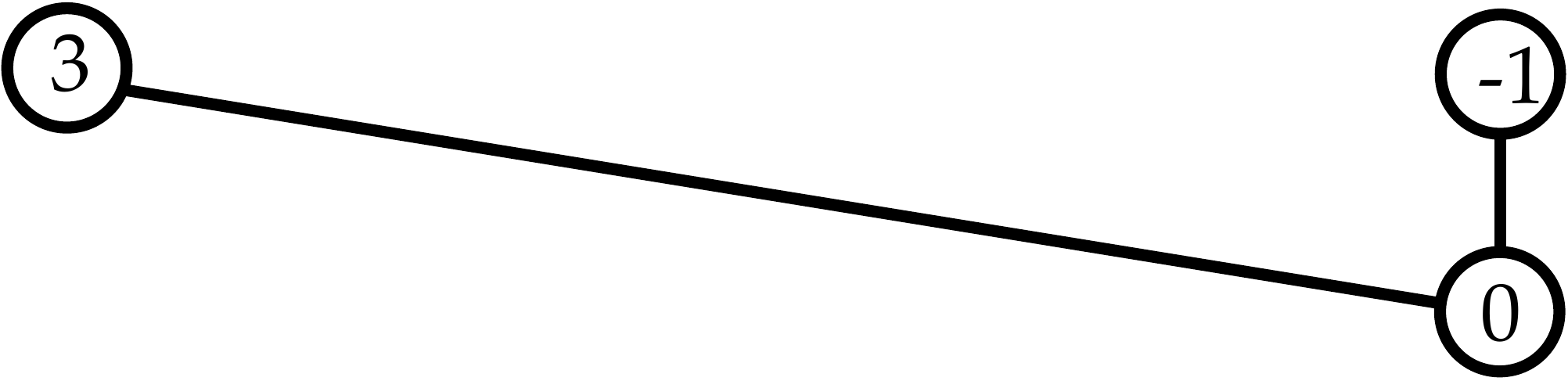} & \includegraphics[height=1cm]{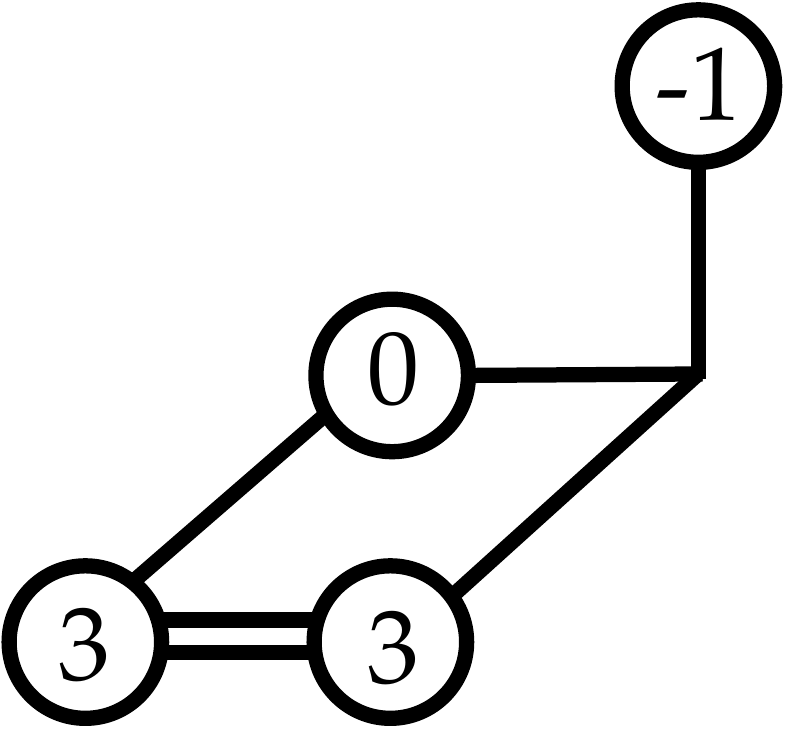} & \includegraphics[height=1cm]{CFD-E8-51.pdf} &  $U(1)$ & {\small $SU(3)_2$} & & 1\cr\hline

61 & 1 &  \includegraphics[height=.8cm]{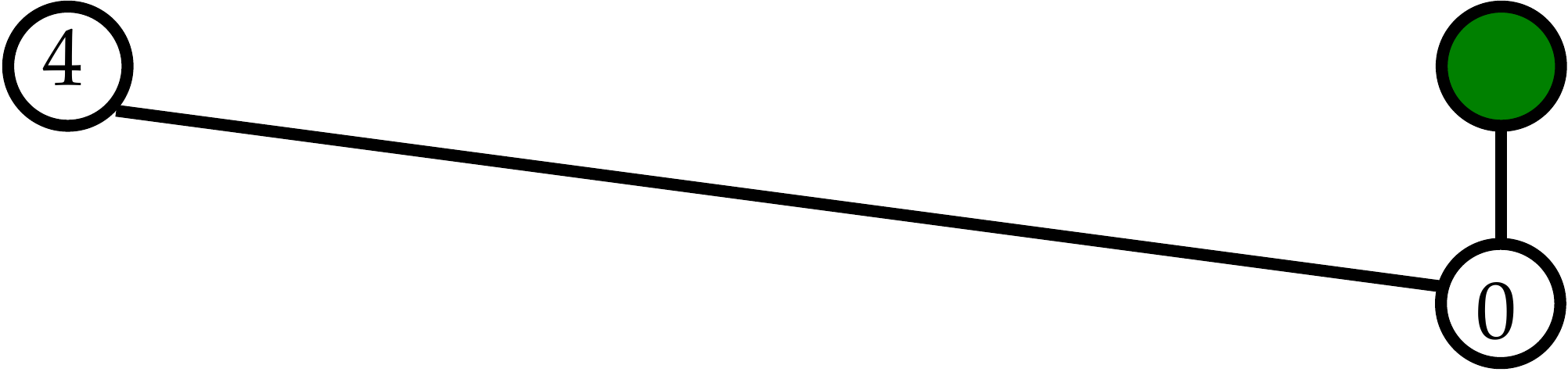} & \includegraphics[height=1cm]{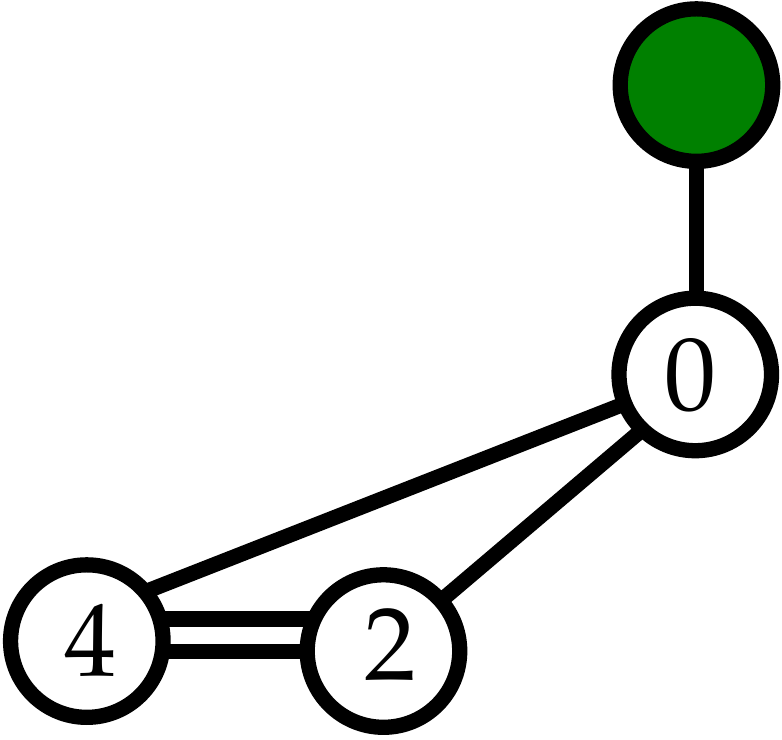} & \includegraphics[height=1cm]{CFD-E8-50.pdf} & {$SU(2)$} & {\small $SU(3)_3$} & & $\mbf{2}$\cr\hline

62 & 1 &  \includegraphics[height=.8cm]{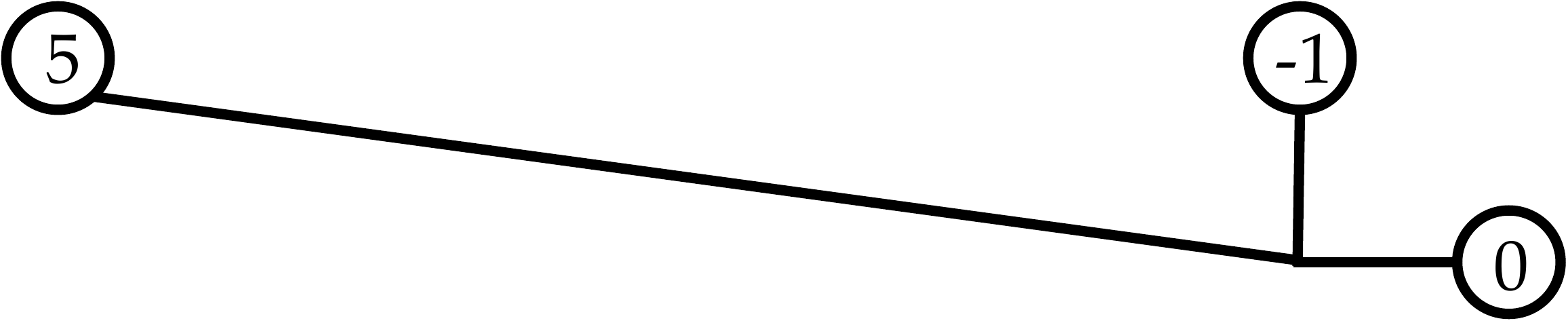} &\includegraphics[height=1cm]{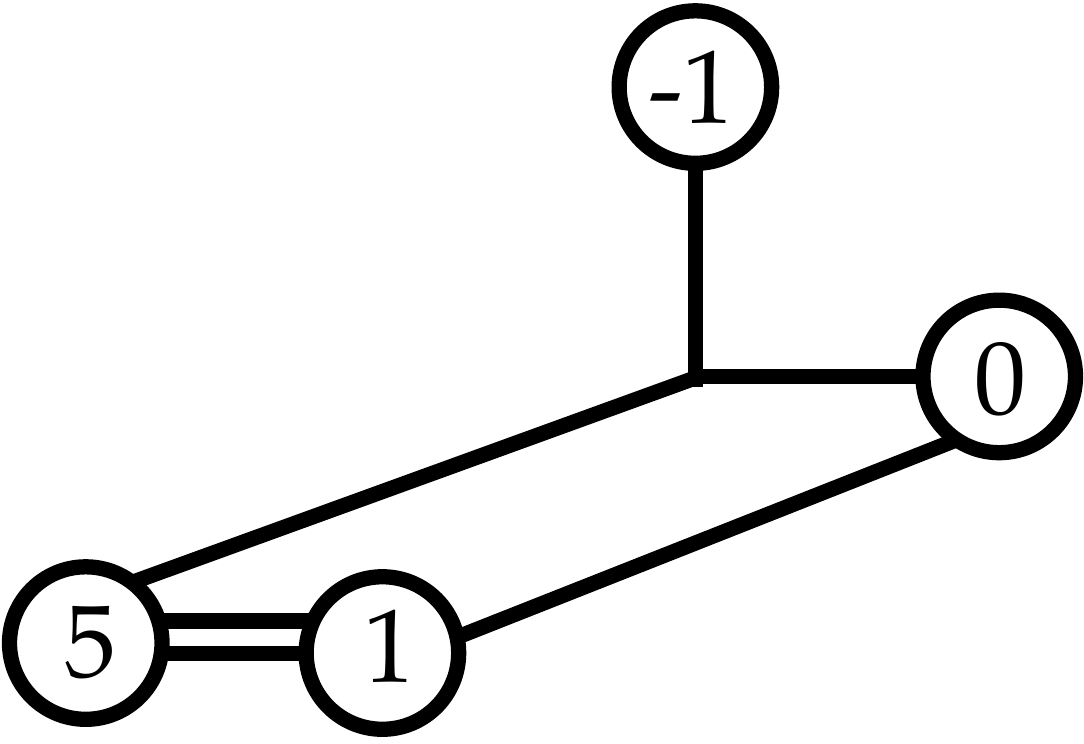} & \includegraphics[height=1cm]{CFD-E8-49.pdf} & $U(1)$ & {\small $SU(3)_4$} & 1 & 1\cr\hline

63 & 1 & \includegraphics[height=.8cm]{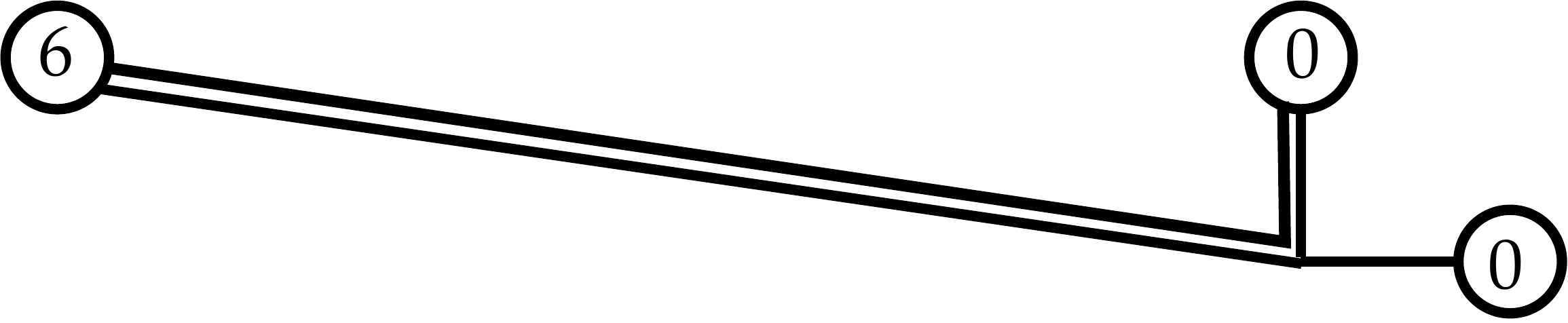} & \includegraphics[height=1cm]{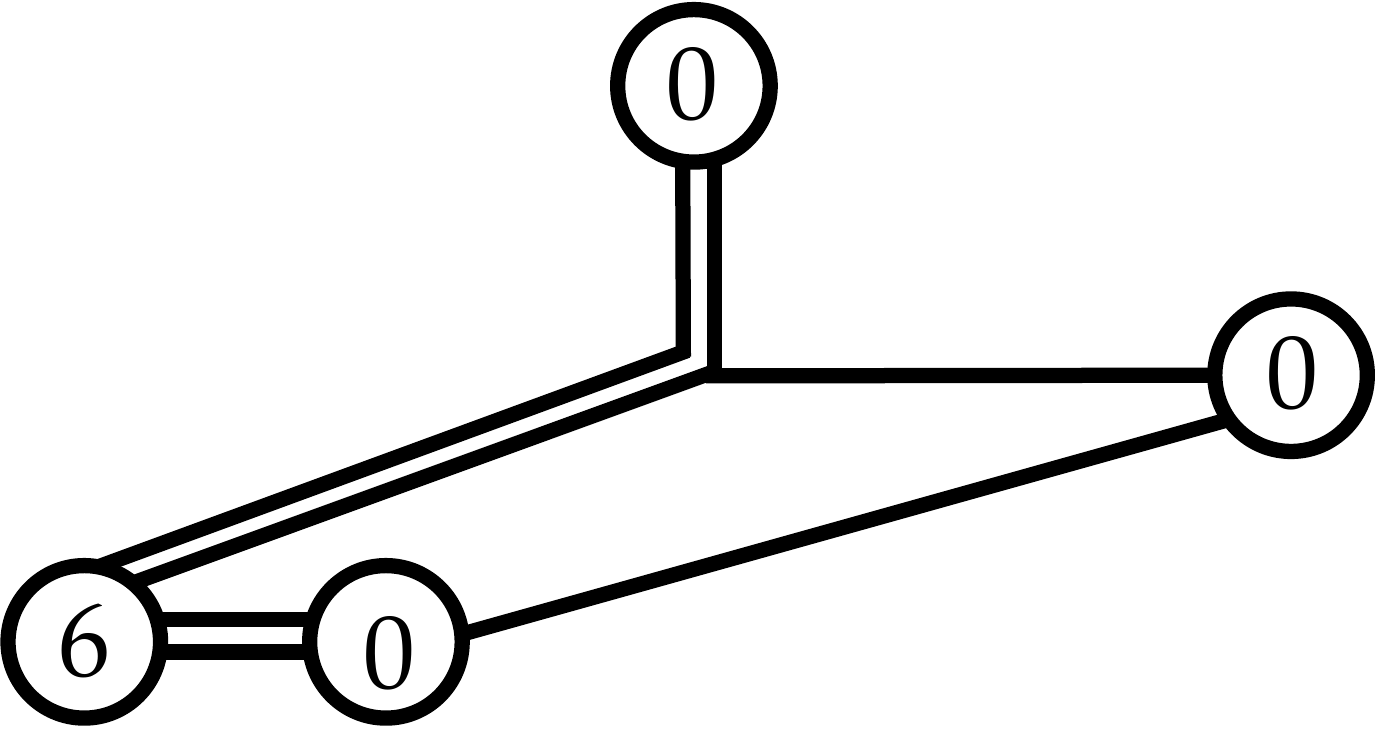} & \includegraphics[height=1cm]{CFD-E8-48.pdf} & $U(1)$ & {\small $SU(3)_5,Sp(2)_\pi$} & & 1,2 \cr \hline

64 & 1 & &\includegraphics[height=1cm]{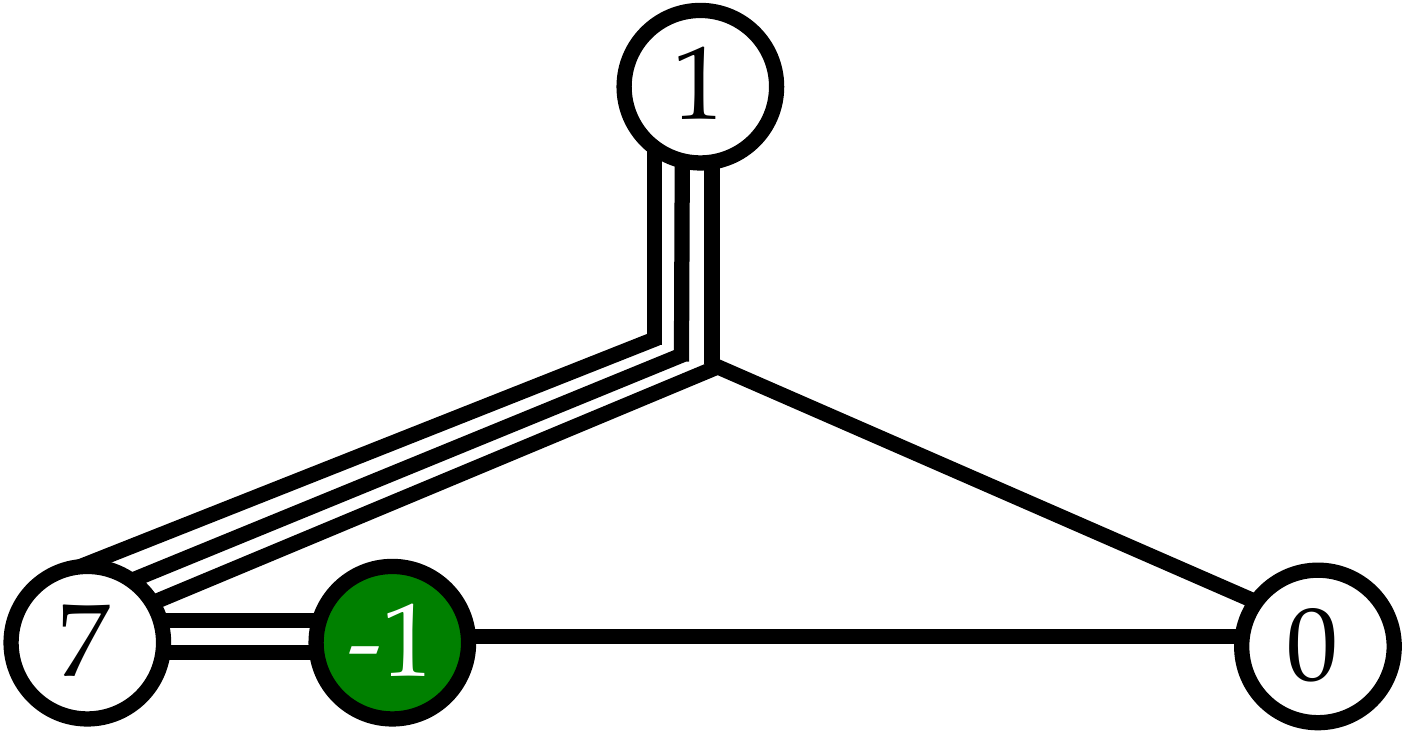} & \includegraphics[height=1cm]{CFD-E8-46.pdf} & $SU(2)$ & {\small $SU(3)_6$} &  & $\mbf{2}$\cr\hline

\multirow{3}{*}{65} & \multirow{3}{*}{1} & & & \multirow{3}{*}{\includegraphics[height=1.2cm]{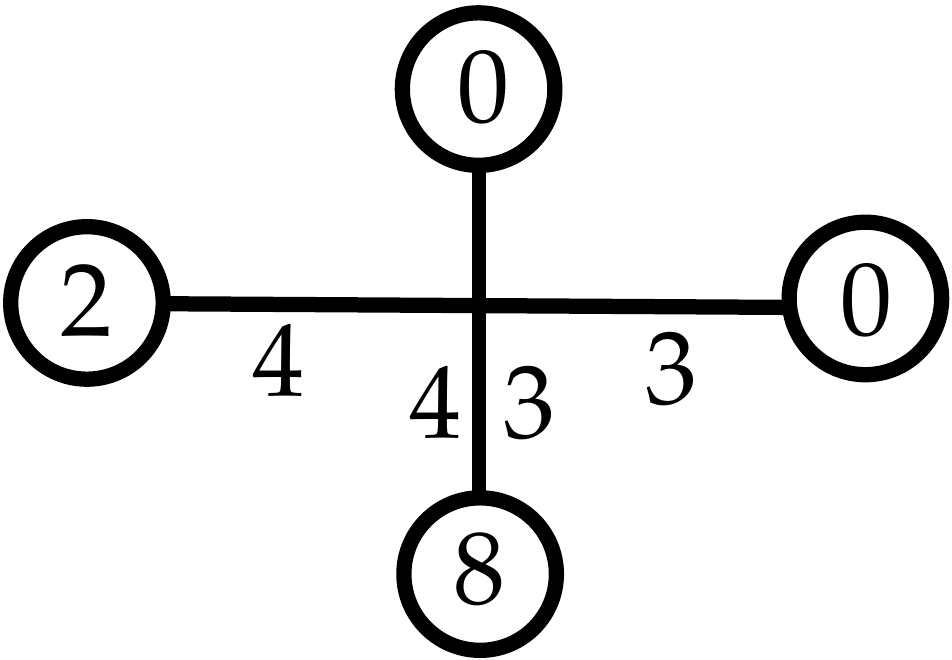}} & \multirow{3}{*}{$U(1)$} & {\small $SU(3)_7$} & & \multirow{3}{*}{$2\cdot\mbf{1}$}\cr
& &  & & & & {\small $G_2$} & & \cr& & & & & & &&\cr\hline

66 & 1 & \includegraphics[height=.6cm]{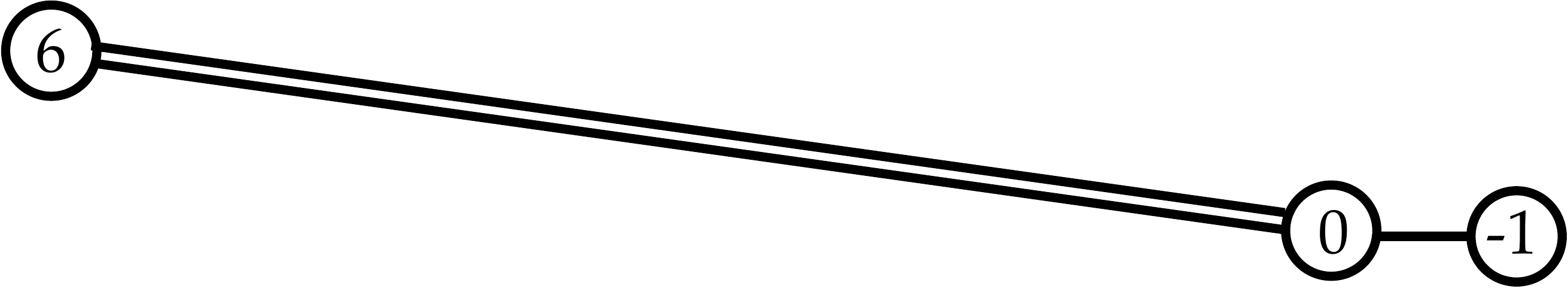} & \includegraphics[height=.8cm]{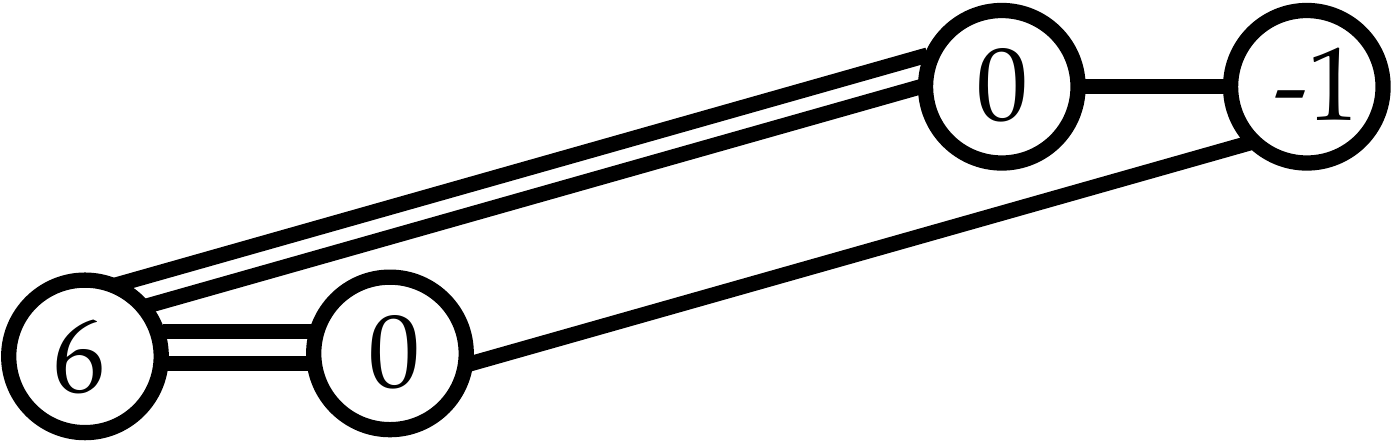} & \includegraphics[height=.8cm]{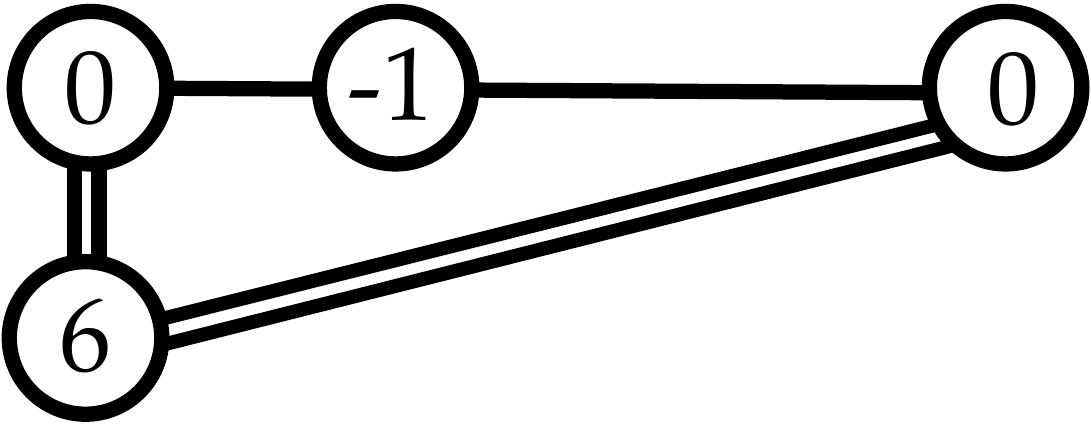} & $U(1)$ & {\small $Sp(2)_0$} & 0 & 1 \cr \hline

67 & 0 & \includegraphics[height=.8cm]{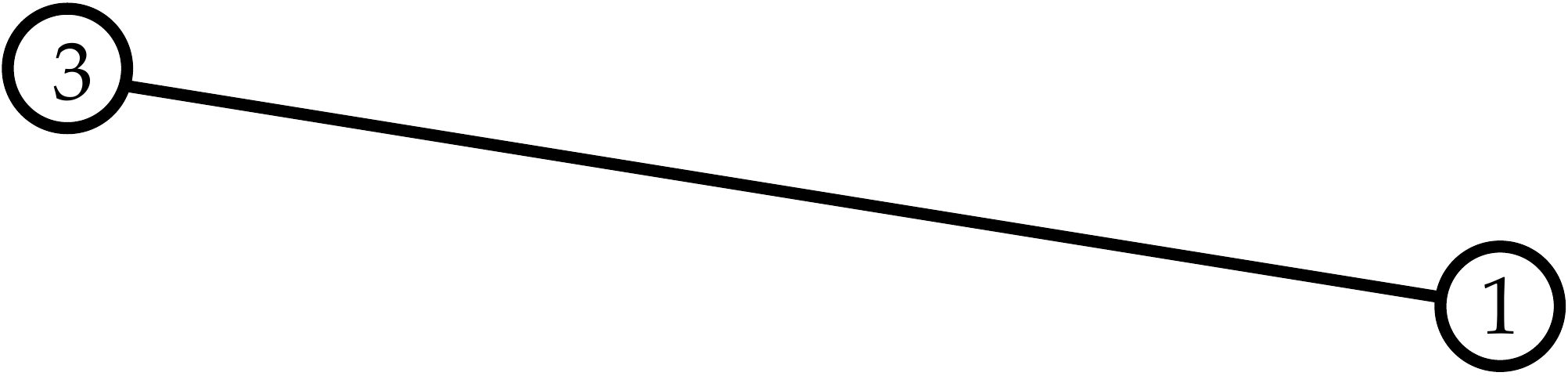} & \includegraphics[height=1cm]{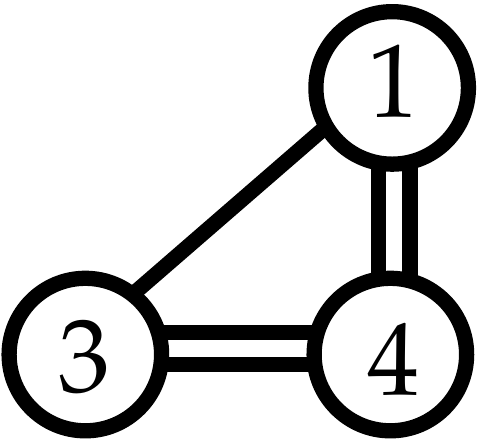} & \includegraphics[height=1cm]{CFD-E8-56.pdf} & $-$ & $-$ & & \cr\hline 

68 & 0 & \includegraphics[height=.6cm]{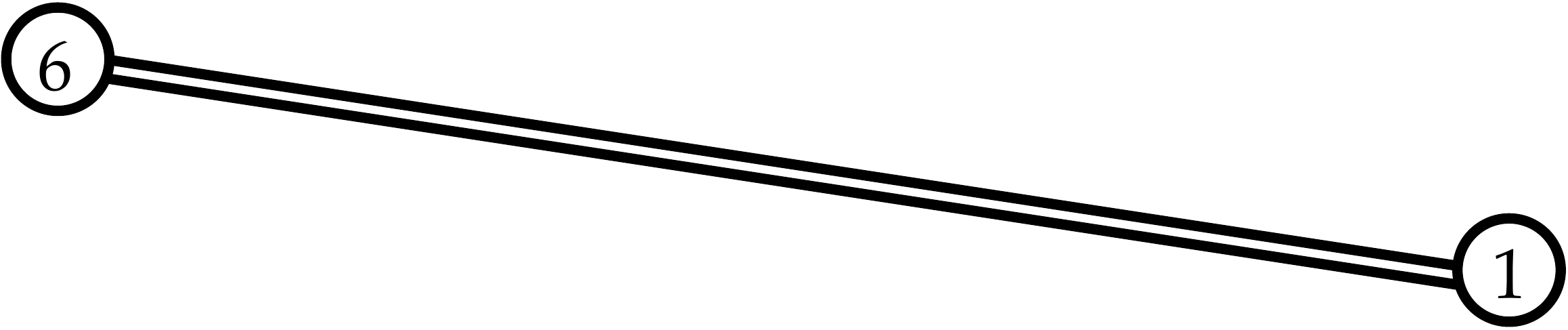} & \includegraphics[height=1cm]{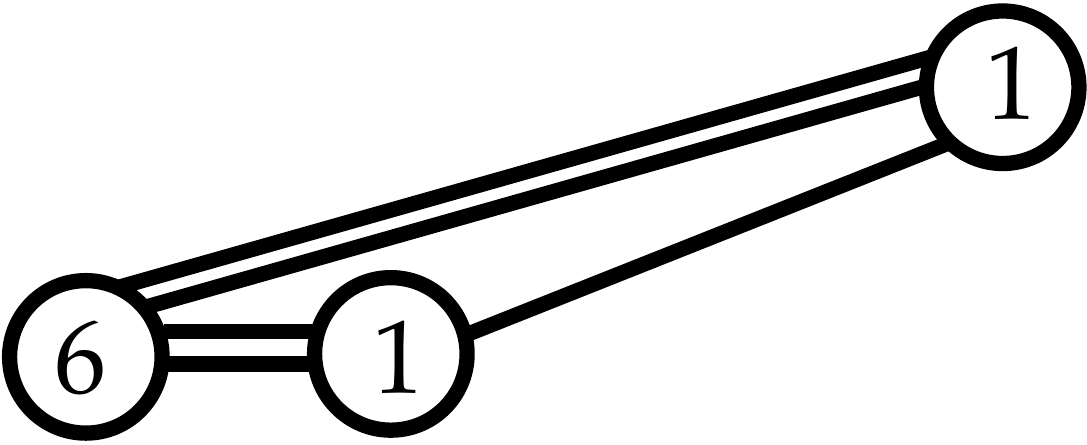} & \includegraphics[height=.8cm]{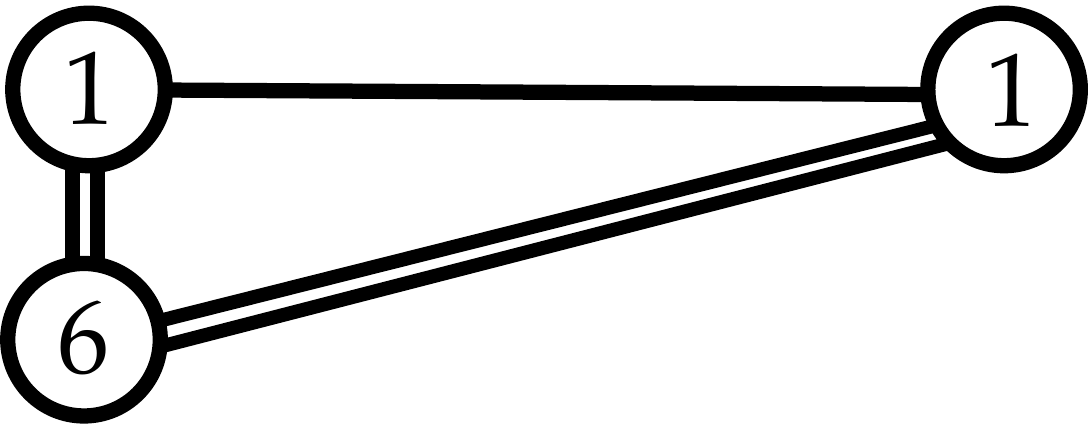} & $-$ & $-$ & & \cr\hline\hline

\end{tabular}\caption{Table summarizing all Rank Two SCFTs (Continued).}
\end{sidewaystable}

\newpage

\section{Rank Two Prepotentials}\label{subsubsec:rank_2_prep}

Using the matching of prepotentials \eqref{eq:matching_prep}, one can easily extract, for a given geometry, the relevant field theory data.
In particular, we are interested in the number $N_{\bf R}$ of hypermultiplets in a representation ${\bf R}$ as well as the discrete Chern--Simons levels.
For a given gauge group $G_\text{gauge}$, in most instances, these determine the effective gauge theory uniquely.

In this appendix we will consider the three types of rank two gauge theories that can be realized in terms of non-flat elliptic fibrations.
These are $G_{\text{gauge}} = SU(3)$ with $N_f$ fundamental hypers and Chern--Simons level $k$, $G_{\text{gauge}} = Sp(2)$ with $N_f$ fundamental and $N_a$ anti-symmetric hypers, and $G_{\text{gauge}} = SU(2)_1 \times SU(2)_2$ with $N_{f_{i}}$ fundamental hypers charged under $SU(2)_i$. These gauge theories are effective low-energy descriptions of most of the rank two descendant 5d SCFTs, which we study in the main text.

There are two furhter gauge theory descriptions for rank two descendants: 
 $G_{\text{gauge}}=G_2$ with $N_7$ hypermultiplets in the 7-dimensional representation of $G_2$, and $G_{\text{gauge}}=SU(3)_{k}$ with $N_f$ fundamental hypers and $N_s$ hypermultiplets in the symmetric representation. These are either alternative dual effective models of the rank two gauge theories that we study here, or they describe the low-energy of the single descendant 5d SCFT outlying our analysis. We will come back to the detailed prepotential analysis of these  theories in \cite{Apruzzi:2019enx}. 

To begin with, note that we can simplify the field theory part on the right-hand side of \eqref{eq:matching_prep} to
\begin{equation} \label{eq:prepr2ls}
	6 \mathcal{F}^{(4)}=  k \, d_{ijl} \phi^i \phi^j \phi^l +  \frac{1}{2} \left( \sum_{\alpha_i \in \Phi_+} 2 (\alpha \cdot \phi)^3 - \sum_f \sum_{ \lambda_f \in \mathbf W_f} \sigma_{\lambda_f}  (w \cdot \phi + m_f)^3\right) \,,
\end{equation}
where $\Phi_+$ are the positive roots, and
\begin{equation}
	\begin{split}
		& \lambda_f \cdot \phi + m_f >0 \ \rightarrow \ \sigma_{\lambda_f}=1 \, ,\\
		& \lambda_f \cdot \phi + m_f <0 \ \rightarrow \ \sigma_{\lambda_f}=-1 \, ,
	\end{split}
\end{equation} 
and $d_{ijl} =0$ for any $G$ but $G=SU(3)$, such that $k$ is only relevant for $SU(3)$ in this case. 
The resulting expression as a function in $\phi^i$ has to be identified with 
\begin{align}\label{eq:prepotential_geometry}
  {\cal F}_\text{geo} = S_1^3\times (\phi^1)^3  + 3 S_1^2\cdot S_2 \times (\phi^1)^2\,\phi^2 + 3 S_1\cdot S_2^2 \times \phi^1\,(\phi^2)^2  + S_2^3 \times (\phi^2)^3\, \equiv (\phi^1 \, S_1+ \phi^2 \, S_2)^3 \, .
\end{align}

\paragraph{\boldmath{$SU(3)$}} $\,$ \\
The cubic part of the prepotential of $SU(3)$ gauge theory with general Chern-Simons level $k$ and $N_f$ hypermultiplets in the fundamental is
\begin{align} \label{eq:gensu3prepls}
\begin{split}
 6 \mathcal  F^{(3)}_{SU(3)} & = 3 k((\phi^1)^2 (\phi^2) - \phi^2(\phi^1)^2)\\
 & +\left|\begin{bmatrix}2\\-1 \end{bmatrix}^T \begin{bmatrix} \phi^1  \\ \phi^2 \end{bmatrix} \right|^3 + \left|\begin{bmatrix}1\\1 \end{bmatrix}^T \begin{bmatrix} \phi^1 \\ \phi^2\end{bmatrix} \right|^3+ \left|\begin{bmatrix}-1\\2 \end{bmatrix}^T \begin{bmatrix} \phi^1   \\ 
 \phi^2\end{bmatrix} \right|^3 \\ 
 &-\frac{1}{2}\left( \sum_i^{N_f} \sigma_1^i (\phi^1 + m_f^i)^3+ \sigma_2^i (\phi^2 - \phi^1 + m_f^i)^3 +  \sigma_3^i (-\phi^2+ m_f^i)^3\right) \,,
 \end{split}
 \end{align}
 where the $\sigma^i_w$ can be $\pm1$ and $w=1,2,3$ labels the positive weights of the fundamental of $SU(3)$. In particular
we would like to compare this with the \eqref{eq:prepotential_geometry} to obtain the CS-level and fix the sign of the terms in the third line.

We can now assume that $\sigma_1^i$ and $\sigma_3^i$ have opposite signs for all $i$, in particular, $\sigma^i_1 = -\sigma^i_3 = 1$. For fixed $N_f$, this is a justified assumption since different sign combinations will lead to regions of validity for $\{\phi^1,\phi^2\}$ outside the Weyl chamber 
\begin{equation}
2\phi^1 - \phi^2\geq0, \qquad \phi^1 + \phi^2\geq0, \qquad -\phi^1 + 2\phi^2\geq0\, .
\end{equation}
In particular, other sign combinations of $\sigma_1^i$ and $\sigma_3^i$ will lead to a gauge theory with different $N_f$. We will give more details on this using the Box-Graph approach in \cite{Apruzzi:2019enx}. 

The only ambiguity can now come from $\sigma_2^i$.
We parametrize the number of flavors with $\sigma_2^i = -1$ by $a$; having carefully kept track of the signs, we can for now ignore the terms proportional to $m_f^i$ and $(m_f^i)^2$, then \eqref{eq:gensu3prepls} becomes
\begin{align} 
\begin{split}
 6 \mathcal  F^{(3)}_{SU(3)} & = 3 k((\phi^1)^2 \phi^2 - (\phi^2)^2 \phi^1)+ (8 (\phi^1)^3 - 3 (\phi^1)^2 \phi^2 - 3(\phi^2)^2 \phi^1 + 8 (\phi^2)^3) \\
 &-\frac{N_f-a}{2}\left( (\phi^1)^3+ (\phi^2 - \phi^1 )^3   - (-\phi^2)^3 \right)-\frac{a}{2}\left( (\phi^1 )^3 - (\phi^2 - \phi^1 )^3   - (-\phi^2 )^3\right) \, .
 \end{split}
 \end{align}
By comparing the cubic terms in $(\phi^1)$ and $(\phi^2)$ of this expression with \eqref{eq:prepotential_geometry} we get the following equations
\begin{align}\label{NkaSU3}
\left(\begin{array}{cc}
  8 - a &-a + \frac{N_f}{2} - 1 + k \cr
 a - \frac{N_f}{2} - 1 - k  & a - N_f + 8  
\end{array}\right) = 
\left( \begin{array}{cc} 
S_1^3 & S_1 S_2^2 \cr 
S_2 S_1^2 & S_2^3
\end{array} 
\right) \,.
   \end{align}
which fix the values of $N_f,k,a$. Furthermore, \eqref{NkaSU3} and \eqref{eq:genusform} imply that $g(\Sigma_{12})=0$, and finally the total number of mass deformations is 
\begin{equation}
M=N_f+1 \,,
\end{equation}
where the $+1$ accounts for the instanton $U(1)_T$ associated to $G_{\text{gauge}}=SU(3)$.

\paragraph{\boldmath{$SU(2)_1 \times SU(2)_2$}} $\,$\\ 
In this case there is no Chern--Simons level to determine. 
The cubic field theory prepotential for generic $N_{f_1},N_{f_2}$ reads
\begin{equation} 
 6 \mathcal  F^{(3)}_{SU(2)^2} = |2 \phi^1|^3 +|2 \phi^2|^3 -|\phi^1 + \phi^2|^3-|\phi^1 - \phi^2|^3 - \sum_{i=1}^{N_{f_1}} |\phi^1 + m_{f_2}^i|^3 - \sum_{i=1}^{N_{f_2}} |\phi^2 + m_{f_2}^i|^3\, .
\end{equation}
We consider the chamber $\phi^1\geq0, \phi^2\geq0$, and we explicitly write the sign dependence like in \eqref{eq:prepr2ls}, such that 
\begin{equation} \label{eq:su2su2prep}
 6 \mathcal  F^{(3)}_{SU(2)-SU(2)} =8 (\phi^1)^3 +8 (\phi^2)^3  -(\phi^1 + \phi^2)^3- \sigma_0(\phi^1 - \phi^2)^3 - \sum_{i=1}^{N_{f_1}} \sigma_1^i (\phi^1 )^3 - \sum_{i=1}^{N_{f_2}} \sigma_2^i(\phi^2 )^3\, ,
\end{equation}
where we also ignored the terms proportional to $m^i_{f_1},m^i_{f_2}$ and  $(m^i_{f_1})^2,(m^i_{f_2})^2$. Similarly to the $SU(3)$ case for determined values of $N_{f_1}$ and $N_{f_2}$ the conditions $\phi^1\geq0, \phi^2\geq0$ fix $\sigma_1^i=\sigma_2^i=1$. Having set this we can now compare \eqref{eq:su2su2prep} with the geometric quantity \eqref{eq:prepotential_geometry}. We get the following equations
\begin{align} \label{eq:su2su2geovsgauge}
\left(\begin{array}{cc}
7-l_0- N_f & -1-l_0 \cr
l_0-1 & 7+l_0 - N_{f_1}
\end{array}\right) = 
\left( \begin{array}{cc} 
S_1^3 & S_1 S_2^2 \cr 
S_2 S_1^2 & S_2^3
\end{array} 
\right) \,.
\end{align}
These equations completely fix $N_{f_1}, N_{f_2}, \sigma_0$. Moreover,  \eqref{eq:su2su2geovsgauge} and \eqref{eq:genusform} imply again that $g(\Sigma_{12})=0$. The total number of mass deformation is given by
\begin{equation}
M=N_{f_1}+N_{f_1}+3 \, ,
\end{equation}
where the $+3$ accounts for the two instanton $U(1)_{T_{1,2}}$ associated  to $SU(2)_{1,2}$, and for the $SU(2)$ baryonic symmetry rotating the hypermultiplet in the bifundamental of $SU(2)_1 \times SU(2)_2$.

\paragraph{\boldmath{$Sp(2)$}} $\,$\\
The prepotential for $Sp(2) (\equiv SO(5))$ gauge theory with $N_f$ hypermultiplets in the fundamental and $N_a$ in the antisymmetric is
\begin{align} \label{eq:gensp2prep}
\begin{split}
 6 \mathcal  F^{(3)}_{Sp(2)} & =  \left|\begin{bmatrix}2\\0 \end{bmatrix}^T \begin{bmatrix} \phi^1  \\ \phi^2\end{bmatrix} \right|^3 + \left|\begin{bmatrix}0\\1 \end{bmatrix}^T \begin{bmatrix} \phi^1  \\ \phi^2\end{bmatrix} \right|^3+ \left|\begin{bmatrix}-2\\2 \end{bmatrix}^T \begin{bmatrix} \phi^1   \\ 
 \phi^2\end{bmatrix} \right|^3 + \left|\begin{bmatrix}2\\-1 \end{bmatrix}^T \begin{bmatrix} \phi ^1  \\ 
 \phi^2\end{bmatrix} \right|^3 \\ 
 &- \sum_{i=1}^{N_a} \left( \left|\begin{bmatrix}0\\1 \end{bmatrix}^T \begin{bmatrix} \phi ^1  \\ 
 \phi^2\end{bmatrix} + m_a^i \right|^3 +\left|\begin{bmatrix}2\\-1 \end{bmatrix}^T \begin{bmatrix}  \phi ^1  \\ 
 \phi^2\end{bmatrix}  + m_a^i \right|^3 \right) \\
  &- \sum_{i=1}^{N_f} \left( \left|\begin{bmatrix}1\\0 \end{bmatrix}^T \begin{bmatrix}  \phi ^1  \\ 
 \phi^2\end{bmatrix} + m_f^i \right|^3 +\left|\begin{bmatrix}-1\\1 \end{bmatrix}^T \begin{bmatrix} \phi ^1  \\ 
 \phi^2\end{bmatrix}  + m_f^i \right|^3 \right) \, .
 \end{split}
 \end{align}
Also in this case there is no CS-level to determine. The cone defined by 
 \begin{equation} \label{eq:sp2coneroots}
 2\phi^1\geq0, \qquad \phi^2\geq0, \qquad -2\phi ^1+2\phi^2\geq0, \qquad 2\phi^1-\phi^2\geq0 \, ,
 \end{equation}
specifies the signs of some of the absolute values in the second and third line of \eqref{eq:gensp2prep}, such that the prepotential reads
 \begin{align}
 6 \mathcal  F^{(3)}_{Sp(2)}  & = (2 \phi^1)^3 + (\phi^2)^3 + (-2\phi^1+2\phi^2)^3 + (2\phi^1-\phi^2)^3\\
 & -(N_a-a_a) \left( (\phi^2)^3+(2\phi^1-\phi^2)^3\right) - a_a  \left( (\phi^2)^3-(2\phi^1-\phi^2)^3\right) \nonumber \\
  & -(N_f-a_f) \left( (\phi^1)^3+(-\phi^1+\phi^2)^3\right) - a_f  \left( (\phi^1)^3-(-\phi^1+\phi^2)^3\right)\nonumber \, .
 \end{align}
For our purpose, we can ignore the terms proportional to $m^i_f, m^i_a$ and $(m^i_f)^2, (m^i_a)^2$, since we have kept track of the signs of the absolute values. By comparing the cubic expansion of this expression with the geometric quantity \eqref{eq:prepotential_geometry}, we get the following constraints on $N_f, N_a,a_f, a_a$
 \begin{align} \label{eq:sp2geovsgauge}
\left(\begin{array}{cc}
 -2a_f + 16 a_a +8 - 8N_a &-2a_f +4a_a -2 N_a+N_f-6 \cr
2a_f-8a_a+4N_a - N_f +4   &2 a_f-2a_a+ 8- N_f 
\end{array}\right) = 
\left( \begin{array}{cc} 
S_1^3 & S_1 S_2^2 \cr 
S_2 S_1^2 & S_2^3
\end{array} 
\right) \,.
   \end{align}
This fixes $N_a, N_f, a_f, a_a$, where $a_f$ and $a_a$ can be half-integer, since $Sp(2)$ has pseudo-real representations for the hypermultiplets. Although they can in principle take values from $0$ to $N_a$ and $0$ to $N_f$, respectively, only some of them are actually allowed and match the geometric prepotential. Using the Box-Graph approach, the set of positivity conditions of the Coulomb branch makes the bounds on the values of $\{a_f,\,a_a\}$ manifest. We will come back to this in \cite{Apruzzi:2019enx}. At last, from  \eqref{eq:su2su2geovsgauge} and \eqref{eq:genusform} we can compute the genus of the gluing curve between the two surfaces, which reads
\begin{equation}
g(\Sigma_{12})=N_a-2a_a\, ,
\end{equation} 
and the total number of mass deformation is given by
\begin{equation}
M=N_{f}+N_{a}+1 \, ,
\end{equation}
where the $+1$ accounts for the two instanton $U(1)_{T}$ associated  to $Sp(2)$.


\section{Rational Surfaces: dP versus gdP}
\label{app:gdP}

\begin{figure}
\centering
\includegraphics[height=7cm]{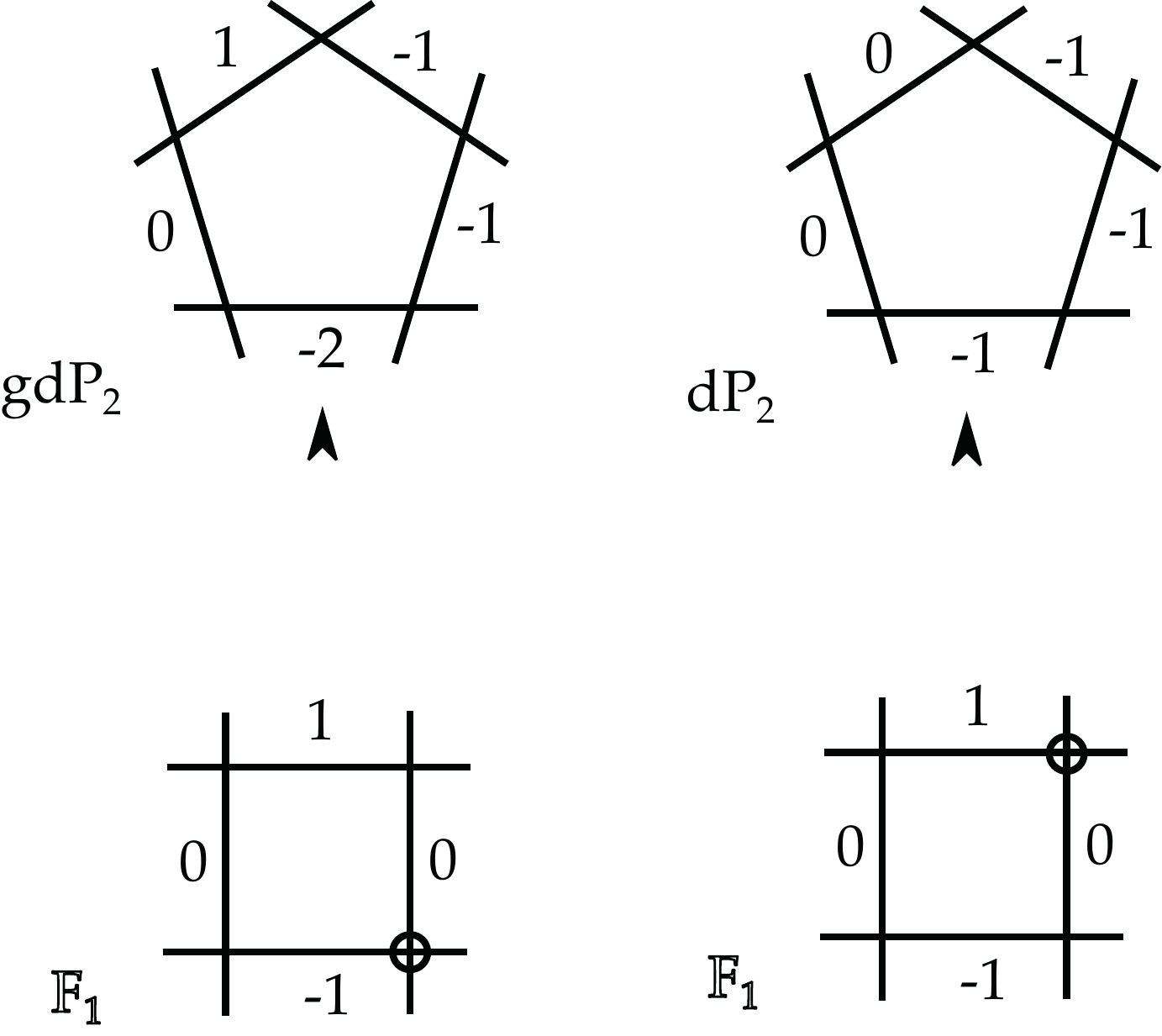}
\caption{The blow-ups from the Hirzebruch surface $\mb{F}_1$ to a generalized del Pezzo surface gdP$_2$ of type $A_1$ and a del Pezzo surface dP$_2$. The circle denotes the points to be blown up on $\mb{F}_1$. Hence gdP$_2$ and dP$_2$ are related by a flop. \label{f:gdPflop}}
\end{figure}

Generalized del Pezzo (gdP) surfaces play an essential part in the geometric description that we choose for 5d SCFTs in this paper. Here we provide some mathematical background on the rational surfaces. More details can be found in~\cite{derenthal2006geometry,derenthal2014singular,Taylor:2015isa}. In the following we always use a smooth rational surface $S$ with an effective anticanonical divisor $-K_S$. 

For a surface $S$ with $h^{1,1}(S)=r>2$, the Picard group of divisors on $S$ is an additive abelian group Pic$(S)=\mb{Z}^r$ with the following generators
\be
\label{rational-Picard}
h\,, e_i\,,\quad i=1,\dots,r-1 \,.
\ee
The intersection numbers between these are
\be\label{heij}
h^2=1,\quad  h\cdot e_i=0,\quad  e_i\cdot e_j=-\delta_{ij} \,.
\ee
A general element of $\mb{Z}^r$ can be written as
\be
C=ah+\sum_{i=1}^{r-1}b_i e_i\,.
\ee
The anticanonical divisor of $S$ is 
\be
-K_S=3h-\sum_{i=1}^{r-1}e_i,.
\ee
If $h^{1,1}(S)=r=2$, then $S$ is a Hirzebruch surface $\mb{F}_n$. If $n$ is even, then we need to use the following set of generators for Pic$(S)$ rather than (\ref{heij}):
\be
s^2=-n,\quad s\cdot f=1,\quad f^2=0 \,.\label{sf}
\ee
For odd $n$, we can use either (\ref{heij}) or (\ref{sf}). The anticanonical divisor of $\mb{F}_n$ is
\be
-K_S=2s+(n+2)f.
\ee

If $h^{1,1}(S)=r=1$, then $S=\mb{P}^2$ and its Picard group is generated by the hyperplane class $h\subset\mb{P}^2$ with $h^2=1$.

The most essential characteristic of a rational surface are the generators of its Mori cone (or the cone of effective divisors as they are equivalent on a complex surface). For any rational surface $S$ with an effective anticanonical divisor $-K_S$ and $r>2$, its Mori cone is generated by a set, Neg$(S)$, of irreducible rational curves $C$ with negative self-intersection. Recall that the genus $g$ of a curve $C$ on $S$ can be computed by the adjunction formula
\be
K\cdot C+C\cdot C=2g-2\,.
\ee
A rational curve is defined to be a curve with genus $g=0$.

If $r=2$, then the Mori cone of $S=\mb{F}_n$ is generated by the $(-n)$-curve $s$ and the 0-curve $f$ in (\ref{sf}). If $r=1$, then the Mori cone of $S=\mb{P}^2$ is generated by $h$.

We can further define the subset of Neg$(S)$: Sing$(S)$ which is the set of irreducible rational curves with self-intersection $(-2)$ or lower. After Sing$(S)$ is fixed, the set of $(-1)$-curves on $S$ is given by all the curves in the form of
\be
C=ah+\sum_{i=1}^{r-1} b_i e_i \,,
\ee
which satisfies
\be
K\cdot C=C\cdot C=-1
\ee
and intersect all the curves in Sing$(S)$ non-negatively \cite{Taylor:2015isa}. This theorem is useful for generating the set of $(-1)$-curves on a rational surface.

Especially, when Sing$(S)$ only contains $(-2)$-curves, $S$ is called a generalized del Pezzo surfaces (or weak-Fano surface). The generalized del Pezzo surfaces are comprehensively classified and studied in~\cite{derenthal2006geometry,derenthal2014singular}. We denote it by 
gdP$_n$ if $n=h^{1,1}(S)-1$, which corresponds to the degree-$(9-n)$ cases in the mathematics literature. They are classified by the configuration of $(-2)$-curves, which form Dynkin diagrams (affine Dynkin diagrams for the cases of gdP$_9$). For example, we call a gdP$_n$ of type $A_1$ if there is only a single $(-2)$-curve on it that form an $A_1$ Dynkin diagram.

If Sing$(S)$ is empty, such that there is no $(-2)$ or lower curve on $S$, then the surface $S$ is a del Pezzo surface $\text{dP}_n$. The del Pezzo surface and generalized del Pezzo surfaces are related by a number of flops, see figure~\ref{f:gdPflop} for the case of $n=2$. One can also think generalized del Pezzo surfaces as special points in the complex structure moduli space of del Pezzo surfaces.


\section{Details of Non-Flat Resolutions}

Various technical details for blow-ups of non-flat resolution of elliptic fibrations are collected in this appendix. 
We begin by deriving the two resolved geometries for the rank two E-string and the $(D_{10},I_1)$ model, which will 
be the starting points for the flop chains for rank two theories, i.e. the marginal geometries, from which we read off the marginal (top) CFDs. 
These are precisely the fibers, where the non-flat surface components $S_i$ contain all the rational curves in the codimension one 
 singular fibers.  We also provide the marginal theories for higher rank conformal matter theories of type $(E_8, SU(n))$ and $(E_n, E_n)$.

\subsection{Non-flat Fiber Resolutions}

\subsubsection{Rank one Theories}
\label{app:Rank1Res}

In this section, all non-flat fiber resolutions of the singular Tate model (\ref{rank-1Tate})  for the rank one E-string are detailed. 
In the main text we already discussed the marginal geometry, where the non-flat fiber is gdP$_9$, see (\ref{gdP9BU}). The descendant theories, which are the rank one 5d SCFTs can either be obtained by flops of the $(-1)$ curve in gdP$_9$, or using the  direct resolutions, which we explain in the following:

\paragraph{gdP\boldmath{$_8$}: }

To get a non-flat fiber with the topology of gdP$_8$, we start with the Tate form (\ref{rank-1Tate}) and do the following sequence of blow-ups
\be
\ba
& \{ \{x,y,u,v,\delta\},\{\left\{x,y,U,u_1\right\},\left\{x,y,u_1,u_2\right\},\left\{y,u_2,u_3\right\},\left\{y,u_1,u_3,u_4\right\} , \cr 
&   \left\{y,u_1,u_5\right\}, \left\{u_1,u_3,u_6\right\},
   \left\{u_1,u_4,u_7\right\},\left\{u_1,u_5,u_8\right\},\left\{u_2,u_3,u_9\right\},  \left\{u_3,u_4,u_{10}\right\},
   \cr 
& \left\{u_4,u_6,u_{11}
   \right\},
   \left\{u_3,u_6,u_{12}\right\},\left\{u_6,u_{10},u_{13}\right\},\left\{u_{10},u_{12},u_{14}\right\},\left\{u_
   {3},u_{12},u_{15}\right\} \} \,.
\ea
\ee
The first blow-up is a weighted blow-up,
\be
x\rightarrow x\delta^3\ ,\ y\rightarrow y\delta^2\ ,\ u\rightarrow U\delta\ ,\ v\rightarrow V\delta \,.
\ee
The exceptional divisor $\delta=0$ is a weighted projective space $\mathbb{P}^{1,1,2,3}$ in the ambient space, but its singular points do not intersect the hypersurface equation. After the replacement of variables, we divide the hypersurface equation by $\delta^6$, which preserves the Calabi--Yau condition of $Y$.

Denote by $S$ the non-flat fiber surface $\delta=0$ inside the resolved Calabi--Yau threefold.  The intersection numbers of $S$ with the Cartan divisors  associated to the affine $E_8$, $D_i$, that are defined in (\ref{CartansRank1}), are 
\be
S \cdot D_i \cdot D_i = (-1,-2,-2,-2,-2,-2,-2,-2,-2)\ ,\ (i=0,\dots,8) \,.
\ee
Hence the flavor symmetry of the corresponding 5d SCFT is $E_8$ from the Dynkin diagram formed by the $(-2)$-curves.

\paragraph{gdP\boldmath{$_7-$}gdP\boldmath{$_3$}: }
To obtain the subsequent geometries corresponds to performing a number of flops, which are essentially a choice of different orderings in the resolution sequences. Consider the sequence
 \be
\ba
&\{\left\{x,y,u,u_1\right\},\left\{x,y,u_1,u_2\right\},\left\{y,u_2,u_3\right\},\underline{\left\{
   u_3,v,\delta   \right\}},\left\{y,u_1,u_3,u_4\right\} , {\bf \Phi_1}, \left\{y,u_1,u_5\right\},\cr 
&    \left\{u_1,u_3,u_6\right\}, {\bf \Phi_2}, 
   \left\{u_1,u_4,u_7\right\},\left\{u_1,u_5,u_8\right\},\left\{u_2,u_3,u_9\right\},  \left\{u_3,u_4,u_{10}\right\}, {\bf \Phi_3}, 
   \cr 
& \left\{u_4,u_6,u_{11}
   \right\},
   \left\{u_3,u_6,u_{12}\right\},{\bf \Phi_4},\left\{u_6,u_{10},u_{13}\right\},\left\{u_{10},u_{12},u_{14}\right\},\left\{u_
   {3},u_{12},u_{15}\right\} \,,
\ea
\ee
 the non-flat fiber surface component $S:\delta=0$ is a gdP$_7$. 
The intersection numbers of $S$ with the Cartan divisors are 
\be
S \cdot D_i \cdot D_i = (0,-1,-2,-2,-2,-2,-2,-2,-2) \qquad \Rightarrow G_\text{F}= E_7 \,.
\ee
For the other cases of $S=$gdP$_6- $gdP$_3$, we move the underlined blow-up $\{u_3,v,\delta\}$ to positions $\Phi_1$, $\Phi_2$,  $\Phi_3$, $\Phi_4$. The intersection numbers of the non-flat fiber $S:\delta=0$ with the Cartan divisors in these cases are 
\be\ba
\Phi_1: \quad S \cdot D_i \cdot D_i &= (0,0,-1,-2,-2,-2,-2,-2,-2) \qquad \Rightarrow G_\text{F}= E_6
\cr 
\Phi_2: \quad S \cdot D_i \cdot D_i &= (0,0,0,-1,-2,-2,-2,-2,-2)\qquad \Rightarrow G_\text{F}= SO(10)
\cr 
\Phi_3: \quad S \cdot D_i \cdot D_i &= (0,0,0,0,-1,-2,-2,-2,-2)\qquad \Rightarrow G_\text{F}= SU(5)
\cr 
\Phi_4: \quad S \cdot D_i \cdot D_i &= (0,0,0,0,0,-1,-2,-2,-2)\qquad \Rightarrow G_\text{F}= SU(3) \times SU(2) \,,
\ea\ee
which also allows us to read off the flavor symmetries $G_\text{F}$ of the strongly coupled SCFT.

\paragraph{gdP\boldmath{$_2$}:}
The resolution
\be
\ba
&\{\left\{x,y,u,u_1\right\},
\left\{x,y,u_1,u_2\right\},
\left\{y,u_2,u_3\right\},
\left\{y,u_1,u_3,u_4\right\},
\left\{y,u_1,u_5\right\}, 
\left\{u_1,u_3,u_6\right\},\left\{u_1,u_4,u_7\right\},\cr 
&
\left\{u_3,u_4,u_{10}\right\},\left\{u_3,u_6,u_{12}\right\},
\left\{u_3,u_{12},u_{15}\right\},
{\left\{u_3,v,\delta \right\}},
\left\{u_6,u_{10},u_{13}\right\},
\left\{u_1,u_5,u_{8}\right\},\cr 
&
\left\{u_2,u_3,u_{9}\right\},
\left\{u_4,u_6,u_{11}\right\},
\left\{u_{10},u_{12},u_{14}\right\}\} \,,
\ea
\ee
results in a non-flat fiber component with geometry gdP$_2$.
The Cartans divisors intersect $S:\delta=0$ as follows
\be
\quad S \cdot D_i \cdot D_i  = (0,0,0,0,0,0,-1,-2,-1)\,,
\ee
wherefore the flavor symmetry is $G_\text{F}=SU(2)\times U(1)$.

\paragraph{dP\boldmath{$_1$}:}
The resolution
\be
\ba
&\{\left\{x,y,u,u_1\right\},
\left\{x,y,u_1,u_2\right\},
\left\{y,u_1,u_2,u_6\right\},
\left\{y,u_1,u_5\right\},
\left\{u_1,u_5,u_8\right\}, 
\left\{y,u_6,u_4\right\},\left\{y,u_2,u_3\right\},\cr 
&
\left\{u_2,u_3,u_9\right\},\left\{u_2,u_4,u_{12}\right\},
\left\{u_3,u_4,u_{10}\right\},
\left\{u_4,u_6,u_{11} \right\},
\left\{u_4,u_9,u_{15}\right\},
\left\{u_4,u_{12},u_{13}\right\},\cr 
&
\left\{u_4,u_{15},u_{14}\right\},
\left\{u_1,u_4,u_7\right\},\left\{u_3,v,\delta\right\}\} \,,
\ea
\ee
leads to
\be
\quad S \cdot D_i \cdot D_i  = (0,0,0,0,0,0,0,-1,0)\,.
\ee
Note that $S$ and $D_{8}$ intersects at a rational curve with self-intersection 0 on $S$.
The flavor symmetry is $G_\text{F}=U(1)$.

\paragraph{gdP\boldmath{$_1 \cong \mb{F}_2$}:}
The resolution
\be
\ba
&\{\left\{x,y,u,u_1\right\},
\left\{x,y,u_1,u_2\right\},
\left\{y,u_1,u_2,u_6\right\},
\left\{y,u_1,u_5\right\},
\left\{y,u_2,u_3\right\},
\left\{u_1,u_5,u_8\right\},
\left\{u_3,u_6,u_{12}\right\},
\cr 
&
\left\{y,u_6,u_4\right\},
\left\{u_4,u_6,u_{11} \right\},
\left\{u_3,u_{12},u_{15}\right\},
\left\{u_2,u_3,u_9\right\},
\left\{u_4,u_{12},u_{13}\right\},
\left\{y,u_{12},u_{10}\right\}\cr 
&
\left\{u_{10},u_{12},u_{14}\right\},
\left\{u_5,u_{6},u_{7}\right\},\left\{u_3,v,\delta\right\}\}
\ea
\ee
results in 
\be
\quad S \cdot D_i \cdot D_i  = (0,0,0,0,0,0,0,-2,0)\,.
\ee
Note that $S$ and $D_{6}$ intersects at a rational curve with self-intersection 0 on $S$.
The flavor symmetry is $G_\text{F}=SU(2)$.


\subsection[Blow-ups for the Rank two E-string \texorpdfstring{$(E_8, SU(2))$}{(E8,SU(2))}]{Blow-ups for the Rank two E-string \boldmath{$(E_8, SU(2))$}}
\label{app:BlowupsRank2E}

We start with the blow-up that generates the geometry associated to the marginal theory in 5d. This is the key input for our subsequent analysis, in particular the CFDs.  We also provide some example blow-ups, which illustrate some of the salient features of non-flat resolutions in higher rank models. 

\subsubsection{Geometry for the Marginal Theory}
\label{app:TopCFDsRank2E}

To derive the blow-up for the marginal theory,  where all of the Dynkin nodes of the affine $E_8\times SU(2)$ are wrapped $(-2)$-curves, we need to blow up the point $U=V=0$ in the base first: $(U,V;\delta_1)$. The Weierstrass model then becomes
\be
y^2=x^3+f_4 U^4 V \delta_1 x+g_6 U^5 V^2 \delta_1 \,,
\ee
while the Tate model is 
\be
y^2+b_1 Uxy+b_3 U^3 V\delta_1 y=x^3+b_2 U^2 x^2+b_4 U^4 V\delta_1 x+b_6 U^5 V^2 \delta_1\,.
\ee
Because of the base blow-up, the locus $U=V=0$ has been removed.
In addition, we now apply the  resolution sequence
\be\label{BUE8SU(2)}
\ba
&\{\left\{x,y,U,u_1\right\},\left\{x,y,V,v_1\right\},\left\{x,y,u_1,u_2\right\},\left\{y,u_2,u_3\right\}, \left\{\delta_1 ,u_3,\delta _2\right\}, \left\{y,u_1,u_3,u_4\right\} , \cr 
&   \left\{y,u_1,u_5\right\}, \left\{u_1,u_3,u_6\right\},
   \left\{u_1,u_4,u_7\right\},\left\{u_1,u_5,u_8\right\},\left\{u_2,u_3,u_9\right\}, \left\{u_3,u_4,u_{10}\right\},
   \cr 
& \left\{u_4,u_6,u_{11}
   \right\}, \left\{u_3,u_6,u_{12}\right\},\left\{u_6,u_{10},u_{13}\right\},\left\{u_{10},u_{12},u_{14}\right\},\left\{u_3,u_{12},u_{15}\right\}\}
\ea
\ee
The Cartan divisors $D_i^{E_8}(i=0,\dots,8)$ of $E_8$ and $D_i^{SU(2)}$ of $SU(2)$ are given by the following hypersurface equations (e. g. $U$ means $U=0$):
\be
\ba
&(U,u_8,u_7,u_{11},u_{13},u_{14},u_{15},u_9,u_{10})\equiv (D_0^{E_8}, D_1^{E_8}, D_2^{E_8}, D_3^{E_8}, D_4^{E_8},D_5^{E_8},D_6^{E_8},D_7^{E_8},D_8^{E_8})\\
&(V,v_1)\equiv (D_0^{SU(2)},D_1^{SU(2)}).
\ea
\ee

The two non-flat fiber components $S_1$ and $S_2$ are given by the hypersurface equations $\delta_1=0$ and $\delta_2=0$ respectively. The full intersection matrices involving the two non-flat surface components are 

{\footnotesize \be
\begin{array}{c|ccccccccccccc}
S_1& D^{E_8}_0 & D^{E_8}_1 & D^{E_8}_2& D^{E_8}_3& D^{E_8}_4& D^{E_8}_5& D^{E_8}_6& D^{E_8}_7& D^{E_8}_8& D^{SU(2)}_0& D_1^{SU(2)} & S_1 & S_2\\\hline
D^{E_8}_0& -2 & 1 & 0 & 0 & 0 & 0 & 0 & 0 & 0 & 0 & 0 & 0 & 0 \\
D^{E_8}_1 &1 & -1 & 0 & 0 & 0 & 0 & 0 & 0 & 0 & 0 & 0 & -1 & 1 \\
D^{E_8}_2& 0 & 0 & 0 & 0 & 0 & 0 & 0 & 0 & 0 & 0 & 0 & 0 & 0 \\
D^{E_8}_3& 0 & 0 & 0 & 0 & 0 & 0 & 0 & 0 & 0 & 0 & 0 & 0 & 0 \\
D^{E_8}_4& 0 & 0 & 0 & 0 & 0 & 0 & 0 & 0 & 0 & 0 & 0 & 0 & 0 \\
D^{E_8}_5& 0 & 0 & 0 & 0 & 0 & 0 & 0 & 0 & 0 & 0 & 0 & 0 & 0 \\
D^{E_8}_6& 0 & 0 & 0 & 0 & 0 & 0 & 0 & 0 & 0 & 0 & 0 & 0 & 0 \\
D^{E_8}_7& 0 & 0 & 0 & 0 & 0 & 0 & 0 & 0 & 0 & 0 & 0 & 0 & 0 \\
D^{E_8}_8& 0 & 0 & 0 & 0 & 0 & 0 & 0 & 0 & 0 & 0 & 0 & 0 & 0 \\
D^{SU(2)}_0& 0 & 0 & 0 & 0 & 0 & 0 & 0 & 0 & 0 & -2 & 2 & 0 & 0 \\
D^{SU(2)}_1& 0 & 0 & 0 & 0 & 0 & 0 & 0 & 0 & 0 & 2 & -2 & 0 & 0 \\
S_1 & 0 & -1 & 0 & 0 & 0 & 0 & 0 & 0 & 0 & 0 & 0 & -2 & 2 \\
S_2 & 0 & 1 & 0 & 0 & 0 & 0 & 0 & 0 & 0 & 0 & 0 & 2 & -2 
\end{array}
\ee

\be
\begin{array}{c|ccccccccccccc}
S_2 & D^{E_8}_0 & D^{E_8}_1 & D^{E_8}_2& D^{E_8}_3& D^{E_8}_4& D^{E_8}_5& D^{E_8}_6& D^{E_8}_7& D^{E_8}_8& D^{SU(2)}_0& D_1^{SU(2)} & S_1 & S_2\\\hline
D^{E_8}_0& 0 & 0 & 0 & 0 & 0 & 0 & 0 & 0 & 0 & 0 & 0 & 0 & 0 \\
D^{E_8}_1 & 0 & -1 & 1 & 0 & 0 & 0 & 0 & 0 & 0 & 0 & 0 & 1 & -1 \\
D^{E_8}_2& 0 & 1 & -2 & 1 & 0 & 0 & 0 & 0 & 0 & 0 & 0 & 0 & 0 \\
D^{E_8}_3& 0 & 0 & 1 & -2 & 1 & 0 & 0 & 0 & 0 & 0 & 0 & 0 & 0 \\
D^{E_8}_4& 0 & 0 & 0 & 1 & -2 & 1 & 0 & 0 & 0 & 0 & 0 & 0 & 0 \\
D^{E_8}_5& 0 & 0 & 0 & 0 & 1 & -2 & 1 & 1 & 0 & 0 & 0 & 0 & 0 \\
D^{E_8}_6& 0 & 0 & 0 & 0 & 0 & 1 & -2 & 0 & 0 & 0 & 0 & 0 & 0 \\
D^{E_8}_7& 0 & 0 & 0 & 0 & 0 & 1 & 0 & -2 & 1 & 0 & 0 & 0 & 0 \\
D^{E_8}_8& 0 & 0 & 0 & 0 & 0 & 0 & 0 & 1 & -2 & 0 & 0 & 0 & 0 \\
D^{SU(2)}_0& 0 & 0 & 0 & 0 & 0 & 0 & 0 & 0 & 0 & 0 & 0 & 0 & 0 \\
D^{SU(2)}_1& 0 & 0 & 0 & 0 & 0 & 0 & 0 & 0 & 0 & 0 & 0 & 0 & 0 \\
S_1 & 0 & 1 & 0 & 0 & 0 & 0 & 0 & 0 & 0 & 0 & 0 & 2 & -2 \\
S_2 & 0 & -1 & 0 & 0 & 0 & 0 & 0 & 0 & 0 & 0 & 0 & -2 & 2 
\end{array}
\ee}

We can also collect all the triple intersection numbers $S_i\cdot D_j^2$ in the following reduced intersection matrix:
\be
\begin{array}{r|ccc|cc|cc}
   D=& D^{E_8}_0 & \cdots & D^{E_8}_8&  D^{SU(2)}_0& D_1^{SU(2)} & S_1 & S_2 \\\hline
S_1 \cdot D_i\cdot D_i & &&&&&& \\
S_2 \cdot D_i\cdot D_i & &&&&&& \\
\end{array}
\ee
From this we can compute the genus of the curve along which the two sections $S_i$ intersect,
\be
2 g -2 = S_1^2 S_2 + S_2^2 S_1 \, .
\ee
For the genus $g=0$ case, we can read off the number of flavors $N_f$ in the $SU(3)$ gauge description and the number of  mass deformations $M=N_f+1$, as in (\ref{NkaSU3}):
\be
M= 17 - (S_1^3 + S_2^3) \, .\label{Mg0}
\ee
For higher genus cases, we can use (\ref{eq:sp2geovsgauge}) with $N_a=g$ and $a_a=0$ if an $Sp(2)$ gauge description exists:
\be
\ba
N_f=16-8g-(S_1^3+S_2^3)\cr
M=17-7g-(S_1^3+S_2^3)\, ,\label{Mgeneralg}
\ea
\ee
where $N_f$ is the number of fundamental flavors in the $Sp(2)+g\mbf{AS}+N_f\mbf{F}$ gauge theory.

\begin{figure}
\centering
\includegraphics[width=12cm]{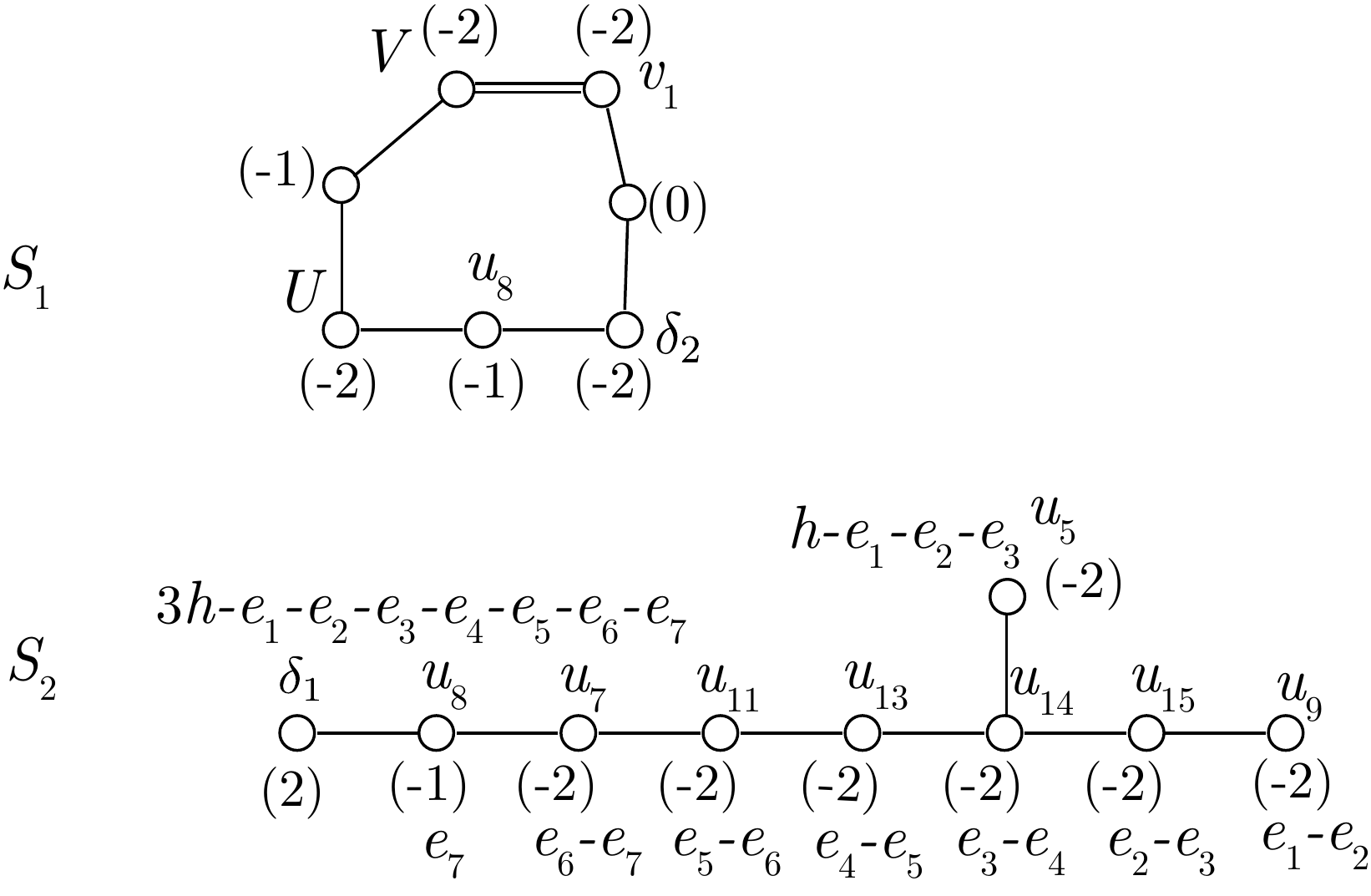}
\caption{The configuration of curves on the two non-flat fiber components $S_1$ and $S_2$ in the geometry~\ref{E8SU(2)-topbf} before the flop. The number in the bracket denotes the self-intersection number of the curve. {The letter denotes an intersection curve with the corresponding divisor in the resolved Calabi--Yau threefold. On $S_2$, we also labeled the curves with the generators of the Picard group on a rational surface (\ref{rational-Picard}). Here $S_1$ is not a rational surface.}}\label{f:E8SU2-topbf}
\end{figure}

In our example, the reduced intersection matrix is
\be\label{E8SU(2)-topbf}
\ba
&
\begin{array}{c|ccccccccc|cc|cc}
 S_i\cdot D_j^2  & D_0^{E_8} & D_1^{E_8} & D_2^{E_8} & D_3^{E_8} & D_4^{E_8} & D_5^{E_8}& D_6^{E_8}& D_7^{E_8}& D_8^{E_8}& D_0^{SU(2)} & D_1^{SU(2)} & S_1 & S_2 \\ \hline
S_1 &  -2 & -1 & 0 & 0 & 0 & 0 & 0 & 0 & 0 & -2 & -2 & -2 & -2 \\
S_2 &  0 & -1 & -2 & -2 & -2 & -2 & -2 & -2 & -2 & 0 & 0 & 2 & 2 \\
\end{array}
\, ,
\ea
\ee
from which we can see that the intersection curve $C=S_1\cdot S_2$ is a genus one curve rather than a rational curve. We draw the configuration of curves on $S_1$ and $S_2$ in figure~\ref{f:E8SU2-topbf}. In fact, the surface $S_1$ is a ruled surface over the genus one curve $C$. From the matrix elements $S_1^3$ and $S_2^3$, we can read off $M=10$ from (\ref{Mgeneralg}).

There exists a geometric transition from this configuration into a combination of two rational surfaces, see Section 3.5 of \cite{Jefferson:2018irk}. We take the limit in the complex structure moduli space of $S_1$, where $C$ becomes a pinched $S^2$ with a double point singularity. After we blow up this double point singularity, the surface $S_1$ will become a blow-up of Hirzebruch surface $S_1'=\mathrm{blp}_{2}\mb{F}_m$ and the surface $S_2$ will be blown up into $S_2'=\mathrm{blp} S_2$. The blow-up of $S_2$ occurs at a double point $P\subset C\subset S_2$. Denoting the exceptional divisor from this blow-up by $e_P$, the proper transform of $C$ on $S_2'$ is then $C'_{S_2'}=C_{S_2}-2e_P$, which is a rational curve. The self-intersection of $C'=S_1'\cdot S_2'$ inside $S_1'$ can then be computed by
\be
C'\cdot_{S_1'} C'=-2-C'\cdot_{S_2'} C'=2-C\cdot_{S_2} C\,.\label{CpS1}
\ee
In figure~\ref{f:E8SU2-topbf}, the genus one curve $C$ has self-intersection $(-2)$ on $S_1$ and 2 on $S_2$. On $S_2$, its representation with the standard Picard group of rational surfaces is
\be
C_{S_2}=3h-\sum_{i=1}^7 e_i\,.
\ee

After the geometric transition, $S_1'$ becomes the blow-up of a Hirzebruch surface and $S_2' = \text{gdP}_8$. The transformed curve $C'_{S_2'}$ is
\be
C'_{S_2'}=3h-\sum_{i=1}^7 e_i-2e_8\,,
\ee
which is a rational $(-2)$-curve (here $e_8\equiv e_p$).

\begin{figure}
\centering
\includegraphics[width=11cm]{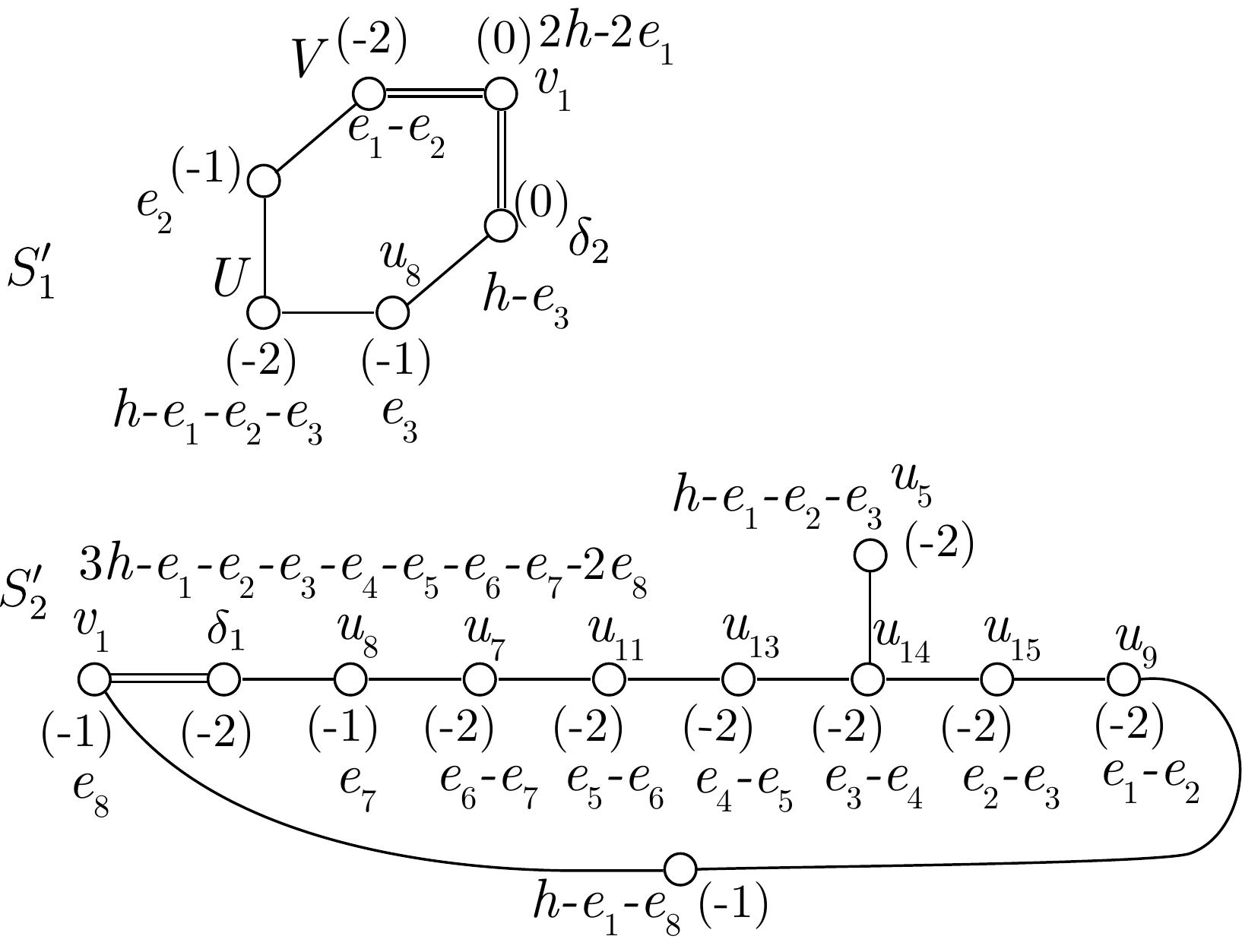}
\caption{The configuration of curves on {$S_1'$} and {$S_2'$} in the $(E_8,SU(2))$ marginal geometry after the flop. The number in the bracket denotes the self-intersection number of the curve. The letter denotes an intersection curve with the corresponding divisor in the resolved Calabi--Yau threefold.}\label{f:E8SU(2)10_7}
\end{figure}

On $S_1'$, from (\ref{CpS1}) we have $C_{S_1'}'^2=0$. The surface $S'_1$ is a gdP$_3$ after the flop, and the curve $D_1^{SU(2)}\cdot S'_1$ is a 0-curve on $S'_1$. Hence the rational $(-2)$-curve on $S_1$ should have been flopped into $S'_2$. On $S'_2=$gdP$_8$, however, there is no room for a new $(-2)$-curve, and $D_1^{SU(2)}\cdot S'_2$ is a $(-1)$-curve on $S'_2$ instead, see figure~\ref{f:E8SU(2)10_7}. The reduced intersection matrix after the flop is
\be\label{E8SU2_10_3/2_7}
\ba
\begin{array}{c|ccccccccc|cc|cc}
 S_i\cdot D_j^2  & D_0^{E_8} & D_1^{E_8} & D_2^{E_8} & D_3^{E_8} & D_4^{E_8} & D_5^{E_8}& D_6^{E_8}& D_7^{E_8}& D_8^{E_8} & D_0^{SU(2)} & D_1^{SU(2)} & S_1 & S_2 \\ \hline
 S_1 & -2 & -1 & 0 & 0 & 0 & 0 & 0 & 0 & 0 & -2 & 0 & 6 & 0 \\
 S_2 & 0 & -1 & -2 & -2 & -2 & -2 & -2 & -2 & -2 & 0 & -1 & -2 & 1 \\
\end{array}\, .
\ea
\ee
Note that the Cartan divisor $D_1^{SU(2)}$ is still considered as fully wrapped inside the non-flat fiber, as we assign a weight 2 to the divisors $D_i^{SU(2)}$ on the surface component $S'_2$. Here we can  confirm that the curve $D_1^{SU(2)}\cdot (S'_1+2S'_2)$ is indeed a rational curve with normal bundle $\mc{O}(0)+\mc{O}(-2)$ in the Calabi--Yau threefold $Y$. The relevant triple intersection numbers can be read off from figure~\ref{f:E8SU(2)10_7}:
\be
\ba
&D_1^{SU(2)2}\cdot S_1^{\prime }=0\,,\ D_1^{SU(2)}\cdot S_1^{\prime 2}=-4\cr
&D_1^{SU(2)2}\cdot S_2'=D_1^{SU(2)}\cdot S_2^{\prime 2}=-1\, ,\ D_1^{SU(2)}\cdot S_1'\cdot S_2'=2\,.
\ea
\ee
Hence we can compute
\be
\ba
D_1^{SU(2)2}\cdot (S_1'+2S_2')&=-2\cr
D_1^{SU(2)}\cdot (S_1'+2S_2')^2&=0\, ,
\ea
\ee
which indeed tells us $D_1^{SU(2)}\cdot (S'_1+2S'_2)$ has the normal bundle $\mc{O}(0)+\mc{O}(-2)$. 

If we contract the $(-1)$-curve $h-e_1-e_8$ on $S'_2$. Then the curve $D_1^{SU(2)}\cdot S_2'$ on $S_2'$ becomes a rational curve with self-intersection 0, which is also reflected in the CFD tree in figure~\ref{fig:E8FibsAll}, see the one with flavor symmetry $SO(16)\times SU(2)$ for example.

Similarly, in any blow down of the figure~\ref{f:E8SU(2)10_7}, the divisor $D_1^{SU(2)}$ has to be considered as fully wrapped and generate the  flavor symmetry if $D_1^{SU(2) 2}\cdot S_2=-1$.

\subsubsection[Example Resolutions: \texorpdfstring{$(E_8, SU(2))$}{(E8,SU(2))}]{Example Resolutions: \boldmath{$(E_8, SU(2))$}}
\label{app:ExamplesE8SU2}

Besides the geometry of marginal theory, the distinct blow-ups are characterized in terms of different orders of resolution of the $U$ and $V$ (and blow-ups thereof). We will provide here a few example resolutions of the rank two E-string Weierstrass model that correspond to descendant 5D SCFTs of the marginal theory.

Consider the first blow-up (we again use the notation for blow-ups introduced in section \ref{sec:ResSing})
\be\label{BU3}
\ba
&BU^{(E_8, SU(2))}_1=\cr
&\big\{\left\{x,y,U,u_1\right\},\left\{x,y,V,v_1\right\},\left\{x,y,u_1,u_2\right\},\left\{y,u_2,u_3\right\},\left\{
   y,u_1,v_1,\delta_1   \right\}, \left\{y,u_1,u_3,u_4\right\} , \cr 
&   \left\{y,u_1,u_5\right\},
 \left\{u_1,u_3,u_6\right\},
   \left\{u_1,u_4,u_7\right\},\left\{u_1,u_5,u_8\right\},\left\{u_2,u_3,u_9\right\},  \left\{u_3,u_4,u_{10}\right\},
   \cr 
& \left\{u_4,u_6,u_{11}
   \right\},
   \left\{u_3,u_6,u_{12}\right\},\left\{u_6,u_{10},u_{13}\right\},\left\{u_{10},u_{12},u_{14}\right\},\left\{u_
   {3},u_{12},u_{15}\right\}{\left\{\delta_1 ,u_3,\delta _2\right\}}\big\} \,.
\ea
\ee

The data that we need to read off from the this resolution are the triple intersection number of the non-flat surface components of the codimension two fibers $S_i$ with the Cartans of both $E_8$ and $SU(2)$, respectively

{\footnotesize
\be\label{BU3Triple1}
\ba&
\begin{array}{c|ccccccccc|cc|cc}
S_1& D^{E_8}_0 & D^{E_8}_1 & D^{E_8}_2& D^{E_8}_3& D^{E_8}_4& D^{E_8}_5& D^{E_8}_6& D^{E_8}_7& D^{E_8}_8& D^{SU(2)}_0& D_1^{SU(2)} & S_1 & S_2\\\hline
D^{E_8}_0 & 0 & 0 & 0 & 0 & 0 & 0 & 0 & 0 & 0 & 0 & 0 & 0 & 0 \\
D^{E_8}_1 & 0 & \boxed{-1} & 1 & 0 & 0 & 0 & 0 & 0 & 0 & 1 & 0 & -1 & 0 \\
D^{E_8}_2 & 0 & 1 & \boxed{-2} & 1 & 0 & 0 & 0 & 0 & 0 & 0 & 0 & 0 & 0 \\
D^{E_8}_3 & 0 & 0 & 1 & \boxed{-2} & 1 & 0 & 0 & 0 & 0 & 0 & 0 & 0 & 0 \\
D^{E_8}_4& 0 & 0 & 0 & 1 & \boxed{-2} & 1 & 0 & 0 & 0 & 0 & 0 & 0 & 0 \\
D^{E_8}_5 & 0 & 0 & 0 & 0 & 1 & \boxed{-1} & 0 & 0 & 0 & 0 & 0 & -1 & 1 \\
D^{E_8}_6 & 0 & 0 & 0 & 0 & 0 & 0 & 0 & 0 & 0 & 0 & 0 & 0 & 0 \\
D^{E_8}_7& 0 & 0 & 0 & 0 & 0 & 0 & 0 & 0 & 0 & 0 & 0 & 0 & 0 \\
D^{E_8}_8& 0 & 0 & 0 & 0 & 0 & 0 & 0 & 0 & 0 & 0 & 0 & 0 & 0 \\
D^{SU(2)}_0 & 0 & 1 & 0 & 0 & 0 & 0 & 0 & 0 & 0 & \boxed{\boxed{0}} & 2 & -2 & 0 \\
D^{SU(2)}_1 & 0 & 0 & 0 & 0 & 0 & 0 & 0 & 0 & 0 & 2 & \boxed{\boxed{0}} & -4 & 2 \\
S_1 & 0 & -1 & 0 & 0 & 0 & -1 & 0 & 0 & 0 & -2 & -4 & 4 & 2 \\
S_2 & 0 & 0 & 0 & 0 & 0 & 1 & 0 & 0 & 0 & 0 & 2 & 2 & -4 \\
\end{array}
\cr &
\begin{array}{c|ccccccccc|cc|cc}
S_2& D^{E_8}_0 & D^{E_8}_1 & D^{E_8}_2& D^{E_8}_3& D^{E_8}_4& D^{E_8}_5& D^{E_8}_6& D^{E_8}_7& D^{E_8}_8& D^{SU(2)}_0& D_1^{SU(2)} & S_1 & S_2\\\hline
D^{E_8}_0&0 & 0 & 0 & 0 & 0 & 0 & 0 & 0 & 0 & 0 & 0 & 0 & 0 \\
D^{E_8}_1 & 0 & 0 & 0 & 0 & 0 & 0 & 0 & 0 & 0 & 0 & 0 & 0 & 0 \\
D^{E_8}_2 & 0 & 0 & 0 & 0 & 0 & 0 & 0 & 0 & 0 & 0 & 0 & 0 & 0 \\
D^{E_8}_3 & 0 & 0 & 0 & 0 & 0 & 0 & 0 & 0 & 0 & 0 & 0 & 0 & 0 \\
D^{E_8}_4 & 0 & 0 & 0 & 0 & 0 & 0 & 0 & 0 & 0 & 0 & 0 & 0 & 0 \\
D^{E_8}_5 & 0 & 0 & 0 & 0 & 0 & \boxed{-1} & 1 & 0 & 1 & 0 & 0 & 1 & -1 \\
D^{E_8}_6 & 0 & 0 & 0 & 0 & 0 & 1 &\boxed{ -1} & 0 & 0 & 0 & 1 & 0 & -1 \\
 D^{E_8}_7 &0 & 0 & 0 & 0 & 0 & 0 & 0 & 0 & 0 & 0 & 0 & 0 & 0 \\
D^{E_8}_8& 0 & 0 & 0 & 0 & 0 & 1 & 0 & 0 & \boxed{-2} & 0 & 0 & 0 & 0 \\
D^{SU(2)}_0& 0 & 0 & 0 & 0 & 0 & 0 & 0 & 0 & 0 & 0 & 0 & 0 & 0 \\
D^{SU(2)}_1& 0 & 0 & 0 & 0 & 0 & 0 & 1 & 0 & 0 & 0 & \boxed{\boxed{1}} & 2 & -3 \\
S_1&0 & 0 & 0 & 0 & 0 & 1 & 0 & 0 & 0 & 0 & 2 & 2 & -4 \\
S_2&  0 & 0 & 0 & 0 & 0 & -1 & -1 & 0 & 0 & 0 & -3 & -4 & 7 
\end{array}
\ea
\ee}

From these intersection matrices we can read off the curves $F_j^{E_8}$ (fibral curves of the Cartan divisors) that are contained also in the surface components $S_i$ (see the entries that are boxed in the above intersection matrices). If again $D_{j}^{E_8 2} \cdot S_i = -2$ then the curve $F_j^{E_8}$ is fully contained in the surface $S_i$, if this is $-1$, then the curve $F_j$ splits in codimension two and one of the split components are contained in $S_i$. 
In the above case, the curves $F_j^{E_8}$ are entirely contained in the surfaces for $j={2,3,4,5,8}$, and $F_j^{E_8}$ for $j=1,6$, the curves split and one of the irreducible components are contained. Note that none of the $F_j^{SU(2)}$ are contained in the surface components in this case. For $j=5$ note that the curve splits, but both irreducible $-1$ curves  are contained in either one of the surfaces $S_i$, so that in the singular, conformal field theory limit, the curve $F_5$ will shrink to zero size and contribute to the flavor symmetry.  

For the blow-up $BU_1^{(E_8, SU(2))}$, we have the reduced intersection matrix 
\be\label{BU1Rank2E}
\ba
&BU_1^{(E_8, SU(2))} : \cr 
&
\begin{array}{r|ccccccccc|cc|cc}
S_i \cdot D_j^2&D_0 & D_1 & D_2& D_3& D_4& D_5& D_6& D_7 &D_8& D^{SU(2)}_0& D_1^{SU(2)} & S_1 & S_2\cr \hline
S_1  & 0 & -1 & -2 & -2 & -2 & -1 & 0 & 0 & 0 & 0 & 0 & 4 & -4 \cr 
S_2  &0 & 0 & 0 & 0 & 0 & -1 & -1 & 0 & -2 & 0 & 1 & 2 & 7 \cr \hline
n(F_j)  & 0 & -1 & -2 & -2 & -2 & -2 & -1 & 0 & -2 & 0 & 2 &  -& -\cr
\end{array}
\cr 
\ea
\ee
In the last line we added the integer $n(F_j)$ defined in (\ref{nFDef}), which determines the non-abelian part of the strongly coupled flavor symmetry. 
From (\ref{BU1Rank2E}) and (\ref{NkaSU3}), we can compute
\be
G_{F,na}= SU(6)\,, \ g=0\,,\ M=6 \,,\ k=\frac{3}{2}\,, \ a=1\,.
\ee
The geometry is shown in figure \ref{fig:ExampleBUE8Rank2}.

We can read off the codimension-two fiber in this case, including how the irreducible fiber components of the codimension one fibers split as well as the wrapping by the non-flat fiber components $S_i$. For the blow-up $BU_1^{(E_8, SU(2))}$ in (\ref{BU3}), the codimension two loci are given by all pair-wise intersections of 
\be
D_i^{E_8} \cdot D_k^{SU(2)}\,,\qquad D_i^{E_8}\cdot S_j, \quad j=1,2\,.
\ee 
From these intersections, we can determine how the $E_8$ fiber components split. 
In the example (\ref{BU3}) the irreducible components are listed in the following table, including the information, indicated by a dash, in which codimension two divisor they are contained in --- $S_j$ or $D_i^{SU(2)}$:
\be\label{BU3CodimTwoFibs}
\begin{array}{|c|c|c|c|c|c|c|}\hline 
\hbox{$E_8$ Root} & \hbox{Irreducible Components} & C\cdot C & S_1 & S_2 & D_0^{SU(2)} & D_1^{SU(2)}  \cr\hline  
\alpha_0^{E_8} & D_0^{E_8} \cdot D_0^{SU(2)} & -2 &  & & -&\cr \hline
\alpha_1^{E_8}& D_1^{E_8} \cdot D_0^{SU(2)} & -1 &  & & -&\cr 
& D_1^{E_8} \cdot S_1& -1 &- & & &   \cr \hline
\alpha_2^{E_8} & D_2^{E_8} \cdot S_1 & -2 & - & & &\cr \hline
\alpha_3^{E_8} & D_3^{E_8}\cdot S_1 & -2 & - & & &\cr \hline
\alpha_4^{E_8} &D_4^{E_8}\cdot S_1 & -2 & - & & &\cr \hline
\alpha_5^{E_8} & D_5^{E_8}\cdot S_1 & -1 & - & & &\cr 
			& D_5^{E_8}\cdot S_2 & -1 &  & - & &\cr \hline
\alpha_6^{E_8} & D_6^{E_8}\cdot S_2 & -1 &  &- & &\cr 
			& D_6^{E_8}\cdot D_1^{SU(2)} & -1 &  &  & &-\cr \hline
\alpha_7^{E_8} &D_7^{E_8}\cdot D_1^{SU(2)} & -2 & & & &-\cr \hline
\alpha_8^{E_8} &D_8^{E_8}\cdot S_2 & -2 & &- & &\cr \hline
\end{array}
\ee
Including the information about the relative intersections of these, which follow almost automatically from the above table, we can read off the codimension two fiber in figure \ref{fig:ExampleBUE8Rank2}.

To count the number of abelian flavor symmetry factors $U(1)^s$, we compute the number of linearly independent curves in form of $D^{E_8}_i\cdot S_j$ and $D^{SU(2)}_i\cdot S_j$, which is the eight given by the rank of the following intersection matrix:
\be
\begin{array}{c|ccccccccc|cc}
 &  D^{E_8}_0 & D^{E_8}_1 & D^{E_8}_2& D^{E_8}_3& D^{E_8}_4& D^{E_8}_5& D^{E_8}_6& D^{E_8}_7& D^{E_8}_8& D^{SU(2)}_0& D_1^{SU(2)}\\ \hline
 D_1^{E_8}\cdot S_1 & 0 & -1 & 1 & 0 & 0 & 0 & 0 & 0 & 0 & 1 & 0\\
 D_2^{E_8}\cdot S_1 & 0 & 1 & -2 & 1 & 0 & 0 & 0 & 0 & 0 & 0 & 0\\
 D_3^{E_8}\cdot S_1 & 0 & 0 & 1 & -2 & 1 & 0 & 0 & 0 & 0 & 0 & 0\\
 D_4^{E_8}\cdot S_1 & 0 & 0 & 0 & 1 & -2 & 1 & 0 & 0 & 0 & 0 & 0\\
 D_5^{E_8}\cdot S_1 & 0 & 0 & 0 & 0 & 1 & -1 & 0 & 0 & 0 & 0 & 0\\
 D_5^{E_8}\cdot S_2 & 0 & 0 & 0 & 0 & 0 & -1 & 1 & 0 & 1 & 0 & 0\\
 D_6^{E_8}\cdot S_2 & 0 & 0 & 0 & 0 & 0 & 1 & -1 & 0 & 0 & 0 & 1\\
 D_8^{E_8}\cdot S_2 & 0 & 0 & 0 & 0 & 0 & 1 & 0 & 0 & -2 & 0 & 0\\
 D_0^{SU(2)}\cdot S_1 & 0 & 1 & 0 & 0 & 0 & 0 & 0 & 0 & 0 & 0 & 2\\
  D_1^{SU(2)}\cdot S_1 & 0 & 0 & 0 & 0 & 0 & 0 & 0 & 0 & 0 & 2 & 0\\
  D_1^{SU(2)}\cdot S_2 & 0 & 0 & 0 & 0 & 0 & 0 & 1 & 0 & 0 & 0 & 1\\
  \end{array}
  \ee
Substracting eight with the rank of gauge group $SU(3)_G$, we get the total rank of rk$(G_\text{F})=6$. Hence we conclude that the number of abelian flavor symmetry factors $s=1$ in this case.

In conclusion, the total flavor symmetry of this SCFT is
\be
G_\text{F}=SU(6)\times U(1)
\ee

A model that is closely related to $BU_1^{(E_8, SU(2))}$ is the following\footnote{The difference between the two models is simply the placement of the resolution step $\left\{\delta_1 ,u_3,\delta _2\right\}$.}
\be\label{BU34}
\ba
&BU^{(E_8, SU(2))}_2=\cr
&\{\left\{x,y,U,u_1\right\},\left\{x,y,V,v_1\right\},\left\{x,y,u_1,u_2\right\},\left\{y,u_2,u_3\right\},\left\{
   y,u_1,v_1,\delta_1   \right\},\left\{\delta_1 ,u_3,\delta _2\right\},\left\{y,u_1,u_3,u_4\right\} , \cr 
&   \left\{y,u_1,u_5\right\}, \left\{u_1,u_3,u_6\right\},
   \left\{u_1,u_4,u_7\right\},\left\{u_1,u_5,u_8\right\},\left\{u_2,u_3,u_9\right\}, \left\{u_3,u_4,u_{10}\right\},
   \cr 
& \left\{u_4,u_6,u_{11}
   \right\},
   \left\{u_3,u_6,u_{12}\right\},\left\{u_6,u_{10},u_{13}\right\},\left\{u_{10},u_{12},u_{14}\right\},\left\{u_
   {3},u_{12},u_{15}\right\}\}
\ea
\ee
The reduced triple intersection matrix is 
\be\ba
&BU^{(E_8, SU(2))}_2: \cr 
& 
\begin{array}{c|ccccccccc|cc|cc}
S_i\cdot D_j^2&D^{E_8}_0 & D^{E_8}_1 & D^{E_8}_2& D^{E_8}_3& D^{E_8}_4& D^{E_8}_5& D^{E_8}_6& D^{E_8}_7 &D^{E_8}_8& D^{SU(2)}_0& D_1^{SU(2)} & S_1 & S_2\cr \hline
S_1 &  0 & 0 & 0 & 0 & 0 & 0 & 0 & 0 & 0 & 0 & 0 & 8 & 0\cr 
S_2   & 0 & -1 & -2 & -2 & -2 & -2 & -1 & 0 & -2 & 0 & 1 & -2 & 3  \cr \hline
n(F_j)  & 0 & -1 & -2 & -2 & -2 & -2 & -1 & 0 & -2 & 0 & 2 &  -& -\cr
\end{array}
\,.
\ea\ee
Note that the numbers $n(F_J)$ is the same as for $BU_1$ in (\ref{BU1Rank2E}), hence it has the same flavor symmetry and $M$:
\be
G_{F}= SU(6)\times U(1)\,, \ g=0\,,\ M=6 \,,\ k=\frac{3}{2}\,, \ a=5\, .
\ee
Thus the only difference is the value of $a$. Geometrically, $BU_2$ can be constructed from $BU_1$ by flopping curves from $S_1$ into $S_2$. Field theoretically, they represent distinct gauge theory phases for the same strongly coupled SCFT. We will systematically study how to characterize all the gauge theory descriptions and associated geometries, that give rise to the same SCFTs in~\cite{Apruzzi:2019enx}. 
Again we can determine the codimension-two fiber explicitly by considering the irreducible curve components:
\be
\label{BU34CodimTwoFibs}
\begin{array}{|c|c|c|c|c|c|c|}
\hline 
\hbox{$E_8$ Root} & \hbox{Irreducible Components} & C\cdot C & S_1 & S_2 & D_0^{SU(2)} & D_1^{SU(2)}  \cr\hline  
\alpha_0^{E_8} & D_0^{E_8} \cdot D_0^{SU(2)} & -2 &  & & -&\cr \hline
\alpha_1^{E_8}& D_1^{E_8} \cdot D_0^{SU(2)} & -1 &  & & -&\cr 
& D_1^{E_8} \cdot S_1& -2 &- & & &   \cr
& D_1^{E_8} \cdot S_2& -1 & &- & &   \cr \hline
\alpha_2^{E_8} & D_2^{E_8} \cdot S_2 & -2 &  &- & &\cr \hline
\alpha_3^{E_8} & D_3^{E_8}\cdot S_2 & -2 & & -& &\cr \hline
\alpha_4^{E_8} &D_4^{E_8}\cdot S_2 & -2 &  & -& &\cr \hline
\alpha_5^{E_8} & D_5^{E_8}\cdot S_2 & -2 &  &- & &\cr \hline
\alpha_6^{E_8} & D_6^{E_8}\cdot S_2 & -1 &  &- & &\cr 
			& D_6^{E_8}\cdot D_1^{SU(2)} & -1 &  &  & &-\cr \hline
\alpha_7^{E_8} &D_7^{E_8}\cdot D_1^{SU(2)} & -2 & & & &-\cr \hline
\alpha_8^{E_8} &D_8^{E_8}\cdot S_2 & -2 & &- & &\cr \hline
\end{array}
\ee
The resulting codimension two fiber is shown in figure \ref{fig:ExampleBUE8Rank2}.

We close with another example blow-up where the 5d SCFT is different from the above models: consider
\be\label{BU105}
\ba
&BU^{(E_8, SU(2))}_3=\cr 
&
\big\{\left\{x,y,U,u_1\right\},\left\{x,y,u_1,u_2\right\},\left\{x,y,V,v_1\right\},\left\{y,u_2,u_3\right\},
\left\{V,u_3,\delta_1  \right\},\left\{y,u_1,u_3,u_4\right\},\left\{y,u_1,u_5\right\},\cr 
&\left\{u_1,u_3,u_6\right\},\left\{u_1,u_4,u_7\right\},\left\{u_1,u_5,u_8\right\},\left\{u_2,u_3,u_9\right\},\left\{u_3,u_4,u_{10}\right\},\left\{u_4,u_6,u_{11}\right\},\cr 
&\left\{u_3,u_6,u_{12}\right\},\left\{u_6,u_{10},u_{   13}\right\},\left\{u_{10},u_{12},u_{14}\right\},\left\{u_{3},u_{12},u_{15}\right\},\left\{\delta_1 ,u_3,\delta _2\right\}\big\}
\ea
\ee
The reduced intersection matrix is 
\be
\ba
&BU^{(E_8, SU(2))}_3: \cr 
&
\begin{array}{c|ccccccccc|cc|cc}
S_i \cdot D_j^2 & D^{E_8}_0 & D^{E_8}_1 & D^{E_8}_2& D^{E_8}_3& D^{E_8}_4& D^{E_8}_5& D^{E_8}_6& D^{E_8}_7 &D^{E_8}_8 & D^{SU(2)}_0& D_1^{SU(2)} & S_1 & S_2\cr \hline
 S_1 &0 & -1 & -2 & -2 & -2 & -1 & 0 & 0 & 0 & 0 & 0 & 4 & -4 \cr 
S_2 & 0 & 0 & 0 & 0 & 0 & -1 & -2 & -1 & -2 & 0 & 0 & 2 & 6 \cr \hline
n(F_j) &0 & -1 & -2 & -2 & -2 & -2 & -2 & -1 & -2 & 0 & 0 & - & -
\end{array}
\ea
\ee

and thus we can read off 
\be
G_{F,na}= SO(12)\,, \ g=0 \,, \ M=7\,,\  k=2\,,\  a=2 \,.
\ee
The irreducible fiber components are likewise obtained from the splitting of the roots as follows:
 \be\label{BU105CodimTwoFibs}
\begin{array}{|c|c|c|c|c|c|c|}
\hline 
\hbox{$E_8$ Root} & \hbox{Irreducible Components} & C\cdot C & S_1 & S_2 & D_0^{SU(2)} & D_1^{SU(2)}  \cr\hline  
\alpha_0^{E_8} & D_0^{E_8} \cdot D_0^{SU(2)} & -2 &  & & -&\cr \hline
\alpha_1^{E_8}& D_1^{E_8} \cdot D_0^{SU(2)} & -1 &  & & -&\cr 
& D_1^{E_8} \cdot S_1& -1 &- & & &   \cr \hline
\alpha_2^{E_8} & D_2^{E_8} \cdot S_1 & -2 & - & & &\cr \hline
\alpha_3^{E_8} & D_3^{E_8}\cdot S_1 & -2 & - & & &\cr \hline
\alpha_4^{E_8} &D_4^{E_8}\cdot S_1 & -2 & - & & &\cr \hline
\alpha_5^{E_8} & D_5^{E_8}\cdot S_1 & -1 & - & & &\cr 
			& D_5^{E_8}\cdot S_2 & -1 &  & - & &\cr \hline
\alpha_6^{E_8} &D_6^{E_8}\cdot D_1^{SU(2)} & -2 & & & &-\cr \hline
\alpha_7^{E_8} & D_7^{E_8}\cdot S_2 & -1 &  &- & &\cr 
			& D_7^{E_8}\cdot D_1^{SU(2)} & -1 &  &  & &-\cr \hline
\alpha_8^{E_8} &D_8^{E_8}\cdot S_2 & -2 & &- & &\cr \hline
\end{array}
\ee
Again the codimension two fiber is shown in figure \ref{fig:ExampleBUE8Rank2}. 

For the abelian part of the flavor symmetry, we compute the rank of the following intersection matrix, which turns out to be nine:
\be
\begin{array}{c|ccccccccc|cc}
 &  D^{E_8}_0 & D^{E_8}_1 & D^{E_8}_2& D^{E_8}_3& D^{E_8}_4& D^{E_8}_5& D^{E_8}_6& D^{E_8}_7& D^{E_8}_8& D^{SU(2)}_0& D_1^{SU(2)}\\ \hline
 D_1^{E_8}\cdot S_1 & 0 & -1 & 1 & 0 & 0 & 0 & 0 & 0 & 0 & 1 & 0\\
 D_2^{E_8}\cdot S_1 & 0 & 1 & -2 & 1 & 0 & 0 & 0 & 0 & 0 & 0 & 0\\
 D_3^{E_8}\cdot S_1 & 0 & 0 & 1 & -2 & 1 & 0 & 0 & 0 & 0 & 0 & 0\\
 D_4^{E_8}\cdot S_1 & 0 & 0 & 0 & 1 & -2 & 1 & 0 & 0 & 0 & 0 & 0\\
 D_5^{E_8}\cdot S_1 & 0 & 0 & 0 & 0 & 1 & -1 & 0 & 0 & 0 & 0 & 0\\
 D_5^{E_8}\cdot S_2 & 0 & 0 & 0 & 0 & 0 & -1 & 1 & 0 & 1 & 0 & 0\\
 D_6^{E_8}\cdot S_2 & 0 & 0 & 0 & 0 & 0 & 1 & -2 & 1 & 0 & 0 & 0\\
 D_7^{E_8}\cdot S_2 & 0 & 0 & 0 & 0 & 0 & 0 & 1 & -1 & 0 & 0 & 1\\
 D_8^{E_8}\cdot S_2 & 0 & 0 & 0 & 0 & 0 & 1 & 0 & 0 & -2 & 0 & 0\\
 D_0^{SU(2)}\cdot S_1 & 0 & 1 & 0 & 0 & 0 & 0 & 0 & 0 & 0 & 0 & 2\\
  D_1^{SU(2)}\cdot S_1 & 0 & 0 & 0 & 0 & 0 & 0 & 0 & 0 & 0 & 2 & 0\\
  D_1^{SU(2)}\cdot S_2 & 0 & 0 & 0 & 0 & 0 & 0 & 0 & 1 & 0 & 0 & 0\\
  \end{array}
  \ee
Substract nine by the rank of $SU(3)_G$, we get the total rank of flavor symmetry $M=7$. Hence the total flavor symmetry is
\be
G_\text{F}=SO(12)\times U(1).
\ee  

This concludes our examples of concrete blow-ups of the rank two E-string. The process is pretty clear from what we have described and many more models can be obtained in this way, by resolving the singularity including the non-minimal one in codimension two. We will next pass to the $D_{10}-I_1$
 starting point and illustrate the blow-ups in this case as well. 
 
 The main observation from this section is that there are multiple resolutions, that will correspond to different gauge theory realizations of the same SCFT. The key to characterizing the distinct 5d SCFTs will be removing this redundant information and extracting the relevant fiber information that uniquely fixes the SCFT, which will be explained in section \ref{sec:CFD}.


\subsection[Blow-ups for the \texorpdfstring{$(D_{10}, I_1)$}{(D10,I1)} Conformal Matter]{Blow-ups for the \boldmath{$(D_{10}, I_1)$} Conformal Matter}
\label{app:BlowupsD10}

\subsubsection{Geometry for the Marginal Theory}
\label{app:TopCFDsD10}

\begin{figure}
\centering
\includegraphics[width=12cm]{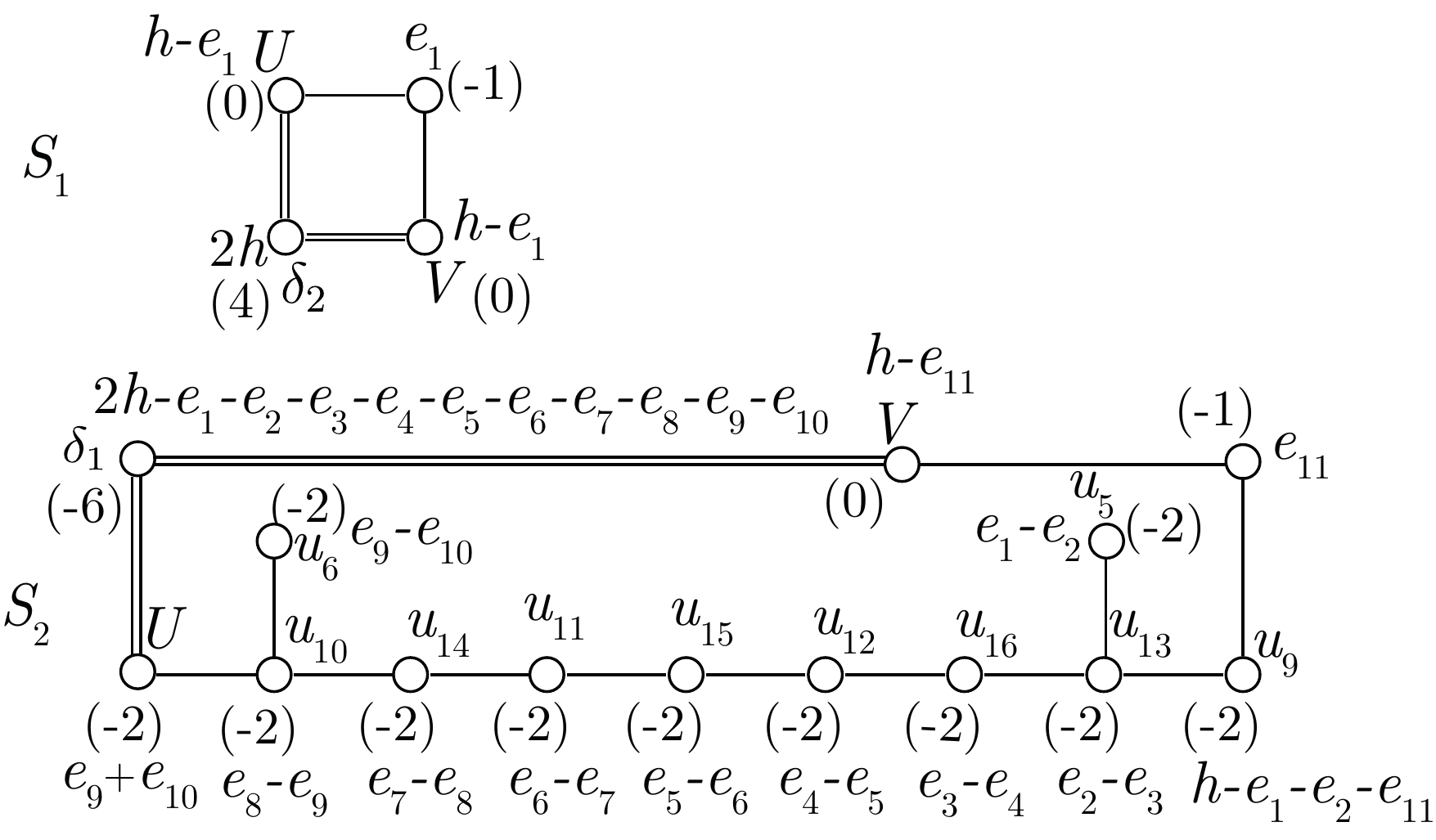}
\caption{The configuration of curves on $S_1$ and $S_2$ in the geometry that has $M=11$ for the $D_{10}-I_1$ collision.  The number in the bracket denotes the self-intersection number of the curve. The letter denotes an intersection curve with the corresponding divisor.}\label{fig:D10topFib}
\end{figure}

To get the marginal geometry with $M=11$, we blow up the base locus $U=V=0$ first: $(U,V;\delta_1)$. The starting point Tate model is then:
\be
y^2+b_1 Uxy+b_3 U^5\delta_1^2=x^3+b_2 UV x^2+b_4 U^5\delta_1 x+b_6 U^{10}\delta_1^4
\ee
We can choose the following blow-up chain:
\be\label{SO20BU11}
\ba
&BU_{M=11}^{(D_{10},I_1)}=\cr
&\{\left\{x,y,\delta_1,\delta_2\}\right\},\{\left\{x,y,U,u_1\right\},\left\{x,y,u_1,u_2\right\},\left\{x,y,u_2,u_3\right\},\left\{x,y,u_3,u_4\right\},\left\{x,y,u_4,u_5\right\},\cr 
&\left\{y,u_1,u_6\right\},\left\{y,u_2,u_7\right\},\left\{y,u_3,u_8\right\}, \left\{y,u_4,u_9\right\}, \left\{u_1,u_6,u_{10}\right\},\left\{u_2,u_7,u_{11}\right\},\left\{u_3,u_8,u_{12}\right\},\cr
&\left\{u_4,u_9,u_{13}\right\},\left\{u_2,u_6,u_{14}\right\},\left\{u_3,u_7,u_{15}\right
   \}, \left\{u_4,u_8,u_{16}\right\}\}
   \ea
\ee
The ordering of the simple roots for $SO(20)$ is as in \cite{Lawrie:2012gg} with the sections 
\be\ba
&\left(D_0, D_1, \cdots, D_{10}\left| D_0^{I_1}\right| S_1, S_2 \right) \cr \equiv& 
\left(U,u_6,u_{10},u_{14},u_{11},u_{15},u_{12},u_{16},u_{13},u_9,u_5| V| \delta _1,\delta _2\right)\,.\ea
\ee
Here $D_i$ is associated to the $i$th simple root of $SO(20)$. 

The configuration of  is shown in figure \ref{fig:D10topFib} and 
\be\label{D10_11_0_10}
\ba
&BU_{M=11}^{(D_{10},I_1)}=\cr
&
\begin{array}{c|ccccccccccc|c|cc}
  S_i\cdot D_j^2  & D_0 & D_1 & D_2 & D_3 & D_4 & D_5 & D_6 & D_7 & D_8 & D_9 & D_{10} & D_0^{I_1} & S_1 & S_2 \\ \hline
S_1 & 0 & 0 & 0 & 0 & 0 & 0 & 0 & 0 & 0 & 0 & 0 & 0 & 8 & 4 \\
 S_2 & -2 & -2 & -2 & -2 & -2 & -2 & -2 & -2 & -2 & -2 & -2 & 0 & -6 & -2\\ \hline
 n(F_j) & -2 & -2 & -2 & -2 & -2 & -2 & -2 & -2 & -2 & -2 & -2 & 0 & - & -\\
\end{array}\,.
\ea
\ee

Note that the curve $D_0\cdot S_2$ is a reducible $(-2)$-curve on $S_2$. Nonetheless, the curve $D_0\cdot (S_1+S_2)$ is still a rational $(-2)$-curve, since we can compute from figure \ref{fig:D10topFib}
\be
D_0^2\cdot(S_1+S_2)=-2\,,\ D_0\cdot(S_1+S_2)^2=0\,.
\ee

\subsubsection[Example Resolution for \texorpdfstring{$(D_{10}, I_1)$}{(D10,I1)}]{Example Resolution for \boldmath{$(D_{10}, I_1)$}}
\label{app:BUD10Ex}

We consider an example resolution of a descendant theory: 
\be\label{SO20BU1}
\ba
&BU_1^{(D_{10}, I_1)}=\cr
&\{\left\{x,y,U,u_1\right\},\left\{x,y,u_1,u_2\right\},\left\{x,y,u_2,u_3\right\},\left\{x,y,u_3,u_4\right\},\left\{x,y,u_4,u_5\right\},\left\{y,u_1,u_6\right\},\cr
&\left\{y,u_2,u_7\right\},\left\{y,u_3,u_8\right\}, \left\{y,u_4,u_9\right\}, \left\{u_1,u_6,u_{10}\right\},\left\{u_2,u_7,u_{11}\right\},\left\{u_3,u_8,u_{12}\right\},\cr
&\left\{u_4,u_9,u_{13}\right\},\left\{u_2,u_6,u_{14}\right\},\left\{u_3,u_7,u_{15}\right
   \}, \left\{u_4,u_8,u_{16}\right\},\left\{V,u_7,\delta _1\right\},\left\{V,u_8,\delta _2\right\}\}\,.
   \ea
\ee

The reduced intersection matrix is then:
\be\label{D10I0BU2}
\begin{array}{r|ccccccccccc|c|cc}
 S_i \cdot D_j^2  &
  D_0 & D_1 & D_2& D_3 & D_4 & D_5 & D_6& D_7 & D_8& D_9 &D_{10} & D_{0}^{I_1}&S _1 & S_2\\ \hline
S_1  & 0 & 0 & 0 & 0 & -1 & -2 & -1 & 0 & 0 & 0 & 0 & 4 & 6 & -2 \\
S_2  & 0 & 0 & 0 & 0 & 0 & 0 & -1 & -2 & -1 & -2 & 0 & 2 & 0 & 6 \\\hline
n(F_j) & 0 & 0 & 0 & 0 & -1 & -2 & -2 & -2 & -1 & -2 & 0 & 6 & - & -\\
\end{array}
\ee
We can read off
\be
G_{F,na}= SU(4) \times SU(2) \,,g=0\,,\ M=5\,,\ k=1\,,\ a=2\,.
\ee
The corresponding codimension two fiber is shown in figure \ref{fig:D10Example}.

We can also compute the total rank of flavor symmetry group $G_\text{F}$ as the rank of the following intersection matrix, which turns out to be seven:
\be
\begin{array}{c|ccccccccccc}
& D_0 & D_1 & D_2 & D_3 & D_4 & D_5 & D_6 & D_7 & D_8 & D_9 & D_{10}\\ \hline
D_3\cdot S_1 & 0 & 0 & 0 & 0 & 1 & 0 & 0 & 0 & 0 & 0 & 0\\ 
D_4\cdot S_1 & 0 & 0 & 0 & 1 & -1 & 1 & 0 & 0 & 0 & 0 & 0\\ 
D_5\cdot S_1 & 0 & 0 & 0 & 0 & 1 & -2 & 1 & 0 & 0 & 0 & 0\\ 
D_6\cdot S_1 & 0 & 0 & 0 & 0 & 0 & 1 & -1 & 0 & 0 & 0 & 0\\ 
D_6\cdot S_2 & 0 & 0 & 0 & 0 & 0 & 0 & -1 & 1 & 0 & 0 & 0\\ 
D_7\cdot S_2 & 0 & 0 & 0 & 0 & 0 & 0 & 1 & -2 & 1 & 0 & 0\\ 
D_8\cdot S_2 & 0 & 0 & 0 & 0 & 0 & 0 & 0 & 1 & -1 & 1 & 0\\
D_9\cdot S_2 & 0 & 0 & 0 & 0 & 0 & 0 & 0 & 0 & 1 & -2 & 0\\  
\end{array}
\ee
Substract seven by the rank of $SU(3)_G$, we confirmed $M=5$ and the total flavor symmetry is
\be
G_\text{F}=SU(4)\times SU(2)\times U(1)\,.
\ee

\subsection[Resolution of \texorpdfstring{$(E_8,SU(3))$}{(E8,SU(3))}]{Resolution of \boldmath{$(E_8, SU(3))$} and \boldmath{$(E_8, G_2)$}}
\label{app:E8SU3}

We give an example of a non-flat resolution of the collision $II^*$ and $I_3$ Kodaira fibers associated to the  $(E_8, SU(3))$ conformal matter. The 5d SCFTs obtained from this have rank four. The sequence of blow-ups that we will consider is
\be\ba
BU^{(E_8, SU(3))}= &\{
\{x,y,u,u_1\},
\{x,y,v,v_1\},
\{x,y,u_1,u_2\},
\{y,v_1,v_2\},\cr 
&\{u_1,v_2,\delta_1\},\{u_2,v_2,\delta _2\},\{y,u_2,u_3\},\{u_3,\delta _1,\delta_3\},\{y,u_1,u_3,u_4\},
\{y,u_1,u_5\},\cr 
&\{u_1,u_3,u_6\},\{u_1,u_4,u_7\},\{u_1,u_5,u_8\},\{u_2,u_3,u_9\},
\{u_3,\delta _3,\delta_4\},\{u_3,u_4,u_{10}\},\cr 
& \{u_4,u_6,u_{11}\},\{u_3,u_6,u_{12}\},\{u_6,u_{10},u_{13}\},\{u_{10},u_{12},u_{14}\},\{u_{3},u_{12},u_{15}\}\}
\ea\ee
The ordering of the $E_8$ Cartan divisors is in the rank one and two examples. The $SU(3)$ affine roots are identified with $V, v_1, v_2$, and the non-flat surface components $S_i$ are given by $\delta_i=0$. The reduced triple intersection matrix of the Cartan divisors $D_i^{\mathfrak{g}}$ of $E_8$ and $SU(3)$, respectively, with the four non-flat fiber components $S_i$ are 

{\small 
\be\label{BUE8SU3}
\ba
&BU^{(E_8, SU(3))}: \cr 
& 
\begin{array}{c|ccccccccc|ccc|cccc}
S_i \cdot D_j^2 &D_0 & D_1 & D_2& D_3& D_4& D_5& D_6& D_7 &D_8& D^{I_3}_0& D_1^{I_3} & D_{2}^{I_3}  & S_1 & S_2 & S_3 & S_4\cr \hline
S_1 & 0 & 0 & 0 & 0 & 0 & 0 & 0 & 0 & 0 & 0 & -1 & -1 & 6 & -2 & -2 & 0 \cr 
S_2 & 0 & 0 & 0 & 0 & 0 & 0 & 0 & -1 & 0 & 0 & -1 & -1 & -2 & 4 & -4 & -2 \cr 
S_3 & 0 & -1 & -2 & -1 & 0 & 0 & 0 & 0 & 0 & 0 & 0 & 0 & 0 & 0 & 6 & -2 \cr 
S_4 & 0 & 0 & 0 & -1 & -2 & -2 & -2 & -1 & -2 & 0 & 0 & 0 & 0 & 0 & 0 & 4 \cr\hline
n(F_j) & 0 & -1 & -2 & -2 & -2 & -2 & -2 & -2 & -2 & 0 & -2 & -2 & - & - & - & -
\end{array}
\ea
\ee
}
The wrapped components of the fiber and codimension two fiber is shown in figure \ref{fig:E8SU3Example}.
The strongly coupled flavor symmetry for the SCFT from this point of view is 
\be
G_\text{F} \supset E_7 \times SU(3) \,.
\ee
With some minor changes we can generalize this to $(E_8, G_2)$, i.e., the collision of $II^*$ with $I_1^{ns}$ (non-split $I_1^*$). 
The vanishing orders  change to 
\be
\hbox{ord}_{v=0} (b_i)=  (0, 1, 1, 2, 3) \,. 
\ee
The same resolution sequence can be applied, except in order to obtain the $G_2$ we need to perform another small resolution
\be
BU^{(E_8, G_2)} = BU^{(E_8, SU(3))} \cup \{v_1, v_2, v_3\} \,.
\ee
The roots of the affine $G_2$ are identified with $\alpha_0 \leftrightarrow V$, $\alpha_1  \leftrightarrow v_3$, $\alpha_2  \leftrightarrow v_2$, so that the intersection matrix for the $G_2$ part of the codimension one singular fibers is 
\be
-C_{ij}^{\widehat{\mathfrak{g}}_2}=
\left(
\begin{array}{ccc}
-2 & 1&	0\cr 
1& 	-2 & 3 \cr 
0& 3 	&	-6 
\end{array}
\right) \,.
\ee
The reduced triple intersection matrix is now
{\small 
\be\label{BUE8G2}
\ba
&BU^{(E_8, G_2)}: \cr 
& 
\begin{array}{c|ccccccccc|ccc|cccc}
S_i \cdot D_j^2 &D_0 & D_1 & D_2 & D_3 & D_4 & D_5 & D_6 & D_7  &D_8 & D^{G_2}_0& D_1^{G_2} & D_{2}^{G_2}  & S_1 & S_2 & S_3 & S_4\cr \hline
S_1 & 0 & 0 & 0 & 0 & 0 & 0 & 0 & 0 & 0 & 0 & -1 & -2 & 6 & -2 & -2 & 0 \\
S_2 & 0 & 0 & 0 & 0 & 0 & 0 & 0 & -1 & 0 & 0 & -1 & -4 & -2 & 4 & -4 & -2 \\
S_3 & 0 & -1 & -2 & -1 & 0 & 0 & 0 & 0 & 0 & 0 & 0 & 0 & 0 & 0 & 6 & -2 \\
S_4 & 0 & 0 & 0 & -1 & -2 & -2 & -2 & -1 & -2 & 0 & 0 & 0 & 0 & 0 & 0 & 4 \\\hline
n(F_j) & 0 & -1 & -2 & -2 & -2 & -2 & -2 & -2 & -2 & 0 & -2 & -6& - & - & - & -
\end{array}
\ea
\ee
}
So here the flavor symmetry at the strongly coupled point is 
\be
G_\text{F} = E_7 \times G_2 \,.
\ee
The fiber is depicted in figure \ref{fig:E8SU3Example}. 

\subsection{Higher Rank Marginal Theories}
\label{app:HigherRankBU}

The $(E_8,SU(2k))$ conformal matter theory is a 6d (1,0) SCFT with rank $2k^2-k+1$. The tensor branch configuration is
\be
[E_8]-\; 1-\; 2 -\; \overset{\mathfrak{su}_2}{2}-\; \overset{\mathfrak{su}_3}{2}\; \cdots -\; \overset{\mathfrak{su}_{2k-1}}{2}\; -[SU(2k)]\,.
\ee
Starting with the Weierstrass model:
\be
y^2+b_1 U xy+b_3 U^3 V^k y=x^3+b_2 U^2 V x^2+b_4 U^4 V^k x+b_6 U^5 V^{2k}\, ,
\ee
we first blow up the base locus $(U,V;\delta_1)$, and then use the following resolution sequence:
\be
\ba
&BU_{\text{marginal}}^{(E_8,SU(2k))}=\cr
&\{\left\{x,y,U,u_1\right\},\left\{x,y,u_1,u_2\right\},\left\{y,u_2,u_3\right\}, \left\{y,u_1,u_3,u_4\right\} ,\left\{y,u_1,u_5\right\}, \left\{u_1,u_3,u_6\right\},\cr
&   \left\{u_1,u_4,u_7\right\},\left\{u_1,u_5,u_8\right\},\left\{u_2,u_3,u_9\right\}, \left\{u_3,u_4,u_{10}\right\},
\left\{u_4,u_6,u_{11}\right\}, \left\{u_3,u_6,u_{12}\right\},\cr
&\left\{u_6,u_{10},u_{13}\right\},\left\{u_{10},u_{12},u_{14}\right\},\left\{u_3,u_{12},u_{15}\right\},\left\{x,y,V,v_1\right\},\left\{x,y,v_i,v_{i+1}\right\}\ (i=1,\dots,k-1),\cr
&\left\{y,v_{i},v_{i+k}\right\}\ (i=1,\dots,k-1)\,,\left\{u_3,\delta_i,\delta_{i+1}\right\}\ (i=1,\dots,2k-1)\,,\cr
&\left[ \left\{x,y,\delta_{2i-1},\delta_{2ik-i^2+i+1}\right\},\left\{x,y,\delta_{2ik-i^2+i+j+1},\delta_{2ik-i^2+i+j+2}\right\}\ (j=0,\cdots,k-i-2), \right. \cr
&\left. \left\{y,\delta_{2ik-i^2+i+j+1},\delta_{2ik-i^2+k+j+1}\right\}\ (j=0,\dots,k-i-1)\; \right] \ (i=1,\cdots,k-1)\,,\cr
&\left[ \left\{x,y,\delta_{2i},\delta_{k^2+2ki-k-i^2+2}\right\},\{x,y,\delta_{k^2+2ki-k-i^2+j+2},\delta_{k^2+2ki-k-i^2+j+3}\}\ (j=0,\cdots,k-i-2) \right. \cr
&\left. \{x,y,\delta_{k^2+2ki-k-i^2+j+2},\delta_{k^2+2ki-i-i^2+j+2}\}\ (j=0,\cdots,k-i-2)\; \right] \ (i=1,\cdots,k-1)\}
\ea
\ee

The $(E_8,SU(2k+1))$ conformal matter theory is a 6d (1,0) SCFT with rank $2k^2+k+1$. The tensor branch configuration is
\be
[E_8]-\; 1-\; 2 -\; \overset{\mathfrak{su}_2}{2}-\; \overset{\mathfrak{su}_3}{2}\; \cdots -\; \overset{\mathfrak{su}_{2k}}{2}\; -[SU(2k+1)]\,.
\ee
Starting with the Weierstrass model:
\be
y^2+b_1 U xy+b_3 U^3 V^k y=x^3+b_2 U^2 V x^2+b_4 U^4 V^{k+1} x+b_6 U^5 V^{2k+1}\, ,
\ee
we first blow up the base locus $(U,V;\delta_1)$, and then use the following resolution sequence:
\be
\ba
&BU_{\text{marginal}}^{(E_8,SU(2k+1))}=\cr
&\{\left\{x,y,U,u_1\right\},\left\{x,y,u_1,u_2\right\},\left\{y,u_2,u_3\right\}, \left\{y,u_1,u_3,u_4\right\} ,\left\{y,u_1,u_5\right\}, \left\{u_1,u_3,u_6\right\},\cr
&   \left\{u_1,u_4,u_7\right\},\left\{u_1,u_5,u_8\right\},\left\{u_2,u_3,u_9\right\}, \left\{u_3,u_4,u_{10}\right\},
\left\{u_4,u_6,u_{11}\right\}, \left\{u_3,u_6,u_{12}\right\},\cr
&\left\{u_6,u_{10},u_{13}\right\},\left\{u_{10},u_{12},u_{14}\right\},\left\{u_3,u_{12},u_{15}\right\},\left\{x,y,V,v_1\right\},\left\{x,y,v_i,v_{i+1}\right\}\ (i=1,\dots,k-1),\cr
&\left\{y,v_{i},v_{i+k}\right\}\ (i=1,\dots,k)\,,\left\{u_3,\delta_i,\delta_{i+1}\right\}\ (i=1,\dots,2k)\,,\cr
&\left[\left\{x,y,\delta_{2i-1},\delta_{2ik-i^2+2i+1}\right\},\left\{x,y,\delta_{2ik-i^2+2i+j+1},\delta_{2ik-i^2+2i+j+2}\right\}\ (j=0,\cdots,k-i-2), \right. \cr
&\left. \left\{y,\delta_{2ik-i^2+2i+j+1},\delta_{2ik-i^2+i+j+k+2}\right\}\ (j=0,\dots,k-i-1)\; \right] \ (i=1,\cdots,k-1)\,,\cr
&\left[ \left\{x,y,\delta_{2i},\delta_{k^2+2ki-i^2+i+2}\right\},\{x,y,\delta_{k^2+2ki-i^2+i+j+2},\delta_{k^2+2ki-i^2+i+j+3}\}\ (j=0,\cdots,k-i-2) \right. \cr
&\left. \{x,y,\delta_{k^2+2ki-i^2+i+j+2},\delta_{k^2+2ki-i^2+j+k+2}\}\ (j=0,\cdots,k-i-2)\; \right] \ (i=1,\cdots,k-1)\}
\ea
\ee

We also present the marginal geometry of $(E_6,E_6)$ conformal matter theory mentioned in \cite{Apruzzi:2019vpe}, with the following the tensor branch configuration is
\be
[E_6]-\; 1-\; \overset{\mathfrak{su}_3}{3}-\; 1-[E_6]\,.
\ee
Starting with the Weierstrass model:
\be
y^2+b_1 UV xy+b_3 U^2 V^2 y=x^3+b_2 U^2 V^2 x^2+b_4 U^3 V^3 x+b_6 U^5 V^5\, ,
\ee
we first blow up the base locus $(U,V;\delta_1)$, and then use the following resolution sequence:
\be
\ba
&BU_{\text{marginal}}^{(E_6,E_6)}=\cr
&\{\left\{x,y,U,u_1\right\},\left\{x,y,u_1,u_2\right\},\left\{y,u_1,u_2,u_3\right\},\left\{y,u_1,u_4\right\},\left\{y,u_2,u_5\right\},\left\{u_3,u_4,u_6\right\},\cr
&\left\{y,u_3,u_7\right\},\left\{u_1,u_4,u_8\right\},\{\left\{x,y,V,v_1\right\},\left\{x,y,v_1,v_2\right\},\left\{y,v_1,v_2,v_3\right\},\left\{y,v_1,v_4\right\},\cr
&\left\{y,v_2,v_5\right\},\left\{v_3,v_4,v_6\right\},\left\{y,v_3,v_7\right\},\left\{v_1,v_4,v_8\right\},\cr
&\left\{u_5,\delta_1,\delta_2\right\},\left\{v_5,\delta_1,\delta_3\right\},\left\{x,y,\delta_1,\delta_4\right\},\left\{y,\delta_4,\delta_5\right\}\}.
\ea
\ee

The $(E_7,E_7)$ conformal matter theory is a 6d (1,0) SCFT with rank 10. The tensor branch configuration is
\be
[E_7]-\; 1-\; \overset{\mathfrak{su}_2}{2}-\; \overset{\mathfrak{so}_7}{3}-\; \overset{\mathfrak{su}_2}{2}-\; 1-[E_7].
\ee
Starting with the Weierstrass model:
\be
y^2+b_1 UV xy+b_3 U^3 V^3 y=x^3+b_2 U^2 V^2 x^2+b_4 U^3 V^3 x+b_6 U^5 V^5\, ,
\ee
we first blow up the base locus $(U,V;\delta_1)$, and then use the following resolution sequence:
\be
\ba
&BU_{\text{marginal}}^{(E_7,E_7)}=\cr
&\{\left\{x,y,U,u_1\right\},\left\{x,y,u_1,u_2\right\},\left\{y,u_1,u_3\right\},\left\{y,u_2,u_4\right\},\left\{u_2,u_3,u_5\right\},\left\{u_1,u_3,u_6\right\},\cr
&\left\{u_2,u_4,u_7\right\},\left\{u_3,u_4,u_8\right\},\left\{u_4,u_5,u_9\right\},\left\{u_5,u_8,u_{10}\right\},\left\{u_3,u_5,u_{11}\right\},\cr
&\{\left\{x,y,V,v_1\right\},\left\{x,y,v_1,v_2\right\},\left\{y,v_1,v_3\right\},\left\{y,v_2,v_4\right\},\left\{v_2,v_3,v_5\right\},\left\{v_1,v_3,v_6\right\},\cr
&\left\{v_2,v_4,v_7\right\},\left\{v_3,v_4,v_8\right\},\left\{v_4,v_5,v_9\right\},\left\{v_5,v_8,v_{10}\right\},\left\{v_3,v_5,v_{11}\right\},\cr
&\left\{u_4,\delta_1,\delta_6\right\},\left\{v_4,\delta_1,\delta_4\right\},\left\{u_4,\delta_6,\delta_{9}\right\},\left\{v_4,\delta_4,\delta_8\right\},\left\{x,y,\delta_1,\epsilon_1\right\},\cr
&\left\{y,\epsilon_1,\delta_2\right\},\left\{\epsilon_1,\delta_2,\delta_3\right\},\left\{x,y,\delta_4,\delta_5\right\},\left\{x,y,\delta_6,\delta_7\right\}\}.
\ea
\ee
Note that the divisor $\epsilon_1=0$ is not present in the Calabi--Yau threefold after the resolution. A subtlety here is that the divisor $\delta_2=0$ is reducible with two components, hence the total number of non-flat surface components is still 10 from $\delta_i$, $(i=1,\cdots,9)$.

The $(E_8,E_8)$ conformal matter theory is a 6d (1,0) SCFT with rank 21. The tensor branch configuration is
\be
[E_8]-\; 1-\; 2-\; \overset{\mathfrak{su}_2}{2}-\; \overset{\mathfrak{g}_2}{3}-\; 1-\; \overset{\mathfrak{f}_4}{5}-\; 1-\; \; \overset{\mathfrak{g}_2}{3}-\; \overset{\mathfrak{su}_2}{2}-\; 2-\; 1-[E_8].
\ee
Starting with the Weierstrass model:
\be
y^2+b_1 UV xy+b_3 U^3 V^3 y=x^3+b_2 U^2 V^2 x^2+b_4 U^4 V^4 x+b_6 U^5 V^5\, ,
\ee
we first blow up the base locus $(U,V;\delta_1)$, and then use the following resolution sequence:
\be
\ba
&BU_{\text{marginal}}^{(E_8,E_8)}=\cr
&\{\left\{x,y,U,u_1\right\},\left\{x,y,u_1,u_2\right\},\left\{y,u_2,u_3\right\}, \left\{y,u_1,u_3,u_4\right\} ,\left\{y,u_1,u_5\right\}, \left\{u_1,u_3,u_6\right\},\cr
&   \left\{u_1,u_4,u_7\right\},\left\{u_1,u_5,u_8\right\},\left\{u_2,u_3,u_9\right\}, \left\{u_3,u_4,u_{10}\right\},
\left\{u_4,u_6,u_{11}\right\}, \left\{u_3,u_6,u_{12}\right\},\cr
&\left\{u_6,u_{10},u_{13}\right\},\left\{u_{10},u_{12},u_{14}\right\},\left\{u_3,u_{12},u_{15}\right\},\{\left\{x,y,V,v_1\right\},\left\{x,y,v_1,v_2\right\},\left\{y,v_2,v_3\right\},\cr
&\left\{y,v_1,v_3,v_4\right\} ,\left\{y,v_1,v_5\right\}, \left\{v_1,v_3,v_6\right\},\left\{v_1,v_4,v_7\right\},\left\{v_1,v_5,v_8\right\},\left\{v_2,v_3,v_9\right\},\cr
&\left\{v_3,v_4,v_{10}\right\},\left\{v_4,v_6,v_{11}\right\}, \left\{v_3,v_6,v_{12}\right\},\left\{v_6,v_{10},v_{13}\right\},\left\{v_{10},v_{12},v_{14}\right\},\left\{v_3,v_{12},v_{15}\right\},\cr
&\left\{u_3,\delta_1,\delta_9\right\},\left\{v_3,\delta_1,\delta_6\right\},\left\{u_3,\delta_9,\delta_{16}\right\},\left\{v_3,\delta_6,\delta_{14}\right\},\left\{u_3,\delta_{16},\delta_{18}\right\},\cr
&\left\{v_3,\delta_{14},\delta_{19}\right\},\left\{u_3,\delta_{18},\delta_{21}\right\},\left\{v_3,\delta_{19},\delta_{20}\right\},\left\{x,y,\delta_1,\epsilon_1\right\},\left\{x,y,\delta_6,\epsilon_3\right\},\cr
&\left\{y,\epsilon_3,\delta_7\right\},\left\{x,y,\delta_9,\epsilon_4\right\},\left\{y,\epsilon_4,\delta_{10}\right\},\left\{\epsilon_1,\delta_7,\delta_{12}\right\},\left\{x,y,\epsilon_1,\delta_2\right\},\cr
&\left\{y,\epsilon_1,\epsilon_2\right\},\left\{\epsilon_1,\delta_{10},\delta_{13}\right\},\left\{\epsilon_4,\delta_{10},\delta_{11}\right\},\left\{x,y,\delta_{14},\delta_{15}\right\},\left\{\epsilon_1,\epsilon_2,\delta_3\right\},\cr
&\left\{\delta_2,\epsilon_2,\delta_4\right\},\left\{\epsilon_2,\delta_4,\delta_5\right\},\left\{\epsilon_3,\delta_7,\delta_8\right\},\left\{x,y,\delta_{16},\delta_{17}\right\}\}.
\ea
\ee

Note that the nine non-flat fiber components are $\delta_i=0$, $(i=1,\cdots,21)$, and the divisors $\epsilon_i=0$, $i=1,\cdots,4$ are not present in the Calabi--Yau threefold after the resolution.


\subsection{Alternative 6d Starting Points}
\label{app:AltStart}

If we start with a 6d SCFT, where the maximal superconformal flavor symmetry is not manifest, the descedant 5d theories will not have manifest superconformal flavor symmetry either. We thus usually start with a 6d model, which has manifestly the maximal $G_\text{F}^{(6d)}$. However it is interesting to see how the flavor symmetry enhancement emerges, when choosing different starting points in 6d. For rank one, an example is $(E_6, SU(3))$ (instead of $(E_8, I_1)$). Likewise in rank two, there is $(D_k,D_k)$ instead of $(D_{2k}, I_1)$. In such cases the flavor symmetry is reconstructed in 5d by considering the BPS states, which reorganize in terms of a larger superconformal flavor symmetry. 

\subsubsection[Alternative Description for Rank one: \texorpdfstring{$(E_6,SU(3))$}{(E6,SU(3))}]{Alternative Description for Rank one: \boldmath{$(E_6, SU(3))$}}
\label{app:Rank1E6SU3}

An alternative description of the rank one E-string theories can be obtained with the following starting point Tate model:
\be
y^2+b_1 U xy+b_3 U^2 Vy=x^3+b_2 U^2 V x^2+b_4 U^3 V^2 x+b_6 U^5 V^3.
\ee
There is an $E_6$ on $U=0$ and $SU(3)$ on $V=0$.

As an example, we can use the following blow-up sequence
\be
\ba
BU^{(E_6,SU(3))}=&\{\left\{x,y,U,u_1\right\},\left\{x,y,V,v_1\right\},\left\{y,v_1,v_2\right\},\left\{u_1,v_2,\delta
   _1\right\},\left\{x,y,u_1,u_2\right\},\cr
   &\left\{y,u_1,u_2,u_3\right\},\left\{y,u_1,u_4\right\},\left\{y,u_2,u_5\right\},\left\{
   u_3,u_4,u_6\right\},\left\{y,u_3,u_7\right\},\left\{u_1,u_4,u_8\right\}\}
\ea\ee
The Cartan divisors $D_i^{E_6}$ and $D_i^{SU(3)}$ are given by
\be
\ba
&(D_0^{E_6},D_1^{E_6},\cdots,D_6^{E_6})\equiv(U,u_2,u_3,u_6,u_7,u_5,u_8)\cr
&(D_0^{SU(3)},D_1^{SU(3)},D_2^{SU(3)})\equiv (V,v_1,v_2)
\ea
\ee

\begin{figure}
  \centering
  \includegraphics[height=7cm]{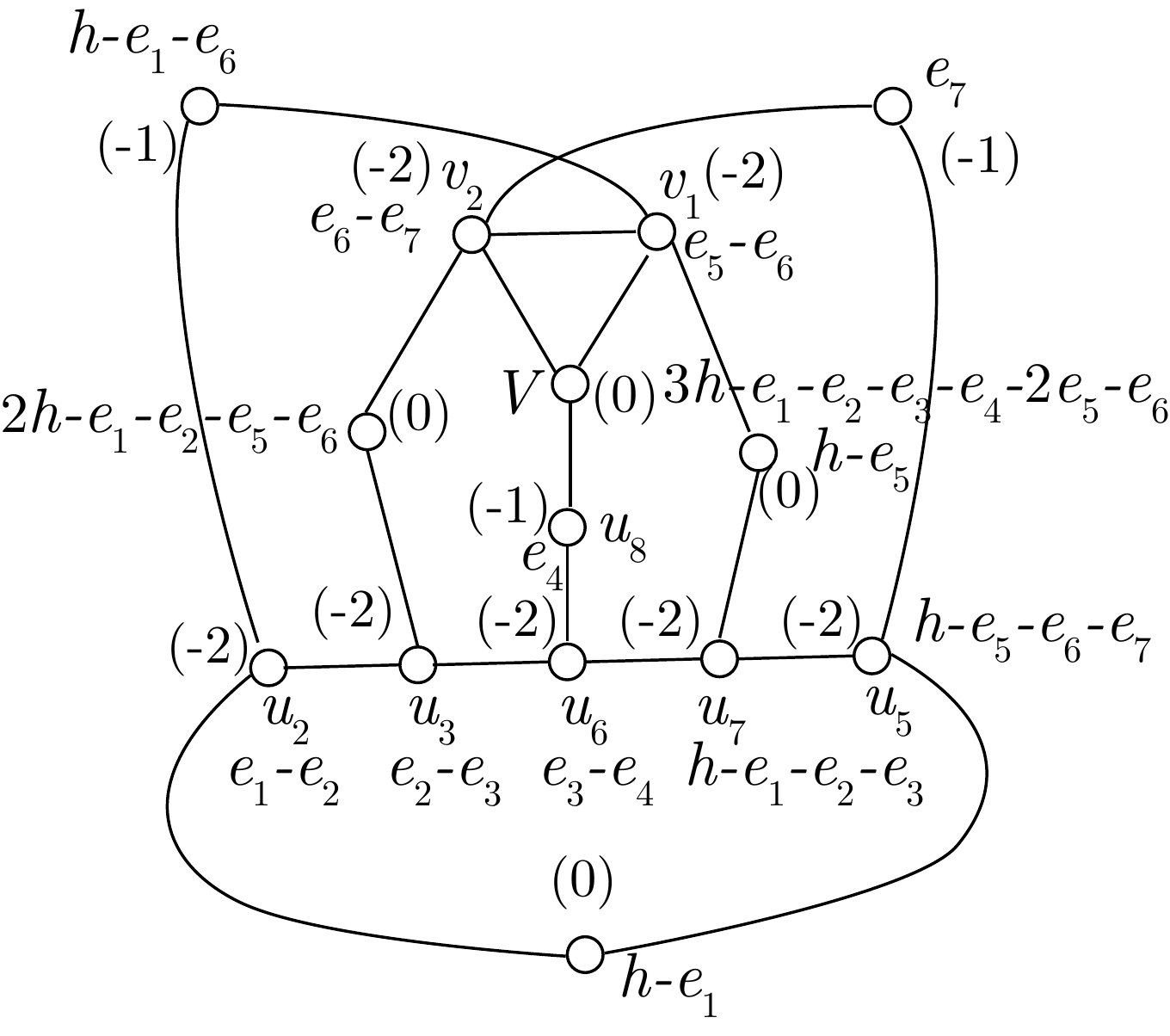}
  \caption{The configuration of curves in the example geometry $E_6\times SU(3)$, including the rational $(-2)$, $(-1)$ and 0-curves.}\label{f:E6SU3}
\end{figure}

We plot the configuration of curves in figure~\ref{f:E6SU3}, and the reduced intersection matrix $S\cdot D_i^2$ is:
\be
\begin{array}{c|ccccccc|ccc|c}
S\cdot D_i^2&  D_0^{E_6} & D_1^{E_6} & D_2^{E_6} & D_3^{E_6} & D_4^{E_6} & D_5^{E_6} & D_6^{E_6} & D_0^{SU(3)} & D_1^{SU(3)} & D_2^{SU(3)} & S \\\hline
D_i & 0 & -2 & -2 & -2 & -2 & -2 & -1 & 0 & -2 & -2 & 2 \\
\end{array}
\ee
The non-flat fiber $S$ is a gdP$_7$, and one can see that the configuration of $(-2)$-curves form the Dynkin diagram of $SU(3)\times SU(6)$. 

From the figure \ref{f:E6SU3}, we read off the following rational $(-1)$-curves corresponding to the highest weight of various representations of $SU(3)\times SU(6)$:
\be
\ba
e_4 &:\ (\mbf{1},\mbf{20})\cr
e_7&:\ (\mbf{3},\mbf{6})\cr
h-e_1-e_6&:\ (\bar{\mbf{3}},\bar{\mbf{6}})\,.
\ea
\ee
They exactly combine into the $\mbf{56}$ representation of $E_7\supset SU(3)\times SU(6)$.

Moreover, the rational 0-curves correspond to the highest weights of the following representations of $SU(3)\times SU(6)$:
\be
\ba
h-e_1 &:\ (\mbf{1},\mbf{35})\cr
h-e_5 &:\ (\mbf{3},\overline{\mbf{15}})\cr
2h-e_1-e_2-e_5-e_6 &:\ (\bar{\mbf{3}},\mbf{15})\cr
3h-e_1-e_2-e_3-e_4-2e_5-e_6 &:\ (\mbf{8},\mbf{1}) \,.
\ea
\ee
These representations can be exactly combined into the adjoint representation $\mbf{133}$ of $E_7\supset SU(3)\times SU(6)$. From this observation, we speculate that the actual flavor symmetry of this geometry is $E_7$. 

More generally, one can read off the actual non-abelian flavor symmetry $G_{F,na}$ with the following steps:

\begin{itemize}
\item{Read off the group $\tilde{G}_{F,na}$ based on the configuration of $(-2)$-curves on the non-flat fiber, which form the Dynkin diagram of $\tilde{G}_{F,na}$. The actual flavor symmetry group should contain $\tilde{G}_{F,na}$.}
\item{Write down all the irreducible (genus 0) curves $C_i$ on the non-flat fiber that gives rise to BPS states of a particular spin, see section~\ref{sec:BPS}, which give rise to highest weight of various representations $R_i$ of $\tilde{G}_{F,na}$.}
\item{Check if the combined representation $\oplus_i R_i$ form an adjoint representation of a larger group $G_{F,na}$. If this is the case, then the actual flavor symmetry group should be $G_{F,na}$.}
\end{itemize} 

\subsubsection[Alternative Description: \texorpdfstring{$(D_5,D_5)$}{(D5,D5)}]{Alternative Description: \boldmath{$(D_5, D_5)$}}
\label{app:D5d5}

In the main text, we considered the starting point $D_{10}-I_1$ instead of the $(D_5, D_5)$ conformal matter. 
In fact it turns out that the former results in a more concise description of the theories, and contains the models descending from the latter. We illustrate now how the models with $(D_5, D_5)$ starting point give rise to the same theories as the $D_{10}$ starting point. 

The starting point Tate model is:
\be
y^2+b_1 UV xy+b_3 U^2 V^2 y=x^3+b_2 U V x^2+b_4 U^3 V^3 x+b_6 U^5 V^5.
\ee

As an example, we can use the following blow-up sequence:
\be\label{BU3D5d5}
\ba
BU^{D_5-D_5}=&\{\left\{x,y,U,V,\delta_2\right\},\left\{x,y,U,u_1\right\},\left\{x,y,u_1,u_2\right\},\left\{y,u_1,u_2,u_3\right\},\left\{y,u_1,u_4\right\},\left\{y,u_2,u_5\right\},
\cr 
&\left\{u_1,u_4,u_6\right\},\left\{x,y,V,v_1\right\},\left\{x,y,v_1,v_2\right\},\left\{
   y,v_1,v_2,v_3\right\},\left\{y,v_1,v_4\right\},\left\{y,v_2,v_5\right\},\cr 
   &\left\{v_1,v_4,v_6\right\},\left\{x,y,\delta_2,\delta _1\right\}\},
   \ea
\ee
where the first blow-up is a weighted blow-up:
\be
x\rightarrow x\delta_2^2\ ,\ y\rightarrow y\delta_2^3\ ,\ U\rightarrow U\delta_2\ ,\ V\rightarrow V\delta_2.
\ee
The Cartan divisors $D_i^{(1)}$ and $D_i^{(2)}$ of the two $SO(10)$ factors are given by
\be
\ba
(D_0^{(1)},D_1^{(1)},D_2^{(1)},D_3^{(1)},D_4^{(1)},D_5^{(1)})&\equiv (U , u_4 , u_6 , u_3 , u_2 , u_5) \cr 
(D_0^{(2)},D_1^{(2)},D_2^{(2)},D_3^{(2)},D_4^{(2)},D_5^{(2)})&\equiv (V , v_4 ,v_6 , v_3 , v_2 , v_5) \,.
\ea\ee
The reduced intersection matrix with the surface components $S_i$ is 
{\footnotesize
\be
\begin{array}{c|cccccc|cccccc|cc}
S_i\cdot D_j^2  & D_0^{(1)} & D_1^{(1)} & D_2^{(1)} & D_3^{(1)} & D_4^{(1)} & D_5^{(1)}& D_0^{(2)} & D_1^{(2)} & D_2^{(2)} & D_3^{(2)} & D_4^{(2)} & D_5^{(2)} & S_1 & S_2 \\ \hline
S _1 & 0 & 0 & 0 & 0 & 0 & 0 & 0 & 0 & 0 & 0 & 0 & 0 & 8 & 4 \\
S_2 & -1 & -2 & -2 & -2 & -2 & -2 & -1 & -2 & -2 & -2 & -2 & -2 & -6 & -1 \\ 
\end{array}
\,.
\ee
}

\begin{figure}
\centering
\includegraphics[height=7cm]{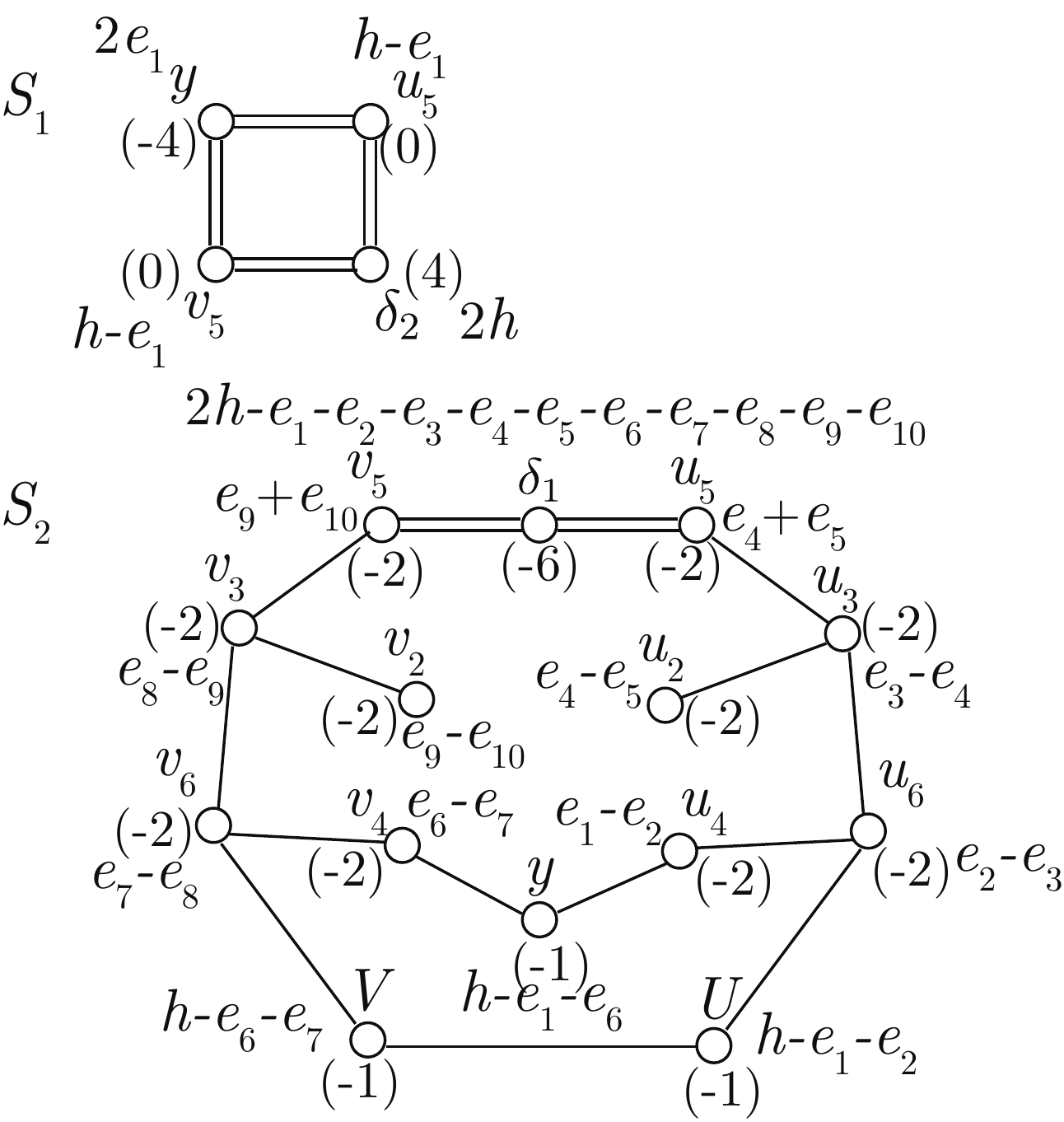}
\caption{ The configuration of curves on $S_1$ and $S_2$ in the geometry (\ref{BU3D5d5}). The number in the bracket denotes the self-intersection number of the curve. The letters $U$, $V$, $u_i$, $v_i$, denote an intersection curve with the corresponding divisor.}\label{f:D5d53}
\end{figure}

We plot the configuration of curves in figure~\ref{f:D5d53}. The configuration of $(-2)$-curves form $H=SO(10)\times SO(10)$. Note that $S_1$ is still an $\mb{F}_1$ as the curve $S_1\cdot S_2$ corresponds to the divisor class $2h$. The geometry only has $Sp(2)$ gauge theory description as any ruling on $S_1$ intersects $S_2\cdot S_1$ at two points. The following $(-1)$-curves on $S_2$ give rise to the highest weight of various representations under $SO(10)\times SO(10)$:
\be
\ba
h-e_1-e_2&:\ (\mbf{45},\mbf{1})\cr
h-e_6-e_7&:\ (\mbf{1},\mbf{45})\cr
h-e_1-e_6&:\ (\mbf{10},\mbf{10}).
\ea
\ee
These representations can be combined into the adjoint representation $\mbf{190}$ of $SO(20)$. Hence we confirmed that the flavor symmetry of this configuration is  $SO(20)$.

\subsubsection{Flavor Symmetry and BPS States}

What we have seen so far in the last subsections is that BPS states can help identify larger flavor symmetries. 
This is particularly important, if the marginal theory from which one starts the CFD-trees does not have the manifest flavor symmetry. 

To determine the superconformal flavor symmetry from the CFD, in the approach
as laid out in the rest of this paper, it is important, nay vital, to begin
with the correct marginal CFD. This marginal CFD is obtained by considering a
particular resolution of singularities of a Weierstrass model which realizes,
over a non-compact locus, a singular fiber associated to the affine Dynkin
diagram of $G_\text{6d}$, the superconformal flavor symmetry in 6d. 

When attempting to determine the descendant 5d $\mathcal{N} = 1$ SCFTs that
arise from 6d $(G, G^\prime)$ conformal matter one naturally begins with a
Weierstrass model that contains two non-compact divisors in the base, which
respectively support singular fibers associated to $G$ and $G^\prime$. However
it is often the case that the superconformal flavour symmetry
\begin{equation}
  G_\text{6d} \supseteq G \times G^\prime \,,
\end{equation}
is not the product, but involves a non-trivial recombination of the two
flavor symmetry factors -- as in the $(E_6, SU(3))$ and $(D_5,D_5)$ examples we studied. 
When this occurs one must first determine this
enhancement before being able to determine the marginal CFD, to which the
full flavor symmetry of the descendants is visible. An example of this that we
have studied in this paper is the rank two theories that come from $(D_5,
D_5)$ minimal conformal matter. In that case the 6d flavor symmetry is not
$SO(10) \times SO(10)$, but, in fact, $SO(20)$, and if one were to write down
the marginal CFD by considering a resolution of the Weierstrass model with $SO(10)
\times SO(10)$, one would not fully capture the flavor symmetry in the CFDs. 

Despite this, the full flavor symmetry is still observable, even if one begins
with a marginal CFD not realizing $G_\text{6d}$, from a study of the BPS
spectra. In this section we will demonstrate this in several examples.

\begin{figure}
  \centering
  \subfloat[ ]{\includegraphics*[width=3cm]{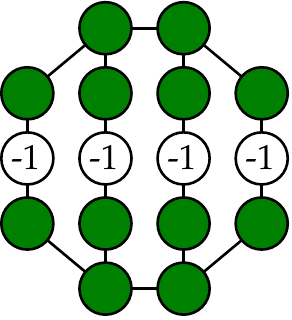}}
  \qquad\qquad\qquad
  \subfloat[ ]{\includegraphics*[width=3cm]{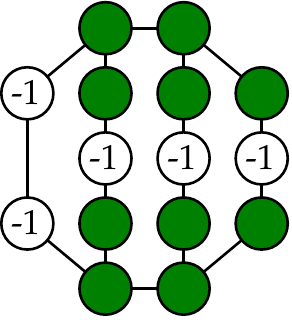}}
  \caption{The marginal CFD for $(D_5, D_5)$ minimal conformal matter as
    determined from the Weierstrass model realizing intersecting $(I_1^*,
    I_1^*)$ singular fibers is shown on the left. On the right is the first
  descendant by taking a single mass deformation of the marginal
CFD.}\label{fig:D5D5falseCFD}
\end{figure}

Let us begin by looking at the aforementioned example of $(D_5, D_5)$ minimal
conformal matter. If one begins with an elliptic fibration with two $I_1^*$
fibers intersecting at a codimension two point then constructing the non-flat
resolution of that fibration would give rise to the marginal CFD as depicted
in figure \ref{fig:D5D5falseCFD} a). Let us consider the first descendant of
this CFD, obtainable by a single (unique) mass deformation from the marginal
theory. From the CFD in figure \ref{fig:D5D5falseCFD} b) one can observe an
$SO(10) \times SO(10)$ flavor symmetry for the 5d theory, however we know that
this theory in fact has an $SO(20)$ flavor symmetry. 

If we determine the BPS states from this descendant CFD we can see that the
spin $0$ BPS states, form representations with the highest weights given by
the vertices labelled with $-1$. In terms of the $SO(10) \times SO(10)$ flavor
group these form the representations
\begin{equation}
  (\bm{10, 10}) \,,\, (\bm{16, 16}) \,,\, (\bm{16^\prime, 16^\prime}) \,,\,
  (\bm{1,45}) \,,\, (\bm{45,1}) \,.
\end{equation}
If we consider the branching 
\begin{equation}
  SO(20) \rightarrow SO(10) \times SO(10) \,,
\end{equation}
then one can see that the relevant representations decompose as
\begin{equation}
  \begin{aligned}
    \bm{190} &\rightarrow (\bm{45,1}) \oplus (\bm{1,45}) \oplus (\bm{10,10})
    \cr
    \bm{512} &\rightarrow (\bm{16, 16}) \oplus (\bm{16^\prime, 16^\prime}) \,.
  \end{aligned}
\end{equation}
In this way one can see that there is, in fact, an enhanced $SO(20)$ flavor
symmetry under which the states transform. One can do a similar analysis for
all descendant states from the $(I_1^*, I_1^*)$ starting point and observe
that the BPS states from $g = 0$ curves at the spin $0$ and spin $1$ level
always form themselves into representations that can be combined into full
representations of the superconformal flavor symmetry as given by the
$(I_{6}^*, I_1)$ marginal CFD.

\providecommand{\href}[2]{#2}\begingroup\raggedright\endgroup

\end{document}